\documentclass[twocolumn,aps,pra,superscriptaddress,tightlines,nofootinbib]{revtex4-2}
\usepackage{hyperref}
\hypersetup{
colorlinks=true,
linkcolor=blue,         
citecolor=blue,         
filecolor=blue,      
urlcolor=blue
}
\usepackage{url}
\usepackage{float}
\usepackage{placeins}
\usepackage{color}
\usepackage{graphicx,epsfig,colordvi,epstopdf}
\usepackage{bm}
\usepackage{mathtools}
\usepackage{mathdots}
\usepackage{physics}

\DeclareMathOperator*{\argmin}{arg\,min}

\usepackage{amsmath}
\usepackage{amsfonts}
\usepackage{amssymb}

\begin{document}

\title{Replacing neural networks by optimal analytical predictors\\ for the detection of phase transitions}
\author{Julian Arnold}
\affiliation{Department of Physics, University of Basel, Klingelbergstrasse 82, 4056 Basel, Switzerland}
\author{Frank Sch\"afer}
\affiliation{Department of Physics, University of Basel, Klingelbergstrasse 82, 4056 Basel, Switzerland}
\affiliation{CSAIL and Department of Mathematics, Massachusetts Institute of Technology, Cambridge, MA
	02139, USA}
\date{\today}

\date{\today}
\begin{abstract}
Identifying phase transitions and classifying phases of matter is central to understanding the properties and behavior of a broad range of material systems. In recent years, machine-learning (ML) techniques have been successfully applied to perform such tasks in a data-driven manner. However, the success of this approach notwithstanding, we still lack a clear understanding of ML methods for detecting phase transitions, particularly of those that utilize neural networks (NNs). In this work, we derive analytical expressions for the optimal output of three widely used NN-based methods for detecting phase transitions. These optimal predictions correspond to the results obtained in the limit of high model capacity. Therefore, in practice they can, for example, be recovered using sufficiently large, well-trained NNs. The inner workings of the considered methods are revealed through the explicit dependence of the optimal output on the input data. By evaluating the analytical expressions, we can identify phase transitions directly from experimentally accessible data without training NNs, which makes this procedure favorable in terms of computation time. Our theoretical results are supported by extensive numerical simulations covering, e.g., topological, quantum, and many-body localization phase transitions. We expect similar analyses to provide a deeper understanding of other classification tasks in condensed matter physics.
\end{abstract}

\maketitle
\section{Introduction}\label{sec_introduction}
\indent In recent years, machine learning (ML) has been used extensively to approach complex physics problems~\cite{dunjko_2018,carleo:2019,dawid:2022}. Among these applications, the task of classifying phases of matter and the identification of phase transitions is particularly exciting~\cite{carrasquilla:2017,van:2017,carrasquilla:2020,carrasquilla:2021}, as it could enable the autonomous discovery of novel phases of matter. Classical ML methods have successfully revealed the phase diagrams of a plethora of systems based on data from experimental measurements~\cite{rem:2019,kaming:2021,bohrdt:2021,miles2:2021,yu:2021} and numerical simulations~\cite{wang:2016,carrasquilla:2017,van:2017,wetzel1:2017,wetzel2:2017,chng:2017,ohtsuki:2017,schindler:2017,zhang:2017,broecker:2017,huembeli:2018,liu:2018,beach:2018,van:2018,zhang:2018,venderley:2018,rodriguez:2019,huembeli:2019,schaefer:2019,scheurer:2020,greplova:2020,kottmann:2020,zvyagintseva:2021,arnold:2021,huang:2021,guo:2022,maskara:2022,patel:2022,zhang:2022}. Many of the most powerful ML methods for detecting phase transitions utilize neural networks (NNs) at their core~\cite{carrasquilla:2017,van:2017,wetzel1:2017,wetzel2:2017,chng:2017,ohtsuki:2017,schindler:2017,zhang:2017,broecker:2017,huembeli:2018,liu:2018,beach:2018,van:2018,zhang:2018,venderley:2018,huembeli:2019,schaefer:2019,greplova:2020,kottmann:2020,zvyagintseva:2021,arnold:2021,guo:2022,maskara:2022,patel:2022,zhang:2022}. Prominent examples are supervised learning~\cite{carrasquilla:2017}, the learning-by-confusion scheme~\cite{van:2017,liu:2018}, and the prediction-based method~\cite{schaefer:2019,greplova:2020,arnold:2021}, which are often applied in conjunction~\cite{beach:2018,greplova:2020,bohrdt:2021}.\\ 

\begin{figure*}[tbh!]
\begin{center}
\includegraphics[width=0.99\textwidth]{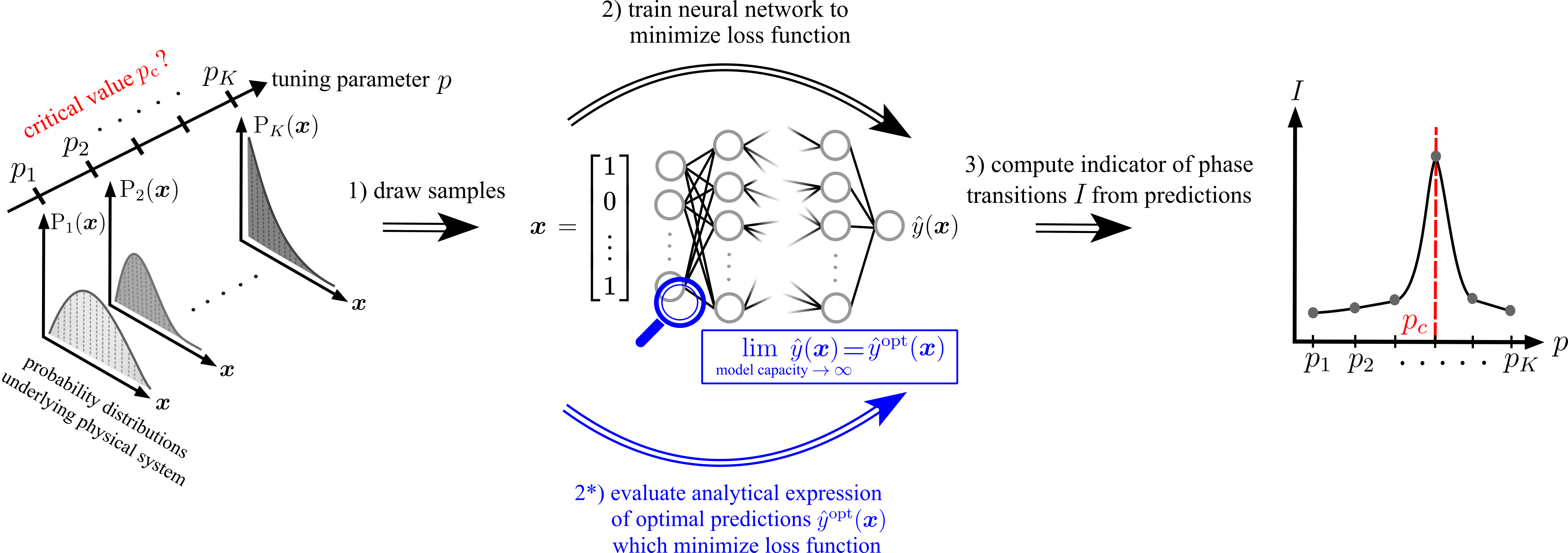}
\caption{Schematic representation of the setup and workflow of supervised learning, the learning-by-confusion scheme, and the prediction-based method for detecting phase transitions from data. The physical system under consideration is characterized by a tuning parameter $p$. The goal is to identify the critical value of the tuning parameter $p_{\rm c}$ at which the system transitions from one phase to another. In a first step (step $1$), the state $\bm{x}$ of the physical system is (repeatedly) sampled at various values of the tuning parameter $\{p_{1},p_{2},\dots , p_{K} \}$, where $\{{\rm P}_{1}(\bm{x}),{\rm P}_{2}(\bm{x}),\dots,{\rm P}_{K}(\bm{x}) \}$ are the corresponding probability distributions. Based on these samples, a neural network (NN) is trained to perform a particular classification or regression task, i.e., its tunable parameters are updated to minimize a particular loss function (step $2$). The three ML methods for detecting phase transitions differ in their formulation of the underlying NN tasks. Having trained the NN, its predictions $\hat{y}$ are used to compute the value of an indicator of phase transitions $I$ at fixed values of the tuning parameter (step $3$). Ideally, the indicator has a local maximum at $p_{\rm c}$ where the largest change in the state of the system occurs. As a result, the ML methods then autonomously highlight phase boundaries along the chosen scanning range of the tuning parameter. Note that the indicators of phase transitions obtained with supervised learning, the learning-by-confusion scheme, and the prediction-based method differ. The contribution of our work is highlighted in blue: We derive analytical expressions for the optimal predictions $\hat{y}^{\rm opt}$ of the NNs used in these three methods. The optimal predictions minimize the corresponding loss function and are thus achieved by NNs whose capacity, i.e., ability to fit a wide variety of functions~\cite{goodfellow:2016,hu:2021}, is sufficiently high. The optimal predictions can solely be expressed in terms of the probability distributions underlying the physical system. Using the optimal predictions $\hat{y}^{\rm opt}$ in place of the NN predictions $\hat{y}$, we further obtain analytical expressions for the optimal indicators of phase transitions $I^{\rm opt}$ (step $2^{*}$). Evaluating these analytical expressions provides an alternative path for computing indicators of phase transitions without \textit{ever} training NNs, see Tab.~\ref{fig_timing} where we compare the computation times of the two approaches.}
\label{fig:workflow}
\end{center}
\end{figure*}

\indent All three methods follow a similar workflow, which is illustrated in Fig.~\ref{fig:workflow} (steps 1-3). They take as input samples that represent the state of a physical system at various values of a tuning parameter. The samples are processed by an NN whose parameters are tuned to minimize a specific loss function. By analyzing the NN predictions, one can compute a scalar quantity that highlights the critical value of the tuning parameter at which the system's state changes most. As such, this quantity highlights phase boundaries and serves as an indicator for phase transitions. The decision whether the change corresponds to a crossover or a phase transition does, however, requires further analysis, such as finite-size scaling. The three methods differ in their choice of loss function, i.e., in the formulation of the underlying classification or regression task, and thus in the resulting indicator for phase transitions.\\

\indent NNs are universal function approximators~\cite{cybenko:1989,hornik:1991,lu:2017,zhou:2020}. This fact makes supervised learning, the learning-by-confusion scheme, and the prediction-based method extremely powerful and has played a central role in the original conception of these methods~\cite{carrasquilla:2017,van:2017,schaefer:2019}. Namely, the use of NNs for detecting phase transitions from data has been inspired by the success of deep NNs (DNNs) in image recognition tasks~\cite{krizhevsky:2012}. The more expressive a ML model~\cite{bengio:2011,goodfellow:2016,raghu:2017}, such as an NN, the more resources are needed to train it, and the more difficult it is to interpret the underlying functional dependence of its predictions on the input~\cite{linardatos:2021,molnar:2022}. Therefore, NNs typically act as black boxes that can correctly highlight phase transitions but whose internal workings remain opaque to the user. Since the proposal of supervised learning, the learning-by-confusion scheme, and the prediction-based method, there have been numerous attempts to understand their working principle, particularly through the extraction of order parameters. As an example, (kernel) support vector machines, which are easier to analyze than NNs due to their inherent linear nature, were used as predictive models~\cite{ponte:2017,zhang:2019,greitemann:2019,liu:2019}. Other approaches to improve interpretability rely on systematic input engineering, such that the objective function that the NN learns is approximately linearly~\cite{zhang:2020}, or on a systematic reduction of the NN expressivity~\cite{wetzel2:2017}. Another set of works~\cite{casert:2019,dawid:2020,blucher:2020,dawid:2021} analyzed trained NNs using standard interpretability tools from ML, which rely on truncated Taylor expansions. Despite these efforts, we still understand little about the working principle of ML methods for the detection of phase transitions based on NNs, when they fail or succeed, and how they differ~\cite{carleo:2019} -- in particular when DNNs are used (i.e., in the limit of high model expressivity). These open questions reflect the general scarcity of rigorous theory in ML~\cite{huang:2021}.\\

\indent Here, we address these gaps in knowledge by pursuing a novel approach based on deriving analytical expressions for the optimal predictions of the NNs underlying supervised learning, learning by confusion, and the prediction-based method. The predictions are optimal in the sense that they minimize the target loss function, i.e., the corresponding model performs the desired task (as specified by the loss function) optimally. Based on the optimal predictions, we find analytical expressions for the optimal indicators of phase transitions of these three methods. The optimal indicators correspond to the output of the methods when using ideal high-capacity~\cite{goodfellow:2016} predictive models, such as well-trained, highly expressive NNs. The inner workings of these methods are revealed through the dependence of the optimal indicators on the input data. Moreover, the analytical expressions make it possible to compute the optimal indicator directly from the input data without training NNs, see step $2^{*})$ in Fig.~\ref{fig:workflow}, manifesting an alternative numerical routine to infer phase transitions. We demonstrate the procedure in a numerical study on a variety of models exhibiting, e.g., symmetry-breaking, topological, quantum, and many-body localization phase transitions.\\

\indent This work is structured as follows: In Sec.~\ref{sec_methods}, we introduce the task of detecting phase transitions from data in an automated fashion, including supervised learning, the learning-by-confusion scheme, and the prediction-based method. Section~\ref{sec_optimal_indicators} discusses the analytical expressions of their optimal indicators of phase transitions, i.e., their output when using well-trained, highly expressive NNs. A numerical study of the optimal predictions and indicators for the Ising model, Ising gauge theory, XY model, XXZ model, Kitaev model, and Bose-Hubbard model is presented in Sec.~\ref{sec_physical_models}. Finally, the results are discussed in Sec.~\ref{sec_discussion} and conclusions are drawn in Sec.~\ref{sec_concl}.

\section{Automated detection of phase transitions from data}\label{sec_methods}
\indent In this section, we will formally introduce the task of automatically detecting phase transitions from data and how supervised learning (SL), learning by confusion (LBC), and the prediction-based method (PBM) approach this problem. We consider the following scenario: The physical system to be analyzed is characterized by a tuning parameter $p$ sampled equidistantly with a grid spacing $\Delta p$. In the following we denote the points at the boundary of the sampled region as $p_{1}$ and $p_{K}$ with $K\in \mathbb{N}$ sampled points in total ($K = \frac{p_{K}-p_{1}}{\Delta p} +1$). At each sampled point $p_{k}$ ($1\leq k \leq K $) we draw $M \in \mathbb{N}$ samples from the system's state $\{ \bm{S}_{jk} \}_{j=1}^{M}$ which constitute our available data. We allow for this data to be pre-processed via a mapping to a representation space $\mathcal{R}:\bm{S}\rightarrow \bm{x}$. At the core of each of the three methods for detecting phase transitions under consideration lies a predictive model $m: \bm{x} \rightarrow \hat{y}$, such as an NN, which takes the pre-processed data $\mathcal{X}=\{\bm{x}_{jk}| 1 \leq j  \leq M, 1 \leq k \leq K \}$ as input. We denote the available data at sampled point $p_{k}$ as $\mathcal{X}_{k}=\{\bm{x}_{jk}\}_{j=1}^{M}$. Note that $\mathcal{X}$ may contain duplicates. Let $\bar{\mathcal{X}}$ be the set of unique inputs obtained from $\mathcal{X}$ by removing all duplicates. We assume that the system is present either in a single phase A or two distinct phases A and B across the sampled range of the tuning parameter $\{ p_{k}\}_{k=1}^{K}$. If a system exhibits multiple distinct phases, the parameter range can (in principle) be analyzed in a piece-wise fashion (for more details on this case, see Appendix~\ref{app_A1} and~\ref{app_A2}). The task is then to compute a scalar indicator $I(p)$, which peaks at the phase boundary if two distinct phases are present, i.e., has a local maximum, and does not exhibit a peak otherwise. More specifically, if the system is in phase A from $p_{1}$ to $p_{\rm c}$ and phase B from $p_{\rm c}$ to $p_{K}$ with critical point $p_{\rm c}$ (not necessarily a sampled point), the indicator $I(p)$ should exhibit a local maximum at the sampled point closest to the critical point $\argmin_{p_{k}} |p_{\rm c}-p_{k}|$.

\subsection{Supervised learning}\label{sec_methods_SL}
\indent In SL, a predictive model $m$ is trained on the data available in regions near the two boundaries of the chosen parameter range denoted by I and II. Region I and II are comprised of the set of sampled points $\{p_{k} | 1 \leq k \leq r_{\rm I} \}$ and $\{p_{k} | l_{\rm II} \leq k \leq K \}$, respectively. Here, $r_{\rm I},l_{\rm II}\in \mathbb{N}$ denote the rightmost and leftmost parameter point in region I and II, respectively. In SL, we assume that there exist two distinct phases A and B, with the regions I and II being located deep within these phases. Without loss of generality we assign the label $y=1$ and $y=0$ to data obtained in region I and II, respectively. The predictive model is trained to minimize a cross-entropy (CE) loss
\begin{align}\label{eq:CE_SL}
\mathcal{L}_{\rm SL} = -\frac{1}{M_{\mathcal{T}}}\sum_{\bm{x} \in {\rm \mathcal{T}}} &[ y(\bm{x})\ln \left(\hat{y}(\bm{x}) \right)\\
&+\left(1-y(\bm{x}) \right)\ln \left(1-\hat{y}(\bm{x}) \right)] \nonumber,
\end{align}
where the sum runs over all $M_{\mathcal{T}}$ data points in the training set $\mathcal{T} \subseteq \mathcal{X}$, $\mathcal{T} = \big\{ \bm{x}_{jk}| 1 \leq j \leq M, k\in \{ 1,\dots, r_{\rm I}\} \cup \{l_{\rm II},\dots,K\} \big\}$. Let us denote the set containing all unique inputs present in $\mathcal{T}$ without repetition as $\bar{\mathcal{T}}$. The output of the predictive model $\hat{y}(\bm{x})\in [0,1]$ corresponds to the probability of input $\bm{x}$ having the label $y=1$, whereas $1-\hat{y}(\bm{x})$ is the probability that the input $\bm{x}$ carries the label $y=0$.\\

\indent After training the predictive model to minimize the loss function in Eq.~\eqref{eq:CE_SL}, it is evaluated on all available data $\mathcal{X}$. Averaging over the predictions $\hat{y}(\bm{x})$ for all data $\mathcal{X}_{k}$ at a given point $p_{k}$ ($1 \leq k \leq K$) yields a prediction as a function of the tuning parameter
\begin{equation}\label{eq:SL_avg_pred}
    \hat{y}_{\rm SL}(p_{k}) = \frac{1}{M} \sum_{\bm{x} \in \mathcal{X}_{k}} \hat{y}(\bm{x}).
\end{equation}
The indicator for phase transitions in SL, $I_{\rm SL}$, is then given by the negative derivative of the prediction with respect to the tuning parameter
\begin{equation}\label{eq:SL_indicator}
    I_{\rm SL}(p_{k}) = - \left. \frac{\partial  \hat{y}_{\rm SL}(p)}{\partial p}\right|_{p_{k}}.
\end{equation}
The estimated critical value of the tuning parameter in SL corresponds to the location of the global maximum in its indicator [Eq.~\eqref{eq:SL_indicator}], which can easily be determined in an automated fashion without human supervision. If one chose to label data obtained in region I with $y=0$ and region II with $y=1$ instead, the same indicator signal can be recovered via a sign change $I_{\rm SL}(p_{k}) \rightarrow - I_{\rm SL}(p_{k})$. Note that it is also common to identify the estimated critical value of the tuning parameter in SL as $\argmin_{p_{k}} |\hat{y}(p_{k})-0.5|$, see Appendix~\ref{app_D1} for a comparison motivating our choice.\\

\indent Intuitively, if there is a transition from one phase to another (phase A to phase B) when varying the tuning parameter $p$, the mean predictions $\hat{y}_{\rm SL}(p)$ should drop from $\hat{y}_{\rm SL}(p_{1})=1$ (deep within phase A) to $\hat{y}_{\rm SL}(p_{N})=0$ (deep within phase B) as $p$ is increased. If the transition is sharp, the predictions should also change abruptly. Such a change results in a peak in the negative derivative of the predictions, i.e., in the indicator for phase transitions. In that case, the predictive model acts as an order parameter that approaches 1/0 deep within phase A/B. In general, one expects the predictions -- and thus the indicator -- to vary most strongly at the critical point $p_{\rm c}$. If there is only a single phase, one expects the predictions to be approximately constant, resulting in a flat indicator $I_{\rm SL}(p)$. Our derivation of the optimal indicator $I_{\rm SL}^{\rm opt}$ will provide a rigorous basis for these heuristic arguments underlying the SL method.

\subsection{Learning by confusion}\label{sec_methods_LBC}
\indent In LBC, predictive models are trained on all available data $\mathcal{X}$. The labels are obtained by performing a split of the sampled parameter range into two neighboring regions labeled I and II. Each input $\bm{x}$ drawn in region I or II carries the label $y=1$ or $y=0$, respectively. The values of the tuning parameters which realize each of the $K+1$ possible bipartitions are given as $p^{\rm bp}_{k} = p_{1}-\Delta p/2+(k-1)\Delta p$, where $1\leq k \leq K+1$. For a given bipartition point $p^{\rm bp}_{k}$, region I and II are then comprised of the sampled points $\{ p_{j}| p_{j}\leq p^{\rm bp}_{k},\, 1 \leq j \leq K \} $ and $\{ p_{j}| p_{j}>p^{\rm bp}_{k},\, 1 \leq j \leq K \} $. Note that for bipartitions $1$ ($p^{\rm bp}_{1}=p_{1}-\Delta p/2$) and $K+1$ ($p^{\rm bp}_{K+1}=p_{K}+\Delta p/2$), region I or II encompasses the entire sampled parameter range and all data is assigned the label 1 or 0, respectively.\\

\indent To each bipartition, i.e., choice of data labeling, we associate a distinct predictive model $m_{k}$ $(1 \leq k \leq K+1)$ which is trained to minimize a CE loss
\begin{align}\label{eq:CE_LBC}
    \mathcal{L}_{\rm LBC} = -\frac{1}{M_{\mathcal{X}}} \sum_{\bm{x} \in \mathcal{X}} &[ y(\bm{x})\ln \left( \hat{y}(\bm{x}) \right) \\
    &+\left( 1-y(\bm{x})\right)\ln \left( 1-\hat{y}(\bm{x})\right)] \nonumber,
\end{align}
where the sum runs over all $M_{\mathcal{X}}=KM$ data points. Again, the output of the predictive model $\hat{y}(\bm{x})\in [0,1]$ corresponds to the probability of input $\bm{x}$ having the label $y=1$, whereas $1-\hat{y}(\bm{x})$ is the probability of the input $\bm{x}$ carrying the label $y=0$.\\

\indent Once a predictive model has been trained to minimize the loss function in Eq.~\eqref{eq:CE_LBC} for a given bipartition, it is evaluated on all available data points. In particular, we can compute the mean classification accuracy as a function of the bipartition parameter $p^{\rm bp}_{k}$ ($1 \leq k \leq K+1$) as
\begin{equation}\label{eq:LBC_error}
    I_{\rm LBC}(p^{\rm bp}_{k}) = 1-\frac{1}{ M_{\mathcal{X}}} \sum_{\bm{x} \in \mathcal{X}} \left|\theta \left(\hat{y}(\bm{x})-0.5\right) - y(\bm{x})\right|,
\end{equation}
where $\theta$ denotes the heaviside step function. The predictions $\hat{y}(\bm{x})$ are obtained from the predictive model $m_{k}$ associated with the bipartition point $p^{\rm bp}_{k}$, and $y(\bm{x})$ are the corresponding labels.\\

\indent Clearly, the mean classification accuracy $I_{\rm LBC}$ will exhibit trivial local maxima at the points $p^{\rm bp}_{1}=p_{1}-\Delta p/2$ and $p^{\rm bp}_{K+1}=p_{1}+\Delta p/2$, where the entire data is assigned the label 0 or 1, respectively. Therefore, a predictive model effortlessly reaches a perfect accuracy of 1, because it simply needs to predict a single label regardless of the input. However, given that the underlying data can be separated into two distinct classes of similar character (i.e., phases) through appropriate bipartitioning of the parameter range at $p_{\rm c}$, one also expects the classification accuracy to have a local maximum at $p_{\rm c}$. At such a point, the predictive model is ``least confused'' by the choice of data labeling. Hence, the mean classification accuracy serves as the indicator for phase transitions within LBC. The estimated critical value of the tuning parameter in LBC corresponds to the location of the largest local maximum (excluding the points $p^{\rm bp}_{1}$ and $p^{\rm bp}_{K+1}$ at the boundary) in its indicator [Eq.~\eqref{eq:LBC_error}].

\subsection{Prediction-based method}\label{sec_methods_PBM}
\indent In PBM, a predictive model $m$ is trained on all available data $\mathcal{X}$ to infer the value of the tuning parameter $p_{k}$ ($1 \leq k \leq K$) at which an input $\bm{x}$ was generated. While SL and LBC constitute supervised \textit{classification} tasks, PBM corresponds to a supervised \textit{regression} task, where the label is given by the tuning parameter itself $y(\bm{x}) = p_{k}\; \forall \bm{x} \in \mathcal{X}_{k}$.\\

\indent We train the predictive model $m$ to minimize a mean-square-error (MSE) loss function
\begin{equation}\label{eq:MSE_PBM}
    \mathcal{L}_{\rm PBM} = \frac{1}{M_{\mathcal{X}}}\sum_{\bm{x} \in \mathcal{X}} \left(\hat{y}(\bm{x}) - y(\bm{x})  \right)^2.
\end{equation}
After training, the predictive model is evaluated on all available data points $\mathcal{X}$. Averaging over the predictions $\hat{y}(\bm{x})$ for all data $\mathcal{X}_{k}$ at a given point $p_{k}$ yields a mean prediction as a function of the tuning parameter
\begin{equation}\label{eq:PBM_avg_pred}
    \hat{y}_{\rm PBM}(p_{k}) = \frac{1}{M} \sum_{\bm{x} \in \mathcal{X}_{k}} \hat{y}(\bm{x}).
\end{equation}
We then compute the deviation of the prediction from the true underlying value of the tuning parameter $\delta y_{\rm PBM}(p_{k}) =\hat{y}_{\rm PBM}(p_{k})-p_{k}$. The indicator for phase transitions of PBM, $I_{\rm PBM}$, is then given by the derivative of this deviation with respect to the tuning parameter
\begin{equation}\label{eq:PBM_indicator}
    I_{\rm PBM}(p_{k}) = \left. \frac{\partial  \delta y_{\rm PBM}(p)}{\partial p}\right|_{p_{k}} = \left. \frac{\partial  \hat{y}_{\rm PBM}(p)}{\partial p}\right|_{p_{k}} - 1.
\end{equation}
The estimated critical value of the tuning parameter in PBM corresponds to the location of the global maximum in its indicator [Eq.~\eqref{eq:PBM_indicator}].\\

\indent Intuitively, if there is only a single phase, in which inputs cannot be distinguished well by the predictive model, one expects the mean predictions to be approximately constant. This results in the deviations $\delta y_{\rm PBM}$ varying approximately linear with the tuning parameter. Hence, the indicator $I_{\rm PBM}$ will be approximately constant. However, if there is a transition from one phase to another as the tuning parameter is varied, the predictions and the corresponding deviations also vary sharply. This results in a peak in the derivative of the deviations, i.e., the indicator for phase transitions $I_{\rm PBM}$. In particular, one expects that the predictions are most susceptible at the phase boundary. Thus, its derivative should vary most strongly at the critical point $p_{\rm c}$.\\

\indent In many standard applications of NNs, it is typical to split the available data into multiple sets, in particular, to avoid overfitting~\cite{goodfellow:2016}. For example, suppose we aim to construct an accurate on-the-fly classifier of individual samples into distinct phases of matter. In this case, it may be beneficial to split the available data into a training and validation set to avoid overfitting if only a limited amount of data is available. In the case of PBM and LBC, we did not explicitly split the data set $\mathcal{X}$ into a training set and test set (as well as a potential validation set). This can be done, e.g., to assess sampling convergence by comparing the predictions obtained on the training set and test set or to perform early stopping with NNs (see Appendix~\ref{app_B2} for concrete examples). Note, however, that the task we consider here is the detection of phase transitions given the data at hand. As such, the data set $\mathcal{X}$ does not \textit{necessarily} need to be split. In particular, in the limit of a sufficient number of samples all splits of a data set coincide, assuming that all samples are drawn independently from the same probability distributions underlying the physical system (see Fig.~\ref{fig:workflow}). Therefore, the predictions and indicators obtained by training NNs using multiple distinct data sets will coincide with the values obtained using the entire data set for training and evaluation up to deviations arising from finite-sample statistics. That is, in the limit of a sufficient number of samples, the results obtained in the two scenarios coincide~\cite{blumer:1989,vapnik:2000,goodfellow:2016}. Moreover, given a fixed amount of data $\mathcal{X}$, better statistics are achieved by utilizing the entire data for training and evaluation.

\section{Optimal indicators of phase transitions}\label{sec_optimal_indicators}
\indent In this section, we discuss the optimal indicators of phase transitions $I^{\rm opt}$ for each of the phase-classification methods presented in Sec.~\ref{sec_methods}. The optimal indicators can be directly calculated given the predictions $\hat{y}^{\rm opt}(\bm{x})$ of an optimal model $m^{\rm opt}$ which minimizes the corresponding loss function. The detailed proofs can be found in Appendix~\ref{app_A1}. In the limit of sufficient data, i.e., given accurate estimates of the probability distributions underlying the physical system $\{{\rm P}_{k}(\bm{x}) \}_{k=1}^{K}$, such a model is also \textit{Bayes optimal}~\cite{devroye2:1996,goodfellow:2016}. Meaning, no other statistical model can outperform it on the classification or regression task at hand (on average). In this case, the optimal loss value it achieves coincides with the \textit{Bayes error}~\cite{devroye2:1996,goodfellow:2016}, i.e., the intrinsic irreducible error inherent to the problem.\\

\indent \textit{Supervised learning}.---In SL (see Sec.~\ref{sec_methods_SL}), the optimal predictions are given as
\begin{equation}\label{eq:SL_derivation_1}
    \hat{y}_{\rm SL}^{\rm opt}(\bm{x}) = \frac{{\rm P}_{\rm I}(\bm {x})}{{\rm P}_{\rm I}(\bm {x}) + {\rm P}_{\rm II}(\bm {x})}\; \forall \bm{x} \in \bar{\mathcal{T}},
\end{equation}
where
\begin{equation}\label{eq:SL_derivation_2}
    {\rm P}_{{\rm I}}(\bm {x}) = \sum_{k= 1}^{r_{\rm I}} {\rm P}_{k}(\bm {x})
\end{equation}
and 
\begin{equation}\label{eq:SL_derivation_3}
    {\rm P}_{{\rm II}}(\bm {x}) = \sum_{k = l_{\rm II}}^{K} {\rm P}_{k}(\bm {x})
\end{equation}
are the (unnormalized) probabilities of drawing an input $\bm{x}$ in region I and II, respectively. Hence, the optimal prediction for a particular input corresponds to the probability of drawing that input in region I compared to region II. Here, ${\rm P}_{k}(\bm {x})$ denotes the (normalized) probability to draw the input $\bm{x}$ at the sampled point $p_{k}$. Given a data set $\mathcal{X}_{k}$, this probability is estimated as ${\rm P}_{k}(\bm {x}) \approx M_{k}(\bm{x})/M$, where $M_{k}(\bm{x})$ is the number of times the input $\bm{x}$ is present in the data set $\mathcal{X}_{k}$. While having access to an analytical expression for the underlying probability distributions $\{{\rm P}_{k}(\bm{x}) \}_{k=1}^{K}$ may ease computation and enable additional insights, it is not required to compute the optimal predictions (see Sec.~\ref{sec_physical_models} for application to physical systems). An expression for the optimal value of the loss in SL, $\mathcal{L}_{\rm SL}^{\rm opt}$, can be obtained by replacing $\hat{y}(\bm{x})$ with $\hat{y}_{\rm SL}^{\rm opt}(\bm{x})$ in Eq.~\eqref{eq:CE_SL}, where, by definition, $\mathcal{L}_{\rm SL}^{\rm opt}\leq \mathcal{L}_{\rm SL}$.\\ 

\indent Assuming that all inputs within the entire data set $\mathcal{X}$ are already present in the training set $\mathcal{T}$, i.e., $\bar{\mathcal{T}} = \bar{\mathcal{X}}$, the mean optimal prediction at a given point $p_{k}$ ($1 \leq k \leq K$) is
\begin{equation}\label{eq:SL_derivation_4}
    \hat{y}_{\rm SL}^{\rm opt}(p_{k}) = \sum_{\bm{x} \in \bar{\mathcal{X}}} {\rm P}_{k}(\bm{x})\hat{y}_{\rm opt}(\bm{x}).
\end{equation}
This corresponds to the probability of finding an input drawn at that point $p_{k}$ in region I compared to region II. We find this assumption to be (approximately) satisfied for all physical systems analyzed in this work and can estimate the errors arising from a violation, see Sec.~\ref{sec_physical_models} and Appendix~\ref{app_A2}. The optimal indicator of phase transitions in SL is then given as 
\begin{equation}\label{eq:SL_derivation_5}
I_{\rm SL}^{\rm opt}(p_{k}) = - \left. \frac{\partial  \hat{y}_{\rm SL}^{\rm opt}(p)}{\partial p}\right|_{p_{k}}.
\end{equation}
In general, there will be a transition point where the probability in Eq.~\eqref{eq:SL_derivation_4} changes most and thus where its derivative, the optimal indicator in Eq.~\eqref{eq:SL_derivation_5}, peaks.\\

\indent \textit{Learning by confusion}.---For a given bipartition of the parameter range into regions I and II, the optimal predictions of LBC (see Sec.~\ref{sec_methods_LBC}) are given as
\begin{equation}\label{eq:LBS_derivation_1}
    \hat{y}_{\rm LBC}^{\rm opt}(\bm{x}) = \frac{{\rm P}_{\rm I}(\bm{x})}{{\rm P}_{\rm I}(\bm{x}) + {\rm P}_{\rm II}(\bm {x})}\; \forall \bm{x} \in \bar{\mathcal{X}},
\end{equation}
which corresponds to the probability of drawing the input in region I compared to region II. This characteristic is inherent to the underlying classification task [compare Eqs.~\eqref{eq:SL_derivation_1} and~\eqref{eq:LBS_derivation_1}]. The mean classification error associated with an input $\bm{x}$ is given by ${\min}\{ \hat{y}_{\rm LBC}^{\rm opt}(\bm{x}), 1- \hat{y}_{\rm LBC}^{\rm opt}(\bm{x})\}$. This classification error arises from a ``confusion'' of the model: different labels can be assigned to the same input due to an overlap of the underlying probability distributions. The mean classification error over the entire parameter range given a particular choice of bipartition, i.e., labeling of the data, then corresponds to
\begin{equation}\label{eq:LBS_derivation_2}
    I_{\rm LBC}^{\rm opt}=1-\frac{1}{K}\sum_{k=1}^{K} \sum_{\bm{x} \in \bar{\mathcal{X}}} {\rm P}_{k}(\bm{x}){\min}\{ \hat{y}_{\rm LBC}^{\rm opt}(\bm{x}), 1- \hat{y}_{\rm LBC}^{\rm opt}(\bm{x})\}.
\end{equation}
This forms the optimal indicator for phase transitions in LBC. An expression for the optimal value of the loss in LBC, $\mathcal{L}_{\rm LBC}^{\rm opt}$, can be obtained by replacing $\hat{y}(\bm{x})$ with $\hat{y}_{\rm LBC}^{\rm opt}(\bm{x})$ in Eq.~\eqref{eq:CE_LBC}. The critical point $p_{\rm c}$ is highlighted by a dip in the mean classification error, i.e., by a peak in the mean classification accuracy [Eq.~\eqref{eq:LBS_derivation_2}]. It corresponds to the bipartition point for which the probability distributions underlying the two regions have the least overlap (on average), resulting in the highest classification accuracy and the least confusion. While confusion can arise due to sub-optimal predictions of models with restricted capacity (see Appendix~\ref{app_B2} for a concrete example), we find that confusion can even persist in the limit of high model capacity if it is inherent to the underlying data. Based on the analytical expressions, we thus gained an intuitive and rigorous understanding of the concept of confusion underlying LBC~\cite{van:2017}.\\ 

\indent \textit{Prediction-based method}.---The optimal predictions within PBM (see Sec.~\ref{sec_methods_PBM}) are given as
\begin{equation}\label{eq:PBM_derivation_1}
\hat{y}_{\rm PBM}^{\rm opt}\left(\bm{x}\right) = \frac{\sum_{k=1}^{K} {\rm P}_{k}\left( \bm{x}\right) p_{k}}{ \sum_{k=1}^{K} {\rm P}_{k}\left(\bm{x}\right)}\; \forall \bm{x} \in \bar{\mathcal{X}}.
\end{equation}
Here, the optimal prediction for a given input is obtained by a weighted sum over each point in the parameter range, where the weight of each point $p_{k}$ corresponds to the probability of obtaining the input at that point along the parameter range compared to all other points. Therefore, the prediction accuracy decreases if the same input can be drawn at multiple values of the tuning parameter, i.e., when the underlying probability distributions overlap. An expression for the optimal value of the loss in PBM, $\mathcal{L}_{\rm PBM}^{\rm opt}$, can be obtained by replacing $\hat{y}(\bm{x})$ by $\hat{y}_{\rm PBM}^{\rm opt}(\bm{x})$ in Eq.~\eqref{eq:MSE_PBM}. The mean prediction of an optimal model $m^{\rm opt}$ at a sampled point $p_{k}$ is given by
\begin{equation}\label{eq:PBM_derivation_2}
\hat{y}_{\rm PBM}^{\rm opt}(p_{k}) = \sum_{\bm{x} \in \bar{\mathcal{X}}} {\rm P}_{k}(\bm{x}) \hat{y}_{\rm PBM}^{\rm opt}(\bm{x}).
\end{equation}
Thus, the optimal indicator for phase transitions is
\begin{equation}\label{eq:PBM_derivation_3}
    I_{\rm PBM}^{\rm opt}(p_{i}) = \left. \frac{\partial  \delta y_{\rm PBM}^{\rm opt}(p)}{\partial p}\right|_{p_{i}},
\end{equation}
where $\delta y_{\rm PBM}^{\rm opt}(p_{k}) = \hat{y}_{\rm PBM}^{\rm opt}(p_{k})-p_{k}$. Recall that in PBM, phase transitions are detected by analyzing the dependence of the prediction error on the tuning parameter. The optimal indicator [Eq.~\eqref{eq:PBM_derivation_3}] highlights the value of the tuning parameter at which the mean predictions change most, i.e., where the overlap of the underlying probability distributions changes most. The optimal predictions and indicators of PBM have previously been derived in Ref.~\cite{arnold:2021} but have neither been utilized in a numerical routine, nor been used to explain previous studies.\\

\begin{figure*}[tbh!]
\centering
\includegraphics[width=\linewidth]{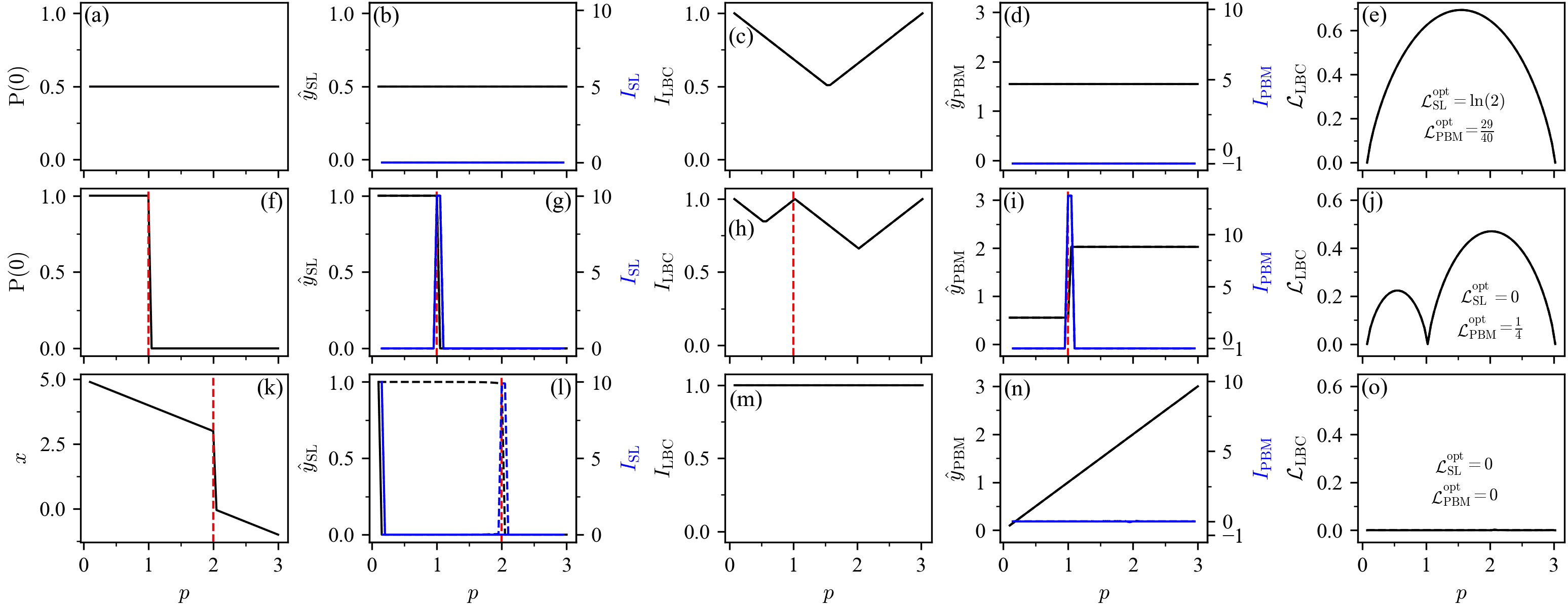}
\caption{Results for prototypical probability distributions in (a)-(e) case 1 with ${\rm P}_{k}(\bm{x}) = {\rm P}(\bm{x})\; \forall k$, where ${\rm P}(0)={\rm P}(1)=0.5$, (f)-(j) case 2 given by Eq.~\eqref{eq:jump_distributions} with 
${\rm P}_{\rm A}(0)=1$, ${\rm P}_{\rm B}(0)=0$ and $p_{\rm c}=1$, and (k)-(o) case 3 with Eqs.~\eqref{eq:continuous}-\eqref{eq:continuous_2}. The tuning parameter ranges from $p_{1}=0.1$ to $p_{K}=3$ with $\Delta p = 0.05$. Critical values of the tuning parameter are highlighted with red-dashed lines. For details on SL, LBC, and PBM using NNs, see Appendix~\ref{app_B}. (a),(f),(k) Illustration of the probability distributions underlying the data. (b),(g),(l) Mean prediction $\hat{y}_{\rm SL}(p)$ obtained using the analytical expression (black, solid) or an NN (black, dashed), as well as the corresponding indicator $I_{\rm SL}(p)$ (blue). Here, we choose $r_{\rm I}=1$ and $l_{\rm II}=K$. (c),(h),(m) The indicator of LBC, $I_{\rm LBC}$, obtained using the analytical expression (black, solid) or an NN (black, dashed). (d),(i),(n) Mean prediction $\hat{y}_{\rm PBM}(p)$ of PBM obtained using the analytical expression (black, solid) or an NN (black, dashed), as well as the corresponding indicator $I_{\rm PBM}(p)$ (blue). (e),(j),(o) Value of the loss function in LBC, $\mathcal{L}_{\rm LBC}$, for each bipartition point $p^{\rm bp}$ obtained using the analytical expression (black, solid) or evaluated after NN training (black, dashed). In addition, the optimal values of the loss function for SL and PBM obtained by evaluating the analytical expressions are reported. Note that, by definition, $\mathcal{L}^{\rm opt}\leq \mathcal{L}$ for all three methods.}
\label{fig:prototypical}
\end{figure*}

\indent The optimal predictions of SL, LBC, and PBM can \textit{solely} be expressed in terms of the probability distributions $\{ {\rm P}_{k}(\bm{x}) \}_{k=1}^{N}$ governing the input data. Crucially, this means that the optimal predictions -- and thus the optimal indicators of phase transitions -- do not depend on the particular nature of an input or how similar it is to other inputs. Such notions of similarity form the basis of a large set of other phase-classification methods, e.g., based on principal component analysis~\cite{wang:2016}, diffusion maps~\cite{rodriguez:2019}, or anomaly detection~\cite{kottmann:2020}. The analytical form of the optimal predictions indicates that SL, LBC, and PBM ultimately gauge changes in the probability distributions governing the data akin to probability metrics~\cite{devroye:1996}. Note that the same optimal predictions and indicators will be obtained for multiple choices of representations $\mathcal{R}$ given that the same probability distributions can still describe the data in the representation space. Consequently, knowledge of the symmetries of the system can be utilized to calculate indicators of phase transitions more efficiently. We will make use of this in Sec.~\ref{sec_physical_models}.
\subsection{Demonstration on prototypical probability distributions}\label{sec_application_trivial_distr}
\indent In this section, we compute the optimal indicators of SL, LBC, and PBM for a set of simple probability distributions governing the input data. As we will see later, the probability distributions governing the data in physical systems can be regarded as generalizations of the special cases discussed in this section. Thus, they serve as a reasonable basis for understanding. We compare these results to the indicators obtained by numerical optimization of NNs. The details on the NN architecture and training, including the corresponding hyperparameters, can be found in Appendix~\ref{app_B}. This first demonstration shows how the analytical expressions can be used to calculate the optimal indicator directly from input data without NNs. Moreover, it confirms that the optimal predictive models can be recovered by training NNs with sufficient expressive power.\\
    
\indent \textit{Case 1}.---Let us first consider the case where the probability distribution governing the data is identical across the parameter range, i.e., ${\rm P}_{k}(\bm{x}) = {\rm P}(\bm{x})\; \forall 1 \leq k \leq K $. Clearly, in this case all three methods should indicate the presence of a single phase. The optimal prediction in SL is
\begin{equation}\label{eq:SL_example_1}
\hat{y}_{\rm SL}^{\rm opt}(p) = \frac{K_{\rm I}}{K_{\rm I} + K_{\rm II}} = {\rm const.},
\end{equation}
corresponding to the relative size of region I compared to region II [see Fig.~\ref{fig:prototypical}(b)]. Here, $K_{\rm I} = r_{\rm I}$ and $K_{\rm II} = K-l_{\rm II}$ correspond to the number of sampled parameter values in region I or II, respectively. Taking the derivative of Eq.~\eqref{eq:SL_example_1} results in a flat indicator signal $I_{\rm SL}^{\rm opt} = 0$. In LBC, the optimal classification accuracy for a particular bipartition is given by $I_{\rm LBC}^{\rm opt} = {\rm max}\{ K_{\rm I}/K, K_{\rm II}/K \}$. This results in a characteristic V-shape~\cite{van:2017}, which has its minimum at the center of the parameter range under consideration, see Fig.~\ref{fig:prototypical}(c). In PBM, the optimal mean prediction is also placed at the center of mass $\hat{y}_{\rm PBM}^{\rm opt}(p_{k}) = 1/K \sum_{k=1}^{K} p_{k}={\rm const.}$, which results in a constant indicator $I_{\rm PBM}^{\rm opt}=-1$ [see Fig.~\ref{fig:prototypical}(d)]. As such, all three methods yield optimal indicators that correctly signal the presence of a single phase, i.e., the absence of two distinct phases. For a concrete numerical demonstration we consider the case of binary inputs $\bar{\mathcal{X}} = \{0,1 \}$ with equal probability ${\rm P}(0)={\rm P}(1)=0.5$. Figure~\ref{fig:prototypical}(a)-(e) shows the results for all three methods using the analytical expressions as well as NNs. Note that the analytical predictions and indicators can be approximated well using NNs as predictive models.\\
    
\indent \textit{Case 2}.---Next, we consider the case where the input data naturally separates into two distinct sets. That is, the underlying probability distributions result in a bipartition of the parameter range into two regions A and B, where each input can only be drawn in one of the two regions. In these regions, we choose the probability distributions to be identical
\begin{equation}\label{eq:jump_distributions}
{\rm P}_{k}(\bm{x}) =
  \begin{cases}
    {\rm P}_{\rm A}(\bm{x}) \; \forall k\leq {\rm c},\\
    {\rm P}_{\rm B}(\bm{x}) \; \forall k> {\rm c},
  \end{cases}
\end{equation}
where $1 \leq k,c \leq K $. This is a prototypical example for the case where the physical system transitions from phase A to B when crossing a critical value of the tuning parameter $p_{\rm c}$. Here, $p_{\rm c}$ corresponds to a sampled value of the tuning parameter, which may, in general, not be the case.\\

\indent Using SL, the optimal strategy corresponds to 
\begin{equation}
\hat{y}_{\rm SL}^{\rm opt}(p_{k}) =
  \begin{cases}
    1 \; \forall  k\leq {\rm c},\\
    0 \; \forall k> {\rm c}.
  \end{cases}
\end{equation}
This results in
\[ I_{\rm SL}^{\rm opt}(p_{k}) =
  \begin{cases}
    0 \; \forall k<{\rm c},\\
    \frac{1}{2 \Delta p} \; \forall k \, \in \{{\rm c}, {\rm c}+1 \},\\
    0 \; \forall k >{\rm c}+1,
  \end{cases}
\]
which diverges as $\Delta p \rightarrow 0$ and exhibits a peak at the two points which constitute the boundary between regions A and B [see Fig.~\ref{fig:prototypical}(g)]. Here, we approximate the derivative in Eq.~\eqref{eq:SL_derivation_5} by a symmetric difference quotient
\begin{equation}\label{eq:SL_indicator_symmetric_difference}
    I_{\rm SL}^{\rm opt}(p_{k}) \approx  \frac{| \hat{y}_{\rm SL}^{\rm opt}(p_{k+1}) - \hat{y}_{\rm SL}^{\rm opt}(p_{k-1})|}{2 \Delta p},
\end{equation}
where $2 \leq k \leq K-1$.\\

\indent In LBC, one can reach a perfect (error-free) classification when matching the natural bipartition present in the data. Let us denote the region between the bipartition point underlying the data, $p_{\rm c}$, and the chosen bipartition point in the LBC scheme, $p_{k}^{\rm bp}$, as III. The number of sampled parameter values within the smallest region between I, II, and III is $K^{\rm m}_{k} = {\rm min}\{K_{\rm I},K_{\rm II},K_{\rm III}\}$. Note that all input data drawn within one of these regions must be misclassified. Thus, the optimal strategy which yields the smallest classification error corresponds to misclassifying all input data drawn within the smallest region. The optimal classification accuracy is then given as
\begin{equation}
    I_{\rm LBC}^{\rm opt}(p_{k}^{\rm bp})=1-\frac{K^{\rm m}_{k}}{K}.
\end{equation}
This results in a characteristic W-shape of the indicator~\cite{van:2017}, see Fig.~\ref{fig:prototypical}(h), where the middle-peak occurs at the bipartition point $p^{\rm bp}_{i}$ closest to $p_{\rm c}$.\\

\indent In PBM, we have
\begin{equation}\label{eq:jump}
\hat{y}_{\rm PBM}^{\rm opt}(p_{k}) =
  \begin{cases}
    \langle p \rangle _{\rm A} = 1/{\rm c} \sum_{j=1}^{\rm c} p_{j} \; \forall k\leq {\rm c},\\
    \langle p \rangle _{\rm B} = 1/(N-{\rm c}) \sum_{j={\rm c}+1}^{N} p_{j} \; \forall k> {\rm c},
  \end{cases}
\end{equation}
where $\langle p \rangle _{\rm A/B}$ denotes the center of region A and B, respectively. This results in
\begin{equation}\label{eq:jump2}
 I_{\rm PBM}^{\rm opt}(p_{k}) =
  \begin{cases}
    -1 \; \forall k<{\rm c},\\
    \frac{\langle p \rangle _{\rm B} -\langle p \rangle _{\rm A}}{2 \Delta p} \; \forall k \in \{{\rm c},{\rm c}+1 \},\\
    -1 \; \forall k> {\rm c}+1,
  \end{cases}
\end{equation}
where we approximated the derivative in Eq.~\eqref{eq:PBM_derivation_3} by a symmetric difference quotient [see Fig.~\ref{fig:prototypical}(i)]. The expression in Eq.~\eqref{eq:jump2} diverges as $\Delta p \rightarrow 0$ for $k\in \{{\rm c},{\rm c}+1 \}$ and results in a peak at the two points which constitute the boundary between regions A and B. As such, the optimal indicators of all three methods correctly indicate the presence of two distinct sets of data, i.e., two distinct phases. The results obtained using the analytical expressions can be approximated well using NNs as predictive models. This is illustrated in Figs.~\ref{fig:prototypical}(f)-(j), where we consider the special case of binary inputs with ${\rm P}_{\rm A}(0)=1$, ${\rm P}_{\rm B}(0)=0$, and $p_{\rm c}=1$.\\
    
\indent \textit{Case 3}.---Lastly, we consider the case where the probability distributions underlying the data do not overlap, i.e., the probability of drawing a given input at two distinct values of the tuning parameter vanishes. In particular, this situation can occur when dealing with large state spaces, which are prone to result in insufficient sampling statistics in practice. That is, even in scenarios where the ground-truth probability distributions underlying the data \textit{do} overlap, the estimated probabilities ${\rm P}_{k}(\bm{x}) \approx M_{k}(\bm{x})/M$ based on the drawn data set $\mathcal{X}$ may not (see Appendix~\ref{app_A5} for a concrete physical example). Many image classification tasks encountered in traditional ML applications~\cite{fei:2004,lecun:2004,griffin:2007,krizhevsky:2009,deng:2009,russakovsky:2015,spanhol:2015} \textit{a priori} fall into this category. In particular, the probability distributions underlying the data are typically not known in these cases. Therefore, constructing optimal models, in particular Bayes optimal models, largely remains conceptual in nature~\cite{devroye2:1996,james:2013}.\\

\indent Here, an optimal predictive model is capable of distinguishing between samples obtained at distinct values of the tuning parameter with perfect accuracy. This results in $I_{\rm LBC}^{\rm opt}(p_{k}^{\rm bp})=1\; \forall 1 \leq k \leq K+1$ for LBC [see Fig.~\ref{fig:prototypical}(m)]. In the case of PBM, we have $\hat{y}_{\rm PBM}^{\rm opt}(p_{k})=p_{k}$ such that $I_{\rm PBM}^{\rm opt}(p_{k})=-1\; \forall 1 \leq k \leq K$, see Fig.~\ref{fig:prototypical}(n). In both cases, the indicator signals the absence of two distinct sets of data, i.e., phases. The optimal predictions of SL for $\bm{x} \in \bar{\mathcal{X}}$ are underdetermined: only the predictions for inputs within the training data $\bm{x} \in \bar{\mathcal{T}}$ are fixed after training and the assumption that $ \bar{\mathcal{X}} = \bar{\mathcal{T}}$ is violated in this particular case [see Fig.~\ref{fig:prototypical}(l)]. Note, however, that the predictions are, in principle, also unconstrained when using SL with NNs. For a simple numerical example, we consider the case where a single unique (scalar) input is drawn at each point along the parameter range
\begin{equation}\label{eq:continuous}
{\rm P}_{k}(x) =
  \begin{cases}
    1 \; {\rm for}\; x=f(p_{k}),\\
    0 \; {\rm otherwise},
  \end{cases}
\end{equation} 
with
\begin{equation}\label{eq:continuous_2}
f(p) =
  \begin{cases}
    5-p \; \forall p\leq 2,\\
    2-p \; \forall p> 2,
  \end{cases}
\end{equation} 
where $1 \leq k \leq K$. The results are shown in Figs.~\ref{fig:prototypical}(k)-(o). In practice, NNs will tend to predict similar outputs for similar inputs. The continuous nature of the NN results in SL highlighting the value of the tuning parameter $p=2$ where a discontinuity in the input data is present. We also observe this tendency for the NNs in LBC and PBM during training. 
\subsection{Computational cost}\label{sec_comp_cost}
We can use the analytical expressions to assess the computational cost associated with the evaluation of the mean optimal predictions and optimal indicators of SL, LBC, and PBM for a given set of input data (see Appendix~\ref{app_A3} for proofs). In our estimation, we neglect the overhead arising from the computation of the probability distributions $\{ {\rm P}_{k}(\bm{x})\}_{k=1}^{K}$ which is identical for all three methods. In the case of SL, the computation of its optimal predictions (as a function of the tuning parameter) and indicator scales as $O(M_{\bar{\mathcal{X}}}K)$. Here, we assume that the number of sampled values of the tuning parameter during training is small compared to the total number of sampled points $K_{\rm I}+K_{\rm II}\ll K$. For PBM and LBC the computation scales as $O(M_{\bar{\mathcal{X}}}K^2)$ and $O(M_{\bar{\mathcal{X}}}K^3)$, respectively. By saving the optimal predictions for each input $\hat{y}_{\rm opt}(\bm{x})$ instead of recomputing it, the computational cost can be reduced and scales as $O(M_{\bar{\mathcal{X}}}K)$, $O(M_{\bar{\mathcal{X}}}K)$, and $O(M_{\bar{\mathcal{X}}}K^2)$, in the case of SL, PBM, and LBC, respectively. Note the appearance of $M_{\bar{\mathcal{X}}}$ which can result in an exponential scaling for quantum problems due to the exponential growth of the Hilbert space $\mathcal{H}$ (and thus the state space $M_{\bar{\mathcal{X}}}$).

\section{Application to physical systems}\label{sec_physical_models}
In this section, we will compute the optimal predictions and indicators of phase transitions of SL, LBC, and PBM, directly from data using the analytical expressions introduced in Sec.~\ref{sec_optimal_indicators} for the Ising model, Ising gauge theory, XY model, XXZ model, Kitaev model, and Bose-Hubbard model. For the classical systems, namely the Ising model, Ising gauge theory, and XY model, spin configurations are sampled from a thermal distribution at various temperatures $T_{k}$ using the Metropolis-Hastings algorithm~\cite{metropolis:1953}. Here, the temperature serves as a tuning parameter. The probability that a system in equilibrium at inverse temperature $\beta_{k} = 1/k_{\rm B}T_{k}$ (where $k_{\rm B}$ is the Boltzmann constant) is found in a state with spin configuration $\bm{\sigma}$ is given by a Boltzmann distribution
\begin{equation}\label{eq:boltzmann}
{\rm P}_{k}(\bm{\sigma}) = \frac{e^{-\beta_{k} H(\bm{\sigma})}}{Z_{k}},
\end{equation}
where $Z_{k} = \sum_{\bm{\sigma}} e^{-\beta_{k} H(\bm{\sigma})}$ is the partition function and $H$ is the respective system Hamiltonian. In principle, one could use the raw spin configurations as input, i.e., estimate the underlying probability distributions as ${\rm P}_{k}(\bm{\sigma}) = M_{k}(\bm{\sigma})/M$. However, the probability of drawing a particular spin configuration only depends on its energy [see Eq.~\eqref{eq:boltzmann}]. One can show that the optimal predictions and indicators remain identical when the energy is used as input instead of the raw configurations, i.e., when the probability distributions governing the data are given by
\begin{equation}\label{eq:boltzmann2}
 {\rm P}_{k}(E) = \frac{g(E) e^{-\beta_{k} E}}{Z_{k}},
\end{equation}
where $g(E)$ is the degeneracy factor (see Appendix~\ref{app_A4} for a proof). Using the energy as input instead of the raw configurations reduces both the input dimension and the size of the associated state space. This, in turn, reduces the cost of computing the optimal predictions and indicators. Similarly, one could take advantage of the symmetries of the system by adopting a symmetry-adapted representation.\\

\begin{figure*}[tbh!]
\centering
\includegraphics[width=\linewidth]{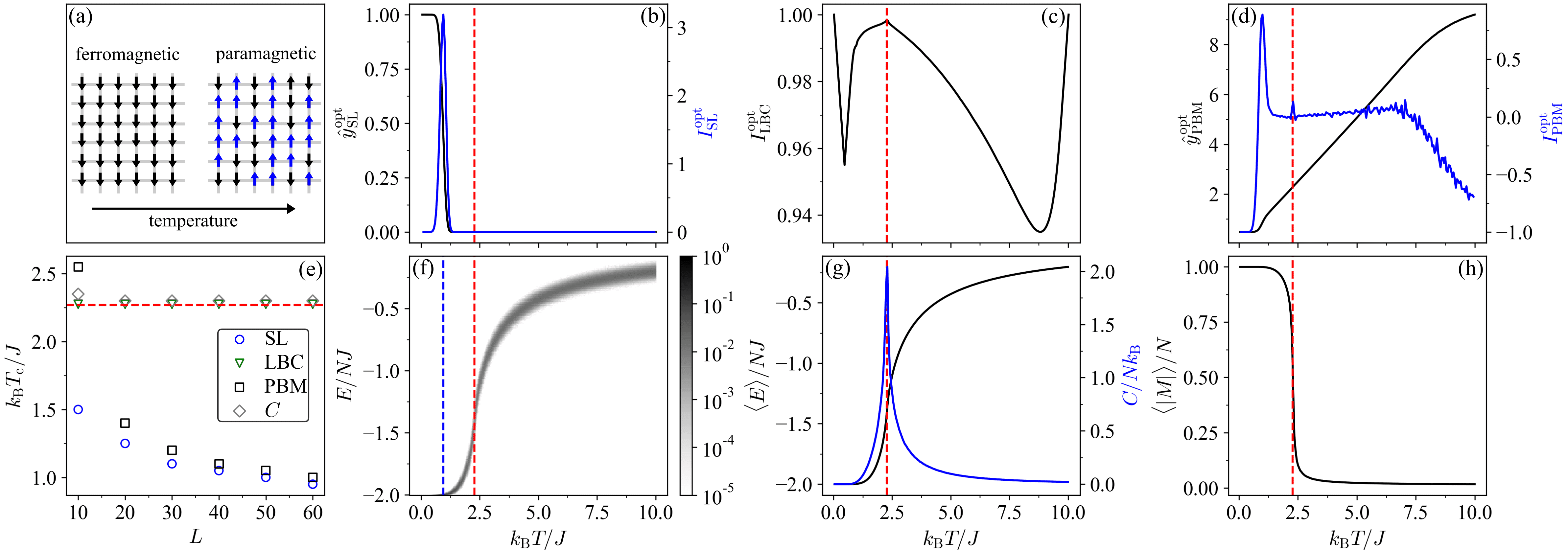}
\caption{Results for the Ising model ($L=60$) with the dimensionless temperature as a tuning parameter $p=k_{\rm B}T/J$, where $p_{1}=0.05$, $p_{K}=10$, and $\Delta p = 0.05$. In SL, the data obtained at $p_{1}$ and $p_{K}$ constitutes our training set, i.e., $r_{\rm I}=1$ and $l_{\rm II}=K$. The critical temperature [Eq.~\eqref{eq:ising_Tc}] is highlighted by a red-dashed line. (a) Illustration of the symmetry-breaking phase transition in the Ising model. (b) Mean optimal prediction $\hat{y}_{\rm SL}^{\rm opt}$ in SL (black) and the corresponding indicator $I_{\rm SL}^{\rm opt}$ (blue). (c) Optimal indicator of LBC, $I_{\rm LBC}^{\rm opt}$ (black). (d) Mean optimal prediction $\hat{y}_{\rm PBM}^{\rm opt}$ in PBM (black) and the corresponding indicator $I_{\rm PBM}^{\rm opt}$ (blue). (e) Estimated critical temperatures based on $I_{\rm SL}^{\rm opt}$ (SL), $I_{\rm LBC}^{\rm opt}$ (LBC), $I_{\rm PBM}^{\rm opt}$ (PBM), and heat capacity ($C$) as a function of the lattice size $L$. The estimated critical temperature based on the heat capacity corresponds to the location of its maximum. (f) Probability distributions governing the input data (here the energy) as a function of the tuning parameter, where the color scale denotes the probability. The blue-dashed line highlights the predicted critical temperature of SL and PBM. (g) Average energy per site (black) and associated heat capacity (blue) as a function of temperature, where $N=L^2$. (h) Average magnetization per site as a function of temperature.}
\label{fig:ising}
\end{figure*}

\indent In the quantum case, we will typically be looking at a state associated with a Hamiltonian $H(p)$ that depends on the tuning parameter $p$. This state could, for example, be the ground state or a state which has undergone unitary time evolution starting from a fixed initial state. Having chosen a complete orthonormal basis $\{ | j\rangle \}_{j=1}^{d}$ to study the system [$d = {\rm dim}(\mathcal{H})$], the relevant quantum state at $p_{k}$ can be written as $| \Psi_{k} \rangle = \sum_{j=1}^{d} c_{jk} | j\rangle$. Thus, the probability distribution ${\rm P}_{k}$ associated with a given value $p_{k}$ of the tuning parameter is ${\rm P}_{k}(j) = |c_{jk}|^2$ with $1 \leq j \leq d$ and $1 \leq k\leq K$. The value of ${\rm P}_{k}(j)$ corresponds to the probability of measuring the system in state $| j \rangle $ given that the value of the tuning parameter is $p_{k}$. This corresponds to using the indices of the basis states $|j \rangle$ ($1 \leq j \leq d$) as inputs, which are governed by the probability distributions $\{ {\rm P}_{k}(j)\}_{k=1}^{K}$. For simplicity, we choose $M_{\bar{\mathcal{X}}} = d$. In the case of spin systems, we use the $S^{z}$ basis, whereas we choose the Fock basis for bosonic and fermionic systems. This choice of bases corresponds to experimentally accessible local
measurements~\cite{simon:2011,bernien:2017,lukin:2019,rispoli:2019,jespen:2020,jespen:2021,ebadi:2021}. In this work, we obtain the ground states through exact diagonalization. Thus, we have direct access to the underlying probability distributions and do not rely on sampling. In Appendix~\ref{app_A5}, we show that the optimal indicators can also be obtained from individual samples, i.e., measurement outcomes (similar to the classical case). As such, the procedure is \textit{in principle} applicable to experimental scenarios.\\

\indent In general, we can consider scenarios where a state $| \Psi_{i} \rangle$ is drawn with probability $a_{k}(i)$ at $p_{k}$. Then, the relevant quantum state is given by a classical probabilistic mixture $\rho_{k} = \sum_{i} a_{k}(i) | \Psi_{i} \rangle \! \langle  \Psi_{i} |, \; i\in \mathbb{N}$. The probability distribution associated with such a state is ${\rm P}_{k}(j) = \sum_{i} a_{k}(i) |c_{ij}|^2$. This case will be particularly relevant for the study of many-body localization phase transitions where disorder is naturally present (see Sec.~\ref{sec_applications_MBL_BH}). Here, the tuning parameter $p_{k}$ itself characterizes a distribution $a_{k}$. Further details on the data generation can be found in Appendix~\ref{app_C}.\\

\indent Clearly, in the quantum case there is an ambiguity in the choice of input, or equivalently, the choice of measurement basis. Changing the measurement basis may change the probability distributions underlying the data, and thus the corresponding optimal predictors and indicators. In turn, the estimated critical value of the tuning parameter may change (which is difficult to assess \textit{a priori} for a given system). In order to avoid an explicit choice of measurement basis, sampling over various classical projections can be performed. Classical representations of quantum states obtained via classical shadow tomography~\cite{huang:2020,huang:2021,huang:2022} are an example of this. Alternatively, measurements given by informationally complete (IC) positive operator-valued measures (POVMs) can be used~\cite{nielsen:2010,carrasquilla:2019}. However, projective measurements in a single basis have been the most common choice, reflecting experimental constraints or prior knowledge of the system~\cite{torlai:2019,greplova:2020,miles:2021,bohrdt:2021,miles2:2021,maskara:2022}.

\subsection{Ising model}\label{sec_ising_model}
\indent The two-dimensional square-lattice ferromagnetic Ising model is described by the following Hamiltonian
\begin{equation}\label{eq:ising_H}
H(\bm{\sigma}) = - J \sum_{\langle ij\rangle} \sigma_{i}\sigma_{j},
\end{equation}
where the sum runs over all nearest-neighboring sites (with periodic boundary conditions) and $J$ is the interaction strength $(J>0)$. At each lattice site $k$, there is a discrete spin variable $\sigma_{i} \in \{+1,-1 \}$. This results in a state space of size $2^{L \times L}$ for a square lattice of linear size $L$. The system is completely characterized by its spin configuration $\bm{\sigma} = (\sigma_{1},\sigma_{2},\dots,\sigma_{L \times L})$. Two example spin configurations of the Ising model at different temperatures are shown in Fig.~\ref{fig:ising}(a). The Ising model exhibits a symmetry-breaking phase transition at a critical temperature of~\cite{onsager:1944}
\begin{equation}\label{eq:ising_Tc}
T_{\rm c} = \frac{2J}{k_{\rm B} \ln(1+\sqrt{2})}.
\end{equation}
The system undergoes a transition between a paramagnetic (disordered) phase at high temperature and a ferromagnetic (ordered) phase at low temperature. Spontaneous magnetization occurs below the critical temperature $T_{\rm c}$, where the interaction is sufficiently strong to cause neighboring spins to align spontaneously. This spontaneous symmetry breaking leads to a non-zero mean magnetization. Above $T_{\rm c}$, thermal fluctuations dominate over spin alignment resulting in a vanishing magnetization. Consequently, the phase transition can be characterized by the magnetization $M(\bm{\sigma}) = \sum_{i=1}^{L^2} \sigma_{i}$ which serves as an order parameter that is zero within the paramagnetic phase and approaches one in the ferromagnetic phase, see Fig.~\ref{fig:ising}(h). The phase transition can also be revealed by the heat capacity
\begin{equation}
    C(T)=\frac{d \langle E\rangle_{T}}{d T} = \frac{\langle E^2 \rangle_{T} - \langle E \rangle_{T}^2}{k_{\rm B}T^2}
\end{equation}
which diverges at $T_{\rm c}$ [see Fig.~\ref{fig:ising}(g)].\\

\indent The results for the Ising model are shown in Fig.~\ref{fig:ising}. Interestingly, SL fails to predict the correct critical temperature even for large lattices [see Figs.~\ref{fig:ising}(b),(e)]. In fact, we can further analyze the special case when the inputs are governed by Boltzmann distributions [Eq.~\eqref{eq:boltzmann2}]: For training data obtained at $T_{1}=0$ (region I) and $T_{K}>0$ (region II), the mean optimal prediction of SL at an intermediate temperature $T_{k}$ is
\begin{equation}\label{eq_SL_boltzmann}
\hat{y}_{\rm SL}^{\rm opt}(T_{k}) = \frac{{\rm P}_{k}(E_{\rm gs})}{1+{\rm P}_{K}(E_{\rm gs})} \propto {\rm P}_{k}(E_{\rm gs}),
\end{equation}
which approaches ${\rm P}_{k}(E_{\rm gs})$ in the thermodynamic limit as $T_{K}\rightarrow \infty$ (see Appendix~\ref{app_A4} for a proof). Here, $E_{\rm gs}$ denotes the ground-state energy. Therefore, in this case, the optimal indicator in SL peaks at the temperature at which the probability of drawing the ground state changes most [see the blue-dashed line in Fig.~\ref{fig:ising}(f)]. The location of the peak tends to zero as one approaches the thermodynamic limit, see Fig.~\ref{fig:ising}(e).\\

\indent The optimal indicator of PBM shows two distinct peaks. One coincides with the peak of the optimal indicator in SL, whereas the other coincides with the critical temperature of the Ising model [see Fig.~\ref{fig:ising}(d)]. This observation suggests a deeper connection between SL and PBM. A similar indicator signal (with two distinct peaks) was observed in Ref.~\cite{schaefer:2019} with NNs after a sufficient number of training epochs. In principle, the finite-size scaling analysis allows one to identify the dominant peak as erroneous without prior knowledge of $T_{\rm c}$, because it shifts towards $T=0$ as the lattice size is increased, whereas the small peak remains stable. In the same fashion, the output of SL can be identified to be erroneous. Note that the fluctuations present in the optimal indicator signal of PBM can be attributed to finite-sample statistics. A detailed study of the effect of finite-sample statistics on the optimal predictions and indicators can be found in Appendix~\ref{app_A5}. Crucially, the analytical expression for the optimal indicator signal allows us to disentangle the stochasticity inherent to the NN training from other sources of noise, which was not rigorously possible in previous works.\\

\indent In Ref.~\cite{carrasquilla:2017}, SL with NNs was able to predict the critical temperature of the Ising model for various lattice sizes correctly. In this case, small NNs with restricted expressive power in combination with $\ell_{2}$ regularization were used. Similarly, using PBM in Ref.~\cite{schaefer:2019} a single, distinct peak at $T_{\rm c}$ was observed after a small number of training epochs with a second peak emerging after longer training. Training time, NN size, and explicit $\ell_{2}$ regularization are all factors which influence the effective capacity of the resulting model and thus determine its ability to approximate the optimal predictive model~\cite{goodfellow:2016,hu:2021}, i.e., to realize the global minimum of the loss function corresponding to the optimal predictions and indicators. We recover the same behavior using NNs as in Ref.~\cite{carrasquilla:2017, schaefer:2019} by restricting the model capacity, e.g., by choosing a small NN, stopping the training early, or using strong $\ell_{2}$ regularization (see Appendix~\ref{app_B1} for details). As these restrictions are lifted, i.e., by choosing a larger NN, training for longer, or reducing the regularization strength, the NN-based predictions and indicators approach the corresponding optimal predictions and indicators displayed in Fig.~\ref{fig:ising}. Thus, our analysis demonstrates that SL and PBM necessarily rely on models with restricted capacity and hyperparameter tuning to correctly predict the critical temperature of the Ising model.\\

\indent Finally, the optimal indicator of LBC correctly highlights the critical temperature of the Ising model for various lattice sizes matching the results of Ref.~\cite{van:2017}, see Figs.~\ref{fig:ising}(c),(e). Overall, the optimal indicators of all three methods show peaks at temperatures where the probability distribution underlying the data varies strongly. Recall the finding from Sec.~\ref{sec_optimal_indicators} that all three methods gauge changes in the probability distributions underlying the data. We have confirmed that the results shown in Fig.~\ref{fig:ising} are stable against small perturbations of the chosen parameter range, including regions I and II in SL.

\begin{figure*}[tbh!]
\centering
\includegraphics[width=\linewidth]{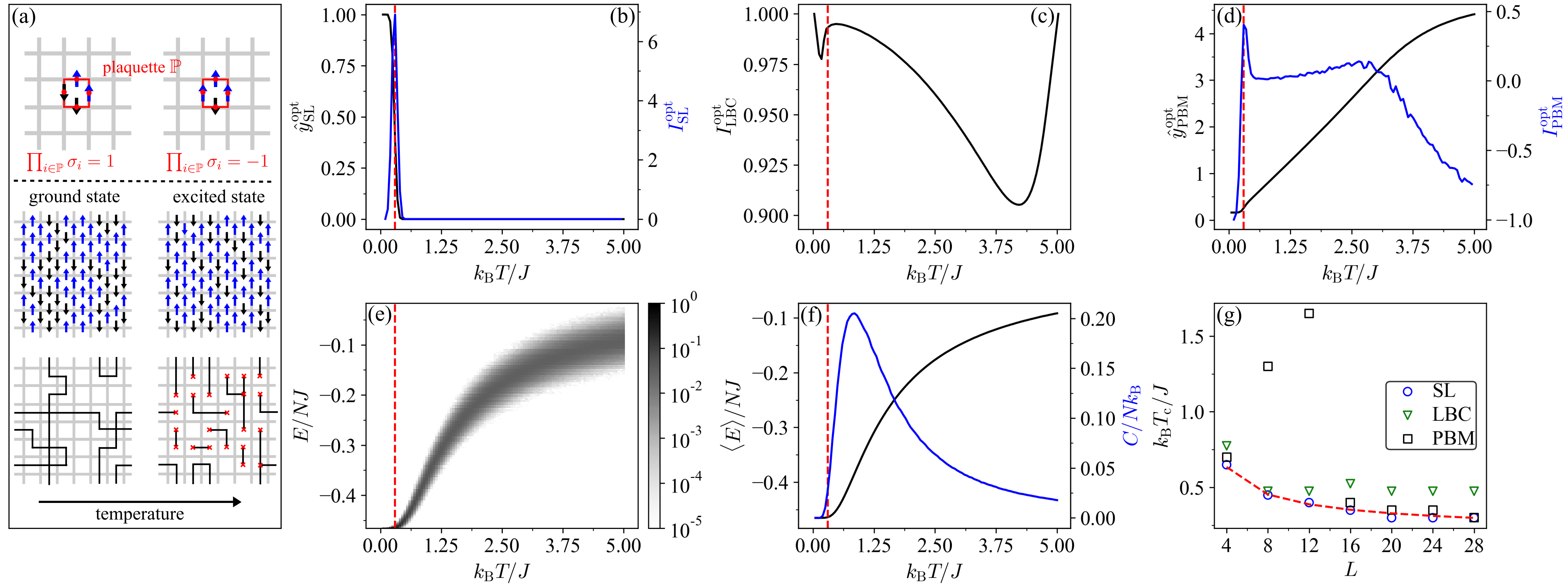}
\caption{Results for the IGT ($L=28$) with the dimensionless temperature as a tuning parameter $p=k_{\rm B}T/J$, where $p_{1}=0.05$, $p_{K}=5$, and $\Delta p = 0.05$. In SL, the data obtained at $p_{1}$ and $p_{K}$ constitutes our training set, i.e., $r_{\rm I}=1$ and $l_{\rm II}=K$. The crossover temperature is highlighted by a red-dashed line and scales as $k_{\rm B}T_{\rm c}/J \propto 1/\ln(2L^2)$~\cite{castelnovo:2007}. (a) Upper panels show examples of plaquettes $\mathbb{P}$ where the topological constraint is met ($\prod_{i \in \mathbb{P}} \sigma_{i}=1$) and violated ($\prod_{i \in \mathbb{P}} \sigma_{i}=-1$). Middle panels show examples of spin configurations within the topological ground-state phase (left) and phase with violated topological constraints at high temperature (right). Lower panels show the corresponding Wilson loops. (b) Mean optimal prediction $\hat{y}_{\rm SL}^{\rm opt}$ in SL (black) and the corresponding indicator $I_{\rm SL}^{\rm opt}$ (blue). (c) Optimal indicator of LBC, $I_{\rm LBC}^{\rm opt}$ (black). (d) Mean optimal prediction $\hat{y}_{\rm PBM}^{\rm opt}$ in PBM (black) and the corresponding indicator $I_{\rm PBM}^{\rm opt}$ (blue). (e) Probability distributions governing the input data (here the energy) as a function of the tuning parameter, where the color scale depicts the probability. (f) Average energy per site (black) and associated heat capacity (blue) as a function of temperature, where $N=2L^2$. Note that the heat capacity does not peak at the crossover temperature. (g) Estimated critical temperature based on $I_{\rm SL}^{\rm opt}$ (SL), $I_{\rm LBC}^{\rm opt}$ (LBC), $I_{\rm PBM}^{\rm opt}$ (PBM) as a function of the lattice size $L$.}
\label{fig:IGT}
\end{figure*}

\subsection{Ising gauge theory}
\indent Wegner's Ising gauge theory (IGT)~\cite{wegner:1971} is described by the following Hamiltonian
\begin{equation}
	H(\bm{\sigma}) = - J \sum_{\mathbb{P}} \prod_{i \in \mathbb{P}} \sigma_{i},
\end{equation}
where $\mathbb{P}$ refers to plaquettes on the lattice, see Fig~\ref{fig:IGT}(a). The IGT is a prototypical example of a classical system that exhibits a topological phase of matter~\cite{kogut:1979}. It is a spin model ($\sigma_{i} \in \{+1,-1 \}$) defined on a square lattice of linear size $L$ (with periodic boundary conditions) where the spins are placed on the lattice bonds [see Fig.~\ref{fig:IGT}(a)]. The IGT ground state is a degenerate manifold made up of all states which fulfill the condition that the product of spins on each plaquette is $\prod_{i \in \mathbb{P}} \sigma_{i}=1$ corresponding to a topological phase. These topological constraints can be violated at finite temperature, where the system leaves its ground state. Note that there is no phase transition at finite temperature: the critical temperature approaches zero in the thermodynamic limit. In finite-sized systems, however, the violations of local constraints are suppressed. Therefore, the system exhibits a crossover from the topological phase at low temperature to a phase with violated topological constraints at high temperature. The crossover temperature $T_{\rm c}$ is defined by the first appearance of a violated local constraint and scales as $T_{\rm c} \propto 1/\ln(2L^2)$~\cite{castelnovo:2007}. Figure~\ref{fig:IGT}(a), which shows typical spin configurations of the IGT, highlights that the phases of the IGT are hard to distinguish visually without prior knowledge of the local constraints or a dual representation~\cite{carrasquilla:2017,greplova:2020}. Note that the heat capacity fails to identify the crossover, see Fig.~\ref{fig:IGT}(f). The topological character of the ground-state phase can be revealed through Wilson loops. These are formed by connecting edges with spins of the same orientation, see Fig.~\ref{fig:IGT}(a). In the ground-state phase, all such loops are closed. The violation of a plaquette constraint breaks a loop.\\ 

\indent Recall that SL, LBC, and PBM are \textit{a priori} sensitive to both phase transitions and crossovers. The results for the crossover in the IGT are shown in Fig.~\ref{fig:IGT}. The optimal indicator of SL [Fig.~\ref{fig:IGT}(b)] shows an appropriate scaling behavior. Moreover, the corresponding estimated critical temperature highlights the first appearance of violated local constraints, see Figs.~\ref{fig:IGT}(e),(f). This can be confirmed explicitly as SL can be shown to measure changes in the probability of drawing the ground state (cf. Sec.~\ref{sec_ising_model}). Observe that the underlying probability distribution undergoes a large change at the crossover temperature, see Fig.~\ref{fig:IGT}(e). SL and PBM were found to correctly highlight the crossover temperature of the IGT using NNs in Refs.~\cite{carrasquilla:2017} and~\cite{greplova:2020}, respectively. In fact, the optimal model underlying PBM for the IGT coincides with the physically motivated density-of-states-based model proposed in Ref.~\cite{greplova:2020}, see Appendix~\ref{app_D2} for details. We find that the optimal indicator of PBM correctly marks the crossover temperature of the IGT except at small lattice sizes. As for the Ising model, the optimal indicator of PBM exhibits two peaks in this case. The peak located at the crossover temperature dominates for large lattice sizes. Note that for the IGT is is not beneficial to reduce the model capacity when using PBM or SL, which leads to an erroneous peak closely matching the specific heat [see Fig.~\ref{fig:IGT}(f)], given that the corresponding optimal indicators correctly highlight the crossover temperature.\\

\begin{figure*}[htb!]
\centering
\includegraphics[width=\linewidth]{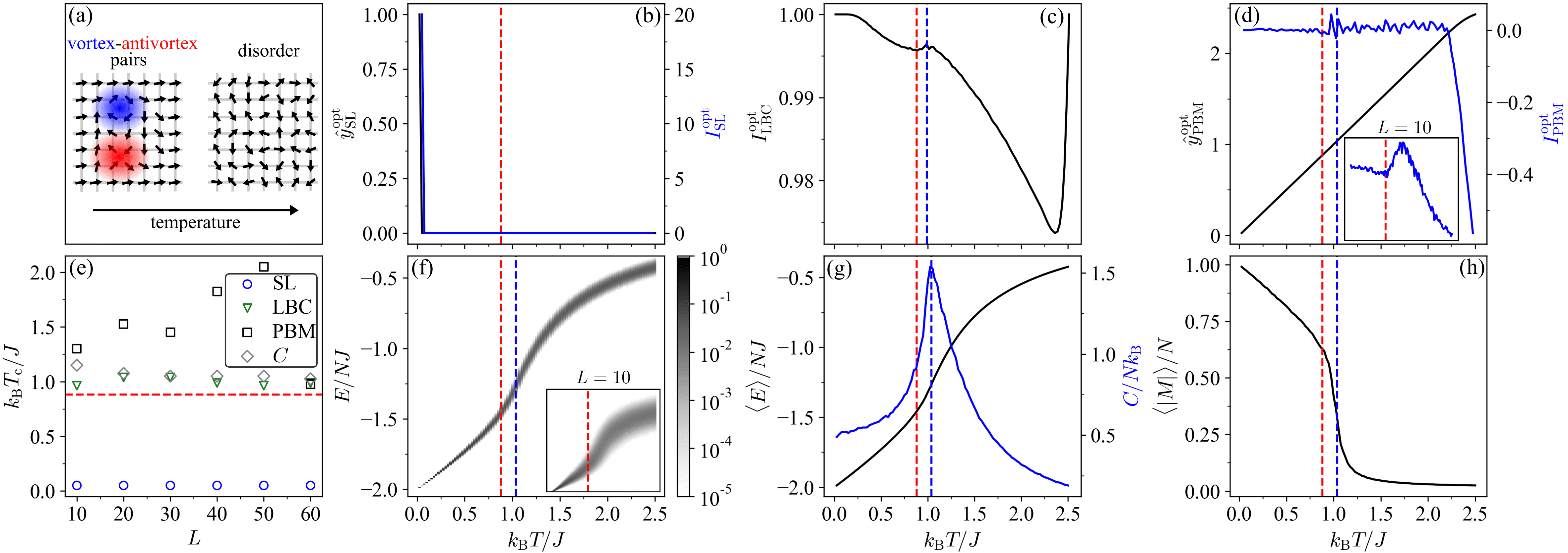}
\caption{Results for the XY model ($L=60$) with the dimensionless temperature as a tuning parameter $p=k_{\rm B}T/J$, where $p_{1}=0.025$, $p_{K}=2.5$, and $\Delta p = 0.025$. In SL, the data obtained at $p_{1}$ and $p_{K}$ constitutes our training set, i.e., $r_{\rm I}=1$ and $l_{\rm II}=K$. The BKT transition temperature $k_{\rm B} T_{\rm c}/J\approx 0.8935$~\cite{hsieh:2013} is highlighted by a red-dashed line. The blue-dashed line highlights the estimated critical temperature using LBC. (a) Illustration of the BKT phase transition in the XY model. (b) Mean optimal prediction $\hat{y}_{\rm SL}^{\rm opt}$ in SL (black) and the corresponding indicator $I_{\rm SL}^{\rm opt}$ (blue). (c) Optimal indicator of LBC, $I_{\rm LBC}^{\rm opt}$ (black). The blue-dashed line highlights the predicted critical temperature of LBC. (d) Mean optimal prediction $\hat{y}_{\rm PBM}^{\rm opt}$ in PBM (black) and the corresponding indicator $I_{\rm PBM}^{\rm opt}$ (blue). The inset shows the optimal indicator signal of PBM for $L=10$, which exhibits a peak near the location of the maximum in the heat capacity. (e) Estimated critical temperature based on $I_{\rm SL}^{\rm opt}$ (SL), $I_{\rm LBC}^{\rm opt}$ (LBC), $I_{\rm PBM}^{\rm opt}$ (PBM), and heat capacity ($C$) as a function of the lattice size $L$. The estimated critical temperature of the heat capacity corresponds to the location of its maximum. (f) Probability distributions governing the input data (here the energy) as a function of the tuning parameter, where the color scale denotes the probability. The inset shows the probability distributions for $L=10$. (g) Average energy per site (black) and associated heat capacity (blue) as a function of temperature, where $N=L^2$. (h) Average magnetization per site as a function of temperature.}
\label{fig:XY}
\end{figure*}

\indent The optimal indicator of LBC correctly highlights the crossover temperature via its local maximum at small lattice sizes, but shows slight deviations from the appropriate scaling behavior for large lattices. In Ref.~\cite{greplova:2020} difficulties were observed to identify the crossover temperature using LBC due to a distorted W-shape of its indicator. Choosing the same range for the tuning parameter, we can qualitatively reproduce their results using our analytical expression for the optimal indicator of LBC, see Appendix~\ref{app_D2}. Using NNs, it is difficult to make concrete statements on whether a method succeeds or fails at identifying a given phase transition due to the inherent stochasticity arising during NN training and the choice of hyperparameters, such as the NN size. Our theoretical analysis allows for rigorous statements to be made about the optimal outcome when applying ML methods for detecting phase transitions to a given system (i.e., data set). In this particular example, the analytical expressions allow us to determine that when training highly expressive NNs for sufficiently long, the indicator signal of LBC is indeed ambiguous (as reported in Ref.~\cite{greplova:2020}). Note that restricting the model capacity is not found to resolve this issue~\cite{greplova:2020}.
\subsection{XY model}\label{sec_xy_model}
Next, we consider the two-dimensional classical XY model that exhibits a Berezinskii–Kosterlitz–Thouless (BKT) transition driven by the emergence of topological defects~\cite{kosterlitz:1973,kosterlitz:1974}. The model is described by the following Hamiltonian
\begin{equation}
    H = -J\sum_{\langle ij\rangle} \cos(\theta_{i} - \theta_{j}),
\end{equation}
where $\langle ij\rangle$ denotes the sum over nearest neighbors (with periodic boundary conditions) of a square lattice of linear size $L$. The angle $\theta_{i}\in [0,2\pi)$ corresponds to the orientation of the spin at site $i$. The formation of topological defects (i.e., vortices and antivortices) results in a quasi-long-range-ordered phase. The transition between the quasi-long-range-ordered phase at low temperature and a disordered phase at high temperature is a BKT transition, and the associated critical temperature is $k_{\rm B} T_{\rm c}/J\approx 0.8935$~\cite{hsieh:2013}. Below $T_{\rm c}$, vortex-antivortex pairs form due to thermal fluctuations, but they remain bound to minimize their total free energy [see Fig.~\ref{fig:XY}(a)]. At $T_{\rm c}$, the entropic contribution to the free energy equals the binding energy of a pair which triggers vortex unbinding. These unbinding events drive the BKT phase transition. Note that the heat capacity has a peak at $T>T_{\rm c}$ which is associated with the entropy released when most vortex pairs unbind~\cite{chaikin:1995,himbergen:1981}, see Fig.~\ref{fig:XY}(g). Moreover, while the XY model has strictly zero magnetization for all $T > 0$ in the thermodynamic limit, a non-zero value is found for systems of finite size~\cite{chung:1999}, see Fig.~\ref{fig:XY}(h). Instead, the critical temperature can, for example, be estimated based on the helicity modulus~\cite{himbergen:1981,minnhagen:2003} (see Appendix~\ref{app_C}).\\
\begin{figure*}[tbh!]
	\centering
	\includegraphics[width=\linewidth]{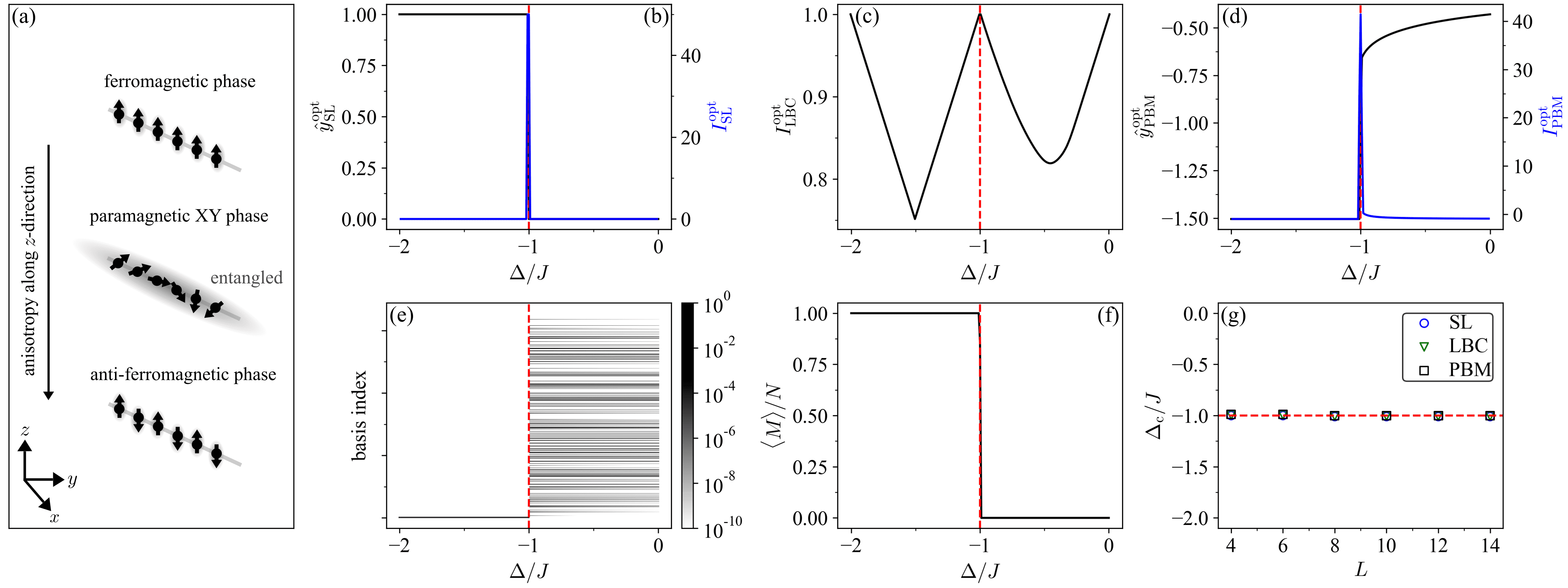}
	\caption{Results for the XXZ chain ($L=14$) with the dimensionless anisotropy strength along the $z$-direction as the tuning parameter $p=\Delta/J$, where $p_{1}=-2$, $p_{K}=0$, and $\Delta p = 0.01$. In SL, the data obtained at $p_{1}$ and $p_{K}$ constitutes our training set, i.e., $r_{\rm I}=1$ and $l_{\rm II}=K$. The critical value of the tuning parameter $\Delta/J=-1$ at which the phase transition between the ferromagnetic phase and paramagnetic XY phase occurs is highlighted by a red-dashed line. (a) Illustration of the quantum phase transitions of the XXZ chain. (b) Mean optimal prediction $\hat{y}_{\rm SL}^{\rm opt}$ in SL (black) and the corresponding indicator $I_{\rm SL}^{\rm opt}$ (blue). (c) Optimal indicator of LBC, $I_{\rm LBC}^{\rm opt}$ (black). (d) Mean optimal prediction $\hat{y}_{\rm PBM}^{\rm opt}$ in PBM (black) and the corresponding indicator $I_{\rm PBM}^{\rm opt}$ (blue). (e) Probability distributions governing the input data (indices of $S^{z}$ basis states) as a function of tuning parameter, where the color scale denotes the probability. The color scale is cut off at $10^{-10}$ to improve visual clarity. (f) Average magnetization per site (black), where $N=L$. (g) Estimated critical value of the tuning parameter based on $I_{\rm SL}^{\rm opt}$ (SL), $I_{\rm LBC}^{\rm opt}$ (LBC), $I_{\rm PBM}^{\rm opt}$ (PBM) as a function of the chain length $L$.}
	\label{fig:XXZ}
\end{figure*}

\indent The results for the XY model are shown in Fig.~\ref{fig:XY}. Here, SL fails to predict the critical temperature correctly. This failure is linked to the fact that the optimal indicator of SL highlights changes in the probability to obtain the ground state (cf. Sec.~\ref{sec_ising_model}), which quickly vanishes with increasing temperature, see Fig.~\ref{fig:XY}(f). In a similar spirit, in Ref.~\cite{beach:2018} it was found that ``naive'' SL (without engineering the features or NN architecture) fails to yield accurate estimates of the critical temperature. Here, we explicitly confirm that a classification based on detecting vortices does not correspond to the most optimal strategy. The peak in the optimal indicator of LBC matches the peak in the heat capacity at $k_{\rm B}T/J \approx 1$, see Figs.~\ref{fig:XY}(c) and (e), and thus overestimates the critical temperature of the XY model. In Ref.~\cite{beach:2018} indicator signals of similar shape were obtained using LBC with NNs for the XY model. The rapid decrease in the optimal indicator of LBC for $k_{\rm B}T/J \gtrsim 1$ can be attributed to the increase in the overlap of the underlying probability distributions [Fig.~\ref{fig:XY}(f)], which results in a higher classification error. Note that the overlap of the probability distributions decreases with increasing lattice size, see Fig.~\ref{fig:XY}(f). Hence, the indicator of PBM [Fig.~\ref{fig:XY}(d)] shows a clear peak close to the location of the peak in the heat capacity for small lattice sizes. For systems of increasing size, the optimal predictions of PBM start to closely match the underlying tuning parameter, resulting in an increasingly linear behavior [see black line in Fig.~\ref{fig:XY}(d)]. This corresponds to an optimal indicator signal close to zero, where the variations in the predicted critical value of the tuning parameter [Fig.~\ref{fig:XY}(e)] are due to small local fluctuations.\\

\indent Overall, the behavior of the optimal indicators of all three methods closely resembles our previous example regarding perfectly distinguishable input data (see case 3 in Sec.~\ref{sec_application_trivial_distr}). This can be traced back to the small overlap of the underlying probability distributions, see Fig.~\ref{fig:XY}(f). The increase in the overlap with increasing temperature results in a decrease in the mean classification accuracy of LBC, i.e., its indicator [see Fig.~\ref{fig:XY}(c)]. Evidently, in such a case, NNs with restricted expressive power and other phase-classification methods based on the similarity of input data~\cite{rodriguez:2019} may provide more valuable insights. In particular, we find that the indicators peak close to the transition temperature, i.e., near the location of the peak in the heat capacity and drop in the magnetization, when restricting the model capacity, e.g., by stopping the NN training early (see Appendix~\ref{app_B}). Recall that this was also observed in the case of the Ising model (see Sec.~\ref{sec_ising_model} and Appendix~\ref{app_B}).

\subsection{XXZ model}\label{sec_methods_xxz}
Having discussed classical models, we move on to the quantum case. First, we consider the spin-1/2 XXZ chain~\cite{schollwock:2008,franchini:2017} with open boundary conditions whose Hamiltonian is given by
\begin{equation}
    H = \sum_{i=1}^{L-1} J(S^{x}_{i+1} S^{x}_{i} + S^{y}_{i+1} S^{y}_{i}) + \Delta S^{z}_{i+1}S^{z}_{i},
\end{equation}
where $J$ is the coupling strength along the $x$- and $y$-direction and $\Delta$ is the coupling strength in the $z$-direction. For $\Delta/J <1$, the XXZ chain is in the ferromagnetic phase, see Fig.~\ref{fig:XXZ}(a). The ground state is spanned by the two product states where all spins point either in the $z$ or $-z$ direction which have a magnetization of $\langle M \rangle = 2\langle S_{\rm tot}^{z}\rangle = \pm L $. The ferromagnetic phase exhibits a broken symmetry: these states do not exhibit the discrete symmetry of spin reflection $S^{z}_{i}\rightarrow -S^{z}_{i}$ under which the Hamiltonian is invariant. For $\Delta/J > 1$, the XXZ chain is in the antiferromagnetic phase with broken symmetry and two degenerate ground states. These are product states with vanishing magnetization. For $-1 <\Delta/J <1$ the XXZ chain is in the paramagnetic XY phase characterized by uni-axial symmetry of the easy-plane type and vanishing magnetization.\\ 

\indent Here, we restrict our analysis to the transition between the ferromagnetic and paramagnetic XY phases. The ground states are obtained through exact diagonalization. Figure~\ref{fig:XXZ} shows the results when the ground state with $\langle S_{\rm tot}^{z} \rangle =+ L/2$ is selected in the ferromagnetic phase and $S^{z}$ is chosen as a measurement basis. The quantum phase transition can be revealed by looking at the magnetization, see Fig.~\ref{fig:XXZ}(f). The optimal indicators of all three methods correctly highlight the phase transition. Looking at the underlying probability distributions [see Fig.~\ref{fig:XXZ}(e)], the problem closely resembles the prototypical case of a bipartitioned data set (see case 2 in Sec.~\ref{sec_application_trivial_distr}). Thus, the optimal predictions and indicators also qualitatively match the results obtained in this case. In particular, the optimal predictions of SL can be described by Eq.~\eqref{eq_SL_boltzmann}, where the ferromagnetic ground state takes the role of the ground state energy (see Appendix~\ref{app_A4} for proof). We verified that the optimal indicators also mark the phase transition when other states from the ground state manifold are selected in the ferromagnetic phase and when measurements are performed in the $S^{x}$ or $S^{y}$ basis.
\subsection{Kitaev model}\label{sec_methods_kitaev}
\begin{figure*}[tbh!]
\centering
\includegraphics[width=\linewidth]{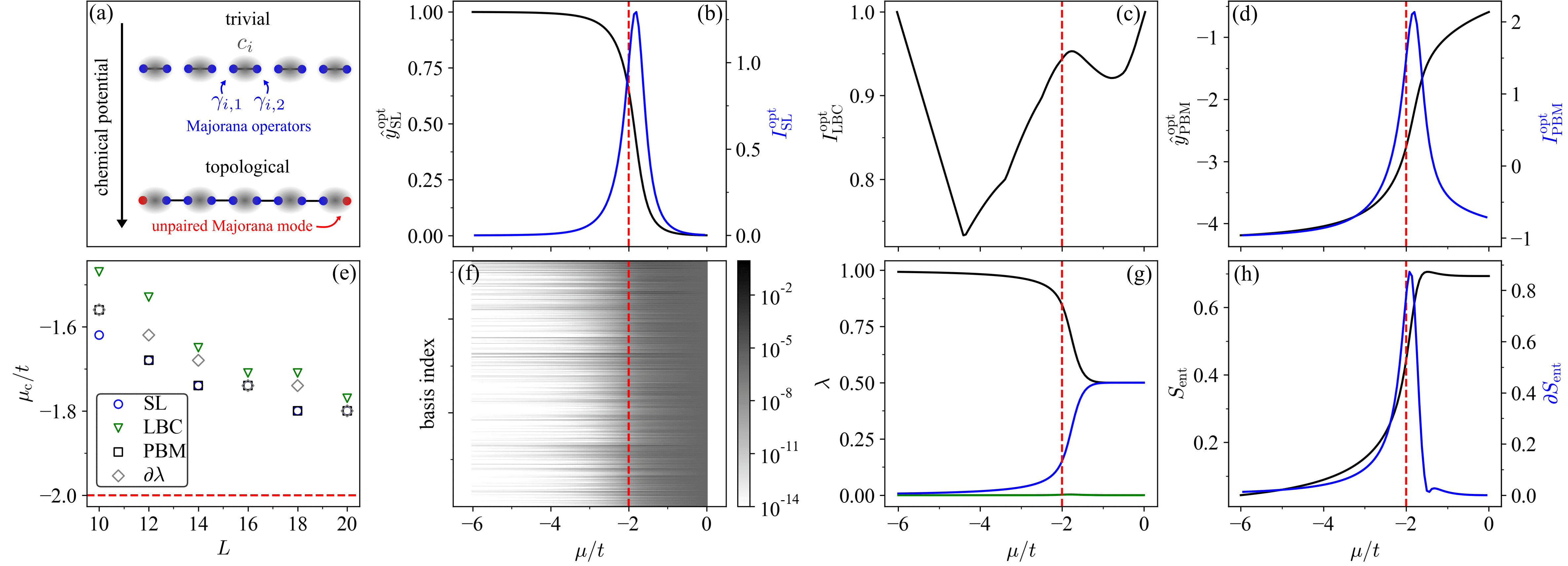}
\caption{Results for the Kitaev chain ($L=20$) with the dimensionless chemical potential as a tuning parameter $p=\mu/t$, where $p_{1}=-6$, $p_{K}=0$, and $\Delta p = 0.06$. In SL, the data obtained at $p_{1}$ and $p_{K}$ constitutes our training set, i.e., $r_{\rm I}=1$ and $l_{\rm II}=K$. The critical value $\mu_{\rm c}/t= - 2$ is highlighted by a red-dashed line. (a) Illustration of the phase transition in the Kitaev chain between a topological and trivial phase, where the Majorana operators $\gamma_{i,1}^{\phantom{\dagger}}$ and $\gamma_{i,2}^{\phantom{\dagger}}$ are defined by $c_{i}^{\phantom{\dagger}} = (\gamma_{i,1}^{\phantom{\dagger}} + i\gamma_{i,2}^{\phantom{\dagger}})/\sqrt{2}$, $c_{i}^{\dagger} = (\gamma_{i,1}^{\phantom{\dagger}} - i\gamma_{i,2}^{\phantom{\dagger}})/\sqrt{2}$. (b) Mean optimal prediction $\hat{y}_{\rm SL}^{\rm opt}$ in SL (black) and the corresponding indicator $I_{\rm SL}^{\rm opt}$ (blue). (c) Optimal indicator of LBC, $I_{\rm LBC}^{\rm opt}$ (black). (d) Mean optimal prediction $\hat{y}_{\rm PBM}^{\rm opt}$ in PBM (black) and the corresponding indicator $I_{\rm PBM}^{\rm opt}$ (blue). (e) Estimated critical value of the tuning parameter based on $I_{\rm SL}^{\rm opt}$ (SL), $I_{\rm LBC}^{\rm opt}$ (LBC), $I_{\rm PBM}^{\rm opt}$ (PBM), and the derivative of the largest eigenvalue of the reduced density matrix [see black line in panel (g)] given by $\partial \lambda /\partial p$ ($\partial \lambda$), as a function of the chain length $L$. The estimated critical value of the tuning parameter denoted by $\partial \lambda$ corresponds to the location of the maximum in $\partial \lambda /\partial p$. (f) Probability distributions governing the input data (indices of Fock basis states) as a function of the tuning parameter, where the color scale denotes the probability. The color scale is cut off at $10^{-14}$ to improve visual clarity. (g) The three largest eigenvalues of $\rho_{\rm A}$ [Eq.~\eqref{eq:reduced_dm}] as a function of the tuning parameter. (h) Entanglement entropy $S_{\rm ent}$ [Eq.~\eqref{eq:entanglement_entropy}] (black) and its derivative with respect to the tuning parameter $\partial S_{\rm ent}/\partial p$ (blue).}
\label{fig:kitaev}
\end{figure*}

\indent The Kitaev chain is a one-dimensional model based on $L$ spinless fermions, which undergoes a quantum phase transition between a topologically trivial and non-trivial phase~\cite{kitaev:2001,alicea:2012}. The Kitaev Hamiltonian is given by
\begin{equation}\label{eq:kitaev_H}
    H=\sum_{i=1}^{L-1} (\Delta c_{i+1}^{\phantom{\dagger}}c_{i}^{\phantom{\dagger}} - tc_{i+1}^{\dagger}c_{i}^{\phantom{\dagger}} + {\rm h.c.}) - \mu \sum_{i=1}^{L} n_{i},
\end{equation}
where we consider open boundary conditions, $\mu$ is the chemical potential, $t$ is the hopping amplitude, and $\Delta$ is the induced superconducting gap. In the following, we set $\Delta=-t$. The ground state of this model features a quantum phase transition from a topologically trivial ($|\mu/t|>2$) to a non-trivial state ($|\mu/t|<2$), see Fig.~\ref{fig:kitaev}(a). In the topological phase, Majorana zero modes~\cite{wilczek:2009} are present. Here we restrict ourselves to $\mu/t\leq 0$. We compute the ground states through exact diagonalization. For results based on individual measurement outcomes (of projective measurements in the Fock basis), see Appendix~\ref{app_A5}.\\

\indent The topologically trivial and non-trivial phase can be distinguished through entanglement spectra and the corresponding entanglement entropy~\cite{amico:2008}. Consider the reduced density matrix $\rho_{\rm A}$ of a system in the pure state $| \Psi \rangle$ obtained by subdividing the Hilbert space $\mathcal{H}$ into two parts, A and B, and tracing out the degrees of freedom of B
\begin{equation}\label{eq:reduced_dm}
\rho_{\rm A} = {\rm Tr}_{\rm B} | \Psi \rangle \!\langle \Psi |,
\end{equation}
with $\{\lambda_{i} \}$ the spectrum of $\rho_{\rm A}$ and $\{-\ln(\lambda_{i}) \}$ the entanglement spectrum. Here, we consider the bipartition of the chain into left and right halves with $L_{\rm A}=L_{\rm B}=L/2$. The entanglement entropy can then be computed as
\begin{equation}\label{eq:entanglement_entropy}
    S_{\rm ent}(\rho_{\rm A}) = -\sum_{i} \lambda_{i} \ln(\lambda_{i}).
\end{equation}
The three largest eigenvalues of $\rho_{A}$ are shown in Fig.~\ref{fig:kitaev}(g) and the resulting entanglement entropy is shown in Fig.~\ref{fig:kitaev}(h). Both the spectrum and entanglement entropy exhibit the largest change close to the critical value $\mu_{\rm c}/t=- 2$. The entanglement entropy approaches zero deep within the topologically trivial phase, signalling that the two halves of the ground state of the chain are not entangled. In the topological phase, the entanglement entropy approaches a value of $\ln(2)$ characteristic of an entangled ground state.\\

\indent Figure~\ref{fig:kitaev} shows the results of SL, LBC, and PBM. The location of the local maxima of the optimal indicators based on all three methods converges to the critical value of $\mu_{\rm c}/t=- 2$ with increasing chain length. Considering the probability distributions governing the input data [see Fig.~\ref{fig:kitaev}(f)], we observe that almost all basis states become occupied with non-negligible probability as the tuning parameter $\mu/t$ is tuned across its critical value. Note that in Ref.~\cite{van:2017}, the phase transition in the Kitaev model was successfully revealed using LBC with NNs where the entanglement spectrum of the ground state served as an input. The scaling behavior of the estimated critical value of the tuning parameter based on the optimal indicators of SL, LBC, and PBM is comparable to standard physical indicators, such as the eigenvalues of the reduced density matrix or the entanglement entropy [see Fig.~\ref{fig:kitaev}(e)]. In the limit $\mu/t \rightarrow -\infty$, the ground state of the Kitaev chain corresponds to the Fock state with each site being occupied. Thus, in the limit $\mu_{1}/t \rightarrow -\infty$, the optimal predictions of SL follow Eq.~\eqref{eq_SL_boltzmann}, where the aforementioned Fock state takes the role of the ground state energy.
\subsection{Bose-Hubbard model}\label{sec_applications_MBL_BH}
\begin{figure*}[tbh!]
\centering
\includegraphics[width=\linewidth]{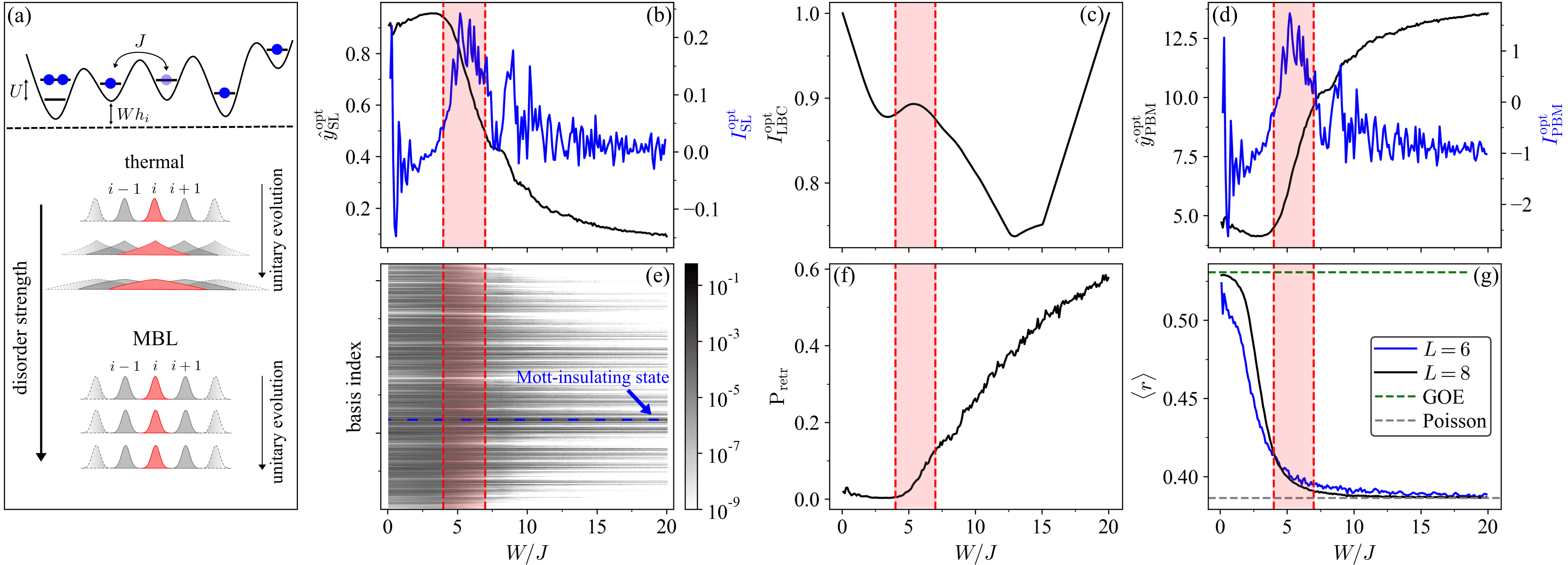}
\caption{Results for the MBL phase transition in the 1D Bose-Hubbard model ($L=8$) with the dimensionless disorder strength as a tuning parameter $p=W/J$ ranging from $p_{1}=0.1$ to $p_{K}=20$ in steps of $\Delta p = 0.1$. Here, $1.1 \times 10^{3}$ different disorder realizations were considered. In SL, the data obtained at $p_{1}$ and $p_{K}$ constitutes our training set, i.e., $r_{\rm I}=1$ and $l_{\rm II}=K$. The reference range for the critical value of the tuning parameter $W_{\rm c}/J\approx 4-7$~\cite{rispoli:2019,bohrdt:2021} at which the phase transition between the thermalizing and MBL phase occurs is highlighted in red. (a) Illustration of the 1D Bose-Hubbard model [Eq.~\eqref{eq:BH_MBL_hamiltonian}] (top) and the MBL phase transition (bottom), where the system is initialized in a Mott-insulating state. (b) Mean optimal prediction $\hat{y}_{\rm SL}^{\rm opt}$ in SL (black) and the corresponding indicator $I_{\rm SL}^{\rm opt}$ (blue). (c) Optimal indicator of LBC $I_{\rm LBC}^{\rm opt}$ (black). (d) Mean optimal prediction $\hat{y}_{\rm PBM}^{\rm opt}$ in PBM (black) and the corresponding indicator $I_{\rm PBM}^{\rm opt}$ (blue). (e) Probability distributions governing the input data (indices of Fock basis states with $N_{b}=8$ particles) as a function of tuning parameter, where the color scale denotes the probability. The color scale is cut off at $10^{-9}$ to improve visual clarity. The blue-dashed line highlights the initial Mott-insulating state. (f) Disorder-average retrieval probability $P_{\rm retr}$ as a function of the tuning parameter corresponding to the line-cut is marked in panel (e). (g) Average ratio of consecutive level spacings $\langle r \rangle$ for a chain of length $L=6$ (blue) and $L=8$ (black) with reference values $r_{\rm GOE} = 0.5307$ (green, dashed) and $r_{\rm Poisson} = 2 \ln(2) - 1  \approx 0.3863$ (grey, dashed). We consider all eigenstates located in the middle one-third of the spectrum~\cite{pal:2010,rispoli:2019} restricted to subspace with $N_{b}=L$ particles and additionally average over multiple disorder realizations ($1\times 10^{4}$ for $L=6$ and $1.1 \times 10^{3}$ for $L=8$).}
\label{fig:MBL_BH_quench}
\end{figure*}

Finally, we consider the many-body localization (MBL) phase transition in the 1D Bose-Hubbard model (with open boundary conditions) following Refs.~\cite{lukin:2019,rispoli:2019,bohrdt:2021}. The system is described by the Hamiltonian
\begin{equation}\label{eq:BH_MBL_hamiltonian}
    H = -J \sum_{i=1}^{L-1} (b_{i+1}^{\dagger}b_{i}^{\phantom{\dagger}}+{\rm h.c.}) + \sum_{i=1}^{L} \frac{U}{2}n_{i}^{\phantom{\dagger}}(n_{i}^{\phantom{\dagger}}-1) + W h_{i} n_{i}^{\phantom{\dagger}},
\end{equation}
where $J$ is the hopping strength and $U$ is the on-site interaction strength [see top panel in Fig.~\ref{fig:MBL_BH_quench}(a)]. Here, we fix $U/J=2.9$. The last term in Eq.~\eqref{eq:BH_MBL_hamiltonian} corresponds to a quasiperiodic potential $h_{i}=\cos(2\pi \beta i + \phi)$ mimicking on-site disorder with amplitude $W$, where we fix $1/\beta  = 1.618$. This system transitions to the MBL phase, where thermalization breaks down as the disorder strength is increased beyond a critical value $W_{\rm c}/J$, see bottom panel in Fig.~\ref{fig:MBL_BH_quench}(a). We analyze the system in the long-time limit $tJ=100$ after unitary time-evolution starting from a Mott-insulating state with one particle per site by solving the Schrödinger equation numerically. We average over different disorder realizations obtained by sampling the phase $\phi \in [0,2\pi)$ of the potential uniformly.\\

\indent A popular way to differentiate between the thermalizing and MBL regimes relies on the study of spectral statistics using tools from random matrix theory~\cite{pal:2010,khemani:2017,alet:2018}. In the thermal regime, the statistical distribution of level spacings is given by a Gaussian orthogonal ensemble (GOE), while a Poisson distribution is expected for localized states. The ratio of consecutive level spacings is
\begin{equation}
r_{i}=\frac{\min(\delta_{i},\delta_{i+1})}{\max(\delta_{i},\delta_{i+1})},
\end{equation}
with $\delta_{i} = E_{i} - E_{i-1}$ at a given eigenenergy $E_{i}$. Averaging over the spectrum and multiple disorder realizations yields $\langle r \rangle$, which varies from $r_{\rm GOE} = 0.5307$ within the thermalizing phase to $r_{\rm Poisson} = 2 \ln(2) - 1  \approx 0.3863$ within the MBL phase, see Fig.~\ref{fig:MBL_BH_quench}(g).\\

\indent The results are shown in Fig.~\ref{fig:MBL_BH_quench}. All three methods correctly identify the MBL phase boundary, where we take $W_{\rm c}/J\approx 4-7$ from Refs.~\cite{rispoli:2019,bohrdt:2021} as a reference. This is in agreement with the spectral analysis: the crossover between the average ratio of consecutive level spacings for systems of size $L=6$ and $L=8$ is located at $W_{\rm c}/J\approx 4$, see Fig.~\ref{fig:MBL_BH_quench}(g). Moreover, the phase boundary marks the range of the tuning parameter in which the most significant change in the underlying probability distribution occurs [see Fig.~\ref{fig:MBL_BH_quench}(e)]. A line-cut along the index corresponding to the initial Mott-insulating state is shown in Fig.~\ref{fig:MBL_BH_quench}(f). It corresponds to the disorder-averaged probability of retrieving the initial state after unitary time evolution. The MBL phase boundary is marked by the sudden increase in ${\rm P}_{\rm retr}$~\cite{lukin:2019} which is correctly picked up by SL, LBC, and PBM.\\ 

\indent Our results are also in agreement with Ref.~\cite{bohrdt:2021}, which examined the MBL phase transition within the same model using SL, PBM, and LBC with NNs on numerical and experimental data. As such, this example highlights the possibility of calculating optimal indicators directly from experimental data. Note that in Ref.~\cite{bohrdt:2021}, the authors attempted to construct a simplified indicator for phase transitions when using LBC by subtracting the V-shaped indicator signal in the case of indistinguishable data (see case 1 in Sec.~\ref{sec_application_trivial_distr}) as a baseline. However, we find that this procedure biases the peak of the optimal indicator signal of LBC towards the center of the parameter range under consideration and is thus not a viable procedure, see Appendix~\ref{app_D3}.
\section{Discussion}\label{sec_discussion}
In the previous section, we have demonstrated that the optimal indicators of SL, LBC, and PBM successfully detect phase transitions and crossovers in a variety of different classical and quantum systems based on numerical data. Recall that the optimal analytical predictors correspond to an optimal model that reaches the global minimum of the loss function. \textit{A priori}, it is unclear if the optimal predictors can be recovered in practice when training NNs, because the employed NNs are of finite size and local optimization techniques are used. In Appendix~\ref{app_B}, we demonstrate that the optimal predictions and indicators of all six systems studied in Sec.~\ref{sec_physical_models} can be recovered by training NNs. This reachability further underpins the practical relevance of our analysis for the case when using SL, LBC, and PBM with NNs.\\

\indent In a traditional NN-based approach, one searches for the optimal model by iteratively updating the parameters of an NN in order to minimize a loss function [see step $2)$ in Fig.~\ref{fig:workflow}]. In contrast, our numerical routine based on the derived analytical expressions allows for the optimal model to be constructed directly from data [see step $2^{*})$ in Fig.~\ref{fig:workflow}]. As such, evaluating the analytical predictors also compares favorably to the NN-based approach in terms of computation time. For each of the three methods and across all six studied physical systems, we find that the time needed to train an NN of \textit{minimal size} (one hidden layer with a single node) for a \textit{single epoch} is of the same order of magnitude as the time needed to compute the optimal predictions, optimal indicator, and optimal loss (see Tab.~\ref{fig_timing}). Therefore, the computation time associated with constructing and evaluating an optimal model is \textit{at worst} comparable with training and evaluating an NN-based model. In practice, however, the latter approach typically requires significantly more computation time because larger NNs need to be used, the training takes many epochs, and hyperparameters need to be adjusted (see Appendix~\ref{app_B} for a detailed discussion). In particular, as the system size increases and the associated state space grows, converging to the global minimum of the loss function can become increasingly difficult. The convergence of the optimal model, on the other hand, is guaranteed \textit{by construction}.\\

\indent We have observed that the optimal indicator of a given method may fail to correctly highlight a phase transition. A failure can, for example, occur if only a limited amount of data is available and finite-sample statistics dominate. In this case, while the ground-truth probability distributions underlying the data show a significant overlap resulting in a peak in the indicator signal, the inferred probability distributions do not (see Appendix~\ref{app_A5} for a concrete example). However, even if the data set is sufficiently large, i.e., the ground-truth probability distributions are well approximated, the optimal model can fail (see classical systems in Sec.~\ref{sec_physical_models} for examples). Both instances of failure can often be resolved by employing non-optimal models. Such a model can be realized by an NN whose capacity, i.e., its ability to fit a wide variety of functions~\cite{goodfellow:2016,hu:2021}, is restricted. This can be achieved, e.g., by reducing the NN size, performing early stopping, or the explicit addition of $\ell_{2}$ regularization (see Appendix~\ref{app_B1} and~\ref{app_B2}). In these instances, other phase-classification methods which are inherently based on the similarity of input data~\cite{wang:2016, rodriguez:2019, kottmann:2020, arnold:2021} are also expected to provide valuable insights. These methods stand in contrast to the optimal predictors of SL, LBC, and PBM, which are not explicitly based on learning order parameters, i.e., recognizing prevalent patterns or orderings. Instead, the optimal predictors gauge changes in the probability distributions governing the data. Contrary to popular opinion, the failure of optimal models, or equivalently high-capacity NNs, does not always correspond to overfitting in the traditional sense~\cite{goodfellow:2016}: the gap between training and test loss vanishes in the limit of a sufficiently large data set (which is available for the examples discussed in Sec.~\ref{sec_physical_models}). Therefore, sub-optimal models, such as NNs with insufficient capacity, are in fact \textit{underfitting} the data. This signals a fundamental mismatch between the classification or regression task underlying a particular ML method, i.e., the corresponding loss function, and the goal of detecting phase transitions. In particular, it raises the intriguing question of whether one can adjust the learning task in SL, PBM, and LBC such that the corresponding optimal models also correctly highlight the phase transition in these problematic cases, e.g., through an appropriate modification of the underlying loss functions or by enforcing explicit constraints.
\section{Conclusion and outlook}\label{sec_concl}
\indent The ML methods for detecting phase transitions from data given by SL, LBC, and PBM can be viewed under a unifying light: all three approaches have predictive models, such as NNs, at their heart which are trained to solve a given classification or regression task. Analyzing their predictions allows us to compute a scalar indicator that highlights phase boundaries. The power and success of these methods is largely attributed to the universal function approximation capabilities of their underlying NNs, which are often sacrificed in practice to regain interpretability~\cite{ponte:2017,wetzel2:2017,zhang:2019,greitemann:2019,greitemann_2:2019,liu:2019,zhang:2020}. Here, we took an alternative approach to cope with the interpretability-expressivity tradeoff: By analyzing the class of predictive models that solve the classification and regression tasks underlying SL, LBC, and PBM optimally, we have derived analytical expressions for the indicators of phase transitions of these three methods.\\

\indent Our work establishes a solid theoretical foundation for SL, LBC, and PBM, based on which we were able to explain and understand the results of a variety of previous studies~\cite{carrasquilla:2017,van:2017,beach:2018,schaefer:2019,greplova:2020,bohrdt:2021}. We anticipate that similar analyses will be useful to gain an understanding of other methods for identifying phase transitions with NNs~\cite{huembeli:2018,huembeli:2019,kottmann:2020,kottmann:2021,guo:2022,patel:2022} and other classification tasks in condensed matter  physics~\cite{bohrdt:2019,zhang2:2019,pilati:2019,ghosh:2020,miles:2021,szoldra:2021,gavreev:2022}. In these cases, the optimal models can also serve as benchmark solutions that enable future studies aimed at investigate the learning process of NNs and improving their design and update routines~\cite{mcclean:2018,vieijra:2020,miles:2021,bukov:2021,miles2:2021,valenti:2022}. For example, in Refs.~\cite{carrasquilla:2017,chng:2017,broecker:2017} it was shown that an NN trained to predict the phase transition in a given model using SL can successfully classify configurations generated from an entirely different Hamiltonian. An exciting prospect is to explore whether the success of this ``transfer learning'' can be rigorously explained based on our results.\\

\indent The analytical expressions not only enable our understanding of the phase-classification methods under consideration -- they also allow for the direct computation of their optimal predictions and indicators based on the input data \textit{without} explicitly training NNs. We have demonstrated that this novel procedure can successfully reveal a broad range of different phase transitions in a numerical setting and is favorable in terms of computation time. Our results suggest a variety of avenues for further explorations. As a next step, one can consider whether tools from ML, especially for density estimation~\cite{bishop:2006,wu:2019,melko:2019,nicoli:2021}, can aid in the computation of the optimal indicators. In the quantum case, classical representations of quantum states obtained via classical shadow tomography~\cite{huang:2020,huang:2021} may help to evade the arising exponential complexity. We believe that optimal predictors will be a valuable tool to detect, interpret, and characterize phases of matter and their transitions from experimental data, particularly in the advent of digital quantum computers~\cite{smith:2019,barratt:2021,satzinger:2021,herrmann:2021,noel:2022} and programmable quantum simulators~\cite{bohrdt:2021,miles2:2021,ebadi:2021,semeghini:2021,scholl:2021,altman:2021}.\\

\indent The code for computing the optimal predictions and indicators of SL, LBC, and PBM utilized in this work is open source~\cite{github}.

\section*{Acknowledgments}
We would like to thank Niels Lörch, Andreas Trabesinger, and Christoph Bruder for helpful suggestions on the manuscript. We thank Niels Lörch, Eliska Greplova, Eugene Demler, Florian Marquardt, and Christoph Bruder for stimulating discussions. We acknowledge financial support from the NCCR QSIT funded by the Swiss National Science Foundation (Grant No. 51NF40-185902). Computation time at sciCORE (scicore.unibas.ch) scientific computing core facility at the University of Basel is gratefully acknowledged. This material is based upon work supported by the National Science Foundation under grant no. OAC-1835443, grant no. SII-2029670, grant no. ECCS-2029670, grant no. OAC-2103804, and grant no. PHY-2021825. We also gratefully acknowledge the U.S. Agency for International Development through Penn State for grant no. S002283-USAID. The information, data, or work presented herein was funded in part by the Advanced Research Projects Agency-Energy (ARPA-E), U.S. Department of Energy, under Award Number DE-AR0001211 and DE-AR0001222. We also gratefully acknowledge the U.S. Agency for International Development through Penn State for grant no. S002283-USAID. The views and opinions of authors expressed herein do not necessarily state or reflect those of the United States Government or any agency thereof. This material was supported by The Research Council of Norway and Equinor ASA through Research Council project "308817 - Digital wells for optimal production and drainage". Research was sponsored by the United States Air Force Research Laboratory and the United States Air Force Artificial Intelligence Accelerator and was accomplished under Cooperative Agreement Number FA8750-19-2-1000. The views and conclusions contained in this document are those of the authors and should not be interpreted as representing the official policies, either expressed or implied, of the United States Air Force or the U.S. Government. The U.S. Government is authorized to reproduce and distribute reprints for Government purposes notwithstanding any copyright notation herein.
\appendix 

\section{Optimal predictions and indicators}\label{app_A}
In this appendix, we provide detailed derivations of the optimal predictions and indicators of SL, LBC, and PBM. In particular, we discuss the assumptions underlying the derivation of the optimal predictions of SL and how the analytical predictors are evaluated in practice. This includes an analysis of the computational cost associated with constructing and evaluating the optimal models and the role of finite-sample statistics.
\subsection{Derivation of optimal predictions and indicators}\label{app_A1}
Here, we derive the form of the optimal predictions and indicators of phase transitions for SL, LBC, and PBM presented in Sec.~\ref{sec_methods} of the main text.\\

\indent \textit{Supervised learning}.---In SL, a predictive model $m$ is trained to minimize the CE loss function given in Eq.~\eqref{eq:CE_SL}. Now, consider a particular input contained within the training set $\tilde{\bm{x}} \in \bar{\mathcal{T}}$. We can determine the optimal model prediction $\hat{y}_{\rm SL}^{\rm opt}(\tilde{\bm{x}})$ for this particular input by minimizing the loss function in Eq.~\eqref{eq:CE_SL} with respect to $\hat{y}(\tilde{\bm{x}})$, i.e., by solving the necessary condition
\begin{equation}\label{eq:app_SL_derivation_0}
\frac{\partial \mathcal{L}_{\rm SL}}{\partial \hat{y}(\tilde{\bm{x}})} = -\frac{1}{M_{\mathcal{T}}} \sum_{\tilde{\bm{x}} \in \mathcal{T}} \left( \frac{y(\tilde{\bm{x}})}{\hat{y}(\tilde{\bm{x}})} - \frac{1-y(\tilde{\bm{x}})}{1-\hat{y}(\tilde{\bm{x}})} \right)=0.
\end{equation}
Using the explicit expressions for the labels ($y=1$ and $y=0$ for all inputs drawn in region I and II, respectively) in Eq.~\eqref{eq:app_SL_derivation_0}, we have
\begin{equation}\label{eq:app_SL_derivation_1}
    \frac{\sum_{k=1}^{r_{\rm I}} M_{k}(\tilde{\bm{x}})}{\sum_{k=l_{\rm II}}^{K} M_{k}(\tilde{\bm{x}})} = \frac{M_{\rm I}(\tilde{\bm{x}})}{M_{\rm II}(\tilde{\bm{x}})} = \frac{\hat{y}(\tilde{\bm{x}})}{1-\hat{y}(\tilde{\bm{x}})}.
\end{equation}
Here, $M_{\rm I/II}(\tilde{\bm{x}})$ denotes the number of times the input $\tilde{\bm{x}}$ is found in region I or II, respectively. In SL, the predictive model must, by definition, satisfy $\hat{y}(\bm{x})\in [0,1] \; \forall \bm{x}$. Thus, Eq.~\eqref{eq:app_SL_derivation_1} is satisfied given predictions of the form
\begin{equation}\label{eq:app_SL_derivation_2}
    \hat{y}_{\rm SL}^{\rm opt}(\tilde{\bm{x}}) = \frac{M_{\rm I}(\tilde{\bm{x}})}{M_{\rm I}(\tilde{\bm{x}}) + M_{\rm II}(\tilde{\bm{x}})}.
\end{equation}
The opposite choice of labeling ($y=0$ and $y=1$ for all inputs drawn in region I and II, respectively) is equally valid and would result in
\begin{equation}
    \hat{y}_{\rm SL}^{\rm opt}(\tilde{\bm{x}}) = \frac{M_{\rm II}(\tilde{\bm{x}})}{M_{\rm I}(\tilde{\bm{x}}) + M_{\rm II}(\tilde{\bm{x}})}.
\end{equation}
That is, the role of $\hat{y}_{\rm SL}^{\rm opt}(\tilde{\bm{x}})$ and $1-\hat{y}_{\rm SL}^{\rm opt}(\tilde{\bm{x}})$ are swapped. In this work, we stick to the former choice [Eq.~\eqref{eq:app_SL_derivation_2}]. The optimality of the predictions in Eq.~\eqref{eq:app_SL_derivation_2} can be confirmed by calculating the second derivative of the loss function
\begin{equation}\label{eq_SL_optimality_condition}
    \frac{\partial^2 \mathcal{L}_{\rm SL}}{\partial \hat{y}(\tilde{\bm{x}})^2} = \frac{M_{\rm I}}{M_{\mathcal{T}}} \frac{1}{\hat{y}(\tilde{\bm{x}})^2} + \frac{M_{\rm II}}{M_{\mathcal{T}}} \frac{1}{  \left(  1-\hat{y}(\tilde{\bm{x}}) \right)^2 } >0.
\end{equation}

\indent The probability distribution governing the input data is denoted as ${\rm P}_{k}(\tilde{\bm{x}})\approx M_{k}(\tilde{\bm{x}}) / M$ ($1 \leq k \leq K$). This allows for Eq.~\eqref{eq:app_SL_derivation_2} to be expressed as
\begin{equation}
    \hat{y}_{\rm SL}^{\rm opt}(\tilde{\bm{x}}) = \frac{{\rm P}_{{\rm I}}(\tilde{\bm{x}})}{{\rm P}_{{\rm I}}(\tilde{\bm{x}}) + {\rm P}_{{\rm II}}(\tilde{\bm{x}})},
\end{equation}
where
\begin{equation}
    {\rm P}_{{\rm I}}(\tilde{\bm{x}}) = \sum_{k= 1}^{r_{\rm I}} {\rm P}_{k}(\tilde{\bm{x}})
\end{equation}
and 
\begin{equation}
    {\rm P}_{{\rm II}}(\tilde{\bm{x}}) = \sum_{k = l_{\rm II}}^{K} {\rm P}_{k}(\tilde{\bm{x}})
\end{equation}
are the (unnormalized) probabilities of drawing the input $\tilde{\bm{x}}$ in region I and II, respectively. Repeating the above procedure for all inputs within the training set $\bar{\mathcal{T}}$, we obtain
\begin{equation}\label{eq:app_SL_derivation_3}
    \hat{y}_{\rm SL}^{\rm opt}(\bm{x}) = \frac{{\rm P}_{\rm I}(\bm {x})}{{\rm P}_{\rm I}(\bm {x}) + {\rm P}_{\rm II}(\bm {x})}\; \forall \bm{x} \in \bar{\mathcal{T}},
\end{equation}
which matches Eq.~\eqref{eq:SL_derivation_1} reported in the main text. Relaxations of the assumption in SL that there are only two distinct phases to be distinguished will be discussed in Appendix~\ref{app_A2}.\\

\indent Note that the same optimal predictions are obtained when training on a MSE loss function
\begin{equation}\label{eq_mse}
    \mathcal{L}_{\rm MSE} = \frac{1}{M_{\mathcal{T}}} \sum_{\bm{x} \in \mathcal{T}} \left(\hat{y}(\bm{x})  - y(\bm{x}) \right)^2,
\end{equation}
instead of a CE loss function. Again, consider a particular input $\tilde{\bm{x}}$ contained within the training set $\bar{\mathcal{T}}$. We can determine the optimal model prediction $\hat{y}_{\rm SL}^{\rm opt}(\tilde{\bm{x}})$ for this input by minimizing the loss function in Eq.~\eqref{eq_mse} with respect to $\hat{y}(\tilde{\bm{x}})$, i.e., by solving
\begin{equation}\label{eq_to_solve}
\frac{\partial \mathcal{L}_{\rm MSE}}{\partial \hat{y}(\tilde{\bm{x}})} = \frac{2}{M_{\mathcal{T}}} \sum_{\tilde{\bm{x}} \in \mathcal{T}} \left(\hat{y}(\tilde{\bm{x}})- y(\tilde{\bm{x}})  \right)=0.
\end{equation}
Plugging the expression for the labels given by a one-hot-encoding in Eq.~\eqref{eq_to_solve}, we have
\begin{equation}\label{eq_xx}
    M_{\rm I}(\tilde{\bm{x}})(1-\hat{y}(\tilde{\bm{x}})) - M_{\rm II}(\tilde{\bm{x}})\hat{y}(\tilde{\bm{x}}) = 0.
\end{equation}
This coincides with the condition for the predictions given in Eq.~\eqref{eq:app_SL_derivation_1} obtained from a CE loss function. Their optimality can be confirmed via
\begin{equation}
    \frac{\partial^2 \mathcal{L}_{\rm MSE}}{\partial \hat{y}(\tilde{\bm{x}})^2} = \frac{2 (M_{\rm I} + M_{\rm II})}{M_{\mathcal{T}}} >0.
\end{equation}
Therefore, in SL, the optimal predictions and indicators associated with optimal models trained on a CE or MSE loss function are identical.\\

\indent \textit{Learning by confusion}.---To reveal the phase transition by means of LBC, we perform several splits of the parameter range into two neighboring regions labeled I and II. For a fixed bipartition, we minimize a CE [Eq.~\eqref{eq:CE_LBC}] or MSE loss function
\begin{equation}\label{eq_mse_LBC}
    \mathcal{L}_{\rm MSE} = \frac{1}{M_{\mathcal{X}}} \sum_{\bm{x} \in \mathcal{X}} \left( \hat{y}(\bm{x})  - y(\bm{x}) \right)^2.
\end{equation}
Following the analysis of SL presented above, we obtain a similar expression for the optimal predictions
\begin{equation}\label{eq:LBC_error_optimal_app}
    \hat{y}_{\rm LBC}^{\rm opt}(\bm{x}) = \frac{{\rm P}_{\rm I}(\bm{x})}{{\rm P}_{\rm I}(\bm{x}) + {\rm P}_{\rm II}(\bm {x})}\; \forall \bm{x} \in \mathcal{X},
\end{equation}
with $\mathcal{T} = \mathcal{X}$ in LBC. Thus, we recover Eq.~\eqref{eq:LBS_derivation_1} of the main text. Their optimality can be confirmed via
\begin{equation}
    \frac{\partial^2 \mathcal{L}_{\rm LBC}}{\partial \hat{y}(\tilde{\bm{x}})^2} = \frac{M_{\rm I}}{M_{\mathcal{X}}} \frac{1}{\hat{y}(\tilde{\bm{x}})^2} + \frac{M_{\rm II}}{M_{\mathcal{X}}} \frac{1}{  \left(  1-\hat{y}(\tilde{\bm{x}}) \right)^2 } >0
\end{equation}
or
\begin{equation}
    \frac{\partial^2 \mathcal{L}_{\rm MSE}}{\partial \hat{y}(\tilde{\bm{x}})^2} = \frac{2 (M_{\rm I} + M_{\rm II})}{M_{\mathcal{X}}} >0,
\end{equation}
in the case of a CE or MSE loss, respectively. The value of the indicator in LBC for a given bipartition corresponds to the mean classification accuracy [Eq.~\eqref{eq:LBC_error}], where the continuous predictions $\hat{y}(\bm{x})\in [0,1]$ are mapped to binary labels via $\theta \left(\hat{y}(\bm{x})-0.5\right)$. Using the optimal prediction in Eq.~\eqref{eq:LBC_error_optimal_app}, the mean classification error for a given input $\bm{x}$ is ${\min}\{ \hat{y}_{\rm LBC}^{\rm opt}(\bm{x}), 1- \hat{y}_{\rm LBC}^{\rm opt}(\bm{x})\}$. Weighting the contribution of each input $\bm{x}$ to the mean classification error by its probability ${\rm P}_{k}(\bm{x})$, we arrive at Eq.~\eqref{eq:LBS_derivation_2} of the main text. Note that, in principle, the assumption in LBC that there are only two phases to be distinguished can be relaxed~\cite{van:2017}. In this case, the optimal indicator may show multiple distinct peaks highlighting the different phase boundaries~\cite{lee:2019}.\\

\indent \textit{Prediction-based method}.---In PBM, a predictive model $m: \bm{x} \rightarrow \hat{y}(\bm{x})$ is trained to minimize the MSE loss function $\mathcal{L}_{\rm PBM}$ specified in Eq.~\eqref{eq:MSE_PBM}. Consider a particular input $\tilde{\bm{x}} \in \bar{\mathcal{X}}$. We can determine the optimal model prediction $\hat{y}_{\rm PBM}^{\rm opt}(\tilde{\bm{x}})$ for this input by minimizing the loss function in Eq.~\eqref{eq:MSE_PBM} with respect to $\hat{y}(\tilde{\bm{x}})$, i.e., by solving
\begin{equation}\label{eq:PBM_derivation_app_1}
\frac{\partial \mathcal{L}_{\rm PBM}}{\partial \hat{y}(\tilde{\bm{x}})} = \frac{2}{KM} \sum_{k=1}^{K} M_{k}(\tilde{\bm{x}}) \left(\hat{y}(\tilde{\bm{x}}) - p_{k}  \right) = 0.
\end{equation}
Solving Eq.~\eqref{eq:PBM_derivation_app_1} yields
\begin{equation}
\hat{y}_{\rm PBM}^{\rm opt}(\tilde{\bm{x}}) = \frac{\sum_{k=1}^{K} {\rm P}_{k}( \tilde{\bm{x}}) p_{k}}{ \sum_{k=1}^{K} {\rm P}_{k}(\tilde{\bm{x}} )}.
\end{equation}
This prediction is indeed optimal, as
\begin{equation}
    \frac{\partial^2 \mathcal{L}_{\rm PBM}}{\partial \hat{y}(\tilde{\bm{x}})^2} = \frac{2}{K} \sum_{k=1}^{K} {\rm P}_{k}(\tilde{\bm{x}})>0.
\end{equation}
Repeating this procedure for all available inputs $\bm{x} \in \bar{\mathcal{X}}$ yields
\begin{equation}
\hat{y}_{\rm PBM}^{\rm opt}(\bm{x}) = \frac{\sum_{k=1}^{K} {\rm P}_{k}( \bm{x}) p_{k}}{ \sum_{k=1}^{K} {\rm P}_{k}(\bm{x} )}\; \forall \bm{x} \in \bar{\mathcal{X}}.
\end{equation}
Thereby, we recover Eq.~\eqref{eq:PBM_derivation_1} of the main text. Note that this derivation can be generalized to higher dimensional parameter spaces (which may host multiple distinct phases) in a straightforward manner (see Ref.~\cite{arnold:2021}), resulting in 
\begin{equation}
    \hat{\bm{y}}_{\rm PBM}^{\rm opt}(\bm{x}) = \frac{\sum_{k} {\rm P}_{k}( \bm{x}) \bm{p}_{k}}{ \sum_{k} {\rm P}_{k}(\bm{x} )}.
\end{equation}
Here, the sum runs over all sampled points $\bm{p}_{k}$ in parameter space. The optimal indicator is then given as a divergence
\begin{equation}
    I_{\rm PBM}^{\rm opt}(\bm{p}) = \nabla_{\bm{p}} \bm{\delta y}^{\rm opt}_{\rm PBM}(\bm{p}),
\end{equation}
where $\bm{\delta y}^{\rm opt}_{\rm PBM}(\bm{p}_{k}) = \sum_{\bm{x} \in \bar{\mathcal{X}}} {\rm P}_{k}(\bm{x}) \hat{\bm{y}}_{\rm PBM}^{\rm opt}(\bm{x}) - \bm{p}_{k}$.

\subsection{Assumptions for supervised learning}\label{app_A2}
\indent Let us we review the assumption of $\bar{\mathcal{X}} = \bar{\mathcal{T}}$ underlying the derivation for the optimal predictions and corresponding indicator of SL. In general, if $\bar{\mathcal{X}} \neq \bar{\mathcal{T}}$ the optimal predictions of SL can be expressed
\begin{equation}\label{eq:assumption_1}
    \hat{y}_{\rm SL}^{\rm opt'}(p_{k}) = \sum_{\bm{x} \in \bar{\mathcal{T}}} {\rm P}_{k}(\bm{x})  \hat{y}_{\rm SL}^{\rm opt}(\bm{x}) + \sum_{\bm{x} \notin \bar{\mathcal{T}}} {\rm P}_{k}(\bm{x})  \hat{y}_{\rm SL}(\bm{x}).
\end{equation}
The first contribution in Eq.~\eqref{eq:assumption_1} comes from predictions for inputs contained in the training data, which are determined through minimization of the corresponding loss function [see Eq.~\eqref{eq:app_SL_derivation_0}]. The second contribution comes from predictions for inputs not contained in the training data, which are \textit{a priori} only restricted to the unit interval $\hat{y}_{\rm SL}(\bm{x}) \in [0,1]$. Therefore, this contribution to Eq.~\eqref{eq:assumption_1} is bounded by the probability of drawing an input at $p_{k}$ that is not present in the training data, which is given by $ \sum_{\bm{x} \notin \bar{\mathcal{T}}} {\rm P}_{k}(\bm{x})$. When using SL with NNs, the predictions for inputs not contained in the training data [second contribution in Eq.~\eqref{eq:assumption_1}] will be most susceptible to noise inherent to NN training and hyperparameter choices. As such, its physical relevance is questionable. It may be possible to obtain better bounds for this second contribution when using SL with NNs, e.g., based on the theory of neural tangent kernels~\cite{jacot:2018}.\\

\indent Let us explicitly discuss the classical systems analyzed in this work, which are governed by Boltzmann distribution [Eqs.~\eqref{eq:boltzmann} and~\eqref{eq:boltzmann2}]. Because the probability of drawing a particular configuration sample (or energy) at any non-zero temperature is non-zero, the assumption of $\bar{\mathcal{X}} = \bar{\mathcal{T}}$ holds given a sufficient number of samples. When computing the optimal indicator of SL numerically, we work with a finite number of samples. Thus, it can happen an input is encountered which is not part of the training data $\bm{x} \not\in \bar{\mathcal{T}}$. In practice, we can verify \textit{on-the-fly} whether this is the case. If so, we set $y_{\rm SL}(\bm{x}) = 0$ in Eq.~\eqref{eq:assumption_1}. Thereby, we effectively ignore the contribution to the predictions of SL from inputs not present in the training data. Note that because these predictions correspond to inputs with low probability, they are also most susceptible to finite-sample statistics. This procedure is further justified by the fact that the optimal predictions $\hat{y}_{\rm SL}^{\rm opt}$ obtained in this manner track the ground-state probability with high accuracy [see Figs.~\ref{fig:ising}(b),~\ref{fig:IGT}(b), and~\ref{fig:XY}(b)]. That is, the optimal predictions closely match the expression in Eq.~\eqref{eq_SL_boltzmann} valid in the case where deviations due to finite-sample statistics vanish.\\

\indent In the quantum case, it is typically not straightforward to determine \textit{a priori} whether the assumption of $\bar{\mathcal{X}} = \bar{\mathcal{T}}$ is met for a given system and choice of basis. Here, when calculating the optimal predictions and indicators numerically, we use the same procedure as described for the classical case. In our study, we only find cases where $\bm{x} \not\in \bar{\mathcal{T}}$ for the XXZ model. The error resulting from neglecting the second contribution in Eq.~\eqref{eq:assumption_1} is marginal, as the probability of drawing such inputs across the parameter range is found to be small. Note that the optimal indicator of SL obtained in such a manner correctly reveals the quantum phase transition in the XXZ (see Fig~\ref{fig:XXZ}). In fact, the optimal predictions calculated via this procedure correspond to the probability of measuring the ferromagnetic ground state (see Sec.~\ref{sec_methods_xxz}). For the above reasons, we expect that the optimal predictions of SL are capable of revealing phase transitions even if $\bar{\mathcal{X}} \neq \bar{\mathcal{T}}$.\\

\indent A relevant scenario in which the assumption that $\bar{\mathcal{X}} = \bar{\mathcal{T}}$ is violated occurs when the system transitions between multiple phases as the tuning parameter is varied. Then, inputs drawn in the phases present in the middle of the sampled range of the tuning parameter may not be present in the two boundary phases. By dropping the second contribution in Eq.~\eqref{eq:assumption_1}, we may still faithfully detect the transition between the first and second phase. However, all subsequent phase boundaries will then likely be missed. In the future, it will be of interest to lift the assumption of $\bar{\mathcal{X}} = \bar{\mathcal{T}}$ underlying the optimal predictions through appropriate interpolation schemes~\cite{jacot:2018,greplova:2020,huang:2021}, which would allow for the generalization capabilities of SL to be explored.

\begin{table*}[thb!]
	\begin{tabular}{l|cccccc@{}}\hline\hline
		& Ising     & IGT & XY & XXZ & Kitaev & Bose-Hubbard \\\hline\hline
		$t_{\rm SL}^{\rm opt}$  & $0.0007 \pm 0.0002$ &  $0.00007 \pm 0.00002$   &  $0.00012 \pm 0.00003$  &  $0.0049 \pm 0.0009$ &   $0.17 \pm 0.02$  &    $0.0044 \pm 0.0009$          \\\hline
		$t_{\rm SL}^{\rm NN}$  & $0.00060 \pm 0.00005$ &  $0.00030 \pm 0.00002$   &  $0.00048 \pm 0.00003$  &  $0.0060\pm 0.0009$ &   $0.14 \pm 0.02$  &    $0.0023 \pm 0.0003$          \\\hline
		$t_{\rm SL}^{\rm NN}/t_{\rm SL}^{\rm opt}$  & $0.9 \pm 0.3$ &  $4.9 \pm 1.3$   &  $4.0 \pm 0.8$  &  $1.2 \pm 0.3$ &   $0.9 \pm 0.2$  &    $0.5\pm 0.1$          \\\hline\hline
		$t_{\rm PBM}^{\rm opt}$ & $0.0016\pm 0.0004$          &  $0.00014 \pm 0.00006$   &  $0.00021\pm 0.00008$  & $0.019 \pm 0.003$  &   $0.42 \pm 0.05 $  &       $0.009 \pm 0.002$       \\\hline
		$t_{\rm PBM}^{\rm NN}$ & $0.0042\pm 0.0007$          &  $0.0005 \pm 0.0001$   &  $0.00084 \pm 0.00004$  & $0.080\pm 0.006$  &   $1.2 \pm 0.1$  &       $0.026 \pm 0.004$       \\\hline
		$t_{\rm PBM}^{\rm NN}/t_{\rm PBM}^{\rm opt}$ & $2.7\pm 0.8$          &  $4.0 \pm 2.1$   &  $4.0 \pm 1.5$  & $4.2 \pm 0.8$  &   $2.8 \pm 0.4$  &       $2.7 \pm 0.6$       \\\hline\hline
		$t_{\rm LBC}^{\rm opt}$ & $0.8\pm 0.1$          &   $0.042 \pm 0.001$  &  $0.041 \pm 0.004$  &   $3.7 \pm 0.4$ &  $32.0 \pm 1.7$  &  $1.4 \pm 0.2$\\\hline
		$t_{\rm LBC}^{\rm NN}$ & $1.11\pm 0.06$          &   $0.09 \pm 0.01$  &  $0.12 \pm 0.01$  &   $12.2 \pm 1.2$ &  $93.9 \pm 3.8$  &  $3.2 \pm 0.4$\\\hline
		$t_{\rm LBC}^{\rm NN}/t_{\rm LBC}^{\rm opt}$ & $1.3\pm 0.2$          &   $2.1 \pm 0.2$  &  $2.8 \pm 0.4$  &   $3.3 \pm 0.5$ &  $3.0 \pm 0.2$  &  $2.4 \pm 0.5$\\\hline\hline
		$t_{\rm PBM}^{\rm opt}$/$t_{\rm SL}^{\rm opt}$ & $2.3 \pm 0.9$          &   $2.0 \pm 1.1$  &  $1.8 \pm 0.7$  &   $3.8 \pm 1.0$ &  $2.7 \pm 0.5$ &  $2.1 \pm 0.6$\\\hline
		$t_{\rm LBC}^{\rm opt}$/$t_{\rm SL}^{\rm opt}$ & $1231 \pm 374$          &   $629 \pm 164$  &  $346 \pm 77$  &   $751 \pm 158$ &  $204 \pm 33$ &  $308 \pm 78$\\\hline\hline
		$L$ & 60   &   28  &  60  &   14 &  20  &  8\\\hline
		$M_{\bar{\mathcal{X}}}$ & 1711          &   353  &  1000  &   16384 &  524288 &  6435\\\hline
		$K$ & 200         &   100  &  100  &   201 &  101  &  200\\\hline
		\hline\hline
	\end{tabular}
	\caption{Measured computations times in seconds associated with constructing and evaluating optimal models, $t^{\rm opt}$, or training an NN of minimal size (one hidden layer with a single node) for a single epoch, $t^{\rm NN}$, for all three methods and six systems discussed in the main text (see Sec.~\ref{sec_physical_models}). The linear system size $L$, the corresponding number of unique samples $M_{\bar{\mathcal{X}}}$, as well as the number of sampled values of the tuning parameter $K$ for each system are also reported. The construction and evaluation of the optimal models yields the optimal predictions, optimal indicator, and optimal loss value. A training epoch is comprised of evaluating the NN at all $M_{\bar{\mathcal{X}}}$ unique samples, calculating the loss function, obtaining the gradient via backpropagation, and performing a single gradient step. For details on the NN architecture and training, see Appendix~\ref{app_B}. Note that in LBC, $t_{\rm LBC}^{\rm NN}$ corresponds to $K+1$ times the computation time of a training epoch for a single NN. All computation times were measured on a single CPU [Intel(R) Xeon(R) CPU E5-2630 v4 @ 2.20GHz] and garbage collection times were subtracted from the total runtime. To gather statistics, for each method and system computations were ran for 20 hours. If $10^5$ independent runs were completed in less than 20 hours, the computations were stopped prematurely. The error corresponds to the observed standard deviation.}
	\label{fig_timing}
\end{table*}

\subsection{Computational cost}\label{app_A3}
Here, we will derive the scaling of the computational cost with the number of unique inputs $M_{\bar{\mathcal{X}}}$ and the number of sampled tuning parameter values $N$ reported in Sec.~\ref{sec_comp_cost} of the main text. Note that we do not consider the overhead associated with computing the probability distributions $\{ {\rm P}_{k}(\bm{x})\}_{k=1}^{K}\; \forall \bm{x} \in \bar{\mathcal{X}}$ from the data at hand (or any other constant overhead). The computation of the optimal predictions and indicators can be approached in two ways: Either the optimal predictions for a given input $\hat{y}^{\rm opt}(\bm{x})$ are recomputed in each function call, or they are cached. We report the required number of floating-point operations in both instances, which can be counted based on the analytical expressions reported in Sec.~\ref{sec_optimal_indicators}. This counting represents a rough, hardware-independent estimate of the required computational cost. In the following, we will assume that the optimal indicators in SL and PBM are computed using a symmetric difference quotient, cf. Eq.~\eqref{eq:SL_indicator_symmetric_difference}.\\

\indent \textit{Supervised learning}.---The computation of $\hat{y}_{\rm SL}^{\rm opt}$ for all $\bm{x} \in \bar{\mathcal{X}}$ requires $M_{\bar{\mathcal{X}}} K_{\mathcal{T}}$ floating-point operations, where $K_{\mathcal{T}}=K_{\rm I} + K_{\rm II}$ is the number of sampled values of the tuning parameter in the training regions I and II. Caching these values, the number of operations required to compute the mean optimal prediction $\hat{y}_{\rm SL}^{\rm opt}$ for all $\{ p_{k} \}_{k=1}^{K}$ is $K(2M_{\bar{\mathcal{X}}}-1) + M_{\bar{\mathcal{X}}}  K_{\mathcal{T}}$. Thus, computing the optimal indicator requires $M_{\bar{\mathcal{X}}}(2K + K_{\mathcal{T}}) + K$ operations. Typically, in SL we have $K_{\mathcal{T}}\ll K$. Under this assumption, the computation of the mean optimal predictions and the optimal indicators each require $O(M_{\bar{\mathcal{X}}}K)$ operations. If the values $\hat{y}_{\rm SL}^{\rm opt}(\bm{x}) \; \forall \bm{x} \in \bar{\mathcal{X}}$ are not cached, computing the mean optimal prediction instead requires $K\left((2M_{\bar{\mathcal{X}}}-1) + M_{\bar{\mathcal{X}}} K_{\mathcal{T}}\right)$ operations. Computing the optimal indicator then requires $M_{\bar{\mathcal{X}}}K (2 + K_{\mathcal{T}}) + K$ operations. For both quantities, this still corresponds to $O(M_{\bar{\mathcal{X}}}K)$ operations.\\

\indent \textit{Learning by confusion}.---The computation of $\hat{y}_{\rm LBC}^{\rm opt}$ for all $\bm{x} \in \bar{\mathcal{X}}$ requires $M_{\bar{\mathcal{X}}} K$ floating-point operations. Caching these values, the number of operations required to compute the optimal indicator is $M_{\bar{\mathcal{X}}}K^2 (F_{\rm min}+2)$, where $F_{\rm min}$ denotes the number of floating-point operations required to compute ${\min}\{ \hat{y}_{\rm LBC}^{\rm opt}(\bm{x}), 1- \hat{y}_{\rm LBC}^{\rm opt}(\bm{x})\}$. This corresponds to $O(M_{\bar{\mathcal{X}}}K^2)$ operations. Without caching, the optimal indicator requires $M_{\bar{\mathcal{X}}}K^3 + M_{\bar{\mathcal{X}}}K^2(F_{\rm min}+2) + K$ operations to compute, resulting in a scaling of $O(M_{\bar{\mathcal{X}}}K^3)$.\\

\indent \textit{Prediction-based method}.---In PBM, the computation of $\hat{y}_{\rm PBM}^{\rm opt}$ for all $\bm{x} \in \bar{\mathcal{X}}$ requires $M_{\bar{\mathcal{X}}}  (3K-1)$ floating-point operations. Caching these values, the number of operations required to compute the mean optimal prediction $\hat{y}_{\rm PBM}^{\rm opt}$ for all $\{ p_{k} \}_{k=1}^{K}$ is $5M_{\bar{\mathcal{X}}}K-K-M_{\bar{\mathcal{X}}}$. Computing the optimal indicator then requires $M_{\bar{\mathcal{X}}}(5K -1)+K$ operations. The computation of the mean optimal predictions and the optimal indicator each require $O(M_{\bar{\mathcal{X}}}K)$ operations. If the values $\hat{y}_{\rm PBM}^{\rm opt}(\bm{x}) \; \forall \bm{x} \in \bar{\mathcal{X}}$ are not cached, computing the mean optimal prediction instead requires $3M_{\bar{\mathcal{X}}}K^2+K(M_{\bar{\mathcal{X}}}-1)$ operations. Computing the optimal indicator then requires $3M_{\bar{\mathcal{X}}}K^2+KM_{\bar{\mathcal{X}}} + K$ operations. For both quantities, this results in a scaling of $O(M_{\bar{\mathcal{X}}}K^2)$.\\

\indent \textit{Numerical implementation.}---The measured computation times associated with calculating the optimal indicators of phase transitions of SL, LBC, and PBM, for all six physical systems discussed in the main text (see Sec.~\ref{sec_physical_models}) are reported in Tab.~\ref{fig_timing}. The corresponding code is open source~\cite{github}. Again, we do not consider the computational cost associated with generating samples and estimating the underlying probability distributions. Overall, the computation times are remarkably low. For all systems, the optimal indicator of SL and PBM can be obtained in under a second, and the optimal indicator of LBC in under a minute. We observe that the computation times of SL and PBM are comparable, with PBM being slightly slower than SL. In contrast, the computations times of LBC are two orders of magnitude larger. Note that these are the evaluation times corresponding to the largest system sizes under consideration. We find that the computation times qualitatively agree with the complexity analysis described above (for the case where caching is performed). An additional speed-up can be gained through parallel execution. In particular, it is straightforward to compute optimal predictions (in the case of SL and PBM) and optimal indicators (in the case of LBC) at discrete values of the tuning parameter in parallel, e.g., via multithreading (which is implemented in~\cite{github}).

\subsection{Boltzmann-distributed inputs}\label{app_A4}
Let us discuss the special case when the drawn inputs $\bm{x}$, such as spin configurations, follow a Boltzmann distribution
\begin{equation}\label{eq:app_b1}
	{\rm P}_{k}(\bm{x}) = \frac{ e^{-H(\bm{x})/k_{\rm B}T_{k}}}{Z_{k}}.
\end{equation}
The probability to draw a sample with energy $E$ is thus given by
\begin{equation}\label{eq:app_b2}
	{\rm P}_{k}(E) = \frac{ g(E) e^{-E/k_{\rm B}T_{k}}}{Z_{k}},
\end{equation}
where $g(E)$ is the corresponding degeneracy factor
\begin{equation}\label{eq:app_bx}
	g(E) = \sum_{\bm{x} \in \mathcal{S}} \delta_{H(\bm{x}),E}.
\end{equation}
Here, $\mathcal{S}$ denotes the state space of the samples $\bm{x}$, i.e., the set of all unique samples without duplicates. Therefore, we have
\begin{equation}\label{eq:app_b3}
	{\rm P}_{k}(\bm{x}) = {\rm P}_{k}\left(H(\bm{x})\right)/g\left(H(\bm{x})\right).
\end{equation}

\indent \textit{Supervised learning}.---Plugging Eq.~\eqref{eq:app_b3} into Eq.~\eqref{eq:SL_derivation_1}, we immediately find that
\begin{equation}\label{eq:app_b4}
    \begin{split}
	 \hat{y}_{\rm SL}^{\rm opt}(\bm{x}) &= \frac{{\rm P}_{\rm I}(H(\bm{x}))}{{\rm P}_{\rm I}(H(\bm{x})) + {\rm P}_{\rm II}(H(\bm{x}))}\\
	 &= \hat{y}_{\rm SL}^{\rm opt}(H(\bm{x})) \; \forall \bm{x} \in \mathcal{S},
	 \end{split}
\end{equation}
where we assume that $\bar{\mathcal{T}} = \bar{\mathcal{X}} = \mathcal{S}$. Using Eq.~\eqref{eq:SL_derivation_4}, we have
\begin{equation}\label{eq:app_b5}
	\begin{split}
		\hat{y}_{\rm SL}^{\rm opt}(p_{k}) &= \sum_{\bm{x} \in \mathcal{S}} {\rm P}_{k}(\bm{x})\hat{y}_{\rm opt}(\bm{x})\\ 
		&= \sum_{\bm{x} \in \mathcal{S}} {\rm P}_{k}(H(\bm{x}))\hat{y}_{\rm opt}(H(\bm{x}))/g(H(\bm{x}))\\ 
		&= \sum_{E \in \mathcal{S}_{E}}  {\rm P}_{k}(E )\hat{y}_{\rm opt}(E ),
	\end{split}
\end{equation}
where $\mathcal{S}_{E}$ is the set of unique energies corresponding to the state space $\mathcal{S}$. To obtain an expression for the optimal loss, we can rewrite Eq.~\eqref{eq:CE_SL} as
\begin{align}\label{eq:app_SL_b6}
\mathcal{L}_{\rm SL} =& -\frac{1}{r_{\rm I} + (K - l_{\rm II})}\sum_{k=1}^{r_{\rm I}} \sum_{k=l_{\rm II}}^{K} \sum_{\bm{x} \in \mathcal{S}} {\rm P}_{k}(\bm{x}) \\&[ y(\bm{x})\ln \left(\hat{y}(\bm{x}) \right) +\left(1-y(\bm{x}) \right)\ln \left(1-\hat{y}(\bm{x}) \right)] \nonumber.
\end{align}
Using Eq.~\eqref{eq:app_b4}, we have
\begin{align}\label{eq:app_SL_b7}
\mathcal{L}_{\rm SL}^{\rm opt} =& -\frac{1}{r_{\rm I} + (K - l_{\rm II}+1)}\sum_{k=1}^{r_{\rm I}} \sum_{k=l_{\rm II}}^{K} \sum_{\bm{x} \in \mathcal{S}} {\rm P}_{k}(H(\bm{x}))\\ &[ y(H(\bm{x}))\ln \left(\hat{y}^{\rm opt}_{\rm SL}(H(\bm{x})) \right)\nonumber \\ &+\left(1-y(H(\bm{x})) \right)\ln \left(1-\hat{y}^{\rm opt}_{\rm SL}(H(\bm{x})) \right)] \nonumber,
\end{align}
where we use the fact that $y(\bm{x})=y(H(\bm{x}))$, i.e., the assigned labels remain identical. Equation~\eqref{eq:app_SL_b7} can be simplified to
\begin{align}\label{eq:app_SL_b8}
\mathcal{L}_{\rm SL}^{\rm opt} =& -\frac{1}{r_{\rm I} + (K - l_{\rm II}+1)}\sum_{k=1}^{r_{\rm I}} \sum_{k=l_{\rm II}}^{K} \sum_{E \in \mathcal{S}_{E}} {\rm P}_{k}(E)\\&[ y(E)\ln \left(\hat{y}^{\rm opt}_{\rm SL}(E) \right) +\left(1-y(E) \right)\ln \left(1-\hat{y}^{\rm opt}_{\rm SL}(E) \right)] \nonumber,
\end{align}
using Eq.~\eqref{eq:app_b3}.\\

\indent \textit{Learning by confusion}.---For a fixed bipartition in LBC, we can proceed in a similar manner. Plugging Eq.~\eqref{eq:app_b3} into Eq.~\eqref{eq:LBS_derivation_1} assuming $\bar{\mathcal{X}} = \mathcal{S}$, we have
\begin{equation}\label{eq:app_b6}
    \begin{split}
	\hat{y}_{\rm LBC}^{\rm opt}(\bm{x}) &= \frac{{\rm P}_{\rm I}(H(\bm{x}))}{{\rm P}_{\rm I}(H(\bm{x})) + {\rm P}_{\rm II}(H(\bm{x}))}\\ 
	&= \hat{y}_{\rm LBC}^{\rm opt}(H(\bm{x})) \; \forall \bm{x} \in \mathcal{S}.
	\end{split}
\end{equation}
Using Eq.~\eqref{eq:LBS_derivation_2}, this yields
\begin{equation}\label{eq:app_b7}
	\begin{split}
	I_{\rm LBC}^{\rm opt} &=1-\frac{1}{K}\sum_{k=1}^{K} \sum_{\bm{x} \in \mathcal{S}} {\rm P}_{k}(\bm{x}){\min}\{ \hat{y}_{\rm LBC}^{\rm opt}(\bm{x}), 1- \hat{y}_{\rm LBC}^{\rm opt}(\bm{x})\}\\
	&= 1-\frac{1}{K}\sum_{k=1}^{K} \sum_{E \in \mathcal{S}_{E}}  {\rm P}_{k}(E ){\min}\{ \hat{y}_{\rm LBC}^{\rm opt}(E), 1- \hat{y}_{\rm LBC}^{\rm opt}(E)\}.
	\end{split}
\end{equation}
To obtain an expression for the optimal loss, we follow the above procedure outlined for SL starting with Eq.~\eqref{eq:CE_LBC} and eventually arrive at
\begin{align}\label{eq:app_LBC_b8}
	\mathcal{L}_{\rm LBC}^{\rm opt} =& -\frac{1}{K} \sum_{k=1}^{K} \sum_{E \in \mathcal{S}_{E}} {\rm P}_{k}(E)\\ &[ y(E)\ln \left( \hat{y}(E) \right) +\left( 1-y(E)\right)\ln \left( 1-\hat{y}(E)\right)] \nonumber.
\end{align}

\indent \textit{Prediction-based method}.---Plugging Eq.~\eqref{eq:app_b3} into Eq.~\eqref{eq:PBM_derivation_1} assuming $\bar{\mathcal{X}} = \mathcal{S}$, we find that
\begin{equation}\label{eq:app_b8}
    \begin{split}
	\hat{y}_{\rm PBM}^{\rm opt}\left(\bm{x}\right) &= \frac{\sum_{k=1}^{K} {\rm P}_{k}\left( H(\bm{x})\right) p_{k}}{ \sum_{k=1}^{K} {\rm P}_{k}\left(H(\bm{x})\right)}\\
	&= \hat{y}_{\rm PBM}^{\rm opt}\left(H(\bm{x})\right) \; \forall \bm{x} \in \mathcal{S}.
	\end{split}
\end{equation}
Using Eq.~\eqref{eq:PBM_derivation_2}, we have
\begin{equation}\label{eq:app_b9}
	\begin{split}
	\hat{y}_{\rm PBM}^{\rm opt}(p_{k}) &= \sum_{\bm{x} \in \mathcal{S}} {\rm P}_{k}(\bm{x}) \hat{y}_{\rm PBM}^{\rm opt}(\bm{x})\\
	&= \sum_{E \in \mathcal{S}_{E}} {\rm P}_{k}(E) \hat{y}_{\rm PBM}^{\rm opt}(E).
\end{split}
\end{equation}
To obtain an expression for the optimal loss, we rewrite Eq.~\eqref{eq:MSE_PBM} as
\begin{equation}\label{eq:app_b10}
\mathcal{L}_{\rm PBM} = \frac{1}{K}\sum_{k=1}^{K} \sum_{\bm{x} \in \mathcal{S}} {\rm P}_{k}(\bm{x}) \left(\hat{y}(\bm{x}) - y(\bm{x})  \right)^2.
\end{equation}
Using Eq.~\eqref{eq:app_b8}, we have
\begin{equation}\label{eq:app_b11}
\mathcal{L}_{\rm PBM}^{\rm opt} = \frac{1}{K}\sum_{k=1}^{K} \sum_{\bm{x} \in \mathcal{S}} {\rm P}_{k}(\bm{x}) \left(\hat{y}_{\rm PBM}^{\rm opt}\left(H(\bm{x})\right) - y(H(\bm{x}))  \right)^2,
\end{equation}
where $y(\bm{x})=y(H(\bm{x}))$. With Eq.~\eqref{eq:app_b3} we finally get
\begin{equation}\label{eq:app_b11}
\mathcal{L}_{\rm PBM}^{\rm opt} = \frac{1}{K}\sum_{k=1}^{K} \sum_{E \in \mathcal{S}_{E}} {\rm P}_{k}(E) \left(\hat{y}_{\rm PBM}^{\rm opt}\left(E\right) - y(E)  \right)^2.
\end{equation}

\indent Thus, we have shown that the optimal predictions, indicators, and loss values of SL, LBC, and PBM remain identical when configuration samples which follow a Boltzmann distribution are used as input, or when the corresponding energies are used as input instead. In practice, given a finite set of samples the inferred probability distribution ${\rm P}_{k}(\bm{x}) \approx M_{k}(\bm{x})/M$ is only approximately Boltzmann, i.e., $\bar{\mathcal{T}}, \bar{\mathcal{X}} \approx \mathcal{S}$, and the two scenarios are only equivalent up to deviations due to finite-sample statistics. In particular, the inferred probability distribution ${\rm P}_{k}(\bm{x}) = M_{k}(\bm{x})/M$ based on raw configuration samples may not correspond to the inferred probability distribution ${\rm P}_{k}(E) = M_{k}(E)/M$ based on the corresponding energy, where the degeneracy factor for the conversion is inferred from the samples as
\begin{equation}
	g(E) = \sum_{\bm{x} \in \mathcal{X}} \delta_{H(\bm{x}),E}.
\end{equation}
However, using the energy as input instead of configuration samples yields a more accurate estimate of the ground-truth distribution. This is because the associated state space $\mathcal{S}_{E}$ is significantly smaller compared to the entire configuration space $\mathcal{S}$, resulting in better statistics given a fixed number of samples. In the 2D Ising model, for example, the size of the configuration space is $2^{L^{2}}$, whereas there are $L^2-1$ unique number of energies (for even $L$). Therefore, the optimal predictions and indicators obtained using the energy as input converge significantly faster compared to the case where raw spin configurations are used. Note that the energy is readily available in numerical studies. However, in principle, one can obtain the same results without having access to the energy given that a sufficient number of raw configurations are sampled. In the future, it will be of interest to employ more elaborate techniques for density estimation~\cite{bishop:2006,wu:2019,melko:2019,nicoli:2021} in order to obtain a more accurate estimate of the underlying distribution given a reduced data set size.\\

\indent Finally, let us continue the analysis of the optimal predictions and indicators of SL in case of Boltzmann-distributed inputs. We take region I to be composed of a single point $T_{1}$. Let $T_{1} \rightarrow 0$ such that
\begin{equation}\label{eq:app_Boltzmann_2}
	{\rm P}_{1}(E) =
	\begin{cases}
		1 \; {\rm if}\; E = E_{\rm gs},\\
		0 \; {\rm otherwise},
	\end{cases}
\end{equation}
where $E_{\rm gs}$ is the ground-state energy. Plugging into Eq.~\eqref{eq:app_SL_derivation_3} yields
\begin{equation}\label{eq:app_Boltzmann_3}
	\hat{y}_{\rm SL}^{\rm opt}(E) =
	\begin{cases}
		\frac{1}{1+{\rm P}_{\rm II}(E_{\rm gs})} \; {\rm if} \; E = E_{\rm gs},\\
		0 \; {\rm otherwise}.
	\end{cases}
\end{equation}
We calculate the mean prediction at a given temperature as
\begin{equation}\label{eq:app_Boltzmann_4}
	\hat{y}_{\rm SL}^{\rm opt}(T_{k}) = \sum_{E \in \mathcal{S}_{E}} {\rm P}_{k}(E) \hat{y}_{\rm SL}^{\rm opt}(E).
\end{equation}
Using Eq.~\eqref{eq:app_Boltzmann_3}, this results in
\begin{equation}\label{eq:app_Boltzmann_5}
	\hat{y}_{\rm SL}^{\rm opt}(T_{k}) = \frac{{\rm P}_{k}(E_{\rm gs})}{1 + {\rm P}_{\rm II}(E_{\rm gs})}.
\end{equation}
Assuming region II is composed of a single point $T_{K}$, we have ${\rm P}_{\rm II}(E_{\rm gs}) = {\rm P}_{K}(E_{\rm gs})$ and recover Eq.~\eqref{eq_SL_boltzmann} of the main text. For $T_{K} \rightarrow \infty$, we have ${\rm P}_{K}(E_{\rm gs}) = g(E_{\rm gs})/M_{\mathcal{S}}$, where $M_{\mathcal{S}}$ is the total number of unique system configurations. For the two-dimensional Ising model, for example, $M_{\mathcal{S}}=2^{L \times L}$. Approaching the thermodynamic limit, this yields $\hat{y}_{\rm SL}^{\rm opt}(T_{k}) \rightarrow {\rm P}_{k}(E_{\rm gs})$.\\

\begin{figure*}[tbh!]
    \centering
    \includegraphics[width=0.9\linewidth]{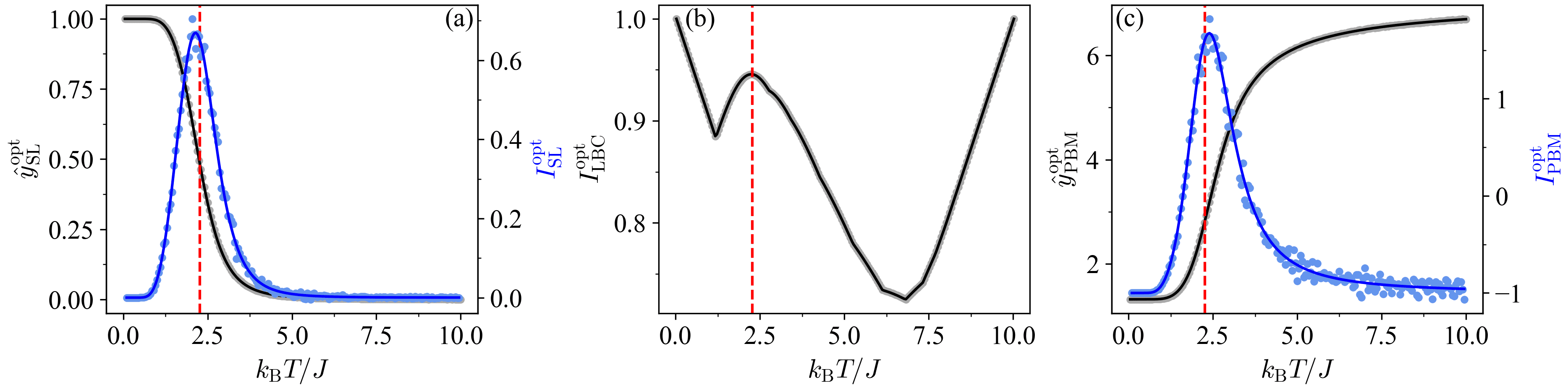}
    \caption{Results for the Ising model ($L=4$) with the dimensionless temperature as a tuning parameter $p=k_{\rm B}T/J$, where $p_{1}=0.05$, $p_{K}=10$, and $\Delta p = 0.05$. The critical temperature [Eq.~\eqref{eq:ising_Tc}] is highlighted by a red-dashed line. In SL, the data obtained at $p_{1}$ and $p_{K}$ constitutes our training set, i.e., $r_{\rm I}=1$ and $l_{\rm II}=K$. The inputs are computed based on spin configurations obtained through exact enumeration (lines) or Monte Carlo sampling (points). (a) Mean optimal prediction $\hat{y}_{\rm SL}^{\rm opt}$ in SL (black) and the corresponding indicator $I_{\rm SL}^{\rm opt}$ (blue). (b) Optimal indicator of LBC, $I_{\rm LBC}^{\rm opt}$ (black). (c) Mean optimal prediction $\hat{y}_{\rm PBM}^{\rm opt}$ in PBM (black) and the corresponding indicator $I_{\rm PBM}^{\rm opt}$ (blue).}
    \label{fig:ising_4x4_exact}
    \end{figure*}
    
\indent Note that these results can be extended to non-Boltzmann distributions: Given that  
\begin{equation}\label{eq:app_Boltzmann_6}
	{\rm P}_{1}(\bm{x}) =
	\begin{cases}
		1 \; {\rm if} \; \bm{x} = \bm{x}^{*},\\
		0 \; {\rm otherwise},
	\end{cases}
\end{equation}
and following the same procedure as above, we have
\begin{equation}\label{eq:app_Boltzmann_7}
	\hat{y}_{\rm SL}^{\rm opt}(p_{k}) = \frac{{\rm P}_{k}(\bm{x}^{*})}{1 + {\rm P}_{\rm II}(\bm{x}^{*})}.
\end{equation}
In particular, Eq.~\eqref{eq:app_Boltzmann_7} can be used to explain the optimal indicator signals of SL in the XXZ chain (Sec.~\ref{sec_methods_xxz}) and Kitaev chain (Sec.~\ref{sec_methods_kitaev}). In this case, $\bm{x}^{*}$ corresponds to a ground state which is one of the chosen basis states.\\
    
\subsection{Finite-sample statistics}\label{app_A5}
Finally, we investigate how the optimal predictions and indicators of SL, LBC, and PBM change as the number of data points $M$ per sampled value of the tuning parameter is varied. Recall that the results for the classical systems displayed in the main text were obtained using the energy from Monte Carlo sampling as input, where $M=10^{5}$ spin configurations are drawn per temperature. For small lattice sizes, however, it is possible to enumerate all spin configurations explicitly. In Fig.~\ref{fig:ising_4x4_exact}, we compare the optimal predictions and indicators for the Ising model on a $4 \times 4$ lattice when enumerating all $2^{16}=65536$ spin configurations explicitly or using Monte Carlo sampling with $10^5$ number of configurations per sampled value of the tuning parameter. The results obtained based on the two distinct data sets are in good agreement, which is to be expected given that there are only 15 unique energies. The noise present in the indicator signals of SL and PBM when using Monte Carlo samples is absent when using exact enumeration. In the latter case, both indicators vary smoothly as a function of temperature. As such, this noise can be attributed to finite-sample statistics.\\

\begin{figure*}[tbh!]
	\begin{center}
		\includegraphics[width=0.99\textwidth]{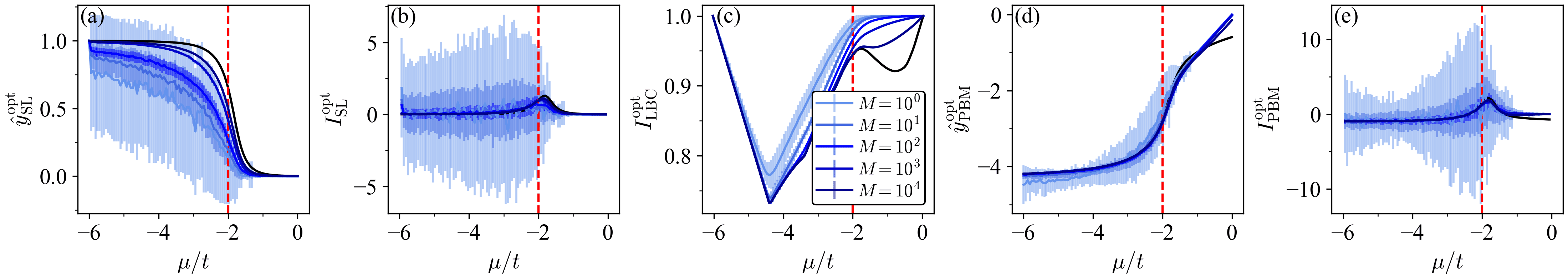}
		\caption{Optimal predictions and indicators of SL, LBC, and PBM for the Kitaev chain ($L=20$) given various number of data points $M$ per sampled value of the tuning parameter $p=\mu /t $, where $p_{1}=-6$, $p_{K}=0$, and $\Delta p = 0.06$. In SL, the data obtained at $p_{1}$ and $p_{K}$ constitutes our training set, i.e., $r_{\rm I}=1$ and $l_{\rm II}=K$. The critical value $\mu_{\rm c}/t= - 2$ is highlighted by a red-dashed line. The optimal predictions and indicators obtained based on the ground-truth probability distributions from exact diagonalization are shown in black. (a) Mean optimal prediction $\hat{y}_{\rm SL}^{\rm opt}$ in SL and (b) the corresponding indicator $I_{\rm SL}^{\rm opt}$. (c) Optimal indicator of LBC, $I_{\rm LBC}^{\rm opt}$. (d) Mean optimal prediction $\hat{y}_{\rm PBM}^{\rm opt}$ in PBM and (f) the corresponding indicator $I_{\rm PBM}^{\rm opt}$. Here, we report results averaged over 100 independent data sets, where the error bars correspond to the standard deviation.}
		\label{fig_Kitaev_finite_samples}
	\end{center}
\end{figure*}

\indent In general, for both the classical and quantum systems we observe that the overlap in the underlying probability distributions leading to a peak in the indicator signals decreases as the number of samples $M$ is decreased. However, meaningful results can already be obtained when only a fraction of the total state space is covered. In the case of the Ising model on a $60 \times 60$ lattice, for example, we observe that the optimal predictions and indicators are already well converged for $M=10^{2}$, i.e., matching the results obtained with $M=10^{5}$. In particular, the key features in the indicators, i.e., the peak locations, can already be identified for $M=10$. Compare this to the unique number of energies given by $3599$.\\

\indent Figure~\ref{fig_Kitaev_finite_samples} shows the optimal predictions and indicators of SL, LBC, and PBM for the Kitaev chain of length $L=20$ given various values of $M$. Recall that the results for the quantum systems displayed in the main text (see Sec.~\ref{sec_physical_models}) were obtained based on the ``ground-truth'' probability distributions from exact diagonalization. Here, we explicitly sample these probability distributions, i.e., perform projective measurements and infer the probability distribution based on the measurement results. In SL and PBM, accurate estimates for the critical value of the tuning parameter can be obtained based on $M=10^3$ samples, whereas $M=10^4$ samples are required for a local maximum to emerge in LBC. This only covers a fraction of the total state space comprised of $M_{\bar{\mathcal{X}}} = 524288$ states. Notice that the indicator of LBC shows a plateau close to one in the topological phase for a small number of samples, which signifies the absence of ``confusion'' inherent to the data (see Fig.~\ref{fig_Kitaev_finite_samples}). Similarly, the optimal prediction of PBM is approximately linear in the topological phase for a small number of samples, corresponding to a model which can perfectly resolve the value of the tuning parameter associated with the input. This demonstrates the fact that while the ground-truth probability distributions may have substantial overlap, estimated probabilities based on a drawn data set may not.\\

\indent The high level of uncertainty in the indicator of SL and PBM compared to LBC can be attributed to the symmetric difference quotient used to approximate the derivative. Moreover, in LBC we associate a distinct optimal predictive model to each bipartition point, whereas the optimal indicator is extracted from a single optimal model in case of SL and PBM. This leads to an additional suppression of fluctuations in case of LBC. In the future, it will be of interest to enhance the quality of the optimal predictions and indicators based on finite data through improved derivative computations in case of SL and PBM~\cite{chartrand:2011}, as well as more elaborate techniques for density estimation~\cite{bishop:2006,wu:2019,melko:2019,nicoli:2021}.

\section{Computation using neural networks}\label{app_B}
In this appendix, we discuss the application of SL, LBC, and PBM to the six physical systems discussed in the main text (see Sec.~\ref{sec_physical_models}) using NNs. First, we show that one can recover the optimal analytical predictions and indicators by training NNs. Next, we discuss the computational cost associated with training NNs compared to constructing and evaluating optimal models. Finally, we investigate the influence of NN size, early stopping, regularization, and finite-sample statistics on the results.\\

\begin{figure*}[htb!]
\centering
\includegraphics[width=0.99\linewidth]{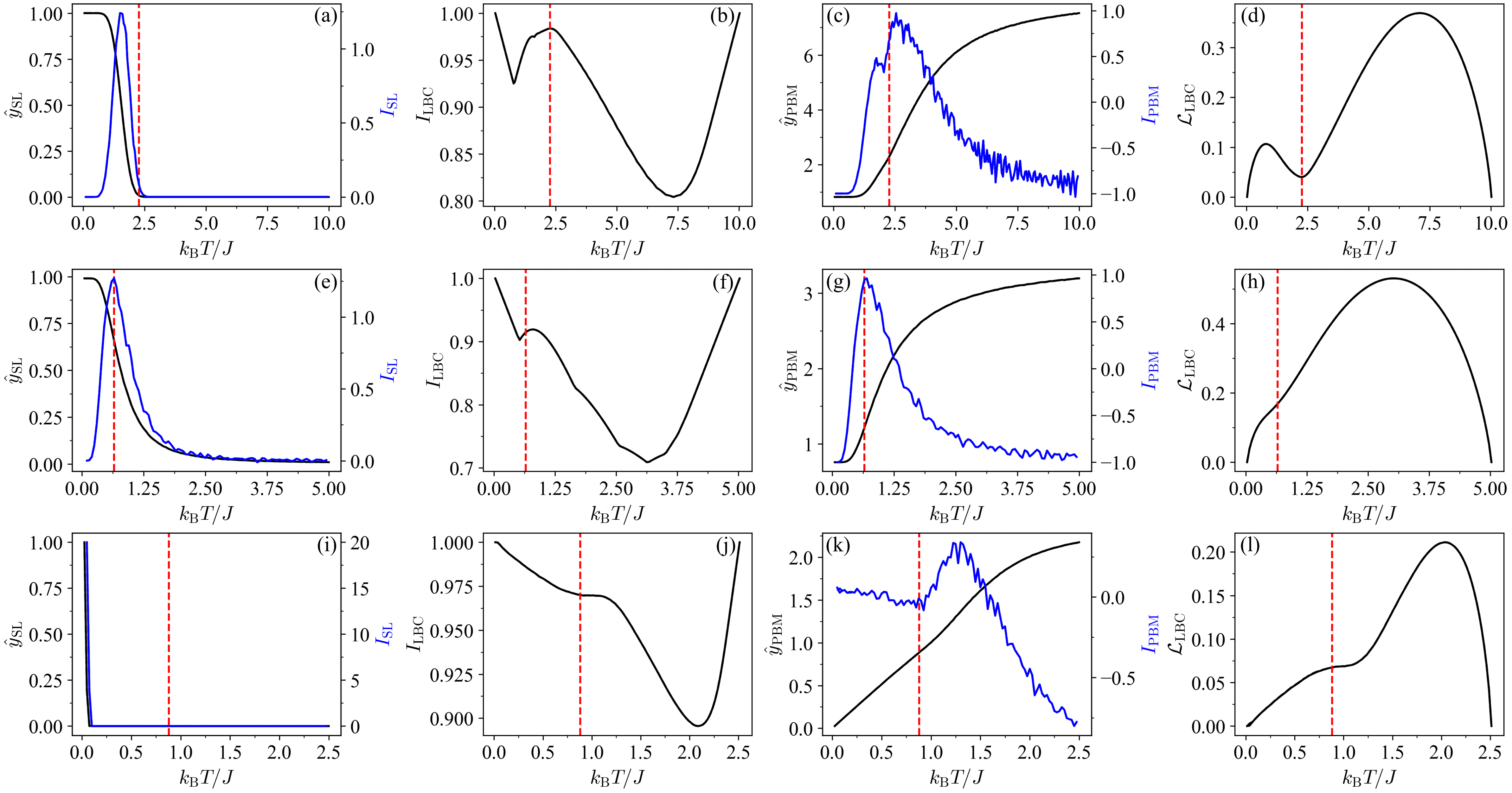}
\caption{(a)-(d) Results for the Ising model ($L=10$) using NNs. The NNs used in SL, LBC, and PBM were trained for 10000, 1000, and 5000 epochs, respectively. The tuning parameter ranges from $p_{1}=0.05$ to $p_{K}=10$ with $\Delta p = 0.05$. (e)-(h) Results for the IGT ($L=4$) using NNs. The NNs used in SL, LBC, and PBM were trained for 10000, 1000, and 5000 epochs, respectively. The tuning parameter ranges from $p_{1}=0.05$ to $p_{K}=5$ with $\Delta p = 0.05$. (i)-(l) Results for the XY model ($L=10$) using NNs. The NNs used in SL, LBC, and PBM were trained for 10000, 1000, and 10000 epochs, respectively. The tuning parameter ranges from $p_{1}=0.025$ to $p_{K}=2.5$ with $\Delta p = 0.025$. The critical value of the tuning parameter $p_{\rm c} = k_{\rm B}T_{\rm c}/J$ is highlighted in red. (a),(e),(i) Mean prediction $\hat{y}_{\rm SL}(p)$ obtained using the analytical expression (black, solid) or an NN (black, dashed), as well as the corresponding indicator $I_{\rm SL}(p)$ (blue). Here, we choose $r_{\rm I}=1$ and $l_{\rm II}=K$. (b),(f),(j) The indicator of LBC, $I_{\rm LBC}$, obtained using the analytical expression (black, solid) or an NN (black, dashed). (c),(g),(k) Mean prediction $\hat{y}_{\rm PBM}(p)$ of PBM obtained using the analytical expression (black, solid) or an NN (black, dashed), as well as the corresponding indicator $I_{\rm PBM}(p)$ (blue). (d),(h),(l) Value of the loss function in LBC, $\mathcal{L}_{\rm LBC}$, for each bipartition point $p^{\rm bp}$ obtained using the analytical expression (black, solid) or evaluated after NN training (black, dashed). In all three models, the NNs were comprised of three hidden layers with 64 nodes each and the learning rate was set to 0.001.}
\label{fig:NN_classical}
\end{figure*}

\begin{figure*}[htb!]
\centering
\includegraphics[width=0.99\linewidth]{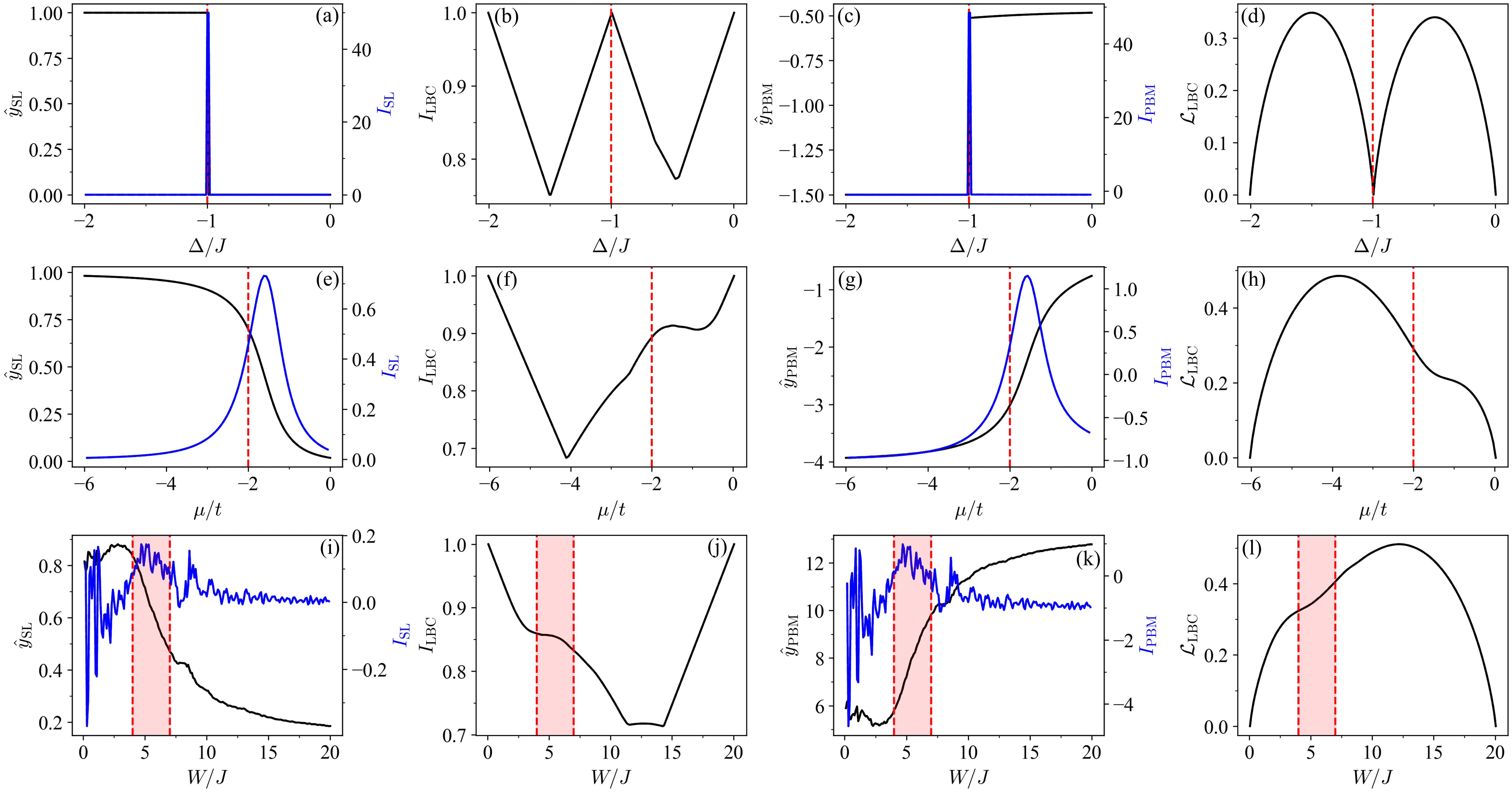}
\caption{(a)-(d) Results for the XXZ chain ($L=4$) using NNs. The NNs used in SL, LBC, and PBM were trained for 10000, 1000, and 5000 epochs, respectively. The tuning parameter ranges from $p_{1}=-2$ to $p_{K}=0$ with $\Delta p = 0.01$. The critical value of the tuning parameter $p_{\rm c} = \Delta_{\rm c}/J$ is highlighted in red. (e)-(h) Results for the Kitaev chain ($L=10$) using NNs. The NNs used in SL, LBC, and PBM were trained for 5000, 500, and 1000 epochs, respectively. The tuning parameter ranges from $p_{1}=-6$ to $p_{K}=0$ with $\Delta p = 0.06$. The critical value of the tuning parameter $p_{\rm c} = \mu_{\rm c}/t$ is highlighted in red. (i)-(l) Results for the many-body localization phase transition in the Bose-Hubbard model ($L=6$) using NNs. The NNs used in SL, LBC, and PBM were trained for 10000, 300, and 1000 epochs, respectively. The tuning parameter ranges from $p_{1}=0.1$ to $p_{K}=20$ with $\Delta p = 0.1$. The critical value of the tuning parameter $p_{\rm c} = W_{\rm c} / J$ is highlighted in red. (a),(e),(i) Mean prediction $\hat{y}_{\rm SL}(p)$ obtained using the analytical expression (black, solid) or an NN (black, dashed), as well as the corresponding indicator $I_{\rm SL}(p)$ (blue). Here, we choose $r_{\rm I}=1$ and $l_{\rm II}=K$. (b),(f),(j) The indicator of LBC, $I_{\rm LBC}$, obtained using the analytical expression (black, solid) or an NN (black, dashed). (c),(g),(k) Mean prediction $\hat{y}_{\rm PBM}(p)$ of PBM obtained using the analytical expression (black, solid) or an NN (black, dashed), as well as the corresponding indicator $I_{\rm PBM}(p)$ (blue). (d),(h),(l) Value of the loss function in LBC, $\mathcal{L}_{\rm LBC}$, for each bipartition point $p^{\rm bp}$ obtained using the analytical expression (black, solid) or evaluated after NN training (black, dashed). For the XXZ model, the NNs were comprised of three hidden layers with 64 nodes each. For the Kitaev chain and Bose-Hubbard model, we use two hidden layers with 128 nodes each, followed by three hidden layers with 64 nodes each. In all three cases, the learning rate was set to 0.001.}
\label{fig:NN_quantum}
\end{figure*}

\indent \textit{Data preparation.}---For the classical systems (Ising model, IGT, and XY model), the energy $H(\bm{\sigma})$ of the spin configurations $\bm{\sigma}$ sampled from Boltzmann distributions at various temperatures serves as an input. To counteract the effect of finite-sample statistics on the predictions in case of SL due to inputs not contained in the training set $\bm{x} \notin \bar{\mathcal{T}}$, i.e., $\bar{\mathcal{X}} \neq \bar{\mathcal{T}}$, we modify the corresponding probability distributions, such that ${\rm P}_{K}(\bm{x}) = 1/(M+M_{\notin \mathcal{T}})$ as opposed to ${\rm P}_{K}(\bm{x}) = 0$. Here, $M_{\notin \mathcal{T}}$ denotes the number of such inputs at $p_{K}$. That is, we add a single instance of each sample which does not appear at the boundary point $p_{K}$ to the corresponding data set $\mathcal{X}_{K}$. Alternatively, we could set these predictions to zero as discussed in Appendix~\ref{app_A2}. While the NN-based indicator can change if no such modifications are performed, this does not resolve the instances where the optimal indicator of SL fails to locate the phase transition (such as in the Ising model or XY model). For the quantum systems (XXZ chain, Kitaev chain, Bose-Hubbard model), the index of the corresponding basis states serves as input. We use a physically-motivated encoding, where the $S^{z}$ eigenstate given by $|\uparrow \downarrow \dots \uparrow \rangle$ and the Fock state $|1 0 \dots 1 \rangle$ are encoded as a bit-string $\bm{x} = (10\dots 1)$.\\

\indent Before training the NNs, each input $\bm{x}= \{x_{i} \}$ is standardized via the following affine transformation
\begin{equation}\label{eq_standardization}
x_{i}' = \frac{x_{i} - \langle x_{i} \rangle}{\sigma_{x_{i}}},
\end{equation}
where $\langle x_{i} \rangle$ and $\sigma_{x_{i}}$ are the mean value and standard deviation of $x_{i}$ across the training data, respectively. Standardization generally leads to a faster rate of convergence when applying gradient-based optimizers~\cite{lecun:2012}. Note that this bijective mapping does not change the probability associated with each input, i.e., ${\rm P}_{k}(\bm{x}) = {\rm P}_{k}(\bm{x}')\; \forall 1 \leq k \leq K $. Therefore, the optimal predictions and indicators remain unchanged.\\

\indent \textit{Neural network architecture.}---For simplicity, the NNs used in this work consist of a series of fully-connected layers, where  rectified linear units (ReLUs), $f\left(z\right) = {\rm max}\left(0,z\right)$, are used as activation functions~\cite{goodfellow:2016}. The NNs for SL and LBC have two output nodes, where a softmax activation function
\begin{equation}
    f_{i}\left(\bm{z}\right) = \frac{e^{z_{i}}}{\sum_{j} e^{z_{j}}}
\end{equation}
is used in the output layer to guarantee that $\hat{y}(\bm{x}') \in [0,1]$. Here, the sum runs over all output nodes, and $\hat{y}$ corresponds to the value of one of the output nodes after application of the softmax activation function. In PBM, no activation function is used for the output layer. The value of the single output node corresponds to $\hat{y}(\bm{x}')$, which is the estimated value of the tuning parameter at which the input $\bm{x}'$ was drawn. For the prototypical probability distributions discussed in Sec.~\ref{sec_application_trivial_distr} in the main text, we use a single hidden layer with 64 nodes. The number of hidden layers and nodes for all other models is reported in the corresponding figure captions.\\
    
\indent \textit{Training.}---The NNs are implemented using Flux in Julia~\cite{innes:2018}, where the weights and biases are optimized via gradient descent with Adam~\cite{kingma:2014} to minimize the loss function over a series of training epochs. In SL and LBC, we train on a CE loss function [Eq.~\eqref{eq:CE_SL} and~\eqref{eq:CE_LBC}, respectively], whereas in PBM we train on a MSE loss function [Eq.~\eqref{eq:MSE_PBM}]. Gradients are calculated using backpropagation~\cite{rumelhart:1986,goodfellow:2016,baydin:2018}. For the prototypical probability distributions discussed in Sec.~\ref{sec_application_trivial_distr}, we train for 10000 epochs with a learning rate of 0.001. The number of training epochs and learning rate for all other models is reported in the corresponding figure captions.\\

\indent \textit{Results.}---Figures~\ref{fig:NN_classical} and~\ref{fig:NN_quantum} show the predictions and indicators of the three methods obtained using NNs (dashed lines) after long training for all six physical systems considered in the main text. Here, we chose the smallest system sizes for convenience. Overall, they are in excellent agreement with the corresponding optimal predictions and indicators (bold lines). As the system size is increased, it becomes increasingly difficult to approximate the corresponding optimal predictions and indicator with high accuracy because the NN size has to be increased systematically, i.e., hyperparameters need to be adjusted more carefully. However, even for the largest system sizes considered in this work qualitative agreement can still be achieved with moderate NN sizes, see Appendix~\ref{app_B1} for an explicit example.\\

\begin{figure*}[tbh!]
	\begin{center}
		\includegraphics[width=0.99\textwidth]{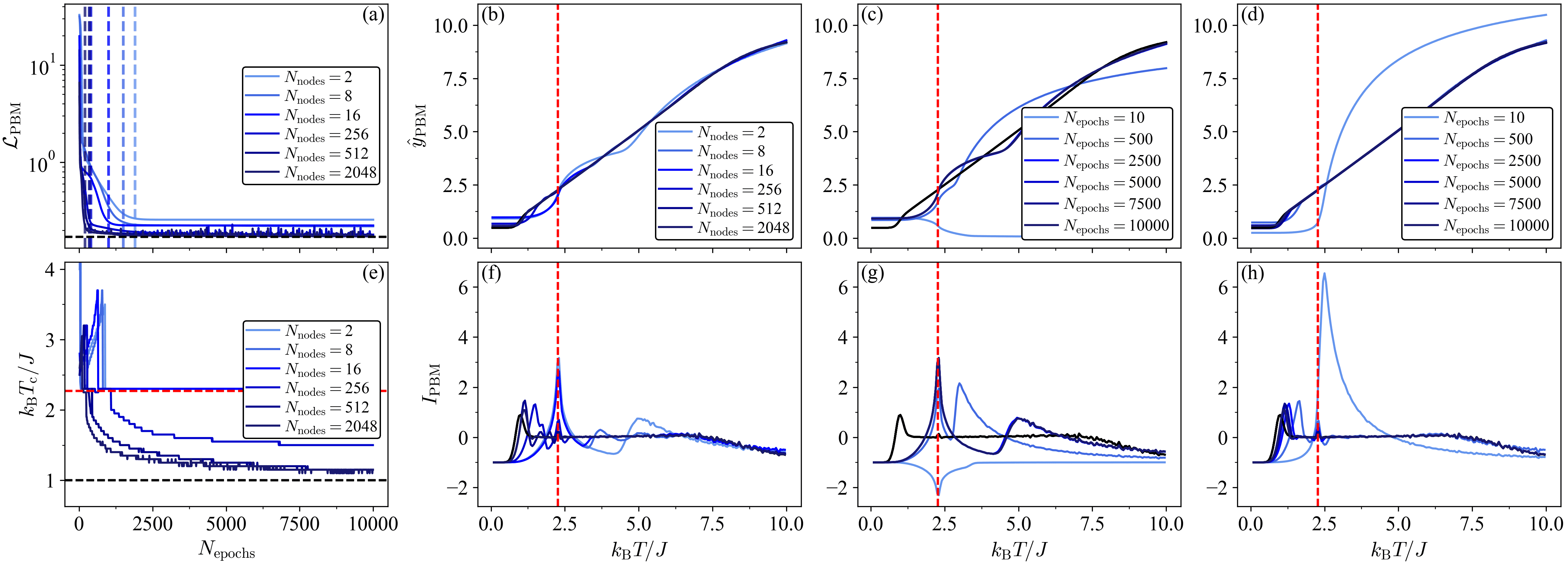}
		\caption{Results for the Ising model ($L=60$) of PBM using NNs with a single hidden layer composed of different number of hidden nodes $N_{\rm nodes}$. The learning rate is set to 0.01. The tuning parameter ranges from $p_{1}=0.05$ to $p_{K}=10$ with $\Delta p = 0.05$. The critical value of the tuning parameter $p_{\rm c} = k_{\rm B}T_{\rm c}/J$ is highlighted in red. The optimal predictions, optimal indicator, optimal loss, and corresponding estimated critical value of the tuning parameter are highlighted in black. (a) Loss $\mathcal{L}_{\rm PBM}$ as a function of the number of training epochs $N_{\rm epochs}$. The location of kinks in the loss are marked by vetical dashed lines. (b),(f) Mean prediction $\hat{y}_{\rm PBM}(p)$ of PBM obtained using NNs after training for 10000 epochs, as well as the corresponding indicator $I_{\rm PBM}(p)$. (c),(g) Mean prediction $\hat{y}_{\rm PBM}(p)$ of PBM obtained using an NN with $N_{\rm nodes}=2$ at various stages during training, as well as the corresponding indicator $I_{\rm PBM}(p)$. (d),(h) Mean prediction $\hat{y}_{\rm PBM}(p)$ of PBM obtained using an NN with $N_{\rm nodes}=2048$ at various stages during training, as well as the corresponding indicator $I_{\rm PBM}(p)$. (e) Estimated critical value of the tuning parameter as a function of the number of training epochs.}
		\label{fig_ising_NN_PBM_systematic}
	\end{center}
\end{figure*}

\indent \textit{Computational cost.}---Finally, let us touch upon the computational cost of training NNs. Table~\ref{fig_timing} reports the measured computation times associated with training an NN with one hidden layer composed of a single node for one epoch. A training epoch is comprised of evaluating the NN (or NNs in the case of LBC) at all $M_{\bar{\mathcal{X}}}$ unique samples (see Tab.~\ref{fig_timing}), calculating the loss function, obtaining the gradient via backpropagation, and performing a single gradient step. This represents a lower bound for the total computation time associated with obtaining NN-based predictions and indicators. In a typical application, however, larger NNs need to be used, the NNs need to be trained for multiple epochs, the NN parameters (or the corresponding predictions and indicator) need to be cached at regular intervals, hyperparameters need to be tuned, and finally the indicator needs to be computed based on the NN predictions. The computation time for a single epoch is also expected to increase if the data is processed in a batchwise fashion (albeit likely at the benefit of requiring less training epochs overall). We find that this lower bound on the training time is comparable with the evaluation time of the corresponding optimal predictions and indicators (and optimal loss) and the two times differ by less than an order of magnitude across all six physical systems studied in the main text. This empirical finding can be explained as follows: To construct the optimal model, the probability of all inputs needs to be evaluated. Similarly, in each training epoch the NN is evaluated at all inputs contained in the training data set. The computation time associated with evaluating a small NN for a given input is comparable with evaluating the corresponding optimal model prediction, and the overhead associated with the gradient computation via backpropagation is of the same order of magnitude as the NN forward pass~\cite{blayo:2014}.\\

\indent Suppose one is interested in the predictions and indicators of SL, PBM, and LBC, in the limit of a perfectly trained, highly expressive NNs. Evidently, based on the discussion above, the evaluation of the analytical expressions is generally more efficient in that case. The precise timings will depend on the particular implementation, as well as the choice of hyperparameters. However, even in the case where small NNs are trained for short times the computation time associated with constructing and evaluating an optimal model is \textit{at worst} comparable. Here, we have neglected any overhead associated with constructing probability distributions based on drawn samples. In principle, when using NN one does not rely on the estimated probability distributions, i.e., one can directly work with the unprocessed dataset. Note, however, that in many scenarios (including this work) the overhead of estimated probability distributions from the dataset is negligible. When studying quantum systems using exact diagonalization, one has direct access to the underlying probability distributions. Similarly, when performing Monte Carlo studies the energy statistics are readily available.
\subsection{Controlling model capacity}\label{app_B1}
Here, we investigate the effect of NN size, training time, and $\ell_{2}$ regularization on the NN-based predictions and indicators and compare them with the corresponding optimal predictions and indicators. All three factors influence the capacity of the resulting model and thus determine its ability to approximate the optimal predictive model realizing the global minimum of the loss function corresponding to the optimal predictions and indicators~\cite{goodfellow:2016,hu:2021}. As pointed out in the main text (see Sec.~\ref{sec_physical_models}), there are instances where the optimal model does not correctly highlight the corresponding phase transition whereas simpler models do.\\

\indent As an example, let us consider the application of PBM to the Ising model. Figure~\ref{fig_ising_NN_PBM_systematic} shows the results for a $60 \times 60$ lattice obtained with NNs composed of a single hidden layer with a variable number of hidden nodes ranging from 2 to 2048. Figs.~\ref{fig_ising_NN_PBM_systematic}(b),(f) show the corresponding NN-based predictions and indicators after training for 10000 epochs. For NNs with 2 and 8 nodes, the indicator shows a clear peak at the critical value of the tuning parameter. As the number of nodes increases, the NN results start to resemble the optimal predictions and indicators (black) more closely. This reflects the fact that, the expressivity of an NN increases as the number of nodes is increased. A similar behavior is also visible in Fig.~\ref{fig_ising_NN_PBM_systematic}(a) which shows the loss over time, where NNs with more than 8 nodes achieve values close to the optimal loss (black), i.e., the global minimum.\\

\begin{figure}[tbh!]
	\begin{center}
		\includegraphics[width=0.99\columnwidth]{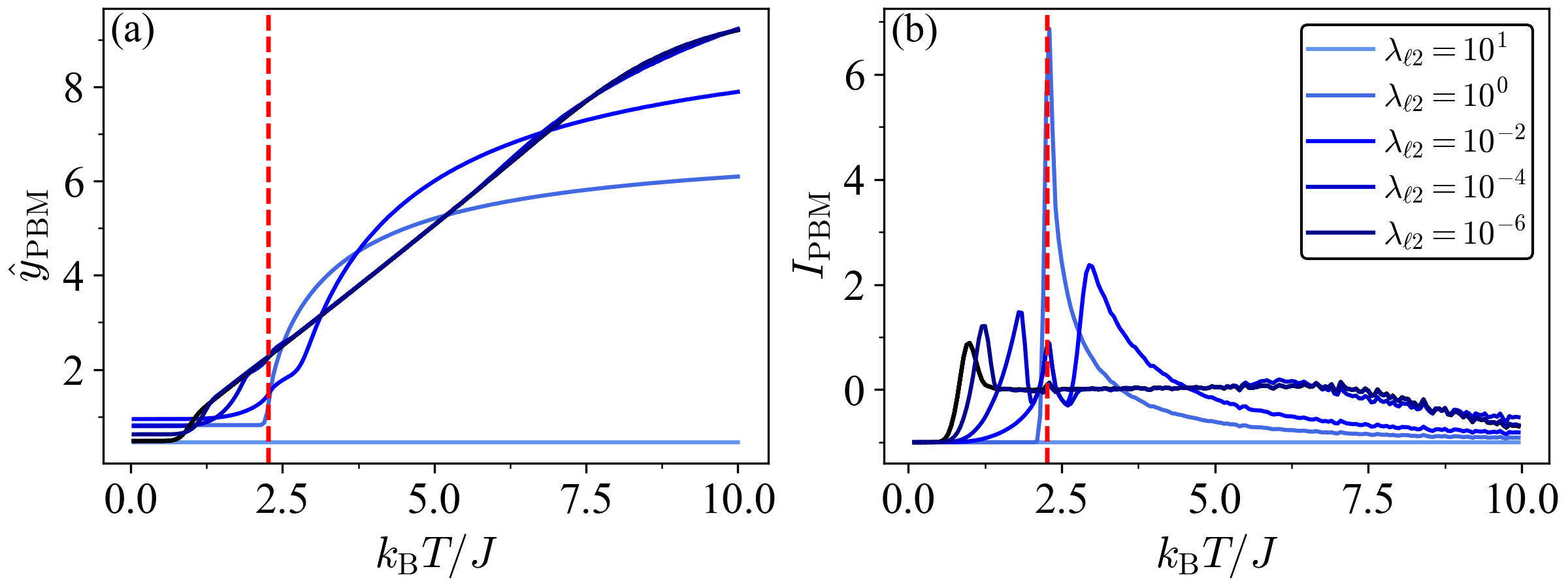}
		\caption{(a) Mean prediction $\hat{y}_{\rm PBM}(p)$ and (b) the corresponding indicator $I_{\rm PBM}(p)$ of PBM for the Ising model ($L=60$) using NNs obtained after long training for various regularization strengths $\lambda_{\ell 2}$. The tuning parameter ranges from $p_{1}=0.05$ to $p_{K}=10$ with $\Delta p = 0.05$. The critical value of the tuning parameter $p_{\rm c} = k_{\rm B}T_{\rm c}/J$ is highlighted in red. The optimal predictions and indicator is highlighted in black. Each NN has a single hidden layer with 2048 nodes and is trained for 10000 epochs with a learning rate of 0.01.}
		\label{fig_ising_NN_PBM_regularization}
	\end{center}
\end{figure}

\indent Figures~\ref{fig_ising_NN_PBM_systematic}(c) and (g) show the predictions and indicators for the smallest NN (2 hidden nodes) evaluated at various training epochs. Here, the indicator gradually converges towards its final form, which exhibits a peak at the critical value of the tuning parameter. Similarly, Figs.~\ref{fig_ising_NN_PBM_systematic}(d) and (h) shows the results for the largest NN (2048 hidden nodes). Here, early on during training the indicator is sharply peaked near the critical value of the tuning parameter. As the training progresses, the indicator signal starts to wash out and converge to the optimal indicator signal. The evolution of the global maximum of the indicator signal as a function of the training epoch for the various NN sizes is shown in Fig.~\ref{fig_ising_NN_PBM_systematic}(e). These results quantify how accurately the estimated critical value of the tuning parameter based on the optimal indicator (black) is reproduced for a given NN size and training time.\\

\indent Figure~\ref{fig_ising_NN_PBM_systematic}(h) shows that even for the large NNs there seems to be an intermediate time period during training where the indicator peaks near the critical value of the tuning parameter correctly highlighting the phase transition. Looking at Fig.~\ref{fig_ising_NN_PBM_systematic}(a), during these intermediate time periods the corresponding loss function starts to saturate and display a kink. This suggests a procedure for early stopping, where the training is stopped once a kink in the loss function is observed~\cite{goodfellow:2016}. Early stopping based on the validation loss will be discussed in the subsequent section (see Sec.~\ref{app_B2}). During training, the model capacity increases as visible by the steady decrease in the corresponding
\begin{figure}[tbh!]
	\begin{center}
		\includegraphics[width=0.99\columnwidth]{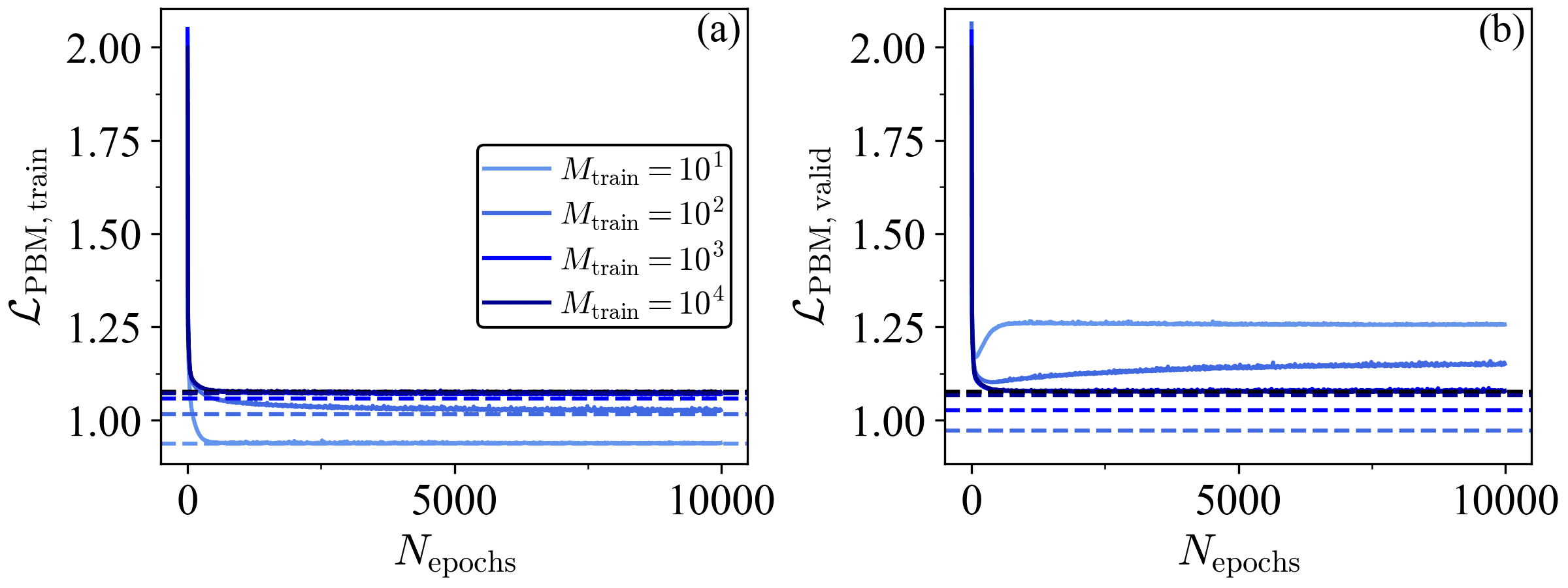}
		\caption{(a) Training loss and (b) validation loss as a function of the number of training epochs of PBM for the Kitaev chain ($L=14$) using an NN composed of a single hidden layer with 128 nodes for various numbers of training samples $M_{\rm train}$ per parameter value, where $M_{\rm valid} = M_{\rm test} = M_{\rm train}/5$. The corresponding optimal loss based on the training or validation data set is highlighted by a colored dashed line. The optimal loss based on the ground-truth probability distributions is highlighted in black. The test loss shows the same behavior as the validation loss. Each NN is trained for 10000 epochs with a learning rate of 0.01. The results averaged over 10 independent data sets.}
		\label{fig_NN_statistics_PBM_Ising}
	\end{center}
\end{figure}
loss~\cite{bishop:1995,sjoberg:1995,goodfellow:2016}: initially the model cannot resolve anything, in the intermediate stages it can resolve between the two phases leading to the sharp peak, and eventually it approaches the optimal predictive model (which, in this case, does not correctly highlight the phase transition). By stopping the training at the intermediate stage (i.e., selecting the corresponding NN parameters after the training is complete) a model of intermediate resolution can be obtained. Thus, early stopping acts as an implicit regularization~\cite{bishop:1995,sjoberg:1995,goodfellow:2016}. In the case of PBM, stopping the training early yields in an NN whose indicator peaks near the critical temperature of the Ising model. However, this is not always the case. In LBC, for example, the estimated critical temperature gradually improves during training, i.e., as the model capacity increases¨. Recall that the optimal indicator of LBC correctly highlights the phase transition. Qualitatively similar results can be obtained for the other methods and systems. In particular, in the Ising model and XY model, we find that the indicators of SL and PBM both show a clear peak near the critical transition temperature early on during training around the epochs marked by a kink in the loss function. The peak locations of the corresponding NN-based indicator signals coincide with the signals of physical indicators, such as the heat capacity or magnetization.\\

\begin{figure*}[tbh!]
	\begin{center}
		\includegraphics[width=0.7\textwidth]{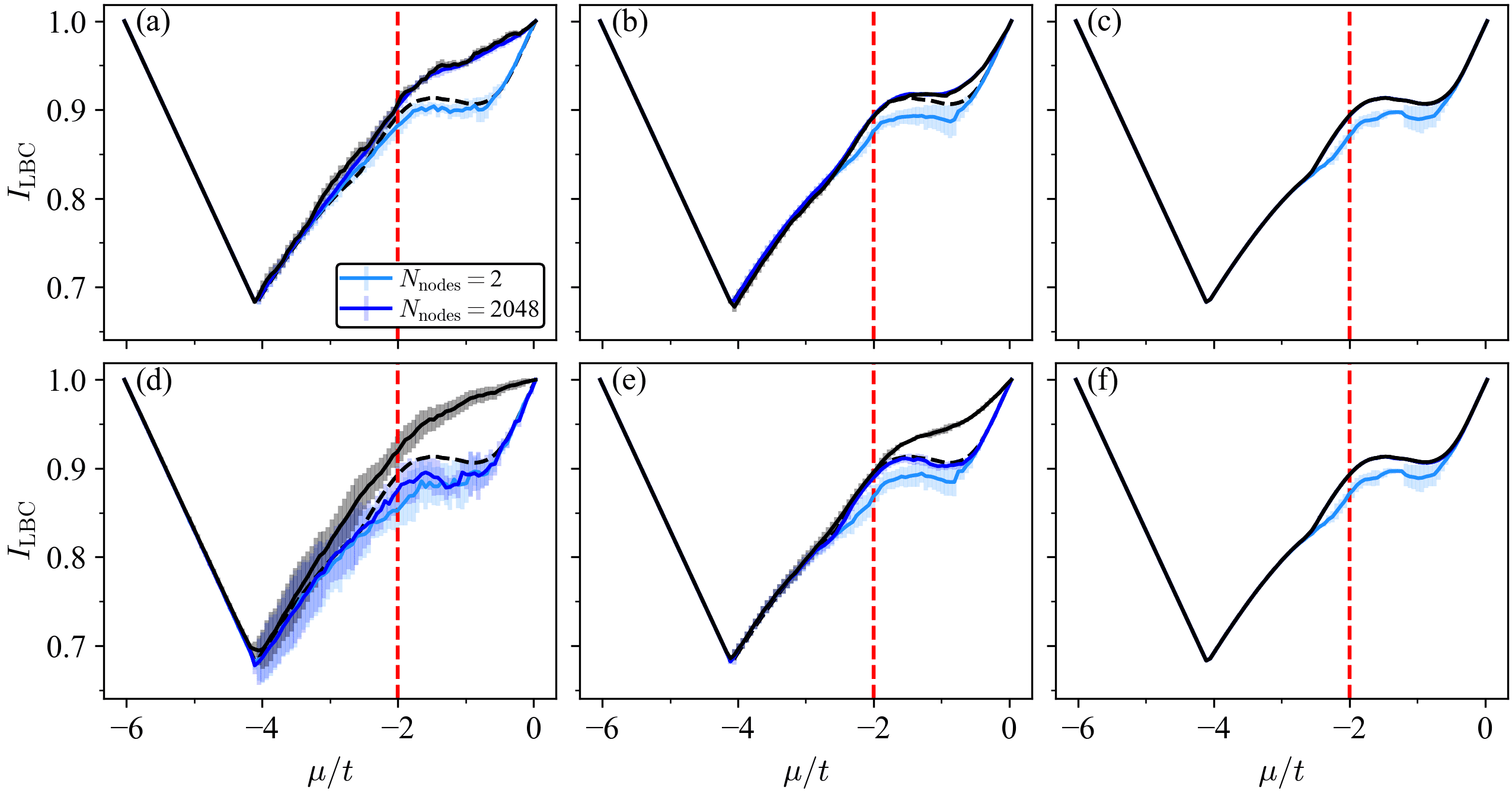}
		\caption{Results of LBC for the Kitaev chain ($L=10$) using NNs composed of a single hidden layer with 2 or 2048 nodes for various numbers of training samples $M_{\rm train}$ per parameter value, where $M_{\rm valid} = M_{\rm test} = M_{\rm train}/5$. The tuning parameter $p=\mu/t$ ranges from $p_{1}=-6$ to $p_{K}=0$ with $\Delta p = 0.06$. The critical value $\mu_{\rm c}/t= - 2$ is highlighted by a red-dashed line. The optimal indicator obtained based on the corresponding data set or the ground-truth probability distributions is highlighted by a black solid or dashed line, respectively. (a)-(c) Indicator $I_{\rm LBC}$ of LBC evaluated on the training set for (a) $M_{\rm train} = 10$, (b) $M_{\rm train} = 10^{2}$, and (c) $M_{\rm train} = 10^{5}$, where the NN-based predictions are obtained after training.  (d)-(f) Indicator $I_{\rm LBC}$ of LBC evaluated on the test set for for (a) $M_{\rm train} = 10$, (b) $M_{\rm train} = 10^{2}$, and (c) $M_{\rm train} = 10^{5}$, where early stopping is performed by minimizing the validation loss. Similar results are obtained when evaluating the NNs at the end of training instead. Each NN is trained for 10000 epochs with a learning rate of 0.005. The results averaged over 10 independent data sets and the error bars are given by the standard deviation.}
		\label{fig_NN_statistics_Kitaev_LBC}
	\end{center}
\end{figure*}

\indent Lastly, we can also control the capacity of our model through explicit $\ell_{2}$ regularization~\cite{goodfellow:2016}
\begin{equation}
    \mathcal{L} \rightarrow \mathcal{L} + \lambda_{\ell 2} \sum_{i} \theta_{i}^2,
\end{equation}
where the sum runs over all tunable parameters $\theta_{i}$ of the NN and $\lambda_{\ell 2}$ is the regularization strength. Figure~\ref{fig_ising_NN_PBM_regularization} shows the NN-based predictions and indicators of PBM for the Ising model after training with various regularization strengths. At large regularization strength, the resulting model cannot resolve any structure leading to a flat indicator signal. At an intermediate regularization strength, the resulting model can distinguish between the two phases leading to a clear peak in the indicator signal at the critical temperature of the Ising model. As the regularization strength is decreased further, the resulting model becomes more complex and converges towards the optimal model that minimizes the loss function in the absence of regularization. Consequently, the predictions and indicators converge towards the optimal predictions and indicator. In the Ising model, we thus find that explicit regularization helps to construct a model of intermediate resolution whose indicator correctly highlights the critical temperature (similarly for SL). However, as mentioned above, models with restricted capacity may not always highlight the critical value of the tuning parameter correctly. In the IGT, for example, the indicator of regularized NNs tends to display an erroneous peak similar to the specific heat, see Fig.~\ref{fig:IGT}. 

\subsection{Finite-sample statistics: Splitting data into training, validation, and test sets}\label{app_B2}
Here, we investigate NN-based predictions and indicators in the case where only a limited amount of data is available. In particular, we discuss the effect of splitting the data into a training, validation, and test set. Recall that in the limit of sufficient data, the training, validation, and test set will coincide as they are all sampled independently from the same probability distribution underlying the physical system, see Sec.~\ref{sec_methods}. Therefore, in the limit of sufficient data the training, validation, and test losses will decrease in lockstep during training. This is illustrated in Fig.~\ref{fig_NN_statistics_PBM_Ising} which shows the training, validation, and test loss of PBM for the Kitaev chain for different data set sizes. For small data sets, the training, validation, and test sets can differ, resulting in differing training, validation, and test losses. In particular, one can observe a characteristic increase of the validation loss after a certain time period attributed to overfitting~\cite{goodfellow:2016}. This allows one to perform early stopping such that the minimum in the validation loss is realized~\cite{goodfellow:2016}. Note that the location of the minima in the validation loss coincides with the kink in the corresponding training loss. The sharp local minimum in the validation loss fades as the data set size is increased further, leaving only the corresponding kinks in the training loss as a signal for early stopping. The latter situation has been discussed in Appendix~\ref{app_B1}. Therefore, a splitting into training, validation, and test set may allow for a clearer signal to perform early stopping given a small data set.\\

\indent Another effect arising when a limited amount of data is available and finite-sample statistics play a role is best illustrated by investigating the Kitaev chain using LBC. Figure~\ref{fig_NN_statistics_Kitaev_LBC} shows the NN-based indicator signal of LBC obtained for training, test, and validation sets of various sizes. For small data set sizes [see Fig.~\ref{fig_NN_statistics_Kitaev_LBC}(a),(d)] the optimal indicator (black, solid) shows no local maximum due to the negligible overlap in the inferred probability distribution. The NN-based indicator of a sufficiently large NN closely matches the optimal indicator on the training set after training [Fig.~\ref{fig_NN_statistics_Kitaev_LBC}(a)], whereas a small NN is incapable of approximating the optimal indicator on the training set. However, interestingly the indicator signal of the small NN qualitatively matches the optimal indicator signal based on the ground-truth probability distributions. In particular, it features a local maximum allowing for an estimate of the critical value of the tuning parameter to be obtained. This is another example illustrating how simple models can lead to sharp indicator signals. While the inferred probability distribution only has a marginal overlap in the topological phase resulting in the absence of a local maximum in the optimal indicator signal (black), the data may be partially indistinguishable to a simple model. This illustrates how ``confusion'' can also arise due to models with restricted expressivity (see Sec.~\ref{sec_optimal_indicators}).
The same phenomenon can also be observed for the indicator signal of the large NN evaluated on the test set (or validation set), see Fig.~\ref{fig_NN_statistics_Kitaev_LBC}(d). Here, the confusion arises because the predictions for the unseen data within the validation and test set are sub-optimal. In the future, it will be of interest to investigate whether this effect can be mimicked through appropriate interpolation of the optimal predictions~\cite{jacot:2018,greplova:2020,huang:2021}. Figures~\ref{fig_NN_statistics_Kitaev_LBC}(b),(e) and Figs.~\ref{fig_NN_statistics_Kitaev_LBC}(c),(f) show how the discrepancy between the optimal indicator signal based on a finite data set and the NN-based indicator vanishes for the large NN as the data set size increases. This arises because eventually the training, validation, and test sets become indistinguishable. Note, however, that the discrepancy persists for the small NN.

\section{Data generation}\label{app_C}
In this appendix, we provide further details on the data-generation process for each of the physical systems analyzed in the main text (see Sec.~\ref{sec_physical_models}). For the classical systems, given by the Ising model, IGT, and XY model, we use the Metropolis-Hastings algorithm~\cite{metropolis:1953} to sample spin configurations from the thermal distribution at a given temperature $T$. The lattice is initialized in a state with all spins pointing up for the Ising model, and a random spin configuration in the case of the IGT and XY model. The lattice is updated by drawing a random spin, which is flipped with probability ${\rm min}(1, e^{-\Delta E}/T)$, where $\Delta E$ is the energy difference resulting from the considered flip. In the XY model, instead of flipping a given spin, we add a perturbation $\Delta \theta \in [-\pi, \pi]$, which is drawn uniformly at random. To ensure that the systems are sufficiently thermalized, we sweep the complete lattice $10^5$ times, where each lattice site is updated once per sweep. After the thermalization period, we collect $10^{5}$ samples, which we find to be sufficient for achieving convergence (see Appendix~\ref{app_A5}). In the Ising model and IGT, we increase the temperature gradually, whereas it is decreased in the XY model.\\
\begin{figure}[t!]
\centering
\includegraphics[width=0.8\linewidth]{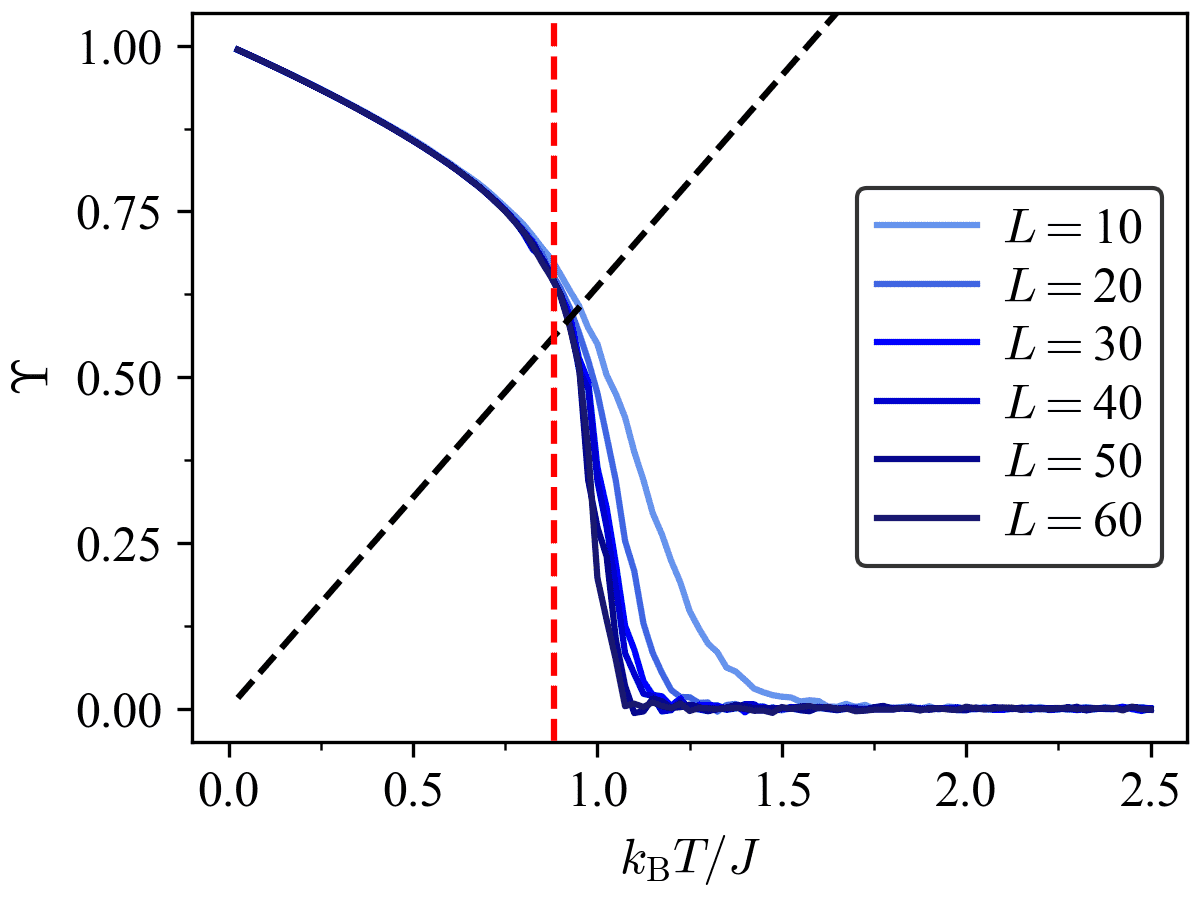}
\caption{Helicity modulus $\Upsilon$ as a function of the tuning parameter $p=k_{\rm B}T/J$ for the two-dimensional XY model for various lattice sizes. The value of the BKT transition point from literature $k_{\rm B} T_{\rm c}/J\approx 0.8935$~\cite{hsieh:2013} is highlighted by a red-dashed line. The estimated transition point based on our Monte Carlo samples at finite size corresponds to the point at which the helicity modulus crosses the line given by $\frac{2k_{\rm B}T}{J\pi}$ (black-dashed line).}
\label{fig:XY_helicity}
\end{figure}
\begin{figure*}[t!]
\centering
\includegraphics[width=\linewidth]{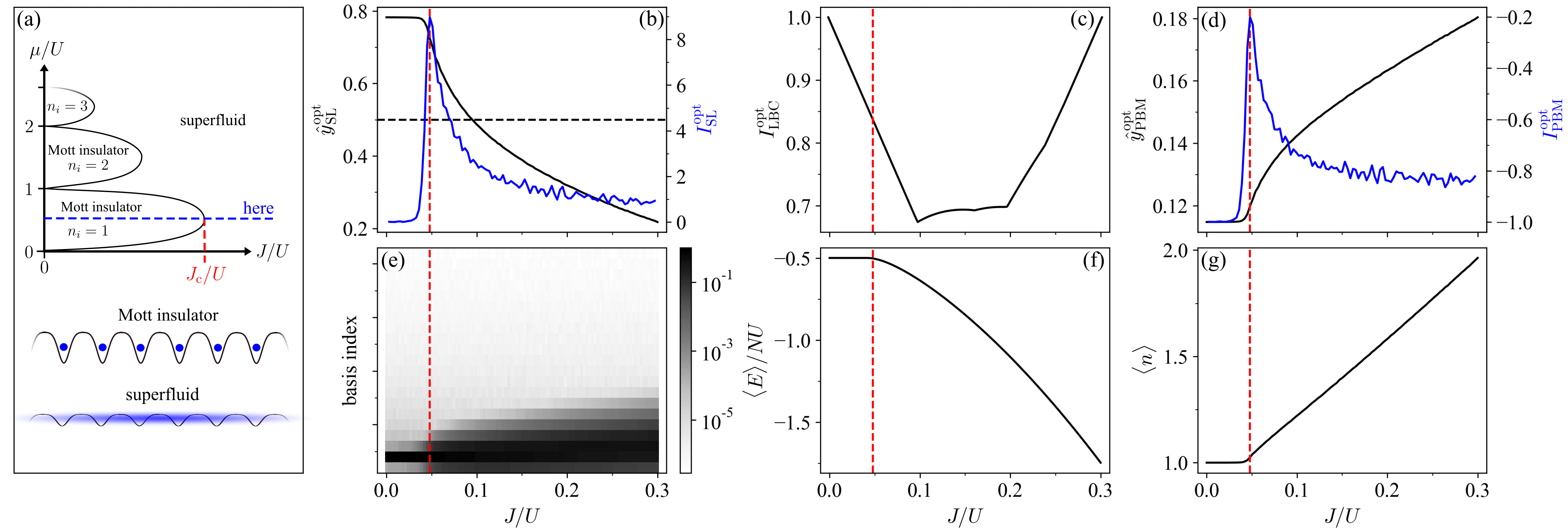}
\caption{Results for the Mott insulating to superfluid phase transition in the (two-dimensional) Bose-Hubbard model with the dimensionless coupling strength as a tuning parameter $p=J/U$ ranging from $p_{1}=0$ to $p_{K}=0.3$ in steps of $\Delta p = 0.03$, where $\mu/U=0.5$. In SL, the data obtained at $p_{1}$ and $p_{K}$ constitutes our training set, i.e., $r_{\rm I}=1$ and $l_{\rm II}=K$. The reference value for the critical value of the tuning parameter $J_{\rm c}/U= 1/(5.8z)$ with $z=4$~\cite{zwerger:2003} is highlighted by a red-dashed line. (a) Illustration of the two-dimensional phase diagram of the Bose-Hubbard model containing three Mott lobes. Here, we analyze the quantum phase transition from a Mott insulating state to a superfluid state occurring at the tip of the first Mott lobe ($\mu/U=0.5$). A sketch of the two distinct phases is shown on the bottom. (b) Mean optimal prediction $\hat{y}_{\rm SL}^{\rm opt}$ in SL (black, solid) and the corresponding indicator $I_{\rm SL}^{\rm opt}$ (blue). The value $\hat{y}_{\rm SL}^{\rm opt}=0.5$ is highlighted by a black-dashed line. (c) Optimal indicator of LBC, $I_{\rm LBC}^{\rm opt}$ (black). (d) Mean optimal prediction $\hat{y}_{\rm PBM}^{\rm opt}$ in PBM (black) and the corresponding indicator $I_{\rm PBM}^{\rm opt}$ (blue). (e) Probability distributions governing the input data (indices of Fock basis states $\{| n_{i} \rangle \}_{i=1}^{n_{\rm max}}$) as a function of tuning parameter, where the color scale denotes the probability. (f) Average energy per site ($N$ sites in total) as a function of the tuning parameter. Notice the drop in the average energy as the system undergoes the quantum phase transition. (g) Average occupation number per site $\langle n \rangle$ as a function of the tuning parameter.}
\label{fig:BH_Mott}
\end{figure*}

\indent In the XY model, we can further validate the quality of the Monte Carlo samples by estimating the BKT transition point. One way to do this is to determine the temperature at which the helicity modulus $\Upsilon$ crosses $2T/\pi$~\cite{himbergen:1981,minnhagen:2003}. The helicity modulus is also referred to as spin stiffness or spin rigidity and measures the response of the system to an in-plane twist of the spins. We find that the estimated BKT transition point based on our samples matches the literature value well, see Fig.~\ref{fig:XY_helicity}. Note that in the XY model, the angle of each spin can take on any value $\theta \in [0,2\pi]$. This results in a continuum of states. Hence, we discretize the energy in practice, which serves as an input for the ML methods. This discretization eases computation and, more crucially, results in overlapping probability distributions given finite-sample statistics (see case 3 in Sec.~\ref{sec_application_trivial_distr}). The discretization is performed through simple histogram-binning using $1000$ bins of equal size. The number of bins was increased systematically until a convergence of the optimal indicator signals was observed. In future works, histogram-binning may be replaced by more elaborate techniques for density estimation~\cite{bishop:2006,wu:2019,melko:2019,nicoli:2021}.\\

\indent Let us move on to the quantum case. To perform exact diagonalization and solve the Schrödinger equation, we use the QuSpin package~\cite{quspin:2017,quspin:2019} in Python. Note that when computing the ground state of the Kitaev chain through exact diagonalization, we restrict ourselves to the even-particle sector whose corresponding ground state has a lower energy within the topologically trivial phase. In the topological phase, the ground state is doubly degenerate, and the two states can be distinguished by their fermionic parity. This is because of the presence of the pairing term in the Kitaev chain Hamiltonian [Eq.~\eqref{eq:kitaev_H}]. As a consequence, $H$ does not conserve the total fermion number $N_{f}=\sum_{i=1}^{L} n_{i}$, i.e., $[H,N_{f}]\neq 0$. However, the fermion number modulo 2 is conserved, $[H,(-1)^{N_{f}} ] = 0$~\cite{katsura:2015}.

\section{Comparison to other works}\label{app_D}
In this appendix, we provide additional material which facilitates the comparison to other
works.
\begin{figure}[h!]
\centering
\includegraphics[width=0.8\linewidth]{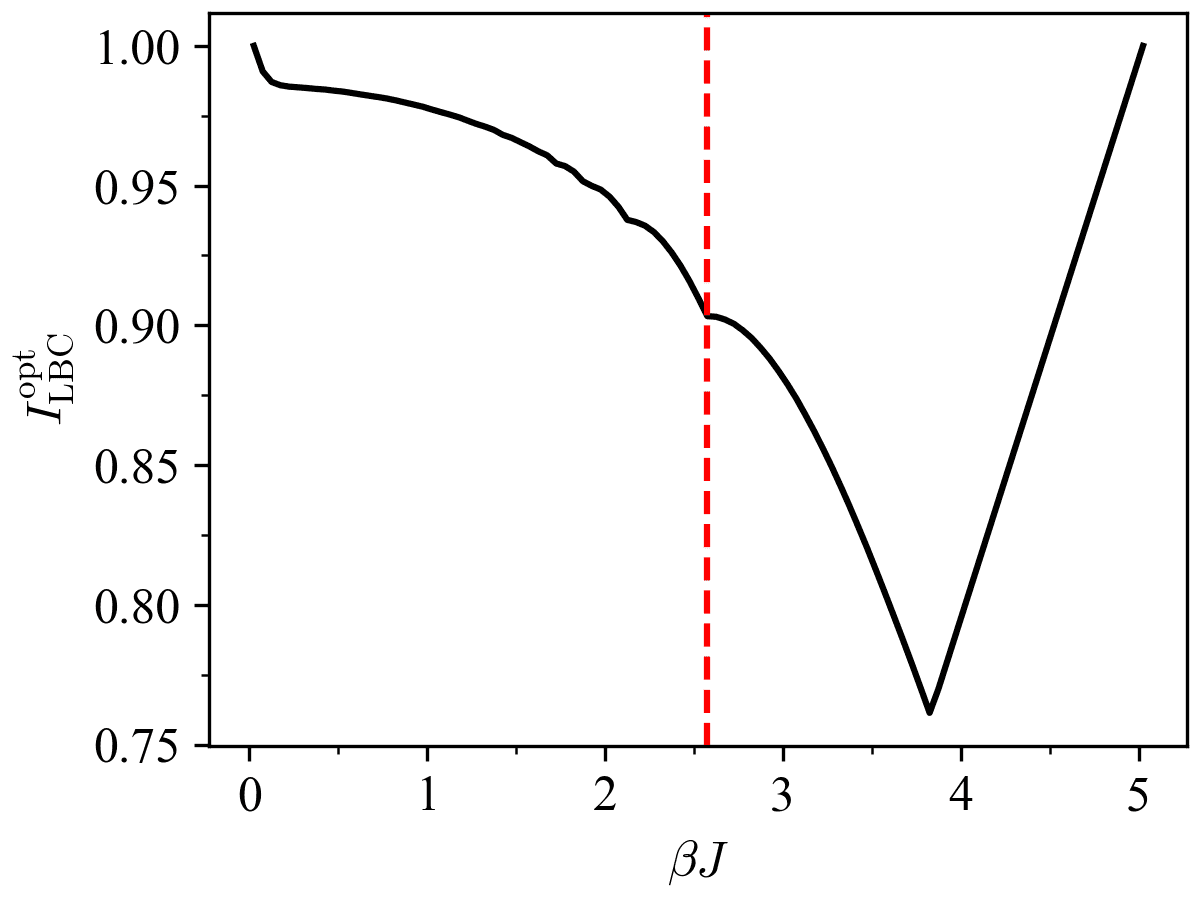}
\caption{Optimal indicator of LBC for the IGT ($L=12$) with dimensionless inverse temperature $p=\beta J $ as a tuning parameter, where $p_{1}=0.05$, $p_{K} = 5$, and $\Delta p = 0.05$. The critical value of the tuning parameter $p_{\rm c}=\beta_{\rm c} J $ from Fig.~\ref{fig:IGT} is highlighted in red.}
\label{fig:IGT_beta}
\end{figure}
\begin{figure*}[bht!]
\centering
\includegraphics[width=0.8\linewidth]{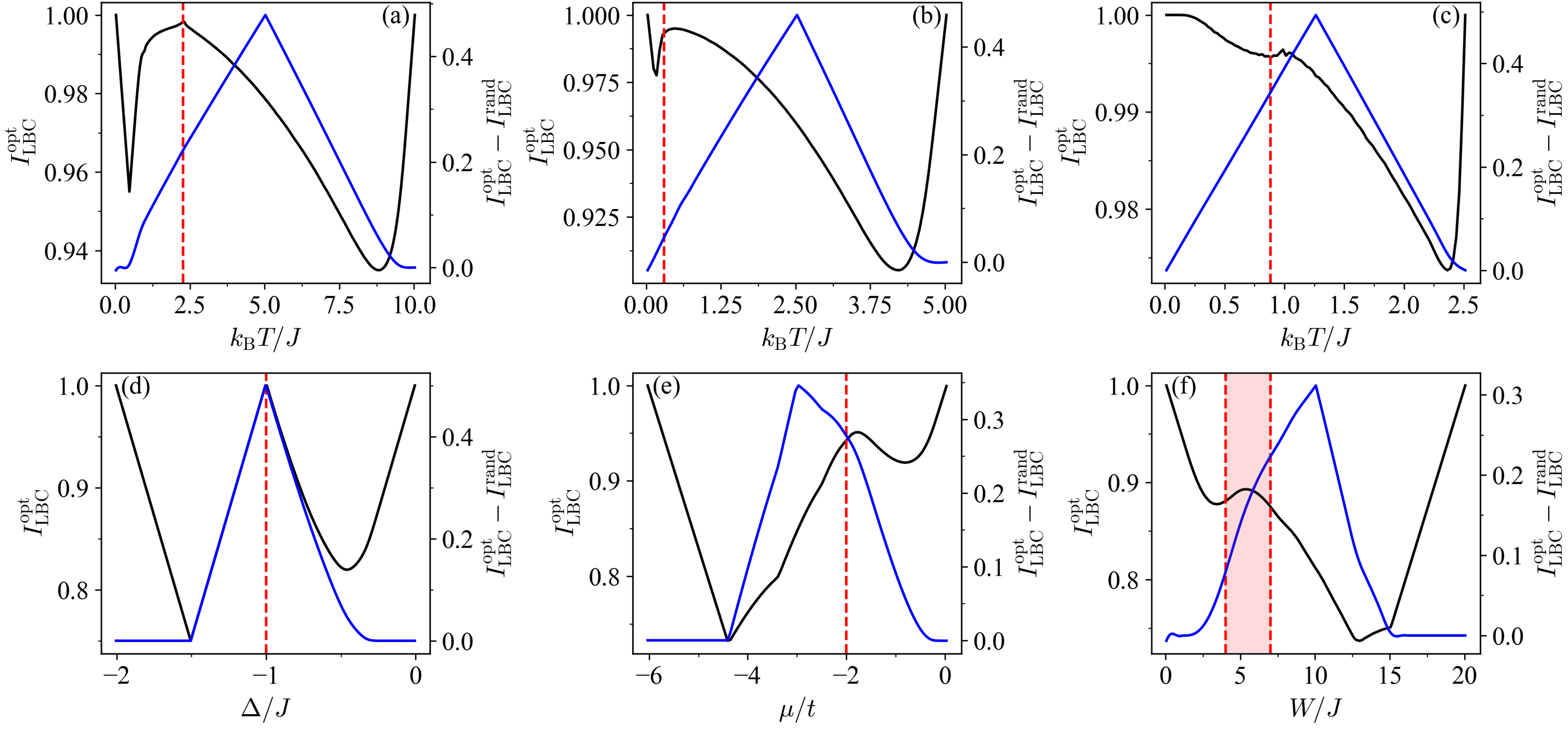}
\caption{Optimal indicator of LBC before (black) and after (blue) background subtraction for the (a) Ising model ($L=60$), (b) IGT ($L=28$), (c) XY model ($L=60$), (d) XXZ chain ($L=14$), (e) Kitaev chain ($L=20$), and (f) Bose-Hubbard model ($L=8$), see Sec.~\ref{sec_physical_models} in main text. The corresponding critical values of the tuning parameters are highlighted in red.}
\label{fig:LBC_background}
\end{figure*}
\subsection{Alternative approach towards supervised learning}\label{app_D1}
Here, we review our approach to SL (see Sec.~\ref{sec_methods_SL}) and put it into context. In Ref.~\cite{carrasquilla:2017}, the authors originally proposed to identify the estimated critical value of the tuning parameter in SL as $\argmin_{p_{k}} |\hat{y}(p_{k})-0.5|$ . In all systems analyzed in the main text (see Sec.~\ref{sec_physical_models}), this yields similar results compared to our approach based on identifying the peak location of the mean prediction's derivative [Eq.~\eqref{eq:SL_indicator}]. Note that the latter approach has, e.g., already been mentioned as an alternative in Ref.~\cite{broecker:2017}. Looking at Fig.~\ref{fig:MBL_BH_quench}(b), we observe that these two procedures would yield slightly different estimated critical values for the MBL phase transition. This discrepancy is even more prominent for the Mott insulator to superfluid transition in the Bose-Hubbard model. Here, we investigate the two-dimensional Bose-Hubbard model whose Hamiltonian is given by
\begin{equation}
    H = -J \sum_{\langle ij\rangle} (b_{i}^{\dagger}b_{j}^{\phantom{\dagger}} + {\rm h.c.}) + \sum_{i} \frac{U}{2} n_{i} (n_{i}-1) - \mu n_{i},
\end{equation}
where $J$ is the nearest-neighbor hopping strength, $U$ is the on-site interaction strength, and $\mu$ is the chemical potential. This model undergoes a quantum phase transition at zero temperature from a Mott insulating phase to a superfluid phase as the tuning parameter $J/U$ is increased at a fixed chemical potential. This gives rise to the characteristic Mott lobes~\cite{fisher:1989,jaksch:1998}, see Fig.~\ref{fig:BH_Mott}(a).\\

\indent We perform mean-field calculations based on a Gutzwiller ansatz in which the ground-state wave function is written as a product state
\begin{equation}
|\Psi_{\rm MF}\rangle = \prod_{i} |\phi_{i}\rangle
\end{equation}
with
\begin{equation}
|\phi_{i}\rangle = \sum_{n=0}^{n_{\rm max}} f_{n}|n_{i} \rangle ,
\end{equation}
where $|n_{i} \rangle$ denotes the Fock state with $n$ bosons at site $i$~\cite{krauth:1992}. We minimize the expectation value of the Hamiltonian with respect to the Gutzwiller coefficients $\{ |f_{n}|^2\}_{n=0}^{n_{\rm max}}$ by means of simulated annealing~\cite{comparin:2017,huembeli:2018} with a maximum number of bosons per site of $n_{\rm max}=20$. As such, the Gutzwiller coefficients $\{ |f_{n}|^2\}_{n=0}^{n_{\rm max}}$ represent the relevant probability distributions governing the data. Note that the simulated annealing algorithm can get stuck in local energy minima. To counteract this noise, we average the Gutzwiller coefficients obtained from 500 independent simulated annealing runs.\\

\indent At the tip of the first Mott lobe ($\mu/U=0.5$) the phase transition occurs at $J_{\rm c}/U = 1/(5.8z)$ [see Fig.~\ref{fig:BH_Mott}(a)], where $z$ is the coordination number (here $z=4$)~\cite{zwerger:2003}. The phase transition can be revealed by looking at the average boson number per site $\langle n\rangle$, see Fig.~\ref{fig:BH_Mott}(g). The Mott insulator is characterized by an integer density enforced by the Mott energy gap $\propto U$. As a result of the energy gap, the Mott insulator is incompressible. In contrast, the superfluid phase is compressible and is characterized by strong number fluctuations (even at low temperature).\\

\indent Figure~\ref{fig:BH_Mott} shows the results of SL, LBC, and PBM. Here, both SL and PBM correctly identify the quantum phase transition, whereas LBC fails. Looking at Fig.~\ref{fig:BH_Mott}(e), we see that a large change in the underlying probability distributions occurs at the quantum phase transition. In Ref.~\cite{liu:2018}, the Mott insulating to superfluid transition in the Bose-Hubbard model was correctly highlighted using LBC with NNs. However, in this case, the Gutzwiller coefficients directly served as input, whereas here the individual Fock basis states (i.e., their indices) constitute the input. Note that the phase transition would not be predicted with a high accuracy using SL if we estimated the predicted critical temperature as the value of the tuning parameter for which $\hat{y}_{\rm SL}^{\rm opt}=0.5$, see black-dashed line in Fig.~\ref{fig:BH_Mott}. This motivates our approach to SL compared to the procedure originally proposed in Ref.~\cite{carrasquilla:2017}. However, both approaches for obtaining estimated critical values are directly applicable given optimal predictions.

\subsection{Analysis of Ising gauge theory}\label{app_D2}
Figure~\ref{fig:IGT_beta} shows the optimal indicator of LBC for the IGT with inverse temperature $\beta$ as a tuning parameter. The signal qualitatively matches the indicator of LBC reported in Fig. C1 of Ref.~\cite{greplova:2020} obtained with NNs, confirming that for high capacity models the indicator signal of LBC is indeed ambiguous in this case.\\ 

\indent In Ref.~\cite{greplova:2020}, the authors also investigated the IGT with PBM using NNs. They empirically find that the NN-based predictions agree well with a physical model based on the underlying density of states, which was proposed in an \textit{ad hoc} fashion guided by physical intuition. In our work, we explicitly confirm this physical intuition on what the NN learns by proving that the optimal prediction of PBM for a given configuration in the IGT corresponds to the most likely tuning parameter value based on the underlying Boltzmann distribution.

\subsection{Background subtraction for learning by confusion}\label{app_D3}
\indent Figure~\ref{fig:LBC_background} shows the optimal indicator in LBC for all physical systems considered in the main text, as well as a modified version where the V-shaped indicator signal characteristic of indistinguishable data is subtracted. Note that this V-shaped indicator signal is computed separately for each system, i.e., parameter range. For all systems, we find that the modified indicator peaks near the center of the parameter range under consideration, whereas the original indicator signal peaks near the phase transition (red-dashed line). This bias arises because the subtracted signal is lowest near the center of the parameter range. As such, the bias can be easily missed if the transition point is indeed located in the center of the chosen parameter range, see Fig.~\ref{fig:LBC_background}(d).

\bibliography{refs}

\begin{thebibliography}{149}%
\makeatletter
\providecommand \@ifxundefined [1]{%
 \@ifx{#1\undefined}
}%
\providecommand \@ifnum [1]{%
 \ifnum #1\expandafter \@firstoftwo
 \else \expandafter \@secondoftwo
 \fi
}%
\providecommand \@ifx [1]{%
 \ifx #1\expandafter \@firstoftwo
 \else \expandafter \@secondoftwo
 \fi
}%
\providecommand \natexlab [1]{#1}%
\providecommand \enquote  [1]{``#1''}%
\providecommand \bibnamefont  [1]{#1}%
\providecommand \bibfnamefont [1]{#1}%
\providecommand \citenamefont [1]{#1}%
\providecommand \href@noop [0]{\@secondoftwo}%
\providecommand \href [0]{\begingroup \@sanitize@url \@href}%
\providecommand \@href[1]{\@@startlink{#1}\@@href}%
\providecommand \@@href[1]{\endgroup#1\@@endlink}%
\providecommand \@sanitize@url [0]{\catcode `\\12\catcode `\$12\catcode
  `\&12\catcode `\#12\catcode `\^12\catcode `\_12\catcode `\%12\relax}%
\providecommand \@@startlink[1]{}%
\providecommand \@@endlink[0]{}%
\providecommand \url  [0]{\begingroup\@sanitize@url \@url }%
\providecommand \@url [1]{\endgroup\@href {#1}{\urlprefix }}%
\providecommand \urlprefix  [0]{URL }%
\providecommand \Eprint [0]{\href }%
\providecommand \doibase [0]{https://doi.org/}%
\providecommand \selectlanguage [0]{\@gobble}%
\providecommand \bibinfo  [0]{\@secondoftwo}%
\providecommand \bibfield  [0]{\@secondoftwo}%
\providecommand \translation [1]{[#1]}%
\providecommand \BibitemOpen [0]{}%
\providecommand \bibitemStop [0]{}%
\providecommand \bibitemNoStop [0]{.\EOS\space}%
\providecommand \EOS [0]{\spacefactor3000\relax}%
\providecommand \BibitemShut  [1]{\csname bibitem#1\endcsname}%
\let\auto@bib@innerbib\@empty
\bibitem [{\citenamefont {Dunjko}\ and\ \citenamefont
  {Briegel}(2018)}]{dunjko_2018}%
  \BibitemOpen
  \bibfield  {author} {\bibinfo {author} {\bibfnamefont {V.}~\bibnamefont
  {Dunjko}}\ and\ \bibinfo {author} {\bibfnamefont {H.~J.}\ \bibnamefont
  {Briegel}},\ }\bibfield  {title} {\bibinfo {title} {Machine learning {\&}
  artificial intelligence in the quantum domain: a review of recent progress},\
  }\href {https://doi.org/10.1088/1361-6633/aab406} {\bibfield  {journal}
  {\bibinfo  {journal} {Rep. Prog. Phys.}\ }\textbf {\bibinfo {volume} {81}},\
  \bibinfo {pages} {074001} (\bibinfo {year} {2018})}\BibitemShut {NoStop}%
\bibitem [{\citenamefont {Carleo}\ \emph {et~al.}(2019)\citenamefont {Carleo},
  \citenamefont {Cirac}, \citenamefont {Cranmer}, \citenamefont {Daudet},
  \citenamefont {Schuld}, \citenamefont {Tishby}, \citenamefont
  {Vogt-Maranto},\ and\ \citenamefont {Zdeborov\'a}}]{carleo:2019}%
  \BibitemOpen
  \bibfield  {author} {\bibinfo {author} {\bibfnamefont {G.}~\bibnamefont
  {Carleo}}, \bibinfo {author} {\bibfnamefont {I.}~\bibnamefont {Cirac}},
  \bibinfo {author} {\bibfnamefont {K.}~\bibnamefont {Cranmer}}, \bibinfo
  {author} {\bibfnamefont {L.}~\bibnamefont {Daudet}}, \bibinfo {author}
  {\bibfnamefont {M.}~\bibnamefont {Schuld}}, \bibinfo {author} {\bibfnamefont
  {N.}~\bibnamefont {Tishby}}, \bibinfo {author} {\bibfnamefont
  {L.}~\bibnamefont {Vogt-Maranto}},\ and\ \bibinfo {author} {\bibfnamefont
  {L.}~\bibnamefont {Zdeborov\'a}},\ }\bibfield  {title} {\bibinfo {title}
  {Machine learning and the physical sciences},\ }\href
  {https://doi.org/10.1103/RevModPhys.91.045002} {\bibfield  {journal}
  {\bibinfo  {journal} {Rev. Mod. Phys.}\ }\textbf {\bibinfo {volume} {91}},\
  \bibinfo {pages} {045002} (\bibinfo {year} {2019})}\BibitemShut {NoStop}%
\bibitem [{\citenamefont {Dawid}\ \emph {et~al.}(2022)\citenamefont {Dawid},
  \citenamefont {Arnold}, \citenamefont {Requena}, \citenamefont {Gresch},
  \citenamefont {P{\l}odzie{\'n}}, \citenamefont {Donatella}, \citenamefont
  {Nicoli}, \citenamefont {Stornati}, \citenamefont {Koch}, \citenamefont
  {B{\"u}ttner} \emph {et~al.}}]{dawid:2022}%
  \BibitemOpen
  \bibfield  {author} {\bibinfo {author} {\bibfnamefont {A.}~\bibnamefont
  {Dawid}}, \bibinfo {author} {\bibfnamefont {J.}~\bibnamefont {Arnold}},
  \bibinfo {author} {\bibfnamefont {B.}~\bibnamefont {Requena}}, \bibinfo
  {author} {\bibfnamefont {A.}~\bibnamefont {Gresch}}, \bibinfo {author}
  {\bibfnamefont {M.}~\bibnamefont {P{\l}odzie{\'n}}}, \bibinfo {author}
  {\bibfnamefont {K.}~\bibnamefont {Donatella}}, \bibinfo {author}
  {\bibfnamefont {K.}~\bibnamefont {Nicoli}}, \bibinfo {author} {\bibfnamefont
  {P.}~\bibnamefont {Stornati}}, \bibinfo {author} {\bibfnamefont
  {R.}~\bibnamefont {Koch}}, \bibinfo {author} {\bibfnamefont {M.}~\bibnamefont
  {B{\"u}ttner}}, \emph {et~al.},\ }\bibfield  {title} {\bibinfo {title}
  {Modern applications of machine learning in quantum sciences},\ }\href
  {https://arxiv.org/abs/2204.04198} {\bibfield  {journal} {\bibinfo  {journal}
  {arXiv:2204.04198}\ } (\bibinfo {year} {2022})}\BibitemShut {NoStop}%
\bibitem [{\citenamefont {Carrasquilla}\ and\ \citenamefont
  {Melko}(2017)}]{carrasquilla:2017}%
  \BibitemOpen
  \bibfield  {author} {\bibinfo {author} {\bibfnamefont {J.}~\bibnamefont
  {Carrasquilla}}\ and\ \bibinfo {author} {\bibfnamefont {R.~G.}\ \bibnamefont
  {Melko}},\ }\bibfield  {title} {\bibinfo {title} {Machine learning phases of
  matter},\ }\href {https://doi.org/10.1038/nphys4035} {\bibfield  {journal}
  {\bibinfo  {journal} {Nat. Phys.}\ }\textbf {\bibinfo {volume} {13}},\
  \bibinfo {pages} {431} (\bibinfo {year} {2017})}\BibitemShut {NoStop}%
\bibitem [{\citenamefont {Van~Nieuwenburg}\ \emph {et~al.}(2017)\citenamefont
  {Van~Nieuwenburg}, \citenamefont {Liu},\ and\ \citenamefont
  {Huber}}]{van:2017}%
  \BibitemOpen
  \bibfield  {author} {\bibinfo {author} {\bibfnamefont {E.~P.}\ \bibnamefont
  {Van~Nieuwenburg}}, \bibinfo {author} {\bibfnamefont {Y.-H.}\ \bibnamefont
  {Liu}},\ and\ \bibinfo {author} {\bibfnamefont {S.~D.}\ \bibnamefont
  {Huber}},\ }\bibfield  {title} {\bibinfo {title} {Learning phase transitions
  by confusion},\ }\href {https://doi.org/10.1038/nphys4037} {\bibfield
  {journal} {\bibinfo  {journal} {Nat. Phys.}\ }\textbf {\bibinfo {volume}
  {13}},\ \bibinfo {pages} {435} (\bibinfo {year} {2017})}\BibitemShut
  {NoStop}%
\bibitem [{\citenamefont {Carrasquilla}(2020)}]{carrasquilla:2020}%
  \BibitemOpen
  \bibfield  {author} {\bibinfo {author} {\bibfnamefont {J.}~\bibnamefont
  {Carrasquilla}},\ }\bibfield  {title} {\bibinfo {title} {Machine learning for
  quantum matter},\ }\href {https://doi.org/10.1080/23746149.2020.1797528}
  {\bibfield  {journal} {\bibinfo  {journal} {Adv. Phys.: X}\ }\textbf
  {\bibinfo {volume} {5}},\ \bibinfo {pages} {1797528} (\bibinfo {year}
  {2020})}\BibitemShut {NoStop}%
\bibitem [{\citenamefont {Carrasquilla}\ and\ \citenamefont
  {Torlai}(2021)}]{carrasquilla:2021}%
  \BibitemOpen
  \bibfield  {author} {\bibinfo {author} {\bibfnamefont {J.}~\bibnamefont
  {Carrasquilla}}\ and\ \bibinfo {author} {\bibfnamefont {G.}~\bibnamefont
  {Torlai}},\ }\bibfield  {title} {\bibinfo {title} {{How To Use Neural
  Networks To Investigate Quantum Many-Body Physics}},\ }\href
  {https://doi.org/10.1103/PRXQuantum.2.040201} {\bibfield  {journal} {\bibinfo
   {journal} {PRX Quantum}\ }\textbf {\bibinfo {volume} {2}},\ \bibinfo {pages}
  {040201} (\bibinfo {year} {2021})}\BibitemShut {NoStop}%
\bibitem [{\citenamefont {Rem}\ \emph {et~al.}(2019)\citenamefont {Rem},
  \citenamefont {K{\"a}ming}, \citenamefont {Tarnowski}, \citenamefont
  {Asteria}, \citenamefont {Fl{\"a}schner}, \citenamefont {Becker},
  \citenamefont {Sengstock},\ and\ \citenamefont {Weitenberg}}]{rem:2019}%
  \BibitemOpen
  \bibfield  {author} {\bibinfo {author} {\bibfnamefont {B.~S.}\ \bibnamefont
  {Rem}}, \bibinfo {author} {\bibfnamefont {N.}~\bibnamefont {K{\"a}ming}},
  \bibinfo {author} {\bibfnamefont {M.}~\bibnamefont {Tarnowski}}, \bibinfo
  {author} {\bibfnamefont {L.}~\bibnamefont {Asteria}}, \bibinfo {author}
  {\bibfnamefont {N.}~\bibnamefont {Fl{\"a}schner}}, \bibinfo {author}
  {\bibfnamefont {C.}~\bibnamefont {Becker}}, \bibinfo {author} {\bibfnamefont
  {K.}~\bibnamefont {Sengstock}},\ and\ \bibinfo {author} {\bibfnamefont
  {C.}~\bibnamefont {Weitenberg}},\ }\bibfield  {title} {\bibinfo {title}
  {Identifying quantum phase transitions using artificial neural networks on
  experimental data},\ }\href {https://doi.org/10.1038/s41567-019-0554-0}
  {\bibfield  {journal} {\bibinfo  {journal} {Nat. Phys.}\ }\textbf {\bibinfo
  {volume} {15}},\ \bibinfo {pages} {917} (\bibinfo {year} {2019})}\BibitemShut
  {NoStop}%
\bibitem [{\citenamefont {K{\"a}ming}\ \emph {et~al.}(2021)\citenamefont
  {K{\"a}ming}, \citenamefont {Dawid}, \citenamefont {Kottmann}, \citenamefont
  {Lewenstein}, \citenamefont {Sengstock}, \citenamefont {Dauphin},\ and\
  \citenamefont {Weitenberg}}]{kaming:2021}%
  \BibitemOpen
  \bibfield  {author} {\bibinfo {author} {\bibfnamefont {N.}~\bibnamefont
  {K{\"a}ming}}, \bibinfo {author} {\bibfnamefont {A.}~\bibnamefont {Dawid}},
  \bibinfo {author} {\bibfnamefont {K.}~\bibnamefont {Kottmann}}, \bibinfo
  {author} {\bibfnamefont {M.}~\bibnamefont {Lewenstein}}, \bibinfo {author}
  {\bibfnamefont {K.}~\bibnamefont {Sengstock}}, \bibinfo {author}
  {\bibfnamefont {A.}~\bibnamefont {Dauphin}},\ and\ \bibinfo {author}
  {\bibfnamefont {C.}~\bibnamefont {Weitenberg}},\ }\bibfield  {title}
  {\bibinfo {title} {Unsupervised machine learning of topological phase
  transitions from experimental data},\ }\href
  {https://doi.org/10.1088/2632-2153/abffe7} {\bibfield  {journal} {\bibinfo
  {journal} {Mach. Learn.: Sci. Technol.}\ } (\bibinfo {year}
  {2021})}\BibitemShut {NoStop}%
\bibitem [{\citenamefont {Bohrdt}\ \emph {et~al.}(2021)\citenamefont {Bohrdt},
  \citenamefont {Kim}, \citenamefont {Lukin}, \citenamefont {Rispoli},
  \citenamefont {Schittko}, \citenamefont {Knap}, \citenamefont {Greiner},\
  and\ \citenamefont {L\'eonard}}]{bohrdt:2021}%
  \BibitemOpen
  \bibfield  {author} {\bibinfo {author} {\bibfnamefont {A.}~\bibnamefont
  {Bohrdt}}, \bibinfo {author} {\bibfnamefont {S.}~\bibnamefont {Kim}},
  \bibinfo {author} {\bibfnamefont {A.}~\bibnamefont {Lukin}}, \bibinfo
  {author} {\bibfnamefont {M.}~\bibnamefont {Rispoli}}, \bibinfo {author}
  {\bibfnamefont {R.}~\bibnamefont {Schittko}}, \bibinfo {author}
  {\bibfnamefont {M.}~\bibnamefont {Knap}}, \bibinfo {author} {\bibfnamefont
  {M.}~\bibnamefont {Greiner}},\ and\ \bibinfo {author} {\bibfnamefont
  {J.}~\bibnamefont {L\'eonard}},\ }\bibfield  {title} {\bibinfo {title}
  {Analyzing {N}onequilibrium {Q}uantum {S}tates through {S}napshots with
  {A}rtificial {N}eural {N}etworks},\ }\href
  {https://doi.org/10.1103/PhysRevLett.127.150504} {\bibfield  {journal}
  {\bibinfo  {journal} {Phys. Rev. Lett.}\ }\textbf {\bibinfo {volume} {127}},\
  \bibinfo {pages} {150504} (\bibinfo {year} {2021})}\BibitemShut {NoStop}%
\bibitem [{\citenamefont {Miles}\ \emph
  {et~al.}(2021{\natexlab{a}})\citenamefont {Miles}, \citenamefont {Samajdar},
  \citenamefont {Ebadi}, \citenamefont {Wang}, \citenamefont {Pichler},
  \citenamefont {Sachdev}, \citenamefont {Lukin}, \citenamefont {Greiner},
  \citenamefont {Weinberger},\ and\ \citenamefont {Kim}}]{miles2:2021}%
  \BibitemOpen
  \bibfield  {author} {\bibinfo {author} {\bibfnamefont {C.}~\bibnamefont
  {Miles}}, \bibinfo {author} {\bibfnamefont {R.}~\bibnamefont {Samajdar}},
  \bibinfo {author} {\bibfnamefont {S.}~\bibnamefont {Ebadi}}, \bibinfo
  {author} {\bibfnamefont {T.~T.}\ \bibnamefont {Wang}}, \bibinfo {author}
  {\bibfnamefont {H.}~\bibnamefont {Pichler}}, \bibinfo {author} {\bibfnamefont
  {S.}~\bibnamefont {Sachdev}}, \bibinfo {author} {\bibfnamefont {M.~D.}\
  \bibnamefont {Lukin}}, \bibinfo {author} {\bibfnamefont {M.}~\bibnamefont
  {Greiner}}, \bibinfo {author} {\bibfnamefont {K.~Q.}\ \bibnamefont
  {Weinberger}},\ and\ \bibinfo {author} {\bibfnamefont {E.-A.}\ \bibnamefont
  {Kim}},\ }\bibfield  {title} {\bibinfo {title} {Machine learning discovery of
  new phases in programmable quantum simulator snapshots},\ }\href
  {https://arxiv.org/abs/2112.10789} {\bibfield  {journal} {\bibinfo  {journal}
  {arXiv:2112.10789}\ } (\bibinfo {year} {2021}{\natexlab{a}})}\BibitemShut
  {NoStop}%
\bibitem [{\citenamefont {Yu}\ \emph {et~al.}(2021)\citenamefont {Yu},
  \citenamefont {Yu}, \citenamefont {Zhang}, \citenamefont {Zhang},
  \citenamefont {Ouyang}, \citenamefont {Liu}, \citenamefont {Deng},\ and\
  \citenamefont {Duan}}]{yu:2021}%
  \BibitemOpen
  \bibfield  {author} {\bibinfo {author} {\bibfnamefont {Y.}~\bibnamefont
  {Yu}}, \bibinfo {author} {\bibfnamefont {L.-W.}\ \bibnamefont {Yu}}, \bibinfo
  {author} {\bibfnamefont {W.}~\bibnamefont {Zhang}}, \bibinfo {author}
  {\bibfnamefont {H.}~\bibnamefont {Zhang}}, \bibinfo {author} {\bibfnamefont
  {X.}~\bibnamefont {Ouyang}}, \bibinfo {author} {\bibfnamefont
  {Y.}~\bibnamefont {Liu}}, \bibinfo {author} {\bibfnamefont {D.-L.}\
  \bibnamefont {Deng}},\ and\ \bibinfo {author} {\bibfnamefont {L.-M.}\
  \bibnamefont {Duan}},\ }\bibfield  {title} {\bibinfo {title} {Experimental
  unsupervised learning of non-hermitian knotted phases with solid-state
  spins},\ }\href {https://arxiv.org/abs/2112.13785} {\bibfield  {journal}
  {\bibinfo  {journal} {arXiv:2112.13785}\ } (\bibinfo {year}
  {2021})}\BibitemShut {NoStop}%
\bibitem [{\citenamefont {Wang}(2016)}]{wang:2016}%
  \BibitemOpen
  \bibfield  {author} {\bibinfo {author} {\bibfnamefont {L.}~\bibnamefont
  {Wang}},\ }\bibfield  {title} {\bibinfo {title} {Discovering phase
  transitions with unsupervised learning},\ }\href
  {https://doi.org/10.1103/PhysRevB.94.195105} {\bibfield  {journal} {\bibinfo
  {journal} {Phys. Rev. B}\ }\textbf {\bibinfo {volume} {94}},\ \bibinfo
  {pages} {195105} (\bibinfo {year} {2016})}\BibitemShut {NoStop}%
\bibitem [{\citenamefont {Wetzel}(2017)}]{wetzel1:2017}%
  \BibitemOpen
  \bibfield  {author} {\bibinfo {author} {\bibfnamefont {S.~J.}\ \bibnamefont
  {Wetzel}},\ }\bibfield  {title} {\bibinfo {title} {Unsupervised learning of
  phase transitions: {F}rom principal component analysis to variational
  autoencoders},\ }\href {https://doi.org/10.1103/PhysRevE.96.022140}
  {\bibfield  {journal} {\bibinfo  {journal} {Phys. Rev. E}\ }\textbf {\bibinfo
  {volume} {96}},\ \bibinfo {pages} {022140} (\bibinfo {year}
  {2017})}\BibitemShut {NoStop}%
\bibitem [{\citenamefont {Wetzel}\ and\ \citenamefont
  {Scherzer}(2017)}]{wetzel2:2017}%
  \BibitemOpen
  \bibfield  {author} {\bibinfo {author} {\bibfnamefont {S.~J.}\ \bibnamefont
  {Wetzel}}\ and\ \bibinfo {author} {\bibfnamefont {M.}~\bibnamefont
  {Scherzer}},\ }\bibfield  {title} {\bibinfo {title} {Machine learning of
  explicit order parameters: {F}rom the {I}sing model to {SU}(2) lattice gauge
  theory},\ }\href {https://doi.org/10.1103/PhysRevB.96.184410} {\bibfield
  {journal} {\bibinfo  {journal} {Phys. Rev. B}\ }\textbf {\bibinfo {volume}
  {96}},\ \bibinfo {pages} {184410} (\bibinfo {year} {2017})}\BibitemShut
  {NoStop}%
\bibitem [{\citenamefont {Ch'ng}\ \emph {et~al.}(2017)\citenamefont {Ch'ng},
  \citenamefont {Carrasquilla}, \citenamefont {Melko},\ and\ \citenamefont
  {Khatami}}]{chng:2017}%
  \BibitemOpen
  \bibfield  {author} {\bibinfo {author} {\bibfnamefont {K.}~\bibnamefont
  {Ch'ng}}, \bibinfo {author} {\bibfnamefont {J.}~\bibnamefont {Carrasquilla}},
  \bibinfo {author} {\bibfnamefont {R.~G.}\ \bibnamefont {Melko}},\ and\
  \bibinfo {author} {\bibfnamefont {E.}~\bibnamefont {Khatami}},\ }\bibfield
  {title} {\bibinfo {title} {Machine {L}earning {P}hases of {S}trongly
  {C}orrelated {F}ermions},\ }\href {https://doi.org/10.1103/PhysRevX.7.031038}
  {\bibfield  {journal} {\bibinfo  {journal} {Phys. Rev. X}\ }\textbf {\bibinfo
  {volume} {7}},\ \bibinfo {pages} {031038} (\bibinfo {year}
  {2017})}\BibitemShut {NoStop}%
\bibitem [{\citenamefont {Ohtsuki}\ and\ \citenamefont
  {Ohtsuki}(2017)}]{ohtsuki:2017}%
  \BibitemOpen
  \bibfield  {author} {\bibinfo {author} {\bibfnamefont {T.}~\bibnamefont
  {Ohtsuki}}\ and\ \bibinfo {author} {\bibfnamefont {T.}~\bibnamefont
  {Ohtsuki}},\ }\bibfield  {title} {\bibinfo {title} {Deep learning the quantum
  phase transitions in random electron systems: {A}pplications to three
  dimensions},\ }\href {https://doi.org/10.7566/JPSJ.86.044708} {\bibfield
  {journal} {\bibinfo  {journal} {J. Phys. Soc. Jpn.}\ }\textbf {\bibinfo
  {volume} {86}},\ \bibinfo {pages} {044708} (\bibinfo {year}
  {2017})}\BibitemShut {NoStop}%
\bibitem [{\citenamefont {Schindler}\ \emph {et~al.}(2017)\citenamefont
  {Schindler}, \citenamefont {Regnault},\ and\ \citenamefont
  {Neupert}}]{schindler:2017}%
  \BibitemOpen
  \bibfield  {author} {\bibinfo {author} {\bibfnamefont {F.}~\bibnamefont
  {Schindler}}, \bibinfo {author} {\bibfnamefont {N.}~\bibnamefont
  {Regnault}},\ and\ \bibinfo {author} {\bibfnamefont {T.}~\bibnamefont
  {Neupert}},\ }\bibfield  {title} {\bibinfo {title} {Probing many-body
  localization with neural networks},\ }\href
  {https://doi.org/10.1103/PhysRevB.95.245134} {\bibfield  {journal} {\bibinfo
  {journal} {Phys. Rev. B}\ }\textbf {\bibinfo {volume} {95}},\ \bibinfo
  {pages} {245134} (\bibinfo {year} {2017})}\BibitemShut {NoStop}%
\bibitem [{\citenamefont {Zhang}\ and\ \citenamefont {Kim}(2017)}]{zhang:2017}%
  \BibitemOpen
  \bibfield  {author} {\bibinfo {author} {\bibfnamefont {Y.}~\bibnamefont
  {Zhang}}\ and\ \bibinfo {author} {\bibfnamefont {E.-A.}\ \bibnamefont
  {Kim}},\ }\bibfield  {title} {\bibinfo {title} {{Q}uantum {L}oop {T}opography
  for {M}achine {L}earning},\ }\href
  {https://doi.org/10.1103/PhysRevLett.118.216401} {\bibfield  {journal}
  {\bibinfo  {journal} {Phys. Rev. Lett.}\ }\textbf {\bibinfo {volume} {118}},\
  \bibinfo {pages} {216401} (\bibinfo {year} {2017})}\BibitemShut {NoStop}%
\bibitem [{\citenamefont {Broecker}\ \emph {et~al.}(2017)\citenamefont
  {Broecker}, \citenamefont {Carrasquilla}, \citenamefont {Melko},\ and\
  \citenamefont {Trebst}}]{broecker:2017}%
  \BibitemOpen
  \bibfield  {author} {\bibinfo {author} {\bibfnamefont {P.}~\bibnamefont
  {Broecker}}, \bibinfo {author} {\bibfnamefont {J.}~\bibnamefont
  {Carrasquilla}}, \bibinfo {author} {\bibfnamefont {R.~G.}\ \bibnamefont
  {Melko}},\ and\ \bibinfo {author} {\bibfnamefont {S.}~\bibnamefont
  {Trebst}},\ }\bibfield  {title} {\bibinfo {title} {Machine learning quantum
  phases of matter beyond the fermion sign problem},\ }\href
  {https://doi.org/10.1038/s41598-017-09098-0} {\bibfield  {journal} {\bibinfo
  {journal} {Sci. Rep.}\ }\textbf {\bibinfo {volume} {7}},\ \bibinfo {pages}
  {1} (\bibinfo {year} {2017})}\BibitemShut {NoStop}%
\bibitem [{\citenamefont {Huembeli}\ \emph {et~al.}(2018)\citenamefont
  {Huembeli}, \citenamefont {Dauphin},\ and\ \citenamefont
  {Wittek}}]{huembeli:2018}%
  \BibitemOpen
  \bibfield  {author} {\bibinfo {author} {\bibfnamefont {P.}~\bibnamefont
  {Huembeli}}, \bibinfo {author} {\bibfnamefont {A.}~\bibnamefont {Dauphin}},\
  and\ \bibinfo {author} {\bibfnamefont {P.}~\bibnamefont {Wittek}},\
  }\bibfield  {title} {\bibinfo {title} {Identifying quantum phase transitions
  with adversarial neural networks},\ }\href
  {https://doi.org/10.1103/PhysRevB.97.134109} {\bibfield  {journal} {\bibinfo
  {journal} {Phys. Rev. B}\ }\textbf {\bibinfo {volume} {97}},\ \bibinfo
  {pages} {134109} (\bibinfo {year} {2018})}\BibitemShut {NoStop}%
\bibitem [{\citenamefont {Liu}\ and\ \citenamefont {van
  Nieuwenburg}(2018)}]{liu:2018}%
  \BibitemOpen
  \bibfield  {author} {\bibinfo {author} {\bibfnamefont {Y.-H.}\ \bibnamefont
  {Liu}}\ and\ \bibinfo {author} {\bibfnamefont {E.~P.~L.}\ \bibnamefont {van
  Nieuwenburg}},\ }\bibfield  {title} {\bibinfo {title} {{D}iscriminative
  {C}ooperative {N}etworks for {D}etecting {P}hase {T}ransitions},\ }\href
  {https://doi.org/10.1103/PhysRevLett.120.176401} {\bibfield  {journal}
  {\bibinfo  {journal} {Phys. Rev. Lett.}\ }\textbf {\bibinfo {volume} {120}},\
  \bibinfo {pages} {176401} (\bibinfo {year} {2018})}\BibitemShut {NoStop}%
\bibitem [{\citenamefont {Beach}\ \emph {et~al.}(2018)\citenamefont {Beach},
  \citenamefont {Golubeva},\ and\ \citenamefont {Melko}}]{beach:2018}%
  \BibitemOpen
  \bibfield  {author} {\bibinfo {author} {\bibfnamefont {M.~J.~S.}\
  \bibnamefont {Beach}}, \bibinfo {author} {\bibfnamefont {A.}~\bibnamefont
  {Golubeva}},\ and\ \bibinfo {author} {\bibfnamefont {R.~G.}\ \bibnamefont
  {Melko}},\ }\bibfield  {title} {\bibinfo {title} {Machine learning vortices
  at the {K}osterlitz-{T}houless transition},\ }\href
  {https://doi.org/10.1103/PhysRevB.97.045207} {\bibfield  {journal} {\bibinfo
  {journal} {Phys. Rev. B}\ }\textbf {\bibinfo {volume} {97}},\ \bibinfo
  {pages} {045207} (\bibinfo {year} {2018})}\BibitemShut {NoStop}%
\bibitem [{\citenamefont {van Nieuwenburg}\ \emph {et~al.}(2018)\citenamefont
  {van Nieuwenburg}, \citenamefont {Bairey},\ and\ \citenamefont
  {Refael}}]{van:2018}%
  \BibitemOpen
  \bibfield  {author} {\bibinfo {author} {\bibfnamefont {E.}~\bibnamefont {van
  Nieuwenburg}}, \bibinfo {author} {\bibfnamefont {E.}~\bibnamefont {Bairey}},\
  and\ \bibinfo {author} {\bibfnamefont {G.}~\bibnamefont {Refael}},\
  }\bibfield  {title} {\bibinfo {title} {Learning phase transitions from
  dynamics},\ }\href {https://doi.org/10.1103/PhysRevB.98.060301} {\bibfield
  {journal} {\bibinfo  {journal} {Phys. Rev. B}\ }\textbf {\bibinfo {volume}
  {98}},\ \bibinfo {pages} {060301} (\bibinfo {year} {2018})}\BibitemShut
  {NoStop}%
\bibitem [{\citenamefont {Zhang}\ \emph {et~al.}(2018)\citenamefont {Zhang},
  \citenamefont {Shen},\ and\ \citenamefont {Zhai}}]{zhang:2018}%
  \BibitemOpen
  \bibfield  {author} {\bibinfo {author} {\bibfnamefont {P.}~\bibnamefont
  {Zhang}}, \bibinfo {author} {\bibfnamefont {H.}~\bibnamefont {Shen}},\ and\
  \bibinfo {author} {\bibfnamefont {H.}~\bibnamefont {Zhai}},\ }\bibfield
  {title} {\bibinfo {title} {Machine {L}earning {T}opological {I}nvariants with
  {N}eural {N}etworks},\ }\href
  {https://doi.org/10.1103/PhysRevLett.120.066401} {\bibfield  {journal}
  {\bibinfo  {journal} {Phys. Rev. Lett.}\ }\textbf {\bibinfo {volume} {120}},\
  \bibinfo {pages} {066401} (\bibinfo {year} {2018})}\BibitemShut {NoStop}%
\bibitem [{\citenamefont {Venderley}\ \emph {et~al.}(2018)\citenamefont
  {Venderley}, \citenamefont {Khemani},\ and\ \citenamefont
  {Kim}}]{venderley:2018}%
  \BibitemOpen
  \bibfield  {author} {\bibinfo {author} {\bibfnamefont {J.}~\bibnamefont
  {Venderley}}, \bibinfo {author} {\bibfnamefont {V.}~\bibnamefont {Khemani}},\
  and\ \bibinfo {author} {\bibfnamefont {E.-A.}\ \bibnamefont {Kim}},\
  }\bibfield  {title} {\bibinfo {title} {Machine learning out-of-equilibrium
  phases of matter},\ }\href {https://doi.org/10.1103/PhysRevLett.120.257204}
  {\bibfield  {journal} {\bibinfo  {journal} {Phys. Rev. Lett.}\ }\textbf
  {\bibinfo {volume} {120}},\ \bibinfo {pages} {257204} (\bibinfo {year}
  {2018})}\BibitemShut {NoStop}%
\bibitem [{\citenamefont {Rodriguez-Nieva}\ and\ \citenamefont
  {Scheurer}(2019)}]{rodriguez:2019}%
  \BibitemOpen
  \bibfield  {author} {\bibinfo {author} {\bibfnamefont {J.~F.}\ \bibnamefont
  {Rodriguez-Nieva}}\ and\ \bibinfo {author} {\bibfnamefont {M.~S.}\
  \bibnamefont {Scheurer}},\ }\bibfield  {title} {\bibinfo {title} {Identifying
  topological order through unsupervised machine learning},\ }\href
  {https://doi.org/10.1038/s41567-019-0512-x} {\bibfield  {journal} {\bibinfo
  {journal} {Nat. Phys.}\ }\textbf {\bibinfo {volume} {15}},\ \bibinfo {pages}
  {790} (\bibinfo {year} {2019})}\BibitemShut {NoStop}%
\bibitem [{\citenamefont {Huembeli}\ \emph {et~al.}(2019)\citenamefont
  {Huembeli}, \citenamefont {Dauphin}, \citenamefont {Wittek},\ and\
  \citenamefont {Gogolin}}]{huembeli:2019}%
  \BibitemOpen
  \bibfield  {author} {\bibinfo {author} {\bibfnamefont {P.}~\bibnamefont
  {Huembeli}}, \bibinfo {author} {\bibfnamefont {A.}~\bibnamefont {Dauphin}},
  \bibinfo {author} {\bibfnamefont {P.}~\bibnamefont {Wittek}},\ and\ \bibinfo
  {author} {\bibfnamefont {C.}~\bibnamefont {Gogolin}},\ }\bibfield  {title}
  {\bibinfo {title} {Automated discovery of characteristic features of phase
  transitions in many-body localization},\ }\href
  {https://doi.org/10.1103/PhysRevB.99.104106} {\bibfield  {journal} {\bibinfo
  {journal} {Phys. Rev. B}\ }\textbf {\bibinfo {volume} {99}},\ \bibinfo
  {pages} {104106} (\bibinfo {year} {2019})}\BibitemShut {NoStop}%
\bibitem [{\citenamefont {Sch\"afer}\ and\ \citenamefont
  {L\"orch}(2019)}]{schaefer:2019}%
  \BibitemOpen
  \bibfield  {author} {\bibinfo {author} {\bibfnamefont {F.}~\bibnamefont
  {Sch\"afer}}\ and\ \bibinfo {author} {\bibfnamefont {N.}~\bibnamefont
  {L\"orch}},\ }\bibfield  {title} {\bibinfo {title} {Vector field divergence
  of predictive model output as indication of phase transitions},\ }\href
  {https://doi.org/10.1103/PhysRevE.99.062107} {\bibfield  {journal} {\bibinfo
  {journal} {Phys. Rev. E}\ }\textbf {\bibinfo {volume} {99}},\ \bibinfo
  {pages} {062107} (\bibinfo {year} {2019})}\BibitemShut {NoStop}%
\bibitem [{\citenamefont {Scheurer}\ and\ \citenamefont
  {Slager}(2020)}]{scheurer:2020}%
  \BibitemOpen
  \bibfield  {author} {\bibinfo {author} {\bibfnamefont {M.~S.}\ \bibnamefont
  {Scheurer}}\ and\ \bibinfo {author} {\bibfnamefont {R.-J.}\ \bibnamefont
  {Slager}},\ }\bibfield  {title} {\bibinfo {title} {Unsupervised {M}achine
  {L}earning and {B}and {T}opology},\ }\href
  {https://doi.org/10.1103/PhysRevLett.124.226401} {\bibfield  {journal}
  {\bibinfo  {journal} {Phys. Rev. Lett.}\ }\textbf {\bibinfo {volume} {124}},\
  \bibinfo {pages} {226401} (\bibinfo {year} {2020})}\BibitemShut {NoStop}%
\bibitem [{\citenamefont {Greplova}\ \emph {et~al.}(2020)\citenamefont
  {Greplova}, \citenamefont {Valenti}, \citenamefont {Boschung}, \citenamefont
  {Schäfer}, \citenamefont {Lörch},\ and\ \citenamefont
  {Huber}}]{greplova:2020}%
  \BibitemOpen
  \bibfield  {author} {\bibinfo {author} {\bibfnamefont {E.}~\bibnamefont
  {Greplova}}, \bibinfo {author} {\bibfnamefont {A.}~\bibnamefont {Valenti}},
  \bibinfo {author} {\bibfnamefont {G.}~\bibnamefont {Boschung}}, \bibinfo
  {author} {\bibfnamefont {F.}~\bibnamefont {Schäfer}}, \bibinfo {author}
  {\bibfnamefont {N.}~\bibnamefont {Lörch}},\ and\ \bibinfo {author}
  {\bibfnamefont {S.~D.}\ \bibnamefont {Huber}},\ }\bibfield  {title} {\bibinfo
  {title} {Unsupervised identification of topological phase transitions using
  predictive models},\ }\href {https://doi.org/10.1088/1367-2630/ab7771}
  {\bibfield  {journal} {\bibinfo  {journal} {New J. Phys.}\ }\textbf {\bibinfo
  {volume} {22}},\ \bibinfo {pages} {045003} (\bibinfo {year}
  {2020})}\BibitemShut {NoStop}%
\bibitem [{\citenamefont {Kottmann}\ \emph {et~al.}(2020)\citenamefont
  {Kottmann}, \citenamefont {Huembeli}, \citenamefont {Lewenstein},\ and\
  \citenamefont {Ac\'{\i}n}}]{kottmann:2020}%
  \BibitemOpen
  \bibfield  {author} {\bibinfo {author} {\bibfnamefont {K.}~\bibnamefont
  {Kottmann}}, \bibinfo {author} {\bibfnamefont {P.}~\bibnamefont {Huembeli}},
  \bibinfo {author} {\bibfnamefont {M.}~\bibnamefont {Lewenstein}},\ and\
  \bibinfo {author} {\bibfnamefont {A.}~\bibnamefont {Ac\'{\i}n}},\ }\bibfield
  {title} {\bibinfo {title} {Unsupervised {P}hase {D}iscovery with {D}eep
  {A}nomaly {D}etection},\ }\href
  {https://doi.org/10.1103/PhysRevLett.125.170603} {\bibfield  {journal}
  {\bibinfo  {journal} {Phys. Rev. Lett.}\ }\textbf {\bibinfo {volume} {125}},\
  \bibinfo {pages} {170603} (\bibinfo {year} {2020})}\BibitemShut {NoStop}%
\bibitem [{\citenamefont {Zvyagintseva}\ \emph {et~al.}(2022)\citenamefont
  {Zvyagintseva}, \citenamefont {Sigurdsson}, \citenamefont {Kozin},
  \citenamefont {Iorsh}, \citenamefont {Shelykh}, \citenamefont {Ulyantsev},\
  and\ \citenamefont {Kyriienko}}]{zvyagintseva:2021}%
  \BibitemOpen
  \bibfield  {author} {\bibinfo {author} {\bibfnamefont {D.}~\bibnamefont
  {Zvyagintseva}}, \bibinfo {author} {\bibfnamefont {H.}~\bibnamefont
  {Sigurdsson}}, \bibinfo {author} {\bibfnamefont {V.}~\bibnamefont {Kozin}},
  \bibinfo {author} {\bibfnamefont {I.}~\bibnamefont {Iorsh}}, \bibinfo
  {author} {\bibfnamefont {I.}~\bibnamefont {Shelykh}}, \bibinfo {author}
  {\bibfnamefont {V.}~\bibnamefont {Ulyantsev}},\ and\ \bibinfo {author}
  {\bibfnamefont {O.}~\bibnamefont {Kyriienko}},\ }\bibfield  {title} {\bibinfo
  {title} {Machine learning of phase transitions in nonlinear polariton
  lattices},\ }\href {https://doi.org/10.1038/s42005-021-00755-5} {\bibfield
  {journal} {\bibinfo  {journal} {Commun. Phys.}\ }\textbf {\bibinfo {volume}
  {5}},\ \bibinfo {pages} {8} (\bibinfo {year} {2022})}\BibitemShut {NoStop}%
\bibitem [{\citenamefont {Arnold}\ \emph {et~al.}(2021)\citenamefont {Arnold},
  \citenamefont {Sch\"afer}, \citenamefont {\ifmmode~\check{Z}\else
  \v{Z}\fi{}onda},\ and\ \citenamefont {Lode}}]{arnold:2021}%
  \BibitemOpen
  \bibfield  {author} {\bibinfo {author} {\bibfnamefont {J.}~\bibnamefont
  {Arnold}}, \bibinfo {author} {\bibfnamefont {F.}~\bibnamefont {Sch\"afer}},
  \bibinfo {author} {\bibfnamefont {M.}~\bibnamefont {\ifmmode~\check{Z}\else
  \v{Z}\fi{}onda}},\ and\ \bibinfo {author} {\bibfnamefont {A.~U.~J.}\
  \bibnamefont {Lode}},\ }\bibfield  {title} {\bibinfo {title} {Interpretable
  and unsupervised phase classification},\ }\href
  {https://doi.org/10.1103/PhysRevResearch.3.033052} {\bibfield  {journal}
  {\bibinfo  {journal} {Phys. Rev. Res.}\ }\textbf {\bibinfo {volume} {3}},\
  \bibinfo {pages} {033052} (\bibinfo {year} {2021})}\BibitemShut {NoStop}%
\bibitem [{\citenamefont {Huang}\ \emph {et~al.}(2021)\citenamefont {Huang},
  \citenamefont {Kueng}, \citenamefont {Torlai}, \citenamefont {Albert},\ and\
  \citenamefont {Preskill}}]{huang:2021}%
  \BibitemOpen
  \bibfield  {author} {\bibinfo {author} {\bibfnamefont {H.-Y.}\ \bibnamefont
  {Huang}}, \bibinfo {author} {\bibfnamefont {R.}~\bibnamefont {Kueng}},
  \bibinfo {author} {\bibfnamefont {G.}~\bibnamefont {Torlai}}, \bibinfo
  {author} {\bibfnamefont {V.~V.}\ \bibnamefont {Albert}},\ and\ \bibinfo
  {author} {\bibfnamefont {J.}~\bibnamefont {Preskill}},\ }\bibfield  {title}
  {\bibinfo {title} {Provably efficient machine learning for quantum many-body
  problems},\ }\href {https://arxiv.org/abs/2106.12627} {\bibfield  {journal}
  {\bibinfo  {journal} {arXiv:2106.12627}\ } (\bibinfo {year}
  {2021})}\BibitemShut {NoStop}%
\bibitem [{\citenamefont {Guo}\ and\ \citenamefont {He}(2022)}]{guo:2022}%
  \BibitemOpen
  \bibfield  {author} {\bibinfo {author} {\bibfnamefont {W.-c.}\ \bibnamefont
  {Guo}}\ and\ \bibinfo {author} {\bibfnamefont {L.}~\bibnamefont {He}},\
  }\bibfield  {title} {\bibinfo {title} {{Learning Phase Transitions from
  Regression Uncertainty}},\ }\href {https://arxiv.org/abs/2203.06455}
  {\bibfield  {journal} {\bibinfo  {journal} {arXiv:2203.06455}\ } (\bibinfo
  {year} {2022})}\BibitemShut {NoStop}%
\bibitem [{\citenamefont {Maskara}\ \emph {et~al.}(2022)\citenamefont
  {Maskara}, \citenamefont {Buchhold}, \citenamefont {Endres},\ and\
  \citenamefont {van Nieuwenburg}}]{maskara:2022}%
  \BibitemOpen
  \bibfield  {author} {\bibinfo {author} {\bibfnamefont {N.}~\bibnamefont
  {Maskara}}, \bibinfo {author} {\bibfnamefont {M.}~\bibnamefont {Buchhold}},
  \bibinfo {author} {\bibfnamefont {M.}~\bibnamefont {Endres}},\ and\ \bibinfo
  {author} {\bibfnamefont {E.}~\bibnamefont {van Nieuwenburg}},\ }\bibfield
  {title} {\bibinfo {title} {Learning algorithm reflecting universal scaling
  behavior near phase transitions},\ }\href
  {https://doi.org/10.1103/PhysRevResearch.4.L022032} {\bibfield  {journal}
  {\bibinfo  {journal} {Phys. Rev. Res.}\ }\textbf {\bibinfo {volume} {4}},\
  \bibinfo {pages} {L022032} (\bibinfo {year} {2022})}\BibitemShut {NoStop}%
\bibitem [{\citenamefont {Patel}\ \emph {et~al.}(2022)\citenamefont {Patel},
  \citenamefont {Merali},\ and\ \citenamefont {Wetzel}}]{patel:2022}%
  \BibitemOpen
  \bibfield  {author} {\bibinfo {author} {\bibfnamefont {Z.}~\bibnamefont
  {Patel}}, \bibinfo {author} {\bibfnamefont {E.}~\bibnamefont {Merali}},\ and\
  \bibinfo {author} {\bibfnamefont {S.~J.}\ \bibnamefont {Wetzel}},\ }\bibfield
   {title} {\bibinfo {title} {{Unsupervised Learning of Rydberg Atom Array
  Phase Diagram with Siamese Neural Networks}},\ }\href
  {https://arxiv.org/abs/2205.04051} {\bibfield  {journal} {\bibinfo  {journal}
  {arXiv:2205.04051}\ } (\bibinfo {year} {2022})}\BibitemShut {NoStop}%
\bibitem [{\citenamefont {Zhang}\ \emph {et~al.}(2022)\citenamefont {Zhang},
  \citenamefont {Yang},\ and\ \citenamefont {Wu}}]{zhang:2022}%
  \BibitemOpen
  \bibfield  {author} {\bibinfo {author} {\bibfnamefont {W.}~\bibnamefont
  {Zhang}}, \bibinfo {author} {\bibfnamefont {H.}~\bibnamefont {Yang}},\ and\
  \bibinfo {author} {\bibfnamefont {N.}~\bibnamefont {Wu}},\ }\bibfield
  {title} {\bibinfo {title} {Neural network topological snake models for
  locating general phase diagrams},\ }\href {https://arxiv.org/abs/2205.09699}
  {\bibfield  {journal} {\bibinfo  {journal} {arXiv:2205.09699}\ } (\bibinfo
  {year} {2022})}\BibitemShut {NoStop}%
\bibitem [{\citenamefont {Goodfellow}\ \emph {et~al.}(2016)\citenamefont
  {Goodfellow}, \citenamefont {Bengio},\ and\ \citenamefont
  {Courville}}]{goodfellow:2016}%
  \BibitemOpen
  \bibfield  {author} {\bibinfo {author} {\bibfnamefont {I.}~\bibnamefont
  {Goodfellow}}, \bibinfo {author} {\bibfnamefont {Y.}~\bibnamefont {Bengio}},\
  and\ \bibinfo {author} {\bibfnamefont {A.}~\bibnamefont {Courville}},\ }\href
  {http://www.deeplearningbook.org} {\emph {\bibinfo {title} {Deep
  {L}earning}}}\ (\bibinfo  {publisher} {MIT Press},\ \bibinfo {year}
  {2016})\BibitemShut {NoStop}%
\bibitem [{\citenamefont {Hu}\ \emph {et~al.}(2021)\citenamefont {Hu},
  \citenamefont {Chu}, \citenamefont {Pei}, \citenamefont {Liu},\ and\
  \citenamefont {Bian}}]{hu:2021}%
  \BibitemOpen
  \bibfield  {author} {\bibinfo {author} {\bibfnamefont {X.}~\bibnamefont
  {Hu}}, \bibinfo {author} {\bibfnamefont {L.}~\bibnamefont {Chu}}, \bibinfo
  {author} {\bibfnamefont {J.}~\bibnamefont {Pei}}, \bibinfo {author}
  {\bibfnamefont {W.}~\bibnamefont {Liu}},\ and\ \bibinfo {author}
  {\bibfnamefont {J.}~\bibnamefont {Bian}},\ }\bibfield  {title} {\bibinfo
  {title} {{Model complexity of deep learning: a survey}},\ }\href
  {https://doi.org/10.1007/s10115-021-01605-0} {\bibfield  {journal} {\bibinfo
  {journal} {Knowl. Inf. Syst.}\ }\textbf {\bibinfo {volume} {63}},\ \bibinfo
  {pages} {2585} (\bibinfo {year} {2021})}\BibitemShut {NoStop}%
\bibitem [{\citenamefont {Cybenko}(1989)}]{cybenko:1989}%
  \BibitemOpen
  \bibfield  {author} {\bibinfo {author} {\bibfnamefont {G.}~\bibnamefont
  {Cybenko}},\ }\bibfield  {title} {\bibinfo {title} {Approximation by
  superpositions of a sigmoidal function},\ }\href
  {https://doi.org/10.1007/BF02551274} {\bibfield  {journal} {\bibinfo
  {journal} {Math. Control Signals Syst.}\ }\textbf {\bibinfo {volume} {2}},\
  \bibinfo {pages} {303} (\bibinfo {year} {1989})}\BibitemShut {NoStop}%
\bibitem [{\citenamefont {Hornik}(1991)}]{hornik:1991}%
  \BibitemOpen
  \bibfield  {author} {\bibinfo {author} {\bibfnamefont {K.}~\bibnamefont
  {Hornik}},\ }\bibfield  {title} {\bibinfo {title} {Approximation capabilities
  of multilayer feedforward networks},\ }\href
  {https://doi.org/https://doi.org/10.1016/0893-6080(91)90009-T} {\bibfield
  {journal} {\bibinfo  {journal} {Neural Netw.}\ }\textbf {\bibinfo {volume}
  {4}},\ \bibinfo {pages} {251} (\bibinfo {year} {1991})}\BibitemShut {NoStop}%
\bibitem [{\citenamefont {Lu}\ \emph {et~al.}(2017)\citenamefont {Lu},
  \citenamefont {Pu}, \citenamefont {Wang}, \citenamefont {Hu},\ and\
  \citenamefont {Wang}}]{lu:2017}%
  \BibitemOpen
  \bibfield  {author} {\bibinfo {author} {\bibfnamefont {Z.}~\bibnamefont
  {Lu}}, \bibinfo {author} {\bibfnamefont {H.}~\bibnamefont {Pu}}, \bibinfo
  {author} {\bibfnamefont {F.}~\bibnamefont {Wang}}, \bibinfo {author}
  {\bibfnamefont {Z.}~\bibnamefont {Hu}},\ and\ \bibinfo {author}
  {\bibfnamefont {L.}~\bibnamefont {Wang}},\ }\bibfield  {title} {\bibinfo
  {title} {The {E}xpressive {P}ower of {N}eural {N}etworks: {A} {V}iew from the
  {W}idth},\ }in\ \href
  {https://proceedings.neurips.cc/paper/2017/hash/32cbf687880eb1674a07bf717761dd3a-Abstract.html}
  {\emph {\bibinfo {booktitle} {Adv. Neural Inf. Process. Syst.}}},\
  Vol.~\bibinfo {volume} {30},\ \bibinfo {editor} {edited by\ \bibinfo {editor}
  {\bibfnamefont {I.}~\bibnamefont {Guyon}}, \bibinfo {editor} {\bibfnamefont
  {U.~V.}\ \bibnamefont {Luxburg}}, \bibinfo {editor} {\bibfnamefont
  {S.}~\bibnamefont {Bengio}}, \bibinfo {editor} {\bibfnamefont
  {H.}~\bibnamefont {Wallach}}, \bibinfo {editor} {\bibfnamefont
  {R.}~\bibnamefont {Fergus}}, \bibinfo {editor} {\bibfnamefont
  {S.}~\bibnamefont {Vishwanathan}},\ and\ \bibinfo {editor} {\bibfnamefont
  {R.}~\bibnamefont {Garnett}}}\ (\bibinfo  {publisher} {Curran Associates,
  Inc.},\ \bibinfo {year} {2017})\BibitemShut {NoStop}%
\bibitem [{\citenamefont {Zhou}(2020)}]{zhou:2020}%
  \BibitemOpen
  \bibfield  {author} {\bibinfo {author} {\bibfnamefont {D.-X.}\ \bibnamefont
  {Zhou}},\ }\bibfield  {title} {\bibinfo {title} {Universality of deep
  convolutional neural networks},\ }\href
  {https://doi.org/https://doi.org/10.1016/j.acha.2019.06.004} {\bibfield
  {journal} {\bibinfo  {journal} {Appl. Comput. Harmon. Anal.}\ }\textbf
  {\bibinfo {volume} {48}},\ \bibinfo {pages} {787} (\bibinfo {year}
  {2020})}\BibitemShut {NoStop}%
\bibitem [{\citenamefont {Krizhevsky}\ \emph {et~al.}(2012)\citenamefont
  {Krizhevsky}, \citenamefont {Sutskever},\ and\ \citenamefont
  {Hinton}}]{krizhevsky:2012}%
  \BibitemOpen
  \bibfield  {author} {\bibinfo {author} {\bibfnamefont {A.}~\bibnamefont
  {Krizhevsky}}, \bibinfo {author} {\bibfnamefont {I.}~\bibnamefont
  {Sutskever}},\ and\ \bibinfo {author} {\bibfnamefont {G.~E.}\ \bibnamefont
  {Hinton}},\ }\bibfield  {title} {\bibinfo {title} {Image{N}et
  {C}lassification with {D}eep {C}onvolutional {N}eural {N}etworks},\ }in\
  \href
  {https://proceedings.neurips.cc/paper/2012/hash/c399862d3b9d6b76c8436e924a68c45b-Abstract.html}
  {\emph {\bibinfo {booktitle} {Adv. Neural Inf. Process. Syst.}}},\
  Vol.~\bibinfo {volume} {25},\ \bibinfo {editor} {edited by\ \bibinfo {editor}
  {\bibfnamefont {F.}~\bibnamefont {Pereira}}, \bibinfo {editor} {\bibfnamefont
  {C.~J.~C.}\ \bibnamefont {Burges}}, \bibinfo {editor} {\bibfnamefont
  {L.}~\bibnamefont {Bottou}},\ and\ \bibinfo {editor} {\bibfnamefont {K.~Q.}\
  \bibnamefont {Weinberger}}}\ (\bibinfo  {publisher} {Curran Associates,
  Inc.},\ \bibinfo {year} {2012})\BibitemShut {NoStop}%
\bibitem [{\citenamefont {Bengio}\ and\ \citenamefont
  {Delalleau}(2011)}]{bengio:2011}%
  \BibitemOpen
  \bibfield  {author} {\bibinfo {author} {\bibfnamefont {Y.}~\bibnamefont
  {Bengio}}\ and\ \bibinfo {author} {\bibfnamefont {O.}~\bibnamefont
  {Delalleau}},\ }\bibfield  {title} {\bibinfo {title} {{On the Expressive
  Power of Deep Architectures}},\ }in\ \href
  {https://doi.org/10.1007/978-3-642-24412-4_3} {\emph {\bibinfo {booktitle}
  {Algorithmic Learning Theory}}},\ \bibinfo {editor} {edited by\ \bibinfo
  {editor} {\bibfnamefont {J.}~\bibnamefont {Kivinen}}, \bibinfo {editor}
  {\bibfnamefont {C.}~\bibnamefont {Szepesv{\'a}ri}}, \bibinfo {editor}
  {\bibfnamefont {E.}~\bibnamefont {Ukkonen}},\ and\ \bibinfo {editor}
  {\bibfnamefont {T.}~\bibnamefont {Zeugmann}}}\ (\bibinfo  {publisher}
  {Springer},\ \bibinfo {address} {Berlin, Heidelberg},\ \bibinfo {year}
  {2011})\ pp.\ \bibinfo {pages} {18--36}\BibitemShut {NoStop}%
\bibitem [{\citenamefont {Raghu}\ \emph {et~al.}(2017)\citenamefont {Raghu},
  \citenamefont {Poole}, \citenamefont {Kleinberg}, \citenamefont {Ganguli},\
  and\ \citenamefont {Sohl-Dickstein}}]{raghu:2017}%
  \BibitemOpen
  \bibfield  {author} {\bibinfo {author} {\bibfnamefont {M.}~\bibnamefont
  {Raghu}}, \bibinfo {author} {\bibfnamefont {B.}~\bibnamefont {Poole}},
  \bibinfo {author} {\bibfnamefont {J.}~\bibnamefont {Kleinberg}}, \bibinfo
  {author} {\bibfnamefont {S.}~\bibnamefont {Ganguli}},\ and\ \bibinfo {author}
  {\bibfnamefont {J.}~\bibnamefont {Sohl-Dickstein}},\ }\bibfield  {title}
  {\bibinfo {title} {{On the Expressive Power of Deep Neural Networks}},\ }in\
  \href {https://proceedings.mlr.press/v70/raghu17a.html} {\emph {\bibinfo
  {booktitle} {Proceedings of the 34th International Conference on Machine
  Learning}}},\ \bibinfo {series} {PMLR}, Vol.~\bibinfo {volume} {70},\
  \bibinfo {editor} {edited by\ \bibinfo {editor} {\bibfnamefont
  {D.}~\bibnamefont {Precup}}\ and\ \bibinfo {editor} {\bibfnamefont {Y.~W.}\
  \bibnamefont {Teh}}}\ (\bibinfo  {publisher} {PMLR},\ \bibinfo {year}
  {2017})\ pp.\ \bibinfo {pages} {2847--2854}\BibitemShut {NoStop}%
\bibitem [{\citenamefont {Linardatos}\ \emph {et~al.}(2021)\citenamefont
  {Linardatos}, \citenamefont {Papastefanopoulos},\ and\ \citenamefont
  {Kotsiantis}}]{linardatos:2021}%
  \BibitemOpen
  \bibfield  {author} {\bibinfo {author} {\bibfnamefont {P.}~\bibnamefont
  {Linardatos}}, \bibinfo {author} {\bibfnamefont {V.}~\bibnamefont
  {Papastefanopoulos}},\ and\ \bibinfo {author} {\bibfnamefont
  {S.}~\bibnamefont {Kotsiantis}},\ }\bibfield  {title} {\bibinfo {title}
  {{Explainable AI: A Review of Machine Learning Interpretability Methods}},\
  }\href {https://doi.org/10.3390/e23010018} {\bibfield  {journal} {\bibinfo
  {journal} {Entropy}\ }\textbf {\bibinfo {volume} {23}},\ \bibinfo {pages}
  {18} (\bibinfo {year} {2021})}\BibitemShut {NoStop}%
\bibitem [{\citenamefont {Molnar}(2022)}]{molnar:2022}%
  \BibitemOpen
  \bibfield  {author} {\bibinfo {author} {\bibfnamefont {C.}~\bibnamefont
  {Molnar}},\ }\href {https://christophm.github.io/interpretable-ml-book}
  {\emph {\bibinfo {title} {{Interpretable Machine Learning}}}},\ \bibinfo
  {edition} {2nd}\ ed.\ (\bibinfo {year} {2022})\BibitemShut {NoStop}%
\bibitem [{\citenamefont {Ponte}\ and\ \citenamefont
  {Melko}(2017)}]{ponte:2017}%
  \BibitemOpen
  \bibfield  {author} {\bibinfo {author} {\bibfnamefont {P.}~\bibnamefont
  {Ponte}}\ and\ \bibinfo {author} {\bibfnamefont {R.~G.}\ \bibnamefont
  {Melko}},\ }\bibfield  {title} {\bibinfo {title} {Kernel methods for
  interpretable machine learning of order parameters},\ }\href
  {https://doi.org/10.1103/PhysRevB.96.205146} {\bibfield  {journal} {\bibinfo
  {journal} {Phys. Rev. B}\ }\textbf {\bibinfo {volume} {96}},\ \bibinfo
  {pages} {205146} (\bibinfo {year} {2017})}\BibitemShut {NoStop}%
\bibitem [{\citenamefont {Zhang}\ \emph
  {et~al.}(2019{\natexlab{a}})\citenamefont {Zhang}, \citenamefont {Wang},\
  and\ \citenamefont {Wang}}]{zhang:2019}%
  \BibitemOpen
  \bibfield  {author} {\bibinfo {author} {\bibfnamefont {W.}~\bibnamefont
  {Zhang}}, \bibinfo {author} {\bibfnamefont {L.}~\bibnamefont {Wang}},\ and\
  \bibinfo {author} {\bibfnamefont {Z.}~\bibnamefont {Wang}},\ }\bibfield
  {title} {\bibinfo {title} {Interpretable machine learning study of the
  many-body localization transition in disordered quantum {I}sing spin
  chains},\ }\href {https://doi.org/10.1103/PhysRevB.99.054208} {\bibfield
  {journal} {\bibinfo  {journal} {Phys. Rev. B}\ }\textbf {\bibinfo {volume}
  {99}},\ \bibinfo {pages} {054208} (\bibinfo {year}
  {2019}{\natexlab{a}})}\BibitemShut {NoStop}%
\bibitem [{\citenamefont {Greitemann}\ \emph
  {et~al.}(2019{\natexlab{a}})\citenamefont {Greitemann}, \citenamefont {Liu},
  \citenamefont {Jaubert}, \citenamefont {Yan}, \citenamefont {Shannon},\ and\
  \citenamefont {Pollet}}]{greitemann:2019}%
  \BibitemOpen
  \bibfield  {author} {\bibinfo {author} {\bibfnamefont {J.}~\bibnamefont
  {Greitemann}}, \bibinfo {author} {\bibfnamefont {K.}~\bibnamefont {Liu}},
  \bibinfo {author} {\bibfnamefont {L.~D.~C.}\ \bibnamefont {Jaubert}},
  \bibinfo {author} {\bibfnamefont {H.}~\bibnamefont {Yan}}, \bibinfo {author}
  {\bibfnamefont {N.}~\bibnamefont {Shannon}},\ and\ \bibinfo {author}
  {\bibfnamefont {L.}~\bibnamefont {Pollet}},\ }\bibfield  {title} {\bibinfo
  {title} {Identification of emergent constraints and hidden order in
  frustrated magnets using tensorial kernel methods of machine learning},\
  }\href {https://doi.org/10.1103/PhysRevB.100.174408} {\bibfield  {journal}
  {\bibinfo  {journal} {Phys. Rev. B}\ }\textbf {\bibinfo {volume} {100}},\
  \bibinfo {pages} {174408} (\bibinfo {year} {2019}{\natexlab{a}})}\BibitemShut
  {NoStop}%
\bibitem [{\citenamefont {Liu}\ \emph {et~al.}(2019)\citenamefont {Liu},
  \citenamefont {Greitemann},\ and\ \citenamefont {Pollet}}]{liu:2019}%
  \BibitemOpen
  \bibfield  {author} {\bibinfo {author} {\bibfnamefont {K.}~\bibnamefont
  {Liu}}, \bibinfo {author} {\bibfnamefont {J.}~\bibnamefont {Greitemann}},\
  and\ \bibinfo {author} {\bibfnamefont {L.}~\bibnamefont {Pollet}},\
  }\bibfield  {title} {\bibinfo {title} {Learning multiple order parameters
  with interpretable machines},\ }\href
  {https://doi.org/10.1103/PhysRevB.99.104410} {\bibfield  {journal} {\bibinfo
  {journal} {Phys. Rev. B}\ }\textbf {\bibinfo {volume} {99}},\ \bibinfo
  {pages} {104410} (\bibinfo {year} {2019})}\BibitemShut {NoStop}%
\bibitem [{\citenamefont {Zhang}\ \emph {et~al.}(2020)\citenamefont {Zhang},
  \citenamefont {Ginsparg},\ and\ \citenamefont {Kim}}]{zhang:2020}%
  \BibitemOpen
  \bibfield  {author} {\bibinfo {author} {\bibfnamefont {Y.}~\bibnamefont
  {Zhang}}, \bibinfo {author} {\bibfnamefont {P.}~\bibnamefont {Ginsparg}},\
  and\ \bibinfo {author} {\bibfnamefont {E.-A.}\ \bibnamefont {Kim}},\
  }\bibfield  {title} {\bibinfo {title} {Interpreting machine learning of
  topological quantum phase transitions},\ }\href
  {https://doi.org/10.1103/PhysRevResearch.2.023283} {\bibfield  {journal}
  {\bibinfo  {journal} {Phys. Rev. Res.}\ }\textbf {\bibinfo {volume} {2}},\
  \bibinfo {pages} {023283} (\bibinfo {year} {2020})}\BibitemShut {NoStop}%
\bibitem [{\citenamefont {Casert}\ \emph {et~al.}(2019)\citenamefont {Casert},
  \citenamefont {Vieijra}, \citenamefont {Nys},\ and\ \citenamefont
  {Ryckebusch}}]{casert:2019}%
  \BibitemOpen
  \bibfield  {author} {\bibinfo {author} {\bibfnamefont {C.}~\bibnamefont
  {Casert}}, \bibinfo {author} {\bibfnamefont {T.}~\bibnamefont {Vieijra}},
  \bibinfo {author} {\bibfnamefont {J.}~\bibnamefont {Nys}},\ and\ \bibinfo
  {author} {\bibfnamefont {J.}~\bibnamefont {Ryckebusch}},\ }\bibfield  {title}
  {\bibinfo {title} {Interpretable machine learning for inferring the phase
  boundaries in a nonequilibrium system},\ }\href
  {https://doi.org/10.1103/PhysRevE.99.023304} {\bibfield  {journal} {\bibinfo
  {journal} {Phys. Rev. E}\ }\textbf {\bibinfo {volume} {99}},\ \bibinfo
  {pages} {023304} (\bibinfo {year} {2019})}\BibitemShut {NoStop}%
\bibitem [{\citenamefont {Dawid}\ \emph {et~al.}(2020)\citenamefont {Dawid},
  \citenamefont {Huembeli}, \citenamefont {Tomza}, \citenamefont {Lewenstein},\
  and\ \citenamefont {Dauphin}}]{dawid:2020}%
  \BibitemOpen
  \bibfield  {author} {\bibinfo {author} {\bibfnamefont {A.}~\bibnamefont
  {Dawid}}, \bibinfo {author} {\bibfnamefont {P.}~\bibnamefont {Huembeli}},
  \bibinfo {author} {\bibfnamefont {M.}~\bibnamefont {Tomza}}, \bibinfo
  {author} {\bibfnamefont {M.}~\bibnamefont {Lewenstein}},\ and\ \bibinfo
  {author} {\bibfnamefont {A.}~\bibnamefont {Dauphin}},\ }\bibfield  {title}
  {\bibinfo {title} {Phase detection with neural networks: interpreting the
  black box},\ }\href {https://doi.org/10.1088/1367-2630/abc463} {\bibfield
  {journal} {\bibinfo  {journal} {New J. Phys.}\ }\textbf {\bibinfo {volume}
  {22}},\ \bibinfo {pages} {115001} (\bibinfo {year} {2020})}\BibitemShut
  {NoStop}%
\bibitem [{\citenamefont {Bl\"ucher}\ \emph {et~al.}(2020)\citenamefont
  {Bl\"ucher}, \citenamefont {Kades}, \citenamefont {Pawlowski}, \citenamefont
  {Strodthoff},\ and\ \citenamefont {Urban}}]{blucher:2020}%
  \BibitemOpen
  \bibfield  {author} {\bibinfo {author} {\bibfnamefont {S.}~\bibnamefont
  {Bl\"ucher}}, \bibinfo {author} {\bibfnamefont {L.}~\bibnamefont {Kades}},
  \bibinfo {author} {\bibfnamefont {J.~M.}\ \bibnamefont {Pawlowski}}, \bibinfo
  {author} {\bibfnamefont {N.}~\bibnamefont {Strodthoff}},\ and\ \bibinfo
  {author} {\bibfnamefont {J.~M.}\ \bibnamefont {Urban}},\ }\bibfield  {title}
  {\bibinfo {title} {Towards novel insights in lattice field theory with
  explainable machine learning},\ }\href
  {https://doi.org/10.1103/PhysRevD.101.094507} {\bibfield  {journal} {\bibinfo
   {journal} {Phys. Rev. D}\ }\textbf {\bibinfo {volume} {101}},\ \bibinfo
  {pages} {094507} (\bibinfo {year} {2020})}\BibitemShut {NoStop}%
\bibitem [{\citenamefont {Dawid}\ \emph {et~al.}(2021)\citenamefont {Dawid},
  \citenamefont {Huembeli}, \citenamefont {Tomza}, \citenamefont {Lewenstein},\
  and\ \citenamefont {Dauphin}}]{dawid:2021}%
  \BibitemOpen
  \bibfield  {author} {\bibinfo {author} {\bibfnamefont {A.}~\bibnamefont
  {Dawid}}, \bibinfo {author} {\bibfnamefont {P.}~\bibnamefont {Huembeli}},
  \bibinfo {author} {\bibfnamefont {M.}~\bibnamefont {Tomza}}, \bibinfo
  {author} {\bibfnamefont {M.}~\bibnamefont {Lewenstein}},\ and\ \bibinfo
  {author} {\bibfnamefont {A.}~\bibnamefont {Dauphin}},\ }\bibfield  {title}
  {\bibinfo {title} {Hessian-based toolbox for reliable and interpretable
  machine learning in physics},\ }\href
  {https://doi.org/10.1088/2632-2153/ac338d} {\bibfield  {journal} {\bibinfo
  {journal} {Mach. Learn.: Sci. Technol.}\ }\textbf {\bibinfo {volume} {3}},\
  \bibinfo {pages} {015002} (\bibinfo {year} {2021})}\BibitemShut {NoStop}%
\bibitem [{\citenamefont {Blumer}\ \emph {et~al.}(1989)\citenamefont {Blumer},
  \citenamefont {Ehrenfeucht}, \citenamefont {Haussler},\ and\ \citenamefont
  {Warmuth}}]{blumer:1989}%
  \BibitemOpen
  \bibfield  {author} {\bibinfo {author} {\bibfnamefont {A.}~\bibnamefont
  {Blumer}}, \bibinfo {author} {\bibfnamefont {A.}~\bibnamefont {Ehrenfeucht}},
  \bibinfo {author} {\bibfnamefont {D.}~\bibnamefont {Haussler}},\ and\
  \bibinfo {author} {\bibfnamefont {M.~K.}\ \bibnamefont {Warmuth}},\
  }\bibfield  {title} {\bibinfo {title} {{Learnability and the
  Vapnik-Chervonenkis Dimension}},\ }\href
  {https://doi.org/10.1145/76359.76371} {\bibfield  {journal} {\bibinfo
  {journal} {J. ACM}\ }\textbf {\bibinfo {volume} {36}},\ \bibinfo {pages}
  {929–965} (\bibinfo {year} {1989})}\BibitemShut {NoStop}%
\bibitem [{\citenamefont {Vapnik}(1999)}]{vapnik:2000}%
  \BibitemOpen
  \bibfield  {author} {\bibinfo {author} {\bibfnamefont {V.~N.}\ \bibnamefont
  {Vapnik}},\ }\href {https://doi.org/10.1007/978-1-4757-3264-1} {\emph
  {\bibinfo {title} {{The Nature of Statistical Learning Theory}}}},\ \bibinfo
  {edition} {2nd}\ ed.\ (\bibinfo  {publisher} {Springer},\ \bibinfo {year}
  {1999})\BibitemShut {NoStop}%
\bibitem [{\citenamefont {Devroye}\ \emph
  {et~al.}(1996{\natexlab{a}})\citenamefont {Devroye}, \citenamefont
  {Gy{\"o}rfi},\ and\ \citenamefont {Lugosi}}]{devroye2:1996}%
  \BibitemOpen
  \bibfield  {author} {\bibinfo {author} {\bibfnamefont {L.}~\bibnamefont
  {Devroye}}, \bibinfo {author} {\bibfnamefont {L.}~\bibnamefont
  {Gy{\"o}rfi}},\ and\ \bibinfo {author} {\bibfnamefont {G.}~\bibnamefont
  {Lugosi}},\ }\bibinfo {title} {{The Bayes Error}},\ in\ \href
  {https://doi.org/10.1007/978-1-4612-0711-5_2} {\emph {\bibinfo {booktitle}
  {{A Probabilistic Theory of Pattern Recognition}}}}\ (\bibinfo  {publisher}
  {Springer},\ \bibinfo {address} {New York, NY},\ \bibinfo {year} {1996})\
  pp.\ \bibinfo {pages} {9--20}\BibitemShut {NoStop}%
\bibitem [{\citenamefont {Devroye}\ \emph
  {et~al.}(1996{\natexlab{b}})\citenamefont {Devroye}, \citenamefont
  {Gy{\"o}rfi},\ and\ \citenamefont {Lugosi}}]{devroye:1996}%
  \BibitemOpen
  \bibfield  {author} {\bibinfo {author} {\bibfnamefont {L.}~\bibnamefont
  {Devroye}}, \bibinfo {author} {\bibfnamefont {L.}~\bibnamefont
  {Gy{\"o}rfi}},\ and\ \bibinfo {author} {\bibfnamefont {G.}~\bibnamefont
  {Lugosi}},\ }\bibinfo {title} {{Inequalities and Alternate Distance
  Measures}},\ in\ \href {https://doi.org/10.1007/978-1-4612-0711-5_3} {\emph
  {\bibinfo {booktitle} {{A Probabilistic Theory of Pattern Recognition}}}}\
  (\bibinfo  {publisher} {Springer},\ \bibinfo {address} {New York, NY},\
  \bibinfo {year} {1996})\ pp.\ \bibinfo {pages} {21--37}\BibitemShut {NoStop}%
\bibitem [{\citenamefont {Fei-Fei}\ \emph {et~al.}(2004)\citenamefont
  {Fei-Fei}, \citenamefont {Fergus},\ and\ \citenamefont {Perona}}]{fei:2004}%
  \BibitemOpen
  \bibfield  {author} {\bibinfo {author} {\bibfnamefont {L.}~\bibnamefont
  {Fei-Fei}}, \bibinfo {author} {\bibfnamefont {R.}~\bibnamefont {Fergus}},\
  and\ \bibinfo {author} {\bibfnamefont {P.}~\bibnamefont {Perona}},\
  }\bibfield  {title} {\bibinfo {title} {{Learning Generative Visual Models
  from Few Training Examples: An Incremental Bayesian Approach Tested on 101
  Object Categories}},\ }in\ \href {https://doi.org/10.1109/CVPR.2004.383}
  {\emph {\bibinfo {booktitle} {2004 Conference on Computer Vision and Pattern
  Recognition Workshop}}}\ (\bibinfo {year} {2004})\ pp.\ \bibinfo {pages}
  {178--178}\BibitemShut {NoStop}%
\bibitem [{\citenamefont {LeCun}\ \emph {et~al.}(2004)\citenamefont {LeCun},
  \citenamefont {Huang},\ and\ \citenamefont {Bottou}}]{lecun:2004}%
  \BibitemOpen
  \bibfield  {author} {\bibinfo {author} {\bibfnamefont {Y.}~\bibnamefont
  {LeCun}}, \bibinfo {author} {\bibfnamefont {F.~J.}\ \bibnamefont {Huang}},\
  and\ \bibinfo {author} {\bibfnamefont {L.}~\bibnamefont {Bottou}},\
  }\bibfield  {title} {\bibinfo {title} {Learning methods for generic object
  recognition with invariance to pose and lighting},\ }in\ \href
  {https://doi.org/10.1109/CVPR.2004.1315150} {\emph {\bibinfo {booktitle}
  {2004 IEEE Computer Society Conference on Computer Vision and Pattern
  Recognition}}},\ Vol.~\bibinfo {volume} {2}\ (\bibinfo {year} {2004})\ pp.\
  \bibinfo {pages} {II--104 Vol.2}\BibitemShut {NoStop}%
\bibitem [{\citenamefont {Griffin}\ \emph {et~al.}(2007)\citenamefont
  {Griffin}, \citenamefont {Holub},\ and\ \citenamefont
  {Perona}}]{griffin:2007}%
  \BibitemOpen
  \bibfield  {author} {\bibinfo {author} {\bibfnamefont {G.}~\bibnamefont
  {Griffin}}, \bibinfo {author} {\bibfnamefont {A.}~\bibnamefont {Holub}},\
  and\ \bibinfo {author} {\bibfnamefont {P.}~\bibnamefont {Perona}},\ }\href
  {https://authors.library.caltech.edu/7694/} {\emph {\bibinfo {title}
  {Caltech-256 object category dataset}}},\ \bibinfo {type} {Tech. Rep.}\
  (\bibinfo {year} {2007})\BibitemShut {NoStop}%
\bibitem [{\citenamefont {Krizhevsky}(2009)}]{krizhevsky:2009}%
  \BibitemOpen
  \bibfield  {author} {\bibinfo {author} {\bibfnamefont {A.}~\bibnamefont
  {Krizhevsky}},\ }\emph {\bibinfo {title} {Learning multiple layers of
  features from tiny images}},\ \href
  {https://www.cs.toronto.edu/~kriz/cifar.html} {Master's thesis},\ \bibinfo
  {school} {University of Toronto}, \bibinfo {address} {Toronto, Ontario}
  (\bibinfo {year} {2009})\BibitemShut {NoStop}%
\bibitem [{\citenamefont {Deng}\ \emph {et~al.}(2009)\citenamefont {Deng},
  \citenamefont {Dong}, \citenamefont {Socher}, \citenamefont {Li},
  \citenamefont {Li},\ and\ \citenamefont {Fei-Fei}}]{deng:2009}%
  \BibitemOpen
  \bibfield  {author} {\bibinfo {author} {\bibfnamefont {J.}~\bibnamefont
  {Deng}}, \bibinfo {author} {\bibfnamefont {W.}~\bibnamefont {Dong}}, \bibinfo
  {author} {\bibfnamefont {R.}~\bibnamefont {Socher}}, \bibinfo {author}
  {\bibfnamefont {L.-J.}\ \bibnamefont {Li}}, \bibinfo {author} {\bibfnamefont
  {K.}~\bibnamefont {Li}},\ and\ \bibinfo {author} {\bibfnamefont
  {L.}~\bibnamefont {Fei-Fei}},\ }\bibfield  {title} {\bibinfo {title}
  {{ImageNet: A large-scale hierarchical image database}},\ }in\ \href
  {https://doi.org/10.1109/CVPR.2009.5206848} {\emph {\bibinfo {booktitle}
  {2009 IEEE Conference on Computer Vision and Pattern Recognition}}}\
  (\bibinfo {year} {2009})\ pp.\ \bibinfo {pages} {248--255}\BibitemShut
  {NoStop}%
\bibitem [{\citenamefont {Russakovsky}\ \emph {et~al.}(2015)\citenamefont
  {Russakovsky}, \citenamefont {Deng}, \citenamefont {Su}, \citenamefont
  {Krause}, \citenamefont {Satheesh}, \citenamefont {Ma}, \citenamefont
  {Huang}, \citenamefont {Karpathy}, \citenamefont {Khosla}, \citenamefont
  {Bernstein} \emph {et~al.}}]{russakovsky:2015}%
  \BibitemOpen
  \bibfield  {author} {\bibinfo {author} {\bibfnamefont {O.}~\bibnamefont
  {Russakovsky}}, \bibinfo {author} {\bibfnamefont {J.}~\bibnamefont {Deng}},
  \bibinfo {author} {\bibfnamefont {H.}~\bibnamefont {Su}}, \bibinfo {author}
  {\bibfnamefont {J.}~\bibnamefont {Krause}}, \bibinfo {author} {\bibfnamefont
  {S.}~\bibnamefont {Satheesh}}, \bibinfo {author} {\bibfnamefont
  {S.}~\bibnamefont {Ma}}, \bibinfo {author} {\bibfnamefont {Z.}~\bibnamefont
  {Huang}}, \bibinfo {author} {\bibfnamefont {A.}~\bibnamefont {Karpathy}},
  \bibinfo {author} {\bibfnamefont {A.}~\bibnamefont {Khosla}}, \bibinfo
  {author} {\bibfnamefont {M.}~\bibnamefont {Bernstein}}, \emph {et~al.},\
  }\bibfield  {title} {\bibinfo {title} {Imagenet large scale visual
  recognition challenge},\ }\href {https://doi.org/10.1007/s11263-015-0816-y}
  {\bibfield  {journal} {\bibinfo  {journal} {Int. J. Comput. Vis.}\ }\textbf
  {\bibinfo {volume} {115}},\ \bibinfo {pages} {211} (\bibinfo {year}
  {2015})}\BibitemShut {NoStop}%
\bibitem [{\citenamefont {Spanhol}\ \emph {et~al.}(2016)\citenamefont
  {Spanhol}, \citenamefont {Oliveira}, \citenamefont {Petitjean},\ and\
  \citenamefont {Heutte}}]{spanhol:2015}%
  \BibitemOpen
  \bibfield  {author} {\bibinfo {author} {\bibfnamefont {F.~A.}\ \bibnamefont
  {Spanhol}}, \bibinfo {author} {\bibfnamefont {L.~S.}\ \bibnamefont
  {Oliveira}}, \bibinfo {author} {\bibfnamefont {C.}~\bibnamefont
  {Petitjean}},\ and\ \bibinfo {author} {\bibfnamefont {L.}~\bibnamefont
  {Heutte}},\ }\bibfield  {title} {\bibinfo {title} {{A Dataset for Breast
  Cancer Histopathological Image Classification}},\ }\href
  {https://doi.org/10.1109/TBME.2015.2496264} {\bibfield  {journal} {\bibinfo
  {journal} {IEEE. Trans. Biomed.}\ }\textbf {\bibinfo {volume} {63}},\
  \bibinfo {pages} {1455} (\bibinfo {year} {2016})}\BibitemShut {NoStop}%
\bibitem [{\citenamefont {James}\ \emph {et~al.}(2013)\citenamefont {James},
  \citenamefont {Witten}, \citenamefont {Hastie},\ and\ \citenamefont
  {Tibshirani}}]{james:2013}%
  \BibitemOpen
  \bibfield  {author} {\bibinfo {author} {\bibfnamefont {G.}~\bibnamefont
  {James}}, \bibinfo {author} {\bibfnamefont {D.}~\bibnamefont {Witten}},
  \bibinfo {author} {\bibfnamefont {T.}~\bibnamefont {Hastie}},\ and\ \bibinfo
  {author} {\bibfnamefont {R.}~\bibnamefont {Tibshirani}},\ }\bibinfo {title}
  {{Statistical Learning}},\ in\ \href
  {https://doi.org/10.1007/978-1-4614-7138-7_2} {\emph {\bibinfo {booktitle}
  {{An Introduction to Statistical Learning: with Applications in R}}}}\
  (\bibinfo  {publisher} {Springer},\ \bibinfo {address} {New York, NY},\
  \bibinfo {year} {2013})\ pp.\ \bibinfo {pages} {15--57}\BibitemShut {NoStop}%
\bibitem [{\citenamefont {Metropolis}\ \emph {et~al.}(1953)\citenamefont
  {Metropolis}, \citenamefont {Rosenbluth}, \citenamefont {Rosenbluth},
  \citenamefont {Teller},\ and\ \citenamefont {Teller}}]{metropolis:1953}%
  \BibitemOpen
  \bibfield  {author} {\bibinfo {author} {\bibfnamefont {N.}~\bibnamefont
  {Metropolis}}, \bibinfo {author} {\bibfnamefont {A.~W.}\ \bibnamefont
  {Rosenbluth}}, \bibinfo {author} {\bibfnamefont {M.~N.}\ \bibnamefont
  {Rosenbluth}}, \bibinfo {author} {\bibfnamefont {A.~H.}\ \bibnamefont
  {Teller}},\ and\ \bibinfo {author} {\bibfnamefont {E.}~\bibnamefont
  {Teller}},\ }\bibfield  {title} {\bibinfo {title} {Equation of state
  calculations by fast computing machines},\ }\href
  {https://doi.org/10.1063/1.1699114} {\bibfield  {journal} {\bibinfo
  {journal} {J. Chem. Phys.}\ }\textbf {\bibinfo {volume} {21}},\ \bibinfo
  {pages} {1087} (\bibinfo {year} {1953})}\BibitemShut {NoStop}%
\bibitem [{\citenamefont {Simon}\ \emph {et~al.}(2011)\citenamefont {Simon},
  \citenamefont {Bakr}, \citenamefont {Ma}, \citenamefont {Tai}, \citenamefont
  {Preiss},\ and\ \citenamefont {Greiner}}]{simon:2011}%
  \BibitemOpen
  \bibfield  {author} {\bibinfo {author} {\bibfnamefont {J.}~\bibnamefont
  {Simon}}, \bibinfo {author} {\bibfnamefont {W.~S.}\ \bibnamefont {Bakr}},
  \bibinfo {author} {\bibfnamefont {R.}~\bibnamefont {Ma}}, \bibinfo {author}
  {\bibfnamefont {M.~E.}\ \bibnamefont {Tai}}, \bibinfo {author} {\bibfnamefont
  {P.~M.}\ \bibnamefont {Preiss}},\ and\ \bibinfo {author} {\bibfnamefont
  {M.}~\bibnamefont {Greiner}},\ }\bibfield  {title} {\bibinfo {title} {Quantum
  simulation of antiferromagnetic spin chains in an optical lattice},\ }\href
  {https://doi.org/10.1038/nature09994} {\bibfield  {journal} {\bibinfo
  {journal} {Nature}\ }\textbf {\bibinfo {volume} {472}},\ \bibinfo {pages}
  {307} (\bibinfo {year} {2011})}\BibitemShut {NoStop}%
\bibitem [{\citenamefont {Bernien}\ \emph {et~al.}(2017)\citenamefont
  {Bernien}, \citenamefont {Schwartz}, \citenamefont {Keesling}, \citenamefont
  {Levine}, \citenamefont {Omran}, \citenamefont {Pichler}, \citenamefont
  {Choi}, \citenamefont {Zibrov}, \citenamefont {Endres}, \citenamefont
  {Greiner} \emph {et~al.}}]{bernien:2017}%
  \BibitemOpen
  \bibfield  {author} {\bibinfo {author} {\bibfnamefont {H.}~\bibnamefont
  {Bernien}}, \bibinfo {author} {\bibfnamefont {S.}~\bibnamefont {Schwartz}},
  \bibinfo {author} {\bibfnamefont {A.}~\bibnamefont {Keesling}}, \bibinfo
  {author} {\bibfnamefont {H.}~\bibnamefont {Levine}}, \bibinfo {author}
  {\bibfnamefont {A.}~\bibnamefont {Omran}}, \bibinfo {author} {\bibfnamefont
  {H.}~\bibnamefont {Pichler}}, \bibinfo {author} {\bibfnamefont
  {S.}~\bibnamefont {Choi}}, \bibinfo {author} {\bibfnamefont {A.~S.}\
  \bibnamefont {Zibrov}}, \bibinfo {author} {\bibfnamefont {M.}~\bibnamefont
  {Endres}}, \bibinfo {author} {\bibfnamefont {M.}~\bibnamefont {Greiner}},
  \emph {et~al.},\ }\bibfield  {title} {\bibinfo {title} {Probing many-body
  dynamics on a 51-atom quantum simulator},\ }\href
  {https://doi.org/doi:10.1038/nature24622} {\bibfield  {journal} {\bibinfo
  {journal} {Nature}\ }\textbf {\bibinfo {volume} {551}},\ \bibinfo {pages}
  {579} (\bibinfo {year} {2017})}\BibitemShut {NoStop}%
\bibitem [{\citenamefont {Lukin}\ \emph {et~al.}(2019)\citenamefont {Lukin},
  \citenamefont {Rispoli}, \citenamefont {Schittko}, \citenamefont {Tai},
  \citenamefont {Kaufman}, \citenamefont {Choi}, \citenamefont {Khemani},
  \citenamefont {L{\'e}onard},\ and\ \citenamefont {Greiner}}]{lukin:2019}%
  \BibitemOpen
  \bibfield  {author} {\bibinfo {author} {\bibfnamefont {A.}~\bibnamefont
  {Lukin}}, \bibinfo {author} {\bibfnamefont {M.}~\bibnamefont {Rispoli}},
  \bibinfo {author} {\bibfnamefont {R.}~\bibnamefont {Schittko}}, \bibinfo
  {author} {\bibfnamefont {M.~E.}\ \bibnamefont {Tai}}, \bibinfo {author}
  {\bibfnamefont {A.~M.}\ \bibnamefont {Kaufman}}, \bibinfo {author}
  {\bibfnamefont {S.}~\bibnamefont {Choi}}, \bibinfo {author} {\bibfnamefont
  {V.}~\bibnamefont {Khemani}}, \bibinfo {author} {\bibfnamefont
  {J.}~\bibnamefont {L{\'e}onard}},\ and\ \bibinfo {author} {\bibfnamefont
  {M.}~\bibnamefont {Greiner}},\ }\bibfield  {title} {\bibinfo {title}
  {{P}robing entanglement in a many-body--localized system},\ }\href
  {https://doi.org/10.1126/science.aau0818} {\bibfield  {journal} {\bibinfo
  {journal} {Science}\ }\textbf {\bibinfo {volume} {364}},\ \bibinfo {pages}
  {256} (\bibinfo {year} {2019})}\BibitemShut {NoStop}%
\bibitem [{\citenamefont {Rispoli}\ \emph {et~al.}(2019)\citenamefont
  {Rispoli}, \citenamefont {Lukin}, \citenamefont {Schittko}, \citenamefont
  {Kim}, \citenamefont {Tai}, \citenamefont {L{\'e}onard},\ and\ \citenamefont
  {Greiner}}]{rispoli:2019}%
  \BibitemOpen
  \bibfield  {author} {\bibinfo {author} {\bibfnamefont {M.}~\bibnamefont
  {Rispoli}}, \bibinfo {author} {\bibfnamefont {A.}~\bibnamefont {Lukin}},
  \bibinfo {author} {\bibfnamefont {R.}~\bibnamefont {Schittko}}, \bibinfo
  {author} {\bibfnamefont {S.}~\bibnamefont {Kim}}, \bibinfo {author}
  {\bibfnamefont {M.~E.}\ \bibnamefont {Tai}}, \bibinfo {author} {\bibfnamefont
  {J.}~\bibnamefont {L{\'e}onard}},\ and\ \bibinfo {author} {\bibfnamefont
  {M.}~\bibnamefont {Greiner}},\ }\bibfield  {title} {\bibinfo {title}
  {{Q}uantum critical behaviour at the many-body localization transition},\
  }\href {https://doi.org/10.1038/s41586-019-1527-2} {\bibfield  {journal}
  {\bibinfo  {journal} {Nature}\ }\textbf {\bibinfo {volume} {573}},\ \bibinfo
  {pages} {385} (\bibinfo {year} {2019})}\BibitemShut {NoStop}%
\bibitem [{\citenamefont {Jepsen}\ \emph {et~al.}(2020)\citenamefont {Jepsen},
  \citenamefont {Amato-Grill}, \citenamefont {Dimitrova}, \citenamefont {Ho},
  \citenamefont {Demler},\ and\ \citenamefont {Ketterle}}]{jespen:2020}%
  \BibitemOpen
  \bibfield  {author} {\bibinfo {author} {\bibfnamefont {P.~N.}\ \bibnamefont
  {Jepsen}}, \bibinfo {author} {\bibfnamefont {J.}~\bibnamefont {Amato-Grill}},
  \bibinfo {author} {\bibfnamefont {I.}~\bibnamefont {Dimitrova}}, \bibinfo
  {author} {\bibfnamefont {W.~W.}\ \bibnamefont {Ho}}, \bibinfo {author}
  {\bibfnamefont {E.}~\bibnamefont {Demler}},\ and\ \bibinfo {author}
  {\bibfnamefont {W.}~\bibnamefont {Ketterle}},\ }\bibfield  {title} {\bibinfo
  {title} {{Spin transport in a tunable Heisenberg model realized with
  ultracold atoms}},\ }\href {https://doi.org/10.1038/s41586-020-3033-y}
  {\bibfield  {journal} {\bibinfo  {journal} {Nature}\ }\textbf {\bibinfo
  {volume} {588}},\ \bibinfo {pages} {403} (\bibinfo {year}
  {2020})}\BibitemShut {NoStop}%
\bibitem [{\citenamefont {Jepsen}\ \emph {et~al.}(2021)\citenamefont {Jepsen},
  \citenamefont {Ho}, \citenamefont {Amato-Grill}, \citenamefont {Dimitrova},
  \citenamefont {Demler},\ and\ \citenamefont {Ketterle}}]{jespen:2021}%
  \BibitemOpen
  \bibfield  {author} {\bibinfo {author} {\bibfnamefont {P.~N.}\ \bibnamefont
  {Jepsen}}, \bibinfo {author} {\bibfnamefont {W.~W.}\ \bibnamefont {Ho}},
  \bibinfo {author} {\bibfnamefont {J.}~\bibnamefont {Amato-Grill}}, \bibinfo
  {author} {\bibfnamefont {I.}~\bibnamefont {Dimitrova}}, \bibinfo {author}
  {\bibfnamefont {E.}~\bibnamefont {Demler}},\ and\ \bibinfo {author}
  {\bibfnamefont {W.}~\bibnamefont {Ketterle}},\ }\bibfield  {title} {\bibinfo
  {title} {{Transverse Spin Dynamics in the Anisotropic Heisenberg Model
  Realized with Ultracold Atoms}},\ }\href
  {https://doi.org/10.1103/PhysRevX.11.041054} {\bibfield  {journal} {\bibinfo
  {journal} {Phys. Rev. X}\ }\textbf {\bibinfo {volume} {11}},\ \bibinfo
  {pages} {041054} (\bibinfo {year} {2021})}\BibitemShut {NoStop}%
\bibitem [{\citenamefont {Ebadi}\ \emph {et~al.}(2021)\citenamefont {Ebadi},
  \citenamefont {Wang}, \citenamefont {Levine}, \citenamefont {Keesling},
  \citenamefont {Semeghini}, \citenamefont {Omran}, \citenamefont {Bluvstein},
  \citenamefont {Samajdar}, \citenamefont {Pichler}, \citenamefont {Ho} \emph
  {et~al.}}]{ebadi:2021}%
  \BibitemOpen
  \bibfield  {author} {\bibinfo {author} {\bibfnamefont {S.}~\bibnamefont
  {Ebadi}}, \bibinfo {author} {\bibfnamefont {T.~T.}\ \bibnamefont {Wang}},
  \bibinfo {author} {\bibfnamefont {H.}~\bibnamefont {Levine}}, \bibinfo
  {author} {\bibfnamefont {A.}~\bibnamefont {Keesling}}, \bibinfo {author}
  {\bibfnamefont {G.}~\bibnamefont {Semeghini}}, \bibinfo {author}
  {\bibfnamefont {A.}~\bibnamefont {Omran}}, \bibinfo {author} {\bibfnamefont
  {D.}~\bibnamefont {Bluvstein}}, \bibinfo {author} {\bibfnamefont
  {R.}~\bibnamefont {Samajdar}}, \bibinfo {author} {\bibfnamefont
  {H.}~\bibnamefont {Pichler}}, \bibinfo {author} {\bibfnamefont {W.~W.}\
  \bibnamefont {Ho}}, \emph {et~al.},\ }\bibfield  {title} {\bibinfo {title}
  {Quantum phases of matter on a 256-atom programmable quantum simulator},\
  }\href {https://doi.org/10.1038/s41586-021-03582-4} {\bibfield  {journal}
  {\bibinfo  {journal} {Nature}\ }\textbf {\bibinfo {volume} {595}},\ \bibinfo
  {pages} {227} (\bibinfo {year} {2021})}\BibitemShut {NoStop}%
\bibitem [{\citenamefont {Huang}\ \emph {et~al.}(2020)\citenamefont {Huang},
  \citenamefont {Kueng},\ and\ \citenamefont {Preskill}}]{huang:2020}%
  \BibitemOpen
  \bibfield  {author} {\bibinfo {author} {\bibfnamefont {H.-Y.}\ \bibnamefont
  {Huang}}, \bibinfo {author} {\bibfnamefont {R.}~\bibnamefont {Kueng}},\ and\
  \bibinfo {author} {\bibfnamefont {J.}~\bibnamefont {Preskill}},\ }\bibfield
  {title} {\bibinfo {title} {Predicting many properties of a quantum system
  from very few measurements},\ }\href
  {https://doi.org/10.1038/s41567-020-0932-7} {\bibfield  {journal} {\bibinfo
  {journal} {Nat. Phys.}\ }\textbf {\bibinfo {volume} {16}},\ \bibinfo {pages}
  {1050} (\bibinfo {year} {2020})}\BibitemShut {NoStop}%
\bibitem [{\citenamefont {Huang}\ \emph {et~al.}(2022)\citenamefont {Huang},
  \citenamefont {Broughton}, \citenamefont {Cotler}, \citenamefont {Chen},
  \citenamefont {Li}, \citenamefont {Mohseni}, \citenamefont {Neven},
  \citenamefont {Babbush}, \citenamefont {Kueng}, \citenamefont {Preskill},\
  and\ \citenamefont {McClean}}]{huang:2022}%
  \BibitemOpen
  \bibfield  {author} {\bibinfo {author} {\bibfnamefont {H.-Y.}\ \bibnamefont
  {Huang}}, \bibinfo {author} {\bibfnamefont {M.}~\bibnamefont {Broughton}},
  \bibinfo {author} {\bibfnamefont {J.}~\bibnamefont {Cotler}}, \bibinfo
  {author} {\bibfnamefont {S.}~\bibnamefont {Chen}}, \bibinfo {author}
  {\bibfnamefont {J.}~\bibnamefont {Li}}, \bibinfo {author} {\bibfnamefont
  {M.}~\bibnamefont {Mohseni}}, \bibinfo {author} {\bibfnamefont
  {H.}~\bibnamefont {Neven}}, \bibinfo {author} {\bibfnamefont
  {R.}~\bibnamefont {Babbush}}, \bibinfo {author} {\bibfnamefont
  {R.}~\bibnamefont {Kueng}}, \bibinfo {author} {\bibfnamefont
  {J.}~\bibnamefont {Preskill}},\ and\ \bibinfo {author} {\bibfnamefont
  {J.~R.}\ \bibnamefont {McClean}},\ }\bibfield  {title} {\bibinfo {title}
  {Quantum advantage in learning from experiments},\ }\href
  {https://doi.org/10.1126/science.abn7293} {\bibfield  {journal} {\bibinfo
  {journal} {Science}\ }\textbf {\bibinfo {volume} {376}},\ \bibinfo {pages}
  {1182} (\bibinfo {year} {2022})}\BibitemShut {NoStop}%
\bibitem [{\citenamefont {Nielsen}\ and\ \citenamefont
  {Chuang}(2010)}]{nielsen:2010}%
  \BibitemOpen
  \bibfield  {author} {\bibinfo {author} {\bibfnamefont {M.~A.}\ \bibnamefont
  {Nielsen}}\ and\ \bibinfo {author} {\bibfnamefont {I.~L.}\ \bibnamefont
  {Chuang}},\ }\href {https://doi.org/10.1017/CBO9780511976667} {\emph
  {\bibinfo {title} {{Quantum Computation and Quantum Information: 10th
  Anniversary Edition}}}}\ (\bibinfo  {publisher} {Cambridge University
  Press},\ \bibinfo {year} {2010})\BibitemShut {NoStop}%
\bibitem [{\citenamefont {Carrasquilla}\ \emph {et~al.}(2019)\citenamefont
  {Carrasquilla}, \citenamefont {Torlai}, \citenamefont {Melko},\ and\
  \citenamefont {Aolita}}]{carrasquilla:2019}%
  \BibitemOpen
  \bibfield  {author} {\bibinfo {author} {\bibfnamefont {J.}~\bibnamefont
  {Carrasquilla}}, \bibinfo {author} {\bibfnamefont {G.}~\bibnamefont
  {Torlai}}, \bibinfo {author} {\bibfnamefont {R.~G.}\ \bibnamefont {Melko}},\
  and\ \bibinfo {author} {\bibfnamefont {L.}~\bibnamefont {Aolita}},\
  }\bibfield  {title} {\bibinfo {title} {Reconstructing quantum states with
  generative models},\ }\href {https://doi.org/10.1038/s42256-019-0028-1}
  {\bibfield  {journal} {\bibinfo  {journal} {Nat. Mach. Intell.}\ }\textbf
  {\bibinfo {volume} {1}},\ \bibinfo {pages} {155} (\bibinfo {year}
  {2019})}\BibitemShut {NoStop}%
\bibitem [{\citenamefont {Torlai}\ \emph {et~al.}(2019)\citenamefont {Torlai},
  \citenamefont {Timar}, \citenamefont {van Nieuwenburg}, \citenamefont
  {Levine}, \citenamefont {Omran}, \citenamefont {Keesling}, \citenamefont
  {Bernien}, \citenamefont {Greiner}, \citenamefont
  {Vuleti\ifmmode~\acute{c}\else \'{c}\fi{}}, \citenamefont {Lukin},
  \citenamefont {Melko},\ and\ \citenamefont {Endres}}]{torlai:2019}%
  \BibitemOpen
  \bibfield  {author} {\bibinfo {author} {\bibfnamefont {G.}~\bibnamefont
  {Torlai}}, \bibinfo {author} {\bibfnamefont {B.}~\bibnamefont {Timar}},
  \bibinfo {author} {\bibfnamefont {E.~P.~L.}\ \bibnamefont {van Nieuwenburg}},
  \bibinfo {author} {\bibfnamefont {H.}~\bibnamefont {Levine}}, \bibinfo
  {author} {\bibfnamefont {A.}~\bibnamefont {Omran}}, \bibinfo {author}
  {\bibfnamefont {A.}~\bibnamefont {Keesling}}, \bibinfo {author}
  {\bibfnamefont {H.}~\bibnamefont {Bernien}}, \bibinfo {author} {\bibfnamefont
  {M.}~\bibnamefont {Greiner}}, \bibinfo {author} {\bibfnamefont
  {V.}~\bibnamefont {Vuleti\ifmmode~\acute{c}\else \'{c}\fi{}}}, \bibinfo
  {author} {\bibfnamefont {M.~D.}\ \bibnamefont {Lukin}}, \bibinfo {author}
  {\bibfnamefont {R.~G.}\ \bibnamefont {Melko}},\ and\ \bibinfo {author}
  {\bibfnamefont {M.}~\bibnamefont {Endres}},\ }\bibfield  {title} {\bibinfo
  {title} {{Integrating Neural Networks with a Quantum Simulator for State
  Reconstruction}},\ }\href {https://doi.org/10.1103/PhysRevLett.123.230504}
  {\bibfield  {journal} {\bibinfo  {journal} {Phys. Rev. Lett.}\ }\textbf
  {\bibinfo {volume} {123}},\ \bibinfo {pages} {230504} (\bibinfo {year}
  {2019})}\BibitemShut {NoStop}%
\bibitem [{\citenamefont {Miles}\ \emph
  {et~al.}(2021{\natexlab{b}})\citenamefont {Miles}, \citenamefont {Bohrdt},
  \citenamefont {Wu}, \citenamefont {Chiu}, \citenamefont {Xu}, \citenamefont
  {Ji}, \citenamefont {Greiner}, \citenamefont {Weinberger}, \citenamefont
  {Demler},\ and\ \citenamefont {Kim}}]{miles:2021}%
  \BibitemOpen
  \bibfield  {author} {\bibinfo {author} {\bibfnamefont {C.}~\bibnamefont
  {Miles}}, \bibinfo {author} {\bibfnamefont {A.}~\bibnamefont {Bohrdt}},
  \bibinfo {author} {\bibfnamefont {R.}~\bibnamefont {Wu}}, \bibinfo {author}
  {\bibfnamefont {C.}~\bibnamefont {Chiu}}, \bibinfo {author} {\bibfnamefont
  {M.}~\bibnamefont {Xu}}, \bibinfo {author} {\bibfnamefont {G.}~\bibnamefont
  {Ji}}, \bibinfo {author} {\bibfnamefont {M.}~\bibnamefont {Greiner}},
  \bibinfo {author} {\bibfnamefont {K.~Q.}\ \bibnamefont {Weinberger}},
  \bibinfo {author} {\bibfnamefont {E.}~\bibnamefont {Demler}},\ and\ \bibinfo
  {author} {\bibfnamefont {E.-A.}\ \bibnamefont {Kim}},\ }\bibfield  {title}
  {\bibinfo {title} {Correlator convolutional neural networks as an
  interpretable architecture for image-like quantum matter data},\ }\href
  {https://doi.org/10.1038/s41467-021-23952-w} {\bibfield  {journal} {\bibinfo
  {journal} {Nat. Commun.}\ }\textbf {\bibinfo {volume} {12}},\ \bibinfo
  {pages} {1} (\bibinfo {year} {2021}{\natexlab{b}})}\BibitemShut {NoStop}%
\bibitem [{\citenamefont {Onsager}(1944)}]{onsager:1944}%
  \BibitemOpen
  \bibfield  {author} {\bibinfo {author} {\bibfnamefont {L.}~\bibnamefont
  {Onsager}},\ }\bibfield  {title} {\bibinfo {title} {Crystal {S}tatistics. i.
  {A} {T}wo-{D}imensional {M}odel with an {O}rder-{D}isorder {T}ransition},\
  }\href {https://doi.org/10.1103/PhysRev.65.117} {\bibfield  {journal}
  {\bibinfo  {journal} {Phys. Rev.}\ }\textbf {\bibinfo {volume} {65}},\
  \bibinfo {pages} {117} (\bibinfo {year} {1944})}\BibitemShut {NoStop}%
\bibitem [{\citenamefont {Castelnovo}\ and\ \citenamefont
  {Chamon}(2007)}]{castelnovo:2007}%
  \BibitemOpen
  \bibfield  {author} {\bibinfo {author} {\bibfnamefont {C.}~\bibnamefont
  {Castelnovo}}\ and\ \bibinfo {author} {\bibfnamefont {C.}~\bibnamefont
  {Chamon}},\ }\bibfield  {title} {\bibinfo {title} {Entanglement and
  topological entropy of the toric code at finite temperature},\ }\href
  {https://doi.org/10.1103/PhysRevB.76.184442} {\bibfield  {journal} {\bibinfo
  {journal} {Phys. Rev. B}\ }\textbf {\bibinfo {volume} {76}},\ \bibinfo
  {pages} {184442} (\bibinfo {year} {2007})}\BibitemShut {NoStop}%
\bibitem [{\citenamefont {Wegner}(1971)}]{wegner:1971}%
  \BibitemOpen
  \bibfield  {author} {\bibinfo {author} {\bibfnamefont {F.~J.}\ \bibnamefont
  {Wegner}},\ }\bibfield  {title} {\bibinfo {title} {Duality in generalized
  {I}sing models and phase transitions without local order parameters},\ }\href
  {https://doi.org/10.1063/1.1665530} {\bibfield  {journal} {\bibinfo
  {journal} {J. Math. Phys.}\ }\textbf {\bibinfo {volume} {12}},\ \bibinfo
  {pages} {2259} (\bibinfo {year} {1971})}\BibitemShut {NoStop}%
\bibitem [{\citenamefont {Kogut}(1979)}]{kogut:1979}%
  \BibitemOpen
  \bibfield  {author} {\bibinfo {author} {\bibfnamefont {J.~B.}\ \bibnamefont
  {Kogut}},\ }\bibfield  {title} {\bibinfo {title} {An introduction to lattice
  gauge theory and spin systems},\ }\href
  {https://doi.org/10.1103/RevModPhys.51.659} {\bibfield  {journal} {\bibinfo
  {journal} {Rev. Mod. Phys.}\ }\textbf {\bibinfo {volume} {51}},\ \bibinfo
  {pages} {659} (\bibinfo {year} {1979})}\BibitemShut {NoStop}%
\bibitem [{\citenamefont {Hsieh}\ \emph {et~al.}(2013)\citenamefont {Hsieh},
  \citenamefont {Kao},\ and\ \citenamefont {Sandvik}}]{hsieh:2013}%
  \BibitemOpen
  \bibfield  {author} {\bibinfo {author} {\bibfnamefont {Y.-D.}\ \bibnamefont
  {Hsieh}}, \bibinfo {author} {\bibfnamefont {Y.-J.}\ \bibnamefont {Kao}},\
  and\ \bibinfo {author} {\bibfnamefont {A.~W.}\ \bibnamefont {Sandvik}},\
  }\bibfield  {title} {\bibinfo {title} {Finite-size scaling method for the
  {B}erezinskii--{K}osterlitz--{T}houless transition},\ }\href
  {https://doi.org/10.1088/1742-5468/2013/09/p09001} {\bibfield  {journal}
  {\bibinfo  {journal} {J. Stat. Mech.}\ }\textbf {\bibinfo {volume} {2013}},\
  \bibinfo {pages} {P09001} (\bibinfo {year} {2013})}\BibitemShut {NoStop}%
\bibitem [{\citenamefont {Kosterlitz}\ and\ \citenamefont
  {Thouless}(1973)}]{kosterlitz:1973}%
  \BibitemOpen
  \bibfield  {author} {\bibinfo {author} {\bibfnamefont {J.~M.}\ \bibnamefont
  {Kosterlitz}}\ and\ \bibinfo {author} {\bibfnamefont {D.~J.}\ \bibnamefont
  {Thouless}},\ }\bibfield  {title} {\bibinfo {title} {Ordering, metastability
  and phase transitions in two-dimensional systems},\ }\href
  {https://doi.org/10.1088/0022-3719/6/7/010} {\bibfield  {journal} {\bibinfo
  {journal} {J. Phys. C: Solid State Phys.}\ }\textbf {\bibinfo {volume} {6}},\
  \bibinfo {pages} {1181} (\bibinfo {year} {1973})}\BibitemShut {NoStop}%
\bibitem [{\citenamefont {Kosterlitz}(1974)}]{kosterlitz:1974}%
  \BibitemOpen
  \bibfield  {author} {\bibinfo {author} {\bibfnamefont {J.}~\bibnamefont
  {Kosterlitz}},\ }\bibfield  {title} {\bibinfo {title} {The critical
  properties of the two-dimensional {XY} model},\ }\href
  {https://doi.org/10.1088/0022-3719/7/6/005} {\bibfield  {journal} {\bibinfo
  {journal} {J. Phys. C: Solid State Phys.}\ }\textbf {\bibinfo {volume} {7}},\
  \bibinfo {pages} {1046} (\bibinfo {year} {1974})}\BibitemShut {NoStop}%
\bibitem [{\citenamefont {Chaikin}\ and\ \citenamefont
  {Lubensky}(1995)}]{chaikin:1995}%
  \BibitemOpen
  \bibfield  {author} {\bibinfo {author} {\bibfnamefont {P.~M.}\ \bibnamefont
  {Chaikin}}\ and\ \bibinfo {author} {\bibfnamefont {T.~C.}\ \bibnamefont
  {Lubensky}},\ }\href {https://doi.org/10.1017/CBO9780511813467} {\emph
  {\bibinfo {title} {{P}rinciples of {C}ondensed {M}atter {P}hysics}}}\
  (\bibinfo  {publisher} {Cambridge University Press},\ \bibinfo {year}
  {1995})\BibitemShut {NoStop}%
\bibitem [{\citenamefont {Van~Himbergen}\ and\ \citenamefont
  {Chakravarty}(1981)}]{himbergen:1981}%
  \BibitemOpen
  \bibfield  {author} {\bibinfo {author} {\bibfnamefont {J.~E.}\ \bibnamefont
  {Van~Himbergen}}\ and\ \bibinfo {author} {\bibfnamefont {S.}~\bibnamefont
  {Chakravarty}},\ }\bibfield  {title} {\bibinfo {title} {Helicity modulus and
  specific heat of classical $\mathrm{XY}$ model in two dimensions},\ }\href
  {https://doi.org/10.1103/PhysRevB.23.359} {\bibfield  {journal} {\bibinfo
  {journal} {Phys. Rev. B}\ }\textbf {\bibinfo {volume} {23}},\ \bibinfo
  {pages} {359} (\bibinfo {year} {1981})}\BibitemShut {NoStop}%
\bibitem [{\citenamefont {Chung}(1999)}]{chung:1999}%
  \BibitemOpen
  \bibfield  {author} {\bibinfo {author} {\bibfnamefont {S.~G.}\ \bibnamefont
  {Chung}},\ }\bibfield  {title} {\bibinfo {title} {Essential finite-size
  effect in the two-dimensional {XY} model},\ }\href
  {https://doi.org/10.1103/PhysRevB.60.11761} {\bibfield  {journal} {\bibinfo
  {journal} {Phys. Rev. B}\ }\textbf {\bibinfo {volume} {60}},\ \bibinfo
  {pages} {11761} (\bibinfo {year} {1999})}\BibitemShut {NoStop}%
\bibitem [{\citenamefont {Minnhagen}\ and\ \citenamefont
  {Kim}(2003)}]{minnhagen:2003}%
  \BibitemOpen
  \bibfield  {author} {\bibinfo {author} {\bibfnamefont {P.}~\bibnamefont
  {Minnhagen}}\ and\ \bibinfo {author} {\bibfnamefont {B.~J.}\ \bibnamefont
  {Kim}},\ }\bibfield  {title} {\bibinfo {title} {{Direct evidence of the
  discontinuous character of the Kosterlitz-Thouless jump}},\ }\href
  {https://doi.org/10.1103/PhysRevB.67.172509} {\bibfield  {journal} {\bibinfo
  {journal} {Phys. Rev. B}\ }\textbf {\bibinfo {volume} {67}},\ \bibinfo
  {pages} {172509} (\bibinfo {year} {2003})}\BibitemShut {NoStop}%
\bibitem [{\citenamefont {Schollw{\"o}ck}\ \emph {et~al.}(2008)\citenamefont
  {Schollw{\"o}ck}, \citenamefont {Richter}, \citenamefont {Farnell},\ and\
  \citenamefont {Bishop}}]{schollwock:2008}%
  \BibitemOpen
  \bibfield  {author} {\bibinfo {author} {\bibfnamefont {U.}~\bibnamefont
  {Schollw{\"o}ck}}, \bibinfo {author} {\bibfnamefont {J.}~\bibnamefont
  {Richter}}, \bibinfo {author} {\bibfnamefont {D.~J.}\ \bibnamefont
  {Farnell}},\ and\ \bibinfo {author} {\bibfnamefont {R.~F.}\ \bibnamefont
  {Bishop}},\ }\href {https://doi.org/10.1007/b96825} {\emph {\bibinfo {title}
  {Quantum {M}agnetism}}},\ Vol.\ \bibinfo {volume} {645}\ (\bibinfo
  {publisher} {Springer},\ \bibinfo {address} {Berlin, Heidelberg},\ \bibinfo
  {year} {2008})\BibitemShut {NoStop}%
\bibitem [{\citenamefont {Franchini}(2017)}]{franchini:2017}%
  \BibitemOpen
  \bibfield  {author} {\bibinfo {author} {\bibfnamefont {F.}~\bibnamefont
  {Franchini}},\ }\href {https://doi.org/10.1007/978-3-319-48487-7} {\emph
  {\bibinfo {title} {An {I}ntroduction to {I}ntegrable {T}echniques for
  {O}ne-{D}imensional {Q}uantum {S}ystems}}}\ (\bibinfo  {publisher}
  {Springer},\ \bibinfo {address} {Cham},\ \bibinfo {year} {2017})\BibitemShut
  {NoStop}%
\bibitem [{\citenamefont {Kitaev}(2001)}]{kitaev:2001}%
  \BibitemOpen
  \bibfield  {author} {\bibinfo {author} {\bibfnamefont {A.~Y.}\ \bibnamefont
  {Kitaev}},\ }\bibfield  {title} {\bibinfo {title} {Unpaired {M}ajorana
  fermions in quantum wires},\ }\href
  {https://doi.org/10.1070/1063-7869/44/10s/s29} {\bibfield  {journal}
  {\bibinfo  {journal} {Phys.-Usp.}\ }\textbf {\bibinfo {volume} {44}},\
  \bibinfo {pages} {131} (\bibinfo {year} {2001})}\BibitemShut {NoStop}%
\bibitem [{\citenamefont {Alicea}(2012)}]{alicea:2012}%
  \BibitemOpen
  \bibfield  {author} {\bibinfo {author} {\bibfnamefont {J.}~\bibnamefont
  {Alicea}},\ }\bibfield  {title} {\bibinfo {title} {New directions in the
  pursuit of {M}ajorana fermions in solid state systems},\ }\href
  {https://doi.org/10.1088/0034-4885/75/7/076501} {\bibfield  {journal}
  {\bibinfo  {journal} {Rep. Prog. Phys.}\ }\textbf {\bibinfo {volume} {75}},\
  \bibinfo {pages} {076501} (\bibinfo {year} {2012})}\BibitemShut {NoStop}%
\bibitem [{\citenamefont {Wilczek}(2009)}]{wilczek:2009}%
  \BibitemOpen
  \bibfield  {author} {\bibinfo {author} {\bibfnamefont {F.}~\bibnamefont
  {Wilczek}},\ }\bibfield  {title} {\bibinfo {title} {Majorana returns},\
  }\href {https://doi.org/10.1038/nphys1380} {\bibfield  {journal} {\bibinfo
  {journal} {Nat. Phys.}\ }\textbf {\bibinfo {volume} {5}},\ \bibinfo {pages}
  {614} (\bibinfo {year} {2009})}\BibitemShut {NoStop}%
\bibitem [{\citenamefont {Amico}\ \emph {et~al.}(2008)\citenamefont {Amico},
  \citenamefont {Fazio}, \citenamefont {Osterloh},\ and\ \citenamefont
  {Vedral}}]{amico:2008}%
  \BibitemOpen
  \bibfield  {author} {\bibinfo {author} {\bibfnamefont {L.}~\bibnamefont
  {Amico}}, \bibinfo {author} {\bibfnamefont {R.}~\bibnamefont {Fazio}},
  \bibinfo {author} {\bibfnamefont {A.}~\bibnamefont {Osterloh}},\ and\
  \bibinfo {author} {\bibfnamefont {V.}~\bibnamefont {Vedral}},\ }\bibfield
  {title} {\bibinfo {title} {Entanglement in many-body systems},\ }\href
  {https://doi.org/10.1103/RevModPhys.80.517} {\bibfield  {journal} {\bibinfo
  {journal} {Rev. Mod. Phys.}\ }\textbf {\bibinfo {volume} {80}},\ \bibinfo
  {pages} {517} (\bibinfo {year} {2008})}\BibitemShut {NoStop}%
\bibitem [{\citenamefont {Pal}\ and\ \citenamefont {Huse}(2010)}]{pal:2010}%
  \BibitemOpen
  \bibfield  {author} {\bibinfo {author} {\bibfnamefont {A.}~\bibnamefont
  {Pal}}\ and\ \bibinfo {author} {\bibfnamefont {D.~A.}\ \bibnamefont {Huse}},\
  }\bibfield  {title} {\bibinfo {title} {Many-body localization phase
  transition},\ }\href {https://doi.org/10.1103/PhysRevB.82.174411} {\bibfield
  {journal} {\bibinfo  {journal} {Phys. Rev. B}\ }\textbf {\bibinfo {volume}
  {82}},\ \bibinfo {pages} {174411} (\bibinfo {year} {2010})}\BibitemShut
  {NoStop}%
\bibitem [{\citenamefont {Khemani}\ \emph {et~al.}(2017)\citenamefont
  {Khemani}, \citenamefont {Lim}, \citenamefont {Sheng},\ and\ \citenamefont
  {Huse}}]{khemani:2017}%
  \BibitemOpen
  \bibfield  {author} {\bibinfo {author} {\bibfnamefont {V.}~\bibnamefont
  {Khemani}}, \bibinfo {author} {\bibfnamefont {S.~P.}\ \bibnamefont {Lim}},
  \bibinfo {author} {\bibfnamefont {D.~N.}\ \bibnamefont {Sheng}},\ and\
  \bibinfo {author} {\bibfnamefont {D.~A.}\ \bibnamefont {Huse}},\ }\bibfield
  {title} {\bibinfo {title} {{C}ritical {P}roperties of the {M}any-{B}ody
  {L}ocalization {T}ransition},\ }\href
  {https://doi.org/10.1103/PhysRevX.7.021013} {\bibfield  {journal} {\bibinfo
  {journal} {Phys. Rev. X}\ }\textbf {\bibinfo {volume} {7}},\ \bibinfo {pages}
  {021013} (\bibinfo {year} {2017})}\BibitemShut {NoStop}%
\bibitem [{\citenamefont {Alet}\ and\ \citenamefont
  {Laflorencie}(2018)}]{alet:2018}%
  \BibitemOpen
  \bibfield  {author} {\bibinfo {author} {\bibfnamefont {F.}~\bibnamefont
  {Alet}}\ and\ \bibinfo {author} {\bibfnamefont {N.}~\bibnamefont
  {Laflorencie}},\ }\bibfield  {title} {\bibinfo {title} {{M}any-body
  localization: {A}n introduction and selected topics},\ }\href
  {https://doi.org/10.1016/j.crhy.2018.03.003} {\bibfield  {journal} {\bibinfo
  {journal} {C. R. Phys.}\ }\textbf {\bibinfo {volume} {19}},\ \bibinfo {pages}
  {498} (\bibinfo {year} {2018})}\BibitemShut {NoStop}%
\bibitem [{\citenamefont {Greitemann}\ \emph
  {et~al.}(2019{\natexlab{b}})\citenamefont {Greitemann}, \citenamefont {Liu},\
  and\ \citenamefont {Pollet}}]{greitemann_2:2019}%
  \BibitemOpen
  \bibfield  {author} {\bibinfo {author} {\bibfnamefont {J.}~\bibnamefont
  {Greitemann}}, \bibinfo {author} {\bibfnamefont {K.}~\bibnamefont {Liu}},\
  and\ \bibinfo {author} {\bibfnamefont {L.}~\bibnamefont {Pollet}},\
  }\bibfield  {title} {\bibinfo {title} {Probing hidden spin order with
  interpretable machine learning},\ }\href
  {https://doi.org/10.1103/PhysRevB.99.060404} {\bibfield  {journal} {\bibinfo
  {journal} {Phys. Rev. B}\ }\textbf {\bibinfo {volume} {99}},\ \bibinfo
  {pages} {060404} (\bibinfo {year} {2019}{\natexlab{b}})}\BibitemShut
  {NoStop}%
\bibitem [{\citenamefont {Kottmann}\ \emph {et~al.}(2021)\citenamefont
  {Kottmann}, \citenamefont {Corboz}, \citenamefont {Lewenstein},\ and\
  \citenamefont {Acín}}]{kottmann:2021}%
  \BibitemOpen
  \bibfield  {author} {\bibinfo {author} {\bibfnamefont {K.}~\bibnamefont
  {Kottmann}}, \bibinfo {author} {\bibfnamefont {P.}~\bibnamefont {Corboz}},
  \bibinfo {author} {\bibfnamefont {M.}~\bibnamefont {Lewenstein}},\ and\
  \bibinfo {author} {\bibfnamefont {A.}~\bibnamefont {Acín}},\ }\bibfield
  {title} {\bibinfo {title} {{Unsupervised mapping of phase diagrams of 2D
  systems from infinite projected entangled-pair states via deep anomaly
  detection}},\ }\href {https://doi.org/10.21468/SciPostPhys.11.2.025}
  {\bibfield  {journal} {\bibinfo  {journal} {SciPost Phys.}\ }\textbf
  {\bibinfo {volume} {11}},\ \bibinfo {pages} {25} (\bibinfo {year}
  {2021})}\BibitemShut {NoStop}%
\bibitem [{\citenamefont {Bohrdt}\ \emph {et~al.}(2019)\citenamefont {Bohrdt},
  \citenamefont {Chiu}, \citenamefont {Ji}, \citenamefont {Xu}, \citenamefont
  {Greif}, \citenamefont {Greiner}, \citenamefont {Demler}, \citenamefont
  {Grusdt},\ and\ \citenamefont {Knap}}]{bohrdt:2019}%
  \BibitemOpen
  \bibfield  {author} {\bibinfo {author} {\bibfnamefont {A.}~\bibnamefont
  {Bohrdt}}, \bibinfo {author} {\bibfnamefont {C.~S.}\ \bibnamefont {Chiu}},
  \bibinfo {author} {\bibfnamefont {G.}~\bibnamefont {Ji}}, \bibinfo {author}
  {\bibfnamefont {M.}~\bibnamefont {Xu}}, \bibinfo {author} {\bibfnamefont
  {D.}~\bibnamefont {Greif}}, \bibinfo {author} {\bibfnamefont
  {M.}~\bibnamefont {Greiner}}, \bibinfo {author} {\bibfnamefont
  {E.}~\bibnamefont {Demler}}, \bibinfo {author} {\bibfnamefont
  {F.}~\bibnamefont {Grusdt}},\ and\ \bibinfo {author} {\bibfnamefont
  {M.}~\bibnamefont {Knap}},\ }\bibfield  {title} {\bibinfo {title}
  {Classifying snapshots of the doped {H}ubbard model with machine learning},\
  }\href {https://doi.org/10.1038/s41567-019-0565-x} {\bibfield  {journal}
  {\bibinfo  {journal} {Nat. Phys.}\ }\textbf {\bibinfo {volume} {15}},\
  \bibinfo {pages} {921} (\bibinfo {year} {2019})}\BibitemShut {NoStop}%
\bibitem [{\citenamefont {Zhang}\ \emph
  {et~al.}(2019{\natexlab{b}})\citenamefont {Zhang}, \citenamefont {Mesaros},
  \citenamefont {Fujita}, \citenamefont {Edkins}, \citenamefont {Hamidian},
  \citenamefont {Ch’ng}, \citenamefont {Eisaki}, \citenamefont {Uchida},
  \citenamefont {Davis}, \citenamefont {Khatami} \emph {et~al.}}]{zhang2:2019}%
  \BibitemOpen
  \bibfield  {author} {\bibinfo {author} {\bibfnamefont {Y.}~\bibnamefont
  {Zhang}}, \bibinfo {author} {\bibfnamefont {A.}~\bibnamefont {Mesaros}},
  \bibinfo {author} {\bibfnamefont {K.}~\bibnamefont {Fujita}}, \bibinfo
  {author} {\bibfnamefont {S.}~\bibnamefont {Edkins}}, \bibinfo {author}
  {\bibfnamefont {M.}~\bibnamefont {Hamidian}}, \bibinfo {author}
  {\bibfnamefont {K.}~\bibnamefont {Ch’ng}}, \bibinfo {author} {\bibfnamefont
  {H.}~\bibnamefont {Eisaki}}, \bibinfo {author} {\bibfnamefont
  {S.}~\bibnamefont {Uchida}}, \bibinfo {author} {\bibfnamefont {J.~S.}\
  \bibnamefont {Davis}}, \bibinfo {author} {\bibfnamefont {E.}~\bibnamefont
  {Khatami}}, \emph {et~al.},\ }\bibfield  {title} {\bibinfo {title} {Machine
  learning in electronic-quantum-matter imaging experiments},\ }\href
  {https://doi.org/10.1038/s41586-019-1319-8} {\bibfield  {journal} {\bibinfo
  {journal} {Nature}\ }\textbf {\bibinfo {volume} {570}},\ \bibinfo {pages}
  {484} (\bibinfo {year} {2019}{\natexlab{b}})}\BibitemShut {NoStop}%
\bibitem [{\citenamefont {Pilati}\ and\ \citenamefont
  {Pieri}(2019)}]{pilati:2019}%
  \BibitemOpen
  \bibfield  {author} {\bibinfo {author} {\bibfnamefont {S.}~\bibnamefont
  {Pilati}}\ and\ \bibinfo {author} {\bibfnamefont {P.}~\bibnamefont {Pieri}},\
  }\bibfield  {title} {\bibinfo {title} {{S}upervised machine learning of
  ultracold atoms with speckle disorder},\ }\href
  {https://doi.org/10.1038/s41598-019-42125-w} {\bibfield  {journal} {\bibinfo
  {journal} {Sci. Rep.}\ }\textbf {\bibinfo {volume} {9}},\ \bibinfo {pages}
  {1} (\bibinfo {year} {2019})}\BibitemShut {NoStop}%
\bibitem [{\citenamefont {Ghosh}\ \emph {et~al.}(2020)\citenamefont {Ghosh},
  \citenamefont {Matty}, \citenamefont {Baumbach}, \citenamefont {Bauer},
  \citenamefont {Modic}, \citenamefont {Shekhter}, \citenamefont {Mydosh},
  \citenamefont {Kim},\ and\ \citenamefont {Ramshaw}}]{ghosh:2020}%
  \BibitemOpen
  \bibfield  {author} {\bibinfo {author} {\bibfnamefont {S.}~\bibnamefont
  {Ghosh}}, \bibinfo {author} {\bibfnamefont {M.}~\bibnamefont {Matty}},
  \bibinfo {author} {\bibfnamefont {R.}~\bibnamefont {Baumbach}}, \bibinfo
  {author} {\bibfnamefont {E.~D.}\ \bibnamefont {Bauer}}, \bibinfo {author}
  {\bibfnamefont {K.~A.}\ \bibnamefont {Modic}}, \bibinfo {author}
  {\bibfnamefont {A.}~\bibnamefont {Shekhter}}, \bibinfo {author}
  {\bibfnamefont {J.}~\bibnamefont {Mydosh}}, \bibinfo {author} {\bibfnamefont
  {E.-A.}\ \bibnamefont {Kim}},\ and\ \bibinfo {author} {\bibfnamefont
  {B.}~\bibnamefont {Ramshaw}},\ }\bibfield  {title} {\bibinfo {title}
  {{O}ne-component order parameter in {UR}u$_{2}${Si}$_{2}$ uncovered by
  resonant ultrasound spectroscopy and machine learning},\ }\href
  {https://doi.org/10.1126/sciadv.aaz4074} {\bibfield  {journal} {\bibinfo
  {journal} {Sci. Adv.}\ }\textbf {\bibinfo {volume} {6}},\ \bibinfo {pages}
  {eaaz4074} (\bibinfo {year} {2020})}\BibitemShut {NoStop}%
\bibitem [{\citenamefont {Szo\l{}dra}\ \emph {et~al.}(2021)\citenamefont
  {Szo\l{}dra}, \citenamefont {Sierant}, \citenamefont {Kottmann},
  \citenamefont {Lewenstein},\ and\ \citenamefont {Zakrzewski}}]{szoldra:2021}%
  \BibitemOpen
  \bibfield  {author} {\bibinfo {author} {\bibfnamefont {T.}~\bibnamefont
  {Szo\l{}dra}}, \bibinfo {author} {\bibfnamefont {P.}~\bibnamefont {Sierant}},
  \bibinfo {author} {\bibfnamefont {K.}~\bibnamefont {Kottmann}}, \bibinfo
  {author} {\bibfnamefont {M.}~\bibnamefont {Lewenstein}},\ and\ \bibinfo
  {author} {\bibfnamefont {J.}~\bibnamefont {Zakrzewski}},\ }\bibfield  {title}
  {\bibinfo {title} {Detecting ergodic bubbles at the crossover to many-body
  localization using neural networks},\ }\href
  {https://doi.org/10.1103/PhysRevB.104.L140202} {\bibfield  {journal}
  {\bibinfo  {journal} {Phys. Rev. B}\ }\textbf {\bibinfo {volume} {104}},\
  \bibinfo {pages} {L140202} (\bibinfo {year} {2021})}\BibitemShut {NoStop}%
\bibitem [{\citenamefont {Gavreev}\ \emph {et~al.}(2022)\citenamefont
  {Gavreev}, \citenamefont {Mastiukova}, \citenamefont {Kiktenko},\ and\
  \citenamefont {Fedorov}}]{gavreev:2022}%
  \BibitemOpen
  \bibfield  {author} {\bibinfo {author} {\bibfnamefont {M.}~\bibnamefont
  {Gavreev}}, \bibinfo {author} {\bibfnamefont {A.}~\bibnamefont {Mastiukova}},
  \bibinfo {author} {\bibfnamefont {E.}~\bibnamefont {Kiktenko}},\ and\
  \bibinfo {author} {\bibfnamefont {A.}~\bibnamefont {Fedorov}},\ }\bibfield
  {title} {\bibinfo {title} {Learning entanglement breakdown as a phase
  transition by confusion},\ }\href {https://arxiv.org/abs/2202.00348}
  {\bibfield  {journal} {\bibinfo  {journal} {arXiv:2202.00348}\ } (\bibinfo
  {year} {2022})}\BibitemShut {NoStop}%
\bibitem [{\citenamefont {McClean}\ \emph {et~al.}(2018)\citenamefont
  {McClean}, \citenamefont {Boixo}, \citenamefont {Smelyanskiy}, \citenamefont
  {Babbush},\ and\ \citenamefont {Neven}}]{mcclean:2018}%
  \BibitemOpen
  \bibfield  {author} {\bibinfo {author} {\bibfnamefont {J.~R.}\ \bibnamefont
  {McClean}}, \bibinfo {author} {\bibfnamefont {S.}~\bibnamefont {Boixo}},
  \bibinfo {author} {\bibfnamefont {V.~N.}\ \bibnamefont {Smelyanskiy}},
  \bibinfo {author} {\bibfnamefont {R.}~\bibnamefont {Babbush}},\ and\ \bibinfo
  {author} {\bibfnamefont {H.}~\bibnamefont {Neven}},\ }\bibfield  {title}
  {\bibinfo {title} {Barren plateaus in quantum neural network training
  landscapes},\ }\href {https://doi.org/10.1038/s41467-018-07090-4} {\bibfield
  {journal} {\bibinfo  {journal} {Nat. Commun.}\ }\textbf {\bibinfo {volume}
  {9}},\ \bibinfo {pages} {1} (\bibinfo {year} {2018})}\BibitemShut {NoStop}%
\bibitem [{\citenamefont {Vieijra}\ \emph {et~al.}(2020)\citenamefont
  {Vieijra}, \citenamefont {Casert}, \citenamefont {Nys}, \citenamefont
  {De~Neve}, \citenamefont {Haegeman}, \citenamefont {Ryckebusch},\ and\
  \citenamefont {Verstraete}}]{vieijra:2020}%
  \BibitemOpen
  \bibfield  {author} {\bibinfo {author} {\bibfnamefont {T.}~\bibnamefont
  {Vieijra}}, \bibinfo {author} {\bibfnamefont {C.}~\bibnamefont {Casert}},
  \bibinfo {author} {\bibfnamefont {J.}~\bibnamefont {Nys}}, \bibinfo {author}
  {\bibfnamefont {W.}~\bibnamefont {De~Neve}}, \bibinfo {author} {\bibfnamefont
  {J.}~\bibnamefont {Haegeman}}, \bibinfo {author} {\bibfnamefont
  {J.}~\bibnamefont {Ryckebusch}},\ and\ \bibinfo {author} {\bibfnamefont
  {F.}~\bibnamefont {Verstraete}},\ }\bibfield  {title} {\bibinfo {title}
  {{Restricted Boltzmann Machines for Quantum States with Non-Abelian or
  Anyonic Symmetries}},\ }\href
  {https://doi.org/10.1103/PhysRevLett.124.097201} {\bibfield  {journal}
  {\bibinfo  {journal} {Phys. Rev. Lett.}\ }\textbf {\bibinfo {volume} {124}},\
  \bibinfo {pages} {097201} (\bibinfo {year} {2020})}\BibitemShut {NoStop}%
\bibitem [{\citenamefont {Bukov}\ \emph {et~al.}(2021)\citenamefont {Bukov},
  \citenamefont {Schmitt},\ and\ \citenamefont {Dupont}}]{bukov:2021}%
  \BibitemOpen
  \bibfield  {author} {\bibinfo {author} {\bibfnamefont {M.}~\bibnamefont
  {Bukov}}, \bibinfo {author} {\bibfnamefont {M.}~\bibnamefont {Schmitt}},\
  and\ \bibinfo {author} {\bibfnamefont {M.}~\bibnamefont {Dupont}},\
  }\bibfield  {title} {\bibinfo {title} {{Learning the ground state of a
  non-stoquastic quantum Hamiltonian in a rugged neural network landscape}},\
  }\href {https://doi.org/10.21468/SciPostPhys.10.6.147} {\bibfield  {journal}
  {\bibinfo  {journal} {SciPost Phys.}\ }\textbf {\bibinfo {volume} {10}},\
  \bibinfo {pages} {147} (\bibinfo {year} {2021})}\BibitemShut {NoStop}%
\bibitem [{\citenamefont {Valenti}\ \emph {et~al.}(2022)\citenamefont
  {Valenti}, \citenamefont {Greplova}, \citenamefont {Lindner},\ and\
  \citenamefont {Huber}}]{valenti:2022}%
  \BibitemOpen
  \bibfield  {author} {\bibinfo {author} {\bibfnamefont {A.}~\bibnamefont
  {Valenti}}, \bibinfo {author} {\bibfnamefont {E.}~\bibnamefont {Greplova}},
  \bibinfo {author} {\bibfnamefont {N.~H.}\ \bibnamefont {Lindner}},\ and\
  \bibinfo {author} {\bibfnamefont {S.~D.}\ \bibnamefont {Huber}},\ }\bibfield
  {title} {\bibinfo {title} {Correlation-enhanced neural networks as
  interpretable variational quantum states},\ }\href
  {https://doi.org/10.1103/PhysRevResearch.4.L012010} {\bibfield  {journal}
  {\bibinfo  {journal} {Phys. Rev. Res.}\ }\textbf {\bibinfo {volume} {4}},\
  \bibinfo {pages} {L012010} (\bibinfo {year} {2022})}\BibitemShut {NoStop}%
\bibitem [{\citenamefont {Bishop}(2006)}]{bishop:2006}%
  \BibitemOpen
  \bibfield  {author} {\bibinfo {author} {\bibfnamefont {C.~M.}\ \bibnamefont
  {Bishop}},\ }\href {https://link.springer.com/book/9780387310732} {\emph
  {\bibinfo {title} {{P}attern {R}ecognition and {M}achine {L}earning
  ({I}nformation {S}cience and {S}tatistics)}}}\ (\bibinfo  {publisher}
  {Springer},\ \bibinfo {address} {Berlin, Heidelberg},\ \bibinfo {year}
  {2006})\BibitemShut {NoStop}%
\bibitem [{\citenamefont {Wu}\ \emph {et~al.}(2019)\citenamefont {Wu},
  \citenamefont {Wang},\ and\ \citenamefont {Zhang}}]{wu:2019}%
  \BibitemOpen
  \bibfield  {author} {\bibinfo {author} {\bibfnamefont {D.}~\bibnamefont
  {Wu}}, \bibinfo {author} {\bibfnamefont {L.}~\bibnamefont {Wang}},\ and\
  \bibinfo {author} {\bibfnamefont {P.}~\bibnamefont {Zhang}},\ }\bibfield
  {title} {\bibinfo {title} {{S}olving {S}tatistical {M}echanics {U}sing
  {V}ariational {A}utoregressive {N}etworks},\ }\href
  {https://doi.org/10.1103/PhysRevLett.122.080602} {\bibfield  {journal}
  {\bibinfo  {journal} {Phys. Rev. Lett.}\ }\textbf {\bibinfo {volume} {122}},\
  \bibinfo {pages} {080602} (\bibinfo {year} {2019})}\BibitemShut {NoStop}%
\bibitem [{\citenamefont {Melko}\ \emph {et~al.}(2019)\citenamefont {Melko},
  \citenamefont {Carleo}, \citenamefont {Carrasquilla},\ and\ \citenamefont
  {Cirac}}]{melko:2019}%
  \BibitemOpen
  \bibfield  {author} {\bibinfo {author} {\bibfnamefont {R.~G.}\ \bibnamefont
  {Melko}}, \bibinfo {author} {\bibfnamefont {G.}~\bibnamefont {Carleo}},
  \bibinfo {author} {\bibfnamefont {J.}~\bibnamefont {Carrasquilla}},\ and\
  \bibinfo {author} {\bibfnamefont {J.~I.}\ \bibnamefont {Cirac}},\ }\bibfield
  {title} {\bibinfo {title} {{R}estricted {B}oltzmann machines in quantum
  physics},\ }\href {https://doi.org/10.1038/s41567-019-0545-1} {\bibfield
  {journal} {\bibinfo  {journal} {Nat. Phys.}\ }\textbf {\bibinfo {volume}
  {15}},\ \bibinfo {pages} {887} (\bibinfo {year} {2019})}\BibitemShut
  {NoStop}%
\bibitem [{\citenamefont {Nicoli}\ \emph {et~al.}(2021)\citenamefont {Nicoli},
  \citenamefont {Anders}, \citenamefont {Funcke}, \citenamefont {Hartung},
  \citenamefont {Jansen}, \citenamefont {Kessel}, \citenamefont {Nakajima},\
  and\ \citenamefont {Stornati}}]{nicoli:2021}%
  \BibitemOpen
  \bibfield  {author} {\bibinfo {author} {\bibfnamefont {K.~A.}\ \bibnamefont
  {Nicoli}}, \bibinfo {author} {\bibfnamefont {C.~J.}\ \bibnamefont {Anders}},
  \bibinfo {author} {\bibfnamefont {L.}~\bibnamefont {Funcke}}, \bibinfo
  {author} {\bibfnamefont {T.}~\bibnamefont {Hartung}}, \bibinfo {author}
  {\bibfnamefont {K.}~\bibnamefont {Jansen}}, \bibinfo {author} {\bibfnamefont
  {P.}~\bibnamefont {Kessel}}, \bibinfo {author} {\bibfnamefont
  {S.}~\bibnamefont {Nakajima}},\ and\ \bibinfo {author} {\bibfnamefont
  {P.}~\bibnamefont {Stornati}},\ }\bibfield  {title} {\bibinfo {title}
  {{E}stimation of {T}hermodynamic {O}bservables in {L}attice {F}ield
  {T}heories with {D}eep {G}enerative {M}odels},\ }\href
  {https://doi.org/10.1103/PhysRevLett.126.032001} {\bibfield  {journal}
  {\bibinfo  {journal} {Phys. Rev. Lett.}\ }\textbf {\bibinfo {volume} {126}},\
  \bibinfo {pages} {032001} (\bibinfo {year} {2021})}\BibitemShut {NoStop}%
\bibitem [{\citenamefont {Smith}\ \emph {et~al.}(2019)\citenamefont {Smith},
  \citenamefont {Kim}, \citenamefont {Pollmann},\ and\ \citenamefont
  {Knolle}}]{smith:2019}%
  \BibitemOpen
  \bibfield  {author} {\bibinfo {author} {\bibfnamefont {A.}~\bibnamefont
  {Smith}}, \bibinfo {author} {\bibfnamefont {M.}~\bibnamefont {Kim}}, \bibinfo
  {author} {\bibfnamefont {F.}~\bibnamefont {Pollmann}},\ and\ \bibinfo
  {author} {\bibfnamefont {J.}~\bibnamefont {Knolle}},\ }\bibfield  {title}
  {\bibinfo {title} {Simulating quantum many-body dynamics on a current digital
  quantum computer},\ }\href {https://doi.org/10.1038/s41534-019-0217-0}
  {\bibfield  {journal} {\bibinfo  {journal} {Npj Quantum Inf.}\ }\textbf
  {\bibinfo {volume} {5}},\ \bibinfo {pages} {1} (\bibinfo {year}
  {2019})}\BibitemShut {NoStop}%
\bibitem [{\citenamefont {Barratt}\ \emph {et~al.}(2021)\citenamefont
  {Barratt}, \citenamefont {Dborin}, \citenamefont {Bal}, \citenamefont
  {Stojevic}, \citenamefont {Pollmann},\ and\ \citenamefont
  {Green}}]{barratt:2021}%
  \BibitemOpen
  \bibfield  {author} {\bibinfo {author} {\bibfnamefont {F.}~\bibnamefont
  {Barratt}}, \bibinfo {author} {\bibfnamefont {J.}~\bibnamefont {Dborin}},
  \bibinfo {author} {\bibfnamefont {M.}~\bibnamefont {Bal}}, \bibinfo {author}
  {\bibfnamefont {V.}~\bibnamefont {Stojevic}}, \bibinfo {author}
  {\bibfnamefont {F.}~\bibnamefont {Pollmann}},\ and\ \bibinfo {author}
  {\bibfnamefont {A.~G.}\ \bibnamefont {Green}},\ }\bibfield  {title} {\bibinfo
  {title} {{Parallel quantum simulation of large systems on small NISQ
  computers}},\ }\href {https://doi.org/10.1038/s41534-021-00420-3} {\bibfield
  {journal} {\bibinfo  {journal} {Npj Quantum Inf.}\ }\textbf {\bibinfo
  {volume} {7}},\ \bibinfo {pages} {1} (\bibinfo {year} {2021})}\BibitemShut
  {NoStop}%
\bibitem [{\citenamefont {Satzinger}\ \emph {et~al.}(2021)\citenamefont
  {Satzinger}, \citenamefont {Liu}, \citenamefont {Smith}, \citenamefont
  {Knapp}, \citenamefont {Newman}, \citenamefont {Jones}, \citenamefont {Chen},
  \citenamefont {Quintana}, \citenamefont {Mi}, \citenamefont {Dunsworth} \emph
  {et~al.}}]{satzinger:2021}%
  \BibitemOpen
  \bibfield  {author} {\bibinfo {author} {\bibfnamefont {K.}~\bibnamefont
  {Satzinger}}, \bibinfo {author} {\bibfnamefont {Y.-J.}\ \bibnamefont {Liu}},
  \bibinfo {author} {\bibfnamefont {A.}~\bibnamefont {Smith}}, \bibinfo
  {author} {\bibfnamefont {C.}~\bibnamefont {Knapp}}, \bibinfo {author}
  {\bibfnamefont {M.}~\bibnamefont {Newman}}, \bibinfo {author} {\bibfnamefont
  {C.}~\bibnamefont {Jones}}, \bibinfo {author} {\bibfnamefont
  {Z.}~\bibnamefont {Chen}}, \bibinfo {author} {\bibfnamefont {C.}~\bibnamefont
  {Quintana}}, \bibinfo {author} {\bibfnamefont {X.}~\bibnamefont {Mi}},
  \bibinfo {author} {\bibfnamefont {A.}~\bibnamefont {Dunsworth}}, \emph
  {et~al.},\ }\bibfield  {title} {\bibinfo {title} {Realizing topologically
  ordered states on a quantum processor},\ }\href
  {https://doi.org/DOI:10.1126/science.abi8378} {\bibfield  {journal} {\bibinfo
   {journal} {Science}\ }\textbf {\bibinfo {volume} {374}},\ \bibinfo {pages}
  {1237} (\bibinfo {year} {2021})}\BibitemShut {NoStop}%
\bibitem [{\citenamefont {Herrmann}\ \emph {et~al.}(2021)\citenamefont
  {Herrmann}, \citenamefont {Llima}, \citenamefont {Remm}, \citenamefont
  {Zapletal}, \citenamefont {McMahon}, \citenamefont {Scarato}, \citenamefont
  {Swiadek}, \citenamefont {Andersen}, \citenamefont {Hellings}, \citenamefont
  {Krinner} \emph {et~al.}}]{herrmann:2021}%
  \BibitemOpen
  \bibfield  {author} {\bibinfo {author} {\bibfnamefont {J.}~\bibnamefont
  {Herrmann}}, \bibinfo {author} {\bibfnamefont {S.~M.}\ \bibnamefont {Llima}},
  \bibinfo {author} {\bibfnamefont {A.}~\bibnamefont {Remm}}, \bibinfo {author}
  {\bibfnamefont {P.}~\bibnamefont {Zapletal}}, \bibinfo {author}
  {\bibfnamefont {N.~A.}\ \bibnamefont {McMahon}}, \bibinfo {author}
  {\bibfnamefont {C.}~\bibnamefont {Scarato}}, \bibinfo {author} {\bibfnamefont
  {F.}~\bibnamefont {Swiadek}}, \bibinfo {author} {\bibfnamefont {C.~K.}\
  \bibnamefont {Andersen}}, \bibinfo {author} {\bibfnamefont {C.}~\bibnamefont
  {Hellings}}, \bibinfo {author} {\bibfnamefont {S.}~\bibnamefont {Krinner}},
  \emph {et~al.},\ }\bibfield  {title} {\bibinfo {title} {{Realizing Quantum
  Convolutional Neural Networks on a Superconducting Quantum Processor to
  Recognize Quantum Phases}},\ }\href {https://arxiv.org/abs/2109.05909}
  {\bibfield  {journal} {\bibinfo  {journal} {arXiv:2109.05909}\ } (\bibinfo
  {year} {2021})}\BibitemShut {NoStop}%
\bibitem [{\citenamefont {Noel}\ \emph {et~al.}(2022)\citenamefont {Noel},
  \citenamefont {Niroula}, \citenamefont {Zhu}, \citenamefont {Risinger},
  \citenamefont {Egan}, \citenamefont {Biswas}, \citenamefont {Cetina},
  \citenamefont {Gorshkov}, \citenamefont {Gullans}, \citenamefont {Huse} \emph
  {et~al.}}]{noel:2022}%
  \BibitemOpen
  \bibfield  {author} {\bibinfo {author} {\bibfnamefont {C.}~\bibnamefont
  {Noel}}, \bibinfo {author} {\bibfnamefont {P.}~\bibnamefont {Niroula}},
  \bibinfo {author} {\bibfnamefont {D.}~\bibnamefont {Zhu}}, \bibinfo {author}
  {\bibfnamefont {A.}~\bibnamefont {Risinger}}, \bibinfo {author}
  {\bibfnamefont {L.}~\bibnamefont {Egan}}, \bibinfo {author} {\bibfnamefont
  {D.}~\bibnamefont {Biswas}}, \bibinfo {author} {\bibfnamefont
  {M.}~\bibnamefont {Cetina}}, \bibinfo {author} {\bibfnamefont {A.~V.}\
  \bibnamefont {Gorshkov}}, \bibinfo {author} {\bibfnamefont {M.~J.}\
  \bibnamefont {Gullans}}, \bibinfo {author} {\bibfnamefont {D.~A.}\
  \bibnamefont {Huse}}, \emph {et~al.},\ }\bibfield  {title} {\bibinfo {title}
  {Measurement-induced quantum phases realized in a trapped-ion quantum
  computer},\ }\href {https://doi.org/10.1038/s41567-022-01619-7} {\bibfield
  {journal} {\bibinfo  {journal} {Nat. Phys.}\ ,\ \bibinfo {pages} {1}}
  (\bibinfo {year} {2022})}\BibitemShut {NoStop}%
\bibitem [{\citenamefont {Semeghini}\ \emph {et~al.}(2021)\citenamefont
  {Semeghini}, \citenamefont {Levine}, \citenamefont {Keesling}, \citenamefont
  {Ebadi}, \citenamefont {Wang}, \citenamefont {Bluvstein}, \citenamefont
  {Verresen}, \citenamefont {Pichler}, \citenamefont {Kalinowski},
  \citenamefont {Samajdar} \emph {et~al.}}]{semeghini:2021}%
  \BibitemOpen
  \bibfield  {author} {\bibinfo {author} {\bibfnamefont {G.}~\bibnamefont
  {Semeghini}}, \bibinfo {author} {\bibfnamefont {H.}~\bibnamefont {Levine}},
  \bibinfo {author} {\bibfnamefont {A.}~\bibnamefont {Keesling}}, \bibinfo
  {author} {\bibfnamefont {S.}~\bibnamefont {Ebadi}}, \bibinfo {author}
  {\bibfnamefont {T.~T.}\ \bibnamefont {Wang}}, \bibinfo {author}
  {\bibfnamefont {D.}~\bibnamefont {Bluvstein}}, \bibinfo {author}
  {\bibfnamefont {R.}~\bibnamefont {Verresen}}, \bibinfo {author}
  {\bibfnamefont {H.}~\bibnamefont {Pichler}}, \bibinfo {author} {\bibfnamefont
  {M.}~\bibnamefont {Kalinowski}}, \bibinfo {author} {\bibfnamefont
  {R.}~\bibnamefont {Samajdar}}, \emph {et~al.},\ }\bibfield  {title} {\bibinfo
  {title} {Probing topological spin liquids on a programmable quantum
  simulator},\ }\href {https://doi.org/10.1126/science.abi8794} {\bibfield
  {journal} {\bibinfo  {journal} {Science}\ }\textbf {\bibinfo {volume}
  {374}},\ \bibinfo {pages} {1242} (\bibinfo {year} {2021})}\BibitemShut
  {NoStop}%
\bibitem [{\citenamefont {Scholl}\ \emph {et~al.}(2021)\citenamefont {Scholl},
  \citenamefont {Schuler}, \citenamefont {Williams}, \citenamefont
  {Eberharter}, \citenamefont {Barredo}, \citenamefont {Schymik}, \citenamefont
  {Lienhard}, \citenamefont {Henry}, \citenamefont {Lang}, \citenamefont
  {Lahaye} \emph {et~al.}}]{scholl:2021}%
  \BibitemOpen
  \bibfield  {author} {\bibinfo {author} {\bibfnamefont {P.}~\bibnamefont
  {Scholl}}, \bibinfo {author} {\bibfnamefont {M.}~\bibnamefont {Schuler}},
  \bibinfo {author} {\bibfnamefont {H.~J.}\ \bibnamefont {Williams}}, \bibinfo
  {author} {\bibfnamefont {A.~A.}\ \bibnamefont {Eberharter}}, \bibinfo
  {author} {\bibfnamefont {D.}~\bibnamefont {Barredo}}, \bibinfo {author}
  {\bibfnamefont {K.-N.}\ \bibnamefont {Schymik}}, \bibinfo {author}
  {\bibfnamefont {V.}~\bibnamefont {Lienhard}}, \bibinfo {author}
  {\bibfnamefont {L.-P.}\ \bibnamefont {Henry}}, \bibinfo {author}
  {\bibfnamefont {T.~C.}\ \bibnamefont {Lang}}, \bibinfo {author}
  {\bibfnamefont {T.}~\bibnamefont {Lahaye}}, \emph {et~al.},\ }\bibfield
  {title} {\bibinfo {title} {{Quantum simulation of 2D antiferromagnets with
  hundreds of Rydberg atoms}},\ }\href
  {https://doi.org/10.1038/s41586-021-03585-1} {\bibfield  {journal} {\bibinfo
  {journal} {Nature}\ }\textbf {\bibinfo {volume} {595}},\ \bibinfo {pages}
  {233} (\bibinfo {year} {2021})}\BibitemShut {NoStop}%
\bibitem [{\citenamefont {Altman}\ \emph {et~al.}(2021)\citenamefont {Altman},
  \citenamefont {Brown}, \citenamefont {Carleo}, \citenamefont {Carr},
  \citenamefont {Demler}, \citenamefont {Chin}, \citenamefont {DeMarco},
  \citenamefont {Economou}, \citenamefont {Eriksson}, \citenamefont {Fu},
  \citenamefont {Greiner}, \citenamefont {Hazzard}, \citenamefont {Hulet},
  \citenamefont {Koll\'ar}, \citenamefont {Lev}, \citenamefont {Lukin},
  \citenamefont {Ma}, \citenamefont {Mi}, \citenamefont {Misra}, \citenamefont
  {Monroe}, \citenamefont {Murch}, \citenamefont {Nazario}, \citenamefont {Ni},
  \citenamefont {Potter}, \citenamefont {Roushan}, \citenamefont {Saffman},
  \citenamefont {Schleier-Smith}, \citenamefont {Siddiqi}, \citenamefont
  {Simmonds}, \citenamefont {Singh}, \citenamefont {Spielman}, \citenamefont
  {Temme}, \citenamefont {Weiss}, \citenamefont {Vu\ifmmode \check{c}\else
  \v{c}\fi{}kovi\ifmmode~\acute{c}\else \'{c}\fi{}}, \citenamefont
  {Vuleti\ifmmode~\acute{c}\else \'{c}\fi{}}, \citenamefont {Ye},\ and\
  \citenamefont {Zwierlein}}]{altman:2021}%
  \BibitemOpen
  \bibfield  {author} {\bibinfo {author} {\bibfnamefont {E.}~\bibnamefont
  {Altman}}, \bibinfo {author} {\bibfnamefont {K.~R.}\ \bibnamefont {Brown}},
  \bibinfo {author} {\bibfnamefont {G.}~\bibnamefont {Carleo}}, \bibinfo
  {author} {\bibfnamefont {L.~D.}\ \bibnamefont {Carr}}, \bibinfo {author}
  {\bibfnamefont {E.}~\bibnamefont {Demler}}, \bibinfo {author} {\bibfnamefont
  {C.}~\bibnamefont {Chin}}, \bibinfo {author} {\bibfnamefont {B.}~\bibnamefont
  {DeMarco}}, \bibinfo {author} {\bibfnamefont {S.~E.}\ \bibnamefont
  {Economou}}, \bibinfo {author} {\bibfnamefont {M.~A.}\ \bibnamefont
  {Eriksson}}, \bibinfo {author} {\bibfnamefont {K.-M.~C.}\ \bibnamefont {Fu}},
  \bibinfo {author} {\bibfnamefont {M.}~\bibnamefont {Greiner}}, \bibinfo
  {author} {\bibfnamefont {K.~R.}\ \bibnamefont {Hazzard}}, \bibinfo {author}
  {\bibfnamefont {R.~G.}\ \bibnamefont {Hulet}}, \bibinfo {author}
  {\bibfnamefont {A.~J.}\ \bibnamefont {Koll\'ar}}, \bibinfo {author}
  {\bibfnamefont {B.~L.}\ \bibnamefont {Lev}}, \bibinfo {author} {\bibfnamefont
  {M.~D.}\ \bibnamefont {Lukin}}, \bibinfo {author} {\bibfnamefont
  {R.}~\bibnamefont {Ma}}, \bibinfo {author} {\bibfnamefont {X.}~\bibnamefont
  {Mi}}, \bibinfo {author} {\bibfnamefont {S.}~\bibnamefont {Misra}}, \bibinfo
  {author} {\bibfnamefont {C.}~\bibnamefont {Monroe}}, \bibinfo {author}
  {\bibfnamefont {K.}~\bibnamefont {Murch}}, \bibinfo {author} {\bibfnamefont
  {Z.}~\bibnamefont {Nazario}}, \bibinfo {author} {\bibfnamefont {K.-K.}\
  \bibnamefont {Ni}}, \bibinfo {author} {\bibfnamefont {A.~C.}\ \bibnamefont
  {Potter}}, \bibinfo {author} {\bibfnamefont {P.}~\bibnamefont {Roushan}},
  \bibinfo {author} {\bibfnamefont {M.}~\bibnamefont {Saffman}}, \bibinfo
  {author} {\bibfnamefont {M.}~\bibnamefont {Schleier-Smith}}, \bibinfo
  {author} {\bibfnamefont {I.}~\bibnamefont {Siddiqi}}, \bibinfo {author}
  {\bibfnamefont {R.}~\bibnamefont {Simmonds}}, \bibinfo {author}
  {\bibfnamefont {M.}~\bibnamefont {Singh}}, \bibinfo {author} {\bibfnamefont
  {I.}~\bibnamefont {Spielman}}, \bibinfo {author} {\bibfnamefont
  {K.}~\bibnamefont {Temme}}, \bibinfo {author} {\bibfnamefont {D.~S.}\
  \bibnamefont {Weiss}}, \bibinfo {author} {\bibfnamefont {J.}~\bibnamefont
  {Vu\ifmmode \check{c}\else \v{c}\fi{}kovi\ifmmode~\acute{c}\else
  \'{c}\fi{}}}, \bibinfo {author} {\bibfnamefont {V.}~\bibnamefont
  {Vuleti\ifmmode~\acute{c}\else \'{c}\fi{}}}, \bibinfo {author} {\bibfnamefont
  {J.}~\bibnamefont {Ye}},\ and\ \bibinfo {author} {\bibfnamefont
  {M.}~\bibnamefont {Zwierlein}},\ }\bibfield  {title} {\bibinfo {title}
  {{Quantum Simulators: Architectures and Opportunities}},\ }\href
  {https://doi.org/10.1103/PRXQuantum.2.017003} {\bibfield  {journal} {\bibinfo
   {journal} {PRX Quantum}\ }\textbf {\bibinfo {volume} {2}},\ \bibinfo {pages}
  {017003} (\bibinfo {year} {2021})}\BibitemShut {NoStop}%
\bibitem [{\citenamefont {Arnold}\ and\ \citenamefont
  {Sch\"{a}fer}(2022)}]{github}%
  \BibitemOpen
  \bibfield  {author} {\bibinfo {author} {\bibfnamefont {J.}~\bibnamefont
  {Arnold}}\ and\ \bibinfo {author} {\bibfnamefont {F.}~\bibnamefont
  {Sch\"{a}fer}},\ }\href
  {https://github.com/arnoldjulian/Replacing-neural-networks-by-optimal-analytical-predictors-for-the-detection-of-phase-transitions}
  {\bibinfo {title} {{R}eplacing neural networks by optimal analytical
  predictors for the detection of phase transitions}} (\bibinfo {year}
  {2022})\BibitemShut {NoStop}%
\bibitem [{\citenamefont {Lee}\ and\ \citenamefont {Kim}(2019)}]{lee:2019}%
  \BibitemOpen
  \bibfield  {author} {\bibinfo {author} {\bibfnamefont {S.~S.}\ \bibnamefont
  {Lee}}\ and\ \bibinfo {author} {\bibfnamefont {B.~J.}\ \bibnamefont {Kim}},\
  }\bibfield  {title} {\bibinfo {title} {Confusion scheme in machine learning
  detects double phase transitions and quasi-long-range order},\ }\href
  {https://doi.org/10.1103/PhysRevE.99.043308} {\bibfield  {journal} {\bibinfo
  {journal} {Phys. Rev. E}\ }\textbf {\bibinfo {volume} {99}},\ \bibinfo
  {pages} {043308} (\bibinfo {year} {2019})}\BibitemShut {NoStop}%
\bibitem [{\citenamefont {Jacot}\ \emph {et~al.}(2018)\citenamefont {Jacot},
  \citenamefont {Gabriel},\ and\ \citenamefont {Hongler}}]{jacot:2018}%
  \BibitemOpen
  \bibfield  {author} {\bibinfo {author} {\bibfnamefont {A.}~\bibnamefont
  {Jacot}}, \bibinfo {author} {\bibfnamefont {F.}~\bibnamefont {Gabriel}},\
  and\ \bibinfo {author} {\bibfnamefont {C.}~\bibnamefont {Hongler}},\
  }\bibfield  {title} {\bibinfo {title} {{N}eural {T}angent {K}ernel:
  {C}onvergence and {G}eneralization in {N}eural {N}etworks},\ }in\ \href
  {https://proceedings.neurips.cc/paper/2018/file/5a4be1fa34e62bb8a6ec6b91d2462f5a-Paper.pdf}
  {\emph {\bibinfo {booktitle} {Adv. Neural Inf. Process. Syst.}}},\
  Vol.~\bibinfo {volume} {31},\ \bibinfo {editor} {edited by\ \bibinfo {editor}
  {\bibfnamefont {S.}~\bibnamefont {Bengio}}, \bibinfo {editor} {\bibfnamefont
  {H.}~\bibnamefont {Wallach}}, \bibinfo {editor} {\bibfnamefont
  {H.}~\bibnamefont {Larochelle}}, \bibinfo {editor} {\bibfnamefont
  {K.}~\bibnamefont {Grauman}}, \bibinfo {editor} {\bibfnamefont
  {N.}~\bibnamefont {Cesa-Bianchi}},\ and\ \bibinfo {editor} {\bibfnamefont
  {R.}~\bibnamefont {Garnett}}}\ (\bibinfo  {publisher} {Curran Associates,
  Inc.},\ \bibinfo {year} {2018})\BibitemShut {NoStop}%
\bibitem [{\citenamefont {Chartrand}()}]{chartrand:2011}%
  \BibitemOpen
  \bibfield  {author} {\bibinfo {author} {\bibfnamefont {R.}~\bibnamefont
  {Chartrand}},\ }\bibfield  {title} {\bibinfo {title} {{Numerical
  Differentiation of Noisy, Nonsmooth Data}},\ }\bibfield  {journal} {\bibinfo
  {journal} {ISRN Appl. Math.}\ }\href {https://doi.org/10.5402/2011/164564}
  {10.5402/2011/164564}\BibitemShut {NoStop}%
\bibitem [{\citenamefont {LeCun}\ \emph {et~al.}(2012)\citenamefont {LeCun},
  \citenamefont {Bottou}, \citenamefont {Orr},\ and\ \citenamefont
  {M{\"u}ller}}]{lecun:2012}%
  \BibitemOpen
  \bibfield  {author} {\bibinfo {author} {\bibfnamefont {Y.~A.}\ \bibnamefont
  {LeCun}}, \bibinfo {author} {\bibfnamefont {L.}~\bibnamefont {Bottou}},
  \bibinfo {author} {\bibfnamefont {G.~B.}\ \bibnamefont {Orr}},\ and\ \bibinfo
  {author} {\bibfnamefont {K.-R.}\ \bibnamefont {M{\"u}ller}},\ }\bibfield
  {title} {\bibinfo {title} {Efficient backprop},\ }in\ \href
  {https://doi.org/https://doi.org/10.1007/978-3-642-35289-8_3} {\emph
  {\bibinfo {booktitle} {Neural {N}etworks: {T}ricks of the {T}rade}}}\
  (\bibinfo  {publisher} {Springer},\ \bibinfo {year} {2012})\ pp.\ \bibinfo
  {pages} {9--48}\BibitemShut {NoStop}%
\bibitem [{\citenamefont {Innes}(2018)}]{innes:2018}%
  \BibitemOpen
  \bibfield  {author} {\bibinfo {author} {\bibfnamefont {M.}~\bibnamefont
  {Innes}},\ }\bibfield  {title} {\bibinfo {title} {{Flux: Elegant machine
  learning with Julia}},\ }\href {https://doi.org/10.21105/joss.00602}
  {\bibfield  {journal} {\bibinfo  {journal} {J. Open Source Softw.}\ }\textbf
  {\bibinfo {volume} {3}},\ \bibinfo {pages} {602} (\bibinfo {year}
  {2018})}\BibitemShut {NoStop}%
\bibitem [{\citenamefont {Kingma}\ and\ \citenamefont
  {Ba}(2014)}]{kingma:2014}%
  \BibitemOpen
  \bibfield  {author} {\bibinfo {author} {\bibfnamefont {D.}~\bibnamefont
  {Kingma}}\ and\ \bibinfo {author} {\bibfnamefont {J.}~\bibnamefont {Ba}},\
  }\bibfield  {title} {\bibinfo {title} {{A}dam: {A} method for stochastic
  optimization},\ }\href {https://arxiv.org/abs/1412.6980} {\bibfield
  {journal} {\bibinfo  {journal} {arXiv:1412.6980}\ } (\bibinfo {year}
  {2014})}\BibitemShut {NoStop}%
\bibitem [{\citenamefont {Rumelhart}\ \emph {et~al.}(1986)\citenamefont
  {Rumelhart}, \citenamefont {Hinton},\ and\ \citenamefont
  {Williams}}]{rumelhart:1986}%
  \BibitemOpen
  \bibfield  {author} {\bibinfo {author} {\bibfnamefont {D.~E.}\ \bibnamefont
  {Rumelhart}}, \bibinfo {author} {\bibfnamefont {G.~E.}\ \bibnamefont
  {Hinton}},\ and\ \bibinfo {author} {\bibfnamefont {R.~J.}\ \bibnamefont
  {Williams}},\ }\bibfield  {title} {\bibinfo {title} {Learning representations
  by back-propagating errors},\ }\href {https://doi.org/10.1038/323533a0}
  {\bibfield  {journal} {\bibinfo  {journal} {Nature}\ }\textbf {\bibinfo
  {volume} {323}},\ \bibinfo {pages} {533} (\bibinfo {year}
  {1986})}\BibitemShut {NoStop}%
\bibitem [{\citenamefont {Baydin}\ \emph {et~al.}(2018)\citenamefont {Baydin},
  \citenamefont {Pearlmutter}, \citenamefont {Radul},\ and\ \citenamefont
  {Siskind}}]{baydin:2018}%
  \BibitemOpen
  \bibfield  {author} {\bibinfo {author} {\bibfnamefont {A.~G.}\ \bibnamefont
  {Baydin}}, \bibinfo {author} {\bibfnamefont {B.~A.}\ \bibnamefont
  {Pearlmutter}}, \bibinfo {author} {\bibfnamefont {A.~A.}\ \bibnamefont
  {Radul}},\ and\ \bibinfo {author} {\bibfnamefont {J.~M.}\ \bibnamefont
  {Siskind}},\ }\bibfield  {title} {\bibinfo {title} {{Automatic
  Differentiation in Machine Learning: a Survey}},\ }\href
  {http://jmlr.org/papers/v18/17-468.html} {\bibfield  {journal} {\bibinfo
  {journal} {J. Mach. Learn. Res.}\ }\textbf {\bibinfo {volume} {18}},\
  \bibinfo {pages} {1} (\bibinfo {year} {2018})}\BibitemShut {NoStop}%
\bibitem [{\citenamefont {Blayo}\ \emph {et~al.}(2014)\citenamefont {Blayo},
  \citenamefont {Bocquet}, \citenamefont {Cosme},\ and\ \citenamefont
  {Cugliandolo}}]{blayo:2014}%
  \BibitemOpen
  \bibfield  {author} {\bibinfo {author} {\bibfnamefont {{\'E}.}~\bibnamefont
  {Blayo}}, \bibinfo {author} {\bibfnamefont {M.}~\bibnamefont {Bocquet}},
  \bibinfo {author} {\bibfnamefont {E.}~\bibnamefont {Cosme}},\ and\ \bibinfo
  {author} {\bibfnamefont {L.~F.}\ \bibnamefont {Cugliandolo}},\ }\href
  {https://doi.org/DOI:10.1093/acprof:oso/9780198723844.001.0001} {\emph
  {\bibinfo {title} {{Advanced Data Assimilation for Geosciences: Lecture Notes
  of the Les Houches School of Physics: Special Issue, June 2012}}}}\ (\bibinfo
   {publisher} {Oxford University Press},\ \bibinfo {year} {2014})\BibitemShut
  {NoStop}%
\bibitem [{\citenamefont {Bishop}(1995)}]{bishop:1995}%
  \BibitemOpen
  \bibfield  {author} {\bibinfo {author} {\bibfnamefont {C.~M.}\ \bibnamefont
  {Bishop}},\ }\bibfield  {title} {\bibinfo {title} {Regularization and
  complexity control in feed-forward networks},\ }in\ \href
  {https://publications.aston.ac.uk/id/eprint/524/} {\emph {\bibinfo
  {booktitle} {Proceedings International Conference on Artificial Neural
  Networks ICANN'95}}}\ (\bibinfo  {publisher} {EC2 et Cie},\ \bibinfo {year}
  {1995})\ pp.\ \bibinfo {pages} {141--148}\BibitemShut {NoStop}%
\bibitem [{\citenamefont {Sj{\"o}berg}\ and\ \citenamefont
  {Ljung}(1995)}]{sjoberg:1995}%
  \BibitemOpen
  \bibfield  {author} {\bibinfo {author} {\bibfnamefont {J.}~\bibnamefont
  {Sj{\"o}berg}}\ and\ \bibinfo {author} {\bibfnamefont {L.}~\bibnamefont
  {Ljung}},\ }\bibfield  {title} {\bibinfo {title} {Overtraining,
  regularization and searching for a minimum, with application to neural
  networks},\ }\href {https://doi.org/10.1080/00207179508921605} {\bibfield
  {journal} {\bibinfo  {journal} {Int. J. Control}\ }\textbf {\bibinfo {volume}
  {62}},\ \bibinfo {pages} {1391} (\bibinfo {year} {1995})}\BibitemShut
  {NoStop}%
\bibitem [{\citenamefont {Zwerger}(2003)}]{zwerger:2003}%
  \BibitemOpen
  \bibfield  {author} {\bibinfo {author} {\bibfnamefont {W.}~\bibnamefont
  {Zwerger}},\ }\bibfield  {title} {\bibinfo {title} {Mott--{H}ubbard
  transition of cold atoms in optical lattices},\ }\href
  {https://doi.org/10.1088/1464-4266/5/2/352} {\bibfield  {journal} {\bibinfo
  {journal} {J. Opt. B: Quantum Semiclass. Opt.}\ }\textbf {\bibinfo {volume}
  {5}},\ \bibinfo {pages} {S9} (\bibinfo {year} {2003})}\BibitemShut {NoStop}%
\bibitem [{\citenamefont {Weinberg}\ and\ \citenamefont
  {Bukov}(2017)}]{quspin:2017}%
  \BibitemOpen
  \bibfield  {author} {\bibinfo {author} {\bibfnamefont {P.}~\bibnamefont
  {Weinberg}}\ and\ \bibinfo {author} {\bibfnamefont {M.}~\bibnamefont
  {Bukov}},\ }\bibfield  {title} {\bibinfo {title} {{QuSpin: a Python Package
  for Dynamics and Exact Diagonalisation of Quantum Many Body Systems part I:
  spin chains}},\ }\href {https://doi.org/10.21468/SciPostPhys.2.1.003}
  {\bibfield  {journal} {\bibinfo  {journal} {SciPost Phys.}\ }\textbf
  {\bibinfo {volume} {2}},\ \bibinfo {pages} {003} (\bibinfo {year}
  {2017})}\BibitemShut {NoStop}%
\bibitem [{\citenamefont {Weinberg}\ and\ \citenamefont
  {Bukov}(2019)}]{quspin:2019}%
  \BibitemOpen
  \bibfield  {author} {\bibinfo {author} {\bibfnamefont {P.}~\bibnamefont
  {Weinberg}}\ and\ \bibinfo {author} {\bibfnamefont {M.}~\bibnamefont
  {Bukov}},\ }\bibfield  {title} {\bibinfo {title} {{QuSpin: a Python Package
  for Dynamics and Exact Diagonalisation of Quantum Many Body Systems. Part II:
  bosons, fermions and higher spins}},\ }\href
  {https://doi.org/10.21468/SciPostPhys.7.2.020} {\bibfield  {journal}
  {\bibinfo  {journal} {SciPost Phys.}\ }\textbf {\bibinfo {volume} {7}},\
  \bibinfo {pages} {20} (\bibinfo {year} {2019})}\BibitemShut {NoStop}%
\bibitem [{\citenamefont {Katsura}\ \emph {et~al.}(2015)\citenamefont
  {Katsura}, \citenamefont {Schuricht},\ and\ \citenamefont
  {Takahashi}}]{katsura:2015}%
  \BibitemOpen
  \bibfield  {author} {\bibinfo {author} {\bibfnamefont {H.}~\bibnamefont
  {Katsura}}, \bibinfo {author} {\bibfnamefont {D.}~\bibnamefont {Schuricht}},\
  and\ \bibinfo {author} {\bibfnamefont {M.}~\bibnamefont {Takahashi}},\
  }\bibfield  {title} {\bibinfo {title} {Exact ground states and topological
  order in interacting {K}itaev/{M}ajorana chains},\ }\href
  {https://doi.org/10.1103/PhysRevB.92.115137} {\bibfield  {journal} {\bibinfo
  {journal} {Phys. Rev. B}\ }\textbf {\bibinfo {volume} {92}},\ \bibinfo
  {pages} {115137} (\bibinfo {year} {2015})}\BibitemShut {NoStop}%
\bibitem [{\citenamefont {Fisher}\ \emph {et~al.}(1989)\citenamefont {Fisher},
  \citenamefont {Weichman}, \citenamefont {Grinstein},\ and\ \citenamefont
  {Fisher}}]{fisher:1989}%
  \BibitemOpen
  \bibfield  {author} {\bibinfo {author} {\bibfnamefont {M.~P.~A.}\
  \bibnamefont {Fisher}}, \bibinfo {author} {\bibfnamefont {P.~B.}\
  \bibnamefont {Weichman}}, \bibinfo {author} {\bibfnamefont {G.}~\bibnamefont
  {Grinstein}},\ and\ \bibinfo {author} {\bibfnamefont {D.~S.}\ \bibnamefont
  {Fisher}},\ }\bibfield  {title} {\bibinfo {title} {Boson localization and the
  superfluid-insulator transition},\ }\href
  {https://doi.org/10.1103/PhysRevB.40.546} {\bibfield  {journal} {\bibinfo
  {journal} {Phys. Rev. B}\ }\textbf {\bibinfo {volume} {40}},\ \bibinfo
  {pages} {546} (\bibinfo {year} {1989})}\BibitemShut {NoStop}%
\bibitem [{\citenamefont {Jaksch}\ \emph {et~al.}(1998)\citenamefont {Jaksch},
  \citenamefont {Bruder}, \citenamefont {Cirac}, \citenamefont {Gardiner},\
  and\ \citenamefont {Zoller}}]{jaksch:1998}%
  \BibitemOpen
  \bibfield  {author} {\bibinfo {author} {\bibfnamefont {D.}~\bibnamefont
  {Jaksch}}, \bibinfo {author} {\bibfnamefont {C.}~\bibnamefont {Bruder}},
  \bibinfo {author} {\bibfnamefont {J.~I.}\ \bibnamefont {Cirac}}, \bibinfo
  {author} {\bibfnamefont {C.~W.}\ \bibnamefont {Gardiner}},\ and\ \bibinfo
  {author} {\bibfnamefont {P.}~\bibnamefont {Zoller}},\ }\bibfield  {title}
  {\bibinfo {title} {Cold {B}osonic {A}toms in {O}ptical {L}attices},\ }\href
  {https://doi.org/10.1103/PhysRevLett.81.3108} {\bibfield  {journal} {\bibinfo
   {journal} {Phys. Rev. Lett.}\ }\textbf {\bibinfo {volume} {81}},\ \bibinfo
  {pages} {3108} (\bibinfo {year} {1998})}\BibitemShut {NoStop}%
\bibitem [{\citenamefont {Krauth}\ \emph {et~al.}(1992)\citenamefont {Krauth},
  \citenamefont {Caffarel},\ and\ \citenamefont {Bouchaud}}]{krauth:1992}%
  \BibitemOpen
  \bibfield  {author} {\bibinfo {author} {\bibfnamefont {W.}~\bibnamefont
  {Krauth}}, \bibinfo {author} {\bibfnamefont {M.}~\bibnamefont {Caffarel}},\
  and\ \bibinfo {author} {\bibfnamefont {J.-P.}\ \bibnamefont {Bouchaud}},\
  }\bibfield  {title} {\bibinfo {title} {Gutzwiller wave function for a model
  of strongly interacting bosons},\ }\href
  {https://doi.org/10.1103/PhysRevB.45.3137} {\bibfield  {journal} {\bibinfo
  {journal} {Phys. Rev. B}\ }\textbf {\bibinfo {volume} {45}},\ \bibinfo
  {pages} {3137} (\bibinfo {year} {1992})}\BibitemShut {NoStop}%
\bibitem [{\citenamefont {Comparin}(2017)}]{comparin:2017}%
  \BibitemOpen
  \bibfield  {author} {\bibinfo {author} {\bibfnamefont {T.}~\bibnamefont
  {Comparin}},\ }\href {https://doi.org/10.5281/zenodo.1067968} {\bibinfo
  {title} {tcompa/bosehubbardgutzwiller v1.0.2}} (\bibinfo {year}
  {2017})\BibitemShut {NoStop}%
\end{thebibliography}%
\end{document}